\documentclass[aps,reprint,twocolumn,10pt,superscriptaddress,english,showpacs,longbibliography,nofootinbib]{revtex4-1}
\usepackage{amsmath}
\usepackage{amssymb}
\usepackage{amsthm}
\usepackage{amsfonts}
\usepackage{listings}
\usepackage{longtable}
\usepackage{enumerate}
\usepackage{latexsym}
\usepackage{color}
\usepackage{multirow}
\usepackage[figuresright]{rotating}
\usepackage[table]{xcolor}
\usepackage{setspace} 
\usepackage{blindtext}
\usepackage{dcolumn}

\usepackage{array,etoolbox}
\preto\tabular{\setcounter{magicrownumbers}{0}}
\newcounter{magicrownumbers}

\def\ie{{\it i.e.},\ }

\usepackage{bm}
\usepackage{hyperref}
\hypersetup{
 pdfnewwindow=true, colorlinks=true,
 linkcolor=blue, anchorcolor=blue,
 citecolor=blue, filecolor=blue,
 menucolor=blue, urlcolor=blue}
\definecolor{red}{rgb}{1,0,0}
\definecolor{blue}{rgb}{0,0,1}
\definecolor{black}{rgb}{0,0,0}

\newcommand{\rhQ}[1][]{$\rho_{G^Q}^{#1}$}

\usepackage{graphicx}
\usepackage{subfigure}

\usepackage{chemformula}
\let\ce\ch

\usepackage{appendix}

\begin{document}

\newcommand{\msgsymb}[2]{\ifnum#1=1
\ifnum#2=1
$P1$\else
\ifnum#2=2
$P11'$\else
\ifnum#2=3
$P_{S}1$\else
{\color{red}{Invalid MSG number}}
\fi
\fi
\fi
\else
\ifnum#1=2
\ifnum#2=4
$P\bar{1}$\else
\ifnum#2=5
$P\bar{1}1'$\else
\ifnum#2=6
$P\bar{1}'$\else
\ifnum#2=7
$P_{S}\bar{1}$\else
{\color{red}{Invalid MSG number}}
\fi
\fi
\fi
\fi
\else
\ifnum#1=3
\ifnum#2=1
$P2$\else
\ifnum#2=2
$P21'$\else
\ifnum#2=3
$P2'$\else
\ifnum#2=4
$P_{a}2$\else
\ifnum#2=5
$P_{b}2$\else
\ifnum#2=6
$P_{C}2$\else
{\color{red}{Invalid MSG number}}
\fi
\fi
\fi
\fi
\fi
\fi
\else
\ifnum#1=4
\ifnum#2=7
$P2_{1}$\else
\ifnum#2=8
$P2_{1}1'$\else
\ifnum#2=9
$P2_{1}'$\else
\ifnum#2=10
$P_{a}2_{1}$\else
\ifnum#2=11
$P_{b}2_{1}$\else
\ifnum#2=12
$P_{C}2_{1}$\else
{\color{red}{Invalid MSG number}}
\fi
\fi
\fi
\fi
\fi
\fi
\else
\ifnum#1=5
\ifnum#2=13
$C2$\else
\ifnum#2=14
$C21'$\else
\ifnum#2=15
$C2'$\else
\ifnum#2=16
$C_{c}2$\else
\ifnum#2=17
$C_{a}2$\else
{\color{red}{Invalid MSG number}}
\fi
\fi
\fi
\fi
\fi
\else
\ifnum#1=6
\ifnum#2=18
$Pm$\else
\ifnum#2=19
$Pm1'$\else
\ifnum#2=20
$Pm'$\else
\ifnum#2=21
$P_{a}m$\else
\ifnum#2=22
$P_{b}m$\else
\ifnum#2=23
$P_{C}m$\else
{\color{red}{Invalid MSG number}}
\fi
\fi
\fi
\fi
\fi
\fi
\else
\ifnum#1=7
\ifnum#2=24
$Pc$\else
\ifnum#2=25
$Pc1'$\else
\ifnum#2=26
$Pc'$\else
\ifnum#2=27
$P_{a}c$\else
\ifnum#2=28
$P_{c}c$\else
\ifnum#2=29
$P_{b}c$\else
\ifnum#2=30
$P_{C}c$\else
\ifnum#2=31
$P_{A}c$\else
{\color{red}{Invalid MSG number}}
\fi
\fi
\fi
\fi
\fi
\fi
\fi
\fi
\else
\ifnum#1=8
\ifnum#2=32
$Cm$\else
\ifnum#2=33
$Cm1'$\else
\ifnum#2=34
$Cm'$\else
\ifnum#2=35
$C_{c}m$\else
\ifnum#2=36
$C_{a}m$\else
{\color{red}{Invalid MSG number}}
\fi
\fi
\fi
\fi
\fi
\else
\ifnum#1=9
\ifnum#2=37
$Cc$\else
\ifnum#2=38
$Cc1'$\else
\ifnum#2=39
$Cc'$\else
\ifnum#2=40
$C_{c}c$\else
\ifnum#2=41
$C_{a}c$\else
{\color{red}{Invalid MSG number}}
\fi
\fi
\fi
\fi
\fi
\else
\ifnum#1=10
\ifnum#2=42
$P2/m$\else
\ifnum#2=43
$P2/m1'$\else
\ifnum#2=44
$P2'/m$\else
\ifnum#2=45
$P2/m'$\else
\ifnum#2=46
$P2'/m'$\else
\ifnum#2=47
$P_{a}2/m$\else
\ifnum#2=48
$P_{b}2/m$\else
\ifnum#2=49
$P_{C}2/m$\else
{\color{red}{Invalid MSG number}}
\fi
\fi
\fi
\fi
\fi
\fi
\fi
\fi
\else
\ifnum#1=11
\ifnum#2=50
$P2_{1}/m$\else
\ifnum#2=51
$P2_{1}/m1'$\else
\ifnum#2=52
$P2_{1}'/m$\else
\ifnum#2=53
$P2_{1}/m'$\else
\ifnum#2=54
$P2_{1}'/m'$\else
\ifnum#2=55
$P_{a}2_{1}/m$\else
\ifnum#2=56
$P_{b}2_{1}/m$\else
\ifnum#2=57
$P_{C}2_{1}/m$\else
{\color{red}{Invalid MSG number}}
\fi
\fi
\fi
\fi
\fi
\fi
\fi
\fi
\else
\ifnum#1=12
\ifnum#2=58
$C2/m$\else
\ifnum#2=59
$C2/m1'$\else
\ifnum#2=60
$C2'/m$\else
\ifnum#2=61
$C2/m'$\else
\ifnum#2=62
$C2'/m'$\else
\ifnum#2=63
$C_{c}2/m$\else
\ifnum#2=64
$C_{a}2/m$\else
{\color{red}{Invalid MSG number}}
\fi
\fi
\fi
\fi
\fi
\fi
\fi
\else
\ifnum#1=13
\ifnum#2=65
$P2/c$\else
\ifnum#2=66
$P2/c1'$\else
\ifnum#2=67
$P2'/c$\else
\ifnum#2=68
$P2/c'$\else
\ifnum#2=69
$P2'/c'$\else
\ifnum#2=70
$P_{a}2/c$\else
\ifnum#2=71
$P_{b}2/c$\else
\ifnum#2=72
$P_{c}2/c$\else
\ifnum#2=73
$P_{A}2/c$\else
\ifnum#2=74
$P_{C}2/c$\else
{\color{red}{Invalid MSG number}}
\fi
\fi
\fi
\fi
\fi
\fi
\fi
\fi
\fi
\fi
\else
\ifnum#1=14
\ifnum#2=75
$P2_{1}/c$\else
\ifnum#2=76
$P2_{1}/c1'$\else
\ifnum#2=77
$P2_{1}'/c$\else
\ifnum#2=78
$P2_{1}/c'$\else
\ifnum#2=79
$P2_{1}'/c'$\else
\ifnum#2=80
$P_{a}2_{1}/c$\else
\ifnum#2=81
$P_{b}2_{1}/c$\else
\ifnum#2=82
$P_{c}2_{1}/c$\else
\ifnum#2=83
$P_{A}2_{1}/c$\else
\ifnum#2=84
$P_{C}2_{1}/c$\else
{\color{red}{Invalid MSG number}}
\fi
\fi
\fi
\fi
\fi
\fi
\fi
\fi
\fi
\fi
\else
\ifnum#1=15
\ifnum#2=85
$C2/c$\else
\ifnum#2=86
$C2/c1'$\else
\ifnum#2=87
$C2'/c$\else
\ifnum#2=88
$C2/c'$\else
\ifnum#2=89
$C2'/c'$\else
\ifnum#2=90
$C_{c}2/c$\else
\ifnum#2=91
$C_{a}2/c$\else
{\color{red}{Invalid MSG number}}
\fi
\fi
\fi
\fi
\fi
\fi
\fi
\else
\ifnum#1=16
\ifnum#2=1
$P222$\else
\ifnum#2=2
$P2221'$\else
\ifnum#2=3
$P2'2'2$\else
\ifnum#2=4
$P_{a}222$\else
\ifnum#2=5
$P_{C}222$\else
\ifnum#2=6
$P_{I}222$\else
{\color{red}{Invalid MSG number}}
\fi
\fi
\fi
\fi
\fi
\fi
\else
\ifnum#1=17
\ifnum#2=7
$P222_{1}$\else
\ifnum#2=8
$P222_{1}1'$\else
\ifnum#2=9
$P2'2'2_{1}$\else
\ifnum#2=10
$P22'2_{1}'$\else
\ifnum#2=11
$P_{a}222_{1}$\else
\ifnum#2=12
$P_{c}222_{1}$\else
\ifnum#2=13
$P_{B}222_{1}$\else
\ifnum#2=14
$P_{C}222_{1}$\else
\ifnum#2=15
$P_{I}222_{1}$\else
{\color{red}{Invalid MSG number}}
\fi
\fi
\fi
\fi
\fi
\fi
\fi
\fi
\fi
\else
\ifnum#1=18
\ifnum#2=16
$P2_{1}2_{1}2$\else
\ifnum#2=17
$P2_{1}2_{1}21'$\else
\ifnum#2=18
$P2_{1}'2_{1}'2$\else
\ifnum#2=19
$P2_{1}2_{1}'2'$\else
\ifnum#2=20
$P_{b}2_{1}2_{1}2$\else
\ifnum#2=21
$P_{c}2_{1}2_{1}2$\else
\ifnum#2=22
$P_{B}2_{1}2_{1}2$\else
\ifnum#2=23
$P_{C}2_{1}2_{1}2$\else
\ifnum#2=24
$P_{I}2_{1}2_{1}2$\else
{\color{red}{Invalid MSG number}}
\fi
\fi
\fi
\fi
\fi
\fi
\fi
\fi
\fi
\else
\ifnum#1=19
\ifnum#2=25
$P2_{1}2_{1}2_{1}$\else
\ifnum#2=26
$P2_{1}2_{1}2_{1}1'$\else
\ifnum#2=27
$P2_{1}'2_{1}'2_{1}$\else
\ifnum#2=28
$P_{c}2_{1}2_{1}2_{1}$\else
\ifnum#2=29
$P_{C}2_{1}2_{1}2_{1}$\else
\ifnum#2=30
$P_{I}2_{1}2_{1}2_{1}$\else
{\color{red}{Invalid MSG number}}
\fi
\fi
\fi
\fi
\fi
\fi
\else
\ifnum#1=20
\ifnum#2=31
$C222_{1}$\else
\ifnum#2=32
$C222_{1}1'$\else
\ifnum#2=33
$C2'2'2_{1}$\else
\ifnum#2=34
$C22'2_{1}'$\else
\ifnum#2=35
$C_{c}222_{1}$\else
\ifnum#2=36
$C_{a}222_{1}$\else
\ifnum#2=37
$C_{A}222_{1}$\else
{\color{red}{Invalid MSG number}}
\fi
\fi
\fi
\fi
\fi
\fi
\fi
\else
\ifnum#1=21
\ifnum#2=38
$C222$\else
\ifnum#2=39
$C2221'$\else
\ifnum#2=40
$C2'2'2$\else
\ifnum#2=41
$C22'2'$\else
\ifnum#2=42
$C_{c}222$\else
\ifnum#2=43
$C_{a}222$\else
\ifnum#2=44
$C_{A}222$\else
{\color{red}{Invalid MSG number}}
\fi
\fi
\fi
\fi
\fi
\fi
\fi
\else
\ifnum#1=22
\ifnum#2=45
$F222$\else
\ifnum#2=46
$F2221'$\else
\ifnum#2=47
$F2'2'2$\else
\ifnum#2=48
$F_{S}222$\else
{\color{red}{Invalid MSG number}}
\fi
\fi
\fi
\fi
\else
\ifnum#1=23
\ifnum#2=49
$I222$\else
\ifnum#2=50
$I2221'$\else
\ifnum#2=51
$I2'2'2$\else
\ifnum#2=52
$I_{c}222$\else
{\color{red}{Invalid MSG number}}
\fi
\fi
\fi
\fi
\else
\ifnum#1=24
\ifnum#2=53
$I2_{1}2_{1}2_{1}$\else
\ifnum#2=54
$I2_{1}2_{1}2_{1}1'$\else
\ifnum#2=55
$I2_{1}'2_{1}'2_{1}$\else
\ifnum#2=56
$I_{c}2_{1}2_{1}2_{1}$\else
{\color{red}{Invalid MSG number}}
\fi
\fi
\fi
\fi
\else
\ifnum#1=25
\ifnum#2=57
$Pmm2$\else
\ifnum#2=58
$Pmm21'$\else
\ifnum#2=59
$Pm'm2'$\else
\ifnum#2=60
$Pm'm'2$\else
\ifnum#2=61
$P_{c}mm2$\else
\ifnum#2=62
$P_{a}mm2$\else
\ifnum#2=63
$P_{C}mm2$\else
\ifnum#2=64
$P_{A}mm2$\else
\ifnum#2=65
$P_{I}mm2$\else
{\color{red}{Invalid MSG number}}
\fi
\fi
\fi
\fi
\fi
\fi
\fi
\fi
\fi
\else
\ifnum#1=26
\ifnum#2=66
$Pmc2_{1}$\else
\ifnum#2=67
$Pmc2_{1}1'$\else
\ifnum#2=68
$Pm'c2_{1}'$\else
\ifnum#2=69
$Pmc'2_{1}'$\else
\ifnum#2=70
$Pm'c'2_{1}$\else
\ifnum#2=71
$P_{a}mc2_{1}$\else
\ifnum#2=72
$P_{b}mc2_{1}$\else
\ifnum#2=73
$P_{c}mc2_{1}$\else
\ifnum#2=74
$P_{A}mc2_{1}$\else
\ifnum#2=75
$P_{B}mc2_{1}$\else
\ifnum#2=76
$P_{C}mc2_{1}$\else
\ifnum#2=77
$P_{I}mc2_{1}$\else
{\color{red}{Invalid MSG number}}
\fi
\fi
\fi
\fi
\fi
\fi
\fi
\fi
\fi
\fi
\fi
\fi
\else
\ifnum#1=27
\ifnum#2=78
$Pcc2$\else
\ifnum#2=79
$Pcc21'$\else
\ifnum#2=80
$Pc'c2'$\else
\ifnum#2=81
$Pc'c'2$\else
\ifnum#2=82
$P_{c}cc2$\else
\ifnum#2=83
$P_{a}cc2$\else
\ifnum#2=84
$P_{C}cc2$\else
\ifnum#2=85
$P_{A}cc2$\else
\ifnum#2=86
$P_{I}cc2$\else
{\color{red}{Invalid MSG number}}
\fi
\fi
\fi
\fi
\fi
\fi
\fi
\fi
\fi
\else
\ifnum#1=28
\ifnum#2=87
$Pma2$\else
\ifnum#2=88
$Pma21'$\else
\ifnum#2=89
$Pm'a2'$\else
\ifnum#2=90
$Pma'2'$\else
\ifnum#2=91
$Pm'a'2$\else
\ifnum#2=92
$P_{a}ma2$\else
\ifnum#2=93
$P_{b}ma2$\else
\ifnum#2=94
$P_{c}ma2$\else
\ifnum#2=95
$P_{A}ma2$\else
\ifnum#2=96
$P_{B}ma2$\else
\ifnum#2=97
$P_{C}ma2$\else
\ifnum#2=98
$P_{I}ma2$\else
{\color{red}{Invalid MSG number}}
\fi
\fi
\fi
\fi
\fi
\fi
\fi
\fi
\fi
\fi
\fi
\fi
\else
\ifnum#1=29
\ifnum#2=99
$Pca2_{1}$\else
\ifnum#2=100
$Pca2_{1}1'$\else
\ifnum#2=101
$Pc'a2_{1}'$\else
\ifnum#2=102
$Pca'2_{1}'$\else
\ifnum#2=103
$Pc'a'2_{1}$\else
\ifnum#2=104
$P_{a}ca2_{1}$\else
\ifnum#2=105
$P_{b}ca2_{1}$\else
\ifnum#2=106
$P_{c}ca2_{1}$\else
\ifnum#2=107
$P_{A}ca2_{1}$\else
\ifnum#2=108
$P_{B}ca2_{1}$\else
\ifnum#2=109
$P_{C}ca2_{1}$\else
\ifnum#2=110
$P_{I}ca2_{1}$\else
{\color{red}{Invalid MSG number}}
\fi
\fi
\fi
\fi
\fi
\fi
\fi
\fi
\fi
\fi
\fi
\fi
\else
\ifnum#1=30
\ifnum#2=111
$Pnc2$\else
\ifnum#2=112
$Pnc21'$\else
\ifnum#2=113
$Pn'c2'$\else
\ifnum#2=114
$Pnc'2'$\else
\ifnum#2=115
$Pn'c'2$\else
\ifnum#2=116
$P_{a}nc2$\else
\ifnum#2=117
$P_{b}nc2$\else
\ifnum#2=118
$P_{c}nc2$\else
\ifnum#2=119
$P_{A}nc2$\else
\ifnum#2=120
$P_{B}nc2$\else
\ifnum#2=121
$P_{C}nc2$\else
\ifnum#2=122
$P_{I}nc2$\else
{\color{red}{Invalid MSG number}}
\fi
\fi
\fi
\fi
\fi
\fi
\fi
\fi
\fi
\fi
\fi
\fi
\else
\ifnum#1=31
\ifnum#2=123
$Pmn2_{1}$\else
\ifnum#2=124
$Pmn2_{1}1'$\else
\ifnum#2=125
$Pm'n2_{1}'$\else
\ifnum#2=126
$Pmn'2_{1}'$\else
\ifnum#2=127
$Pm'n'2_{1}$\else
\ifnum#2=128
$P_{a}mn2_{1}$\else
\ifnum#2=129
$P_{b}mn2_{1}$\else
\ifnum#2=130
$P_{c}mn2_{1}$\else
\ifnum#2=131
$P_{A}mn2_{1}$\else
\ifnum#2=132
$P_{B}mn2_{1}$\else
\ifnum#2=133
$P_{C}mn2_{1}$\else
\ifnum#2=134
$P_{I}mn2_{1}$\else
{\color{red}{Invalid MSG number}}
\fi
\fi
\fi
\fi
\fi
\fi
\fi
\fi
\fi
\fi
\fi
\fi
\else
\ifnum#1=32
\ifnum#2=135
$Pba2$\else
\ifnum#2=136
$Pba21'$\else
\ifnum#2=137
$Pb'a2'$\else
\ifnum#2=138
$Pb'a'2$\else
\ifnum#2=139
$P_{c}ba2$\else
\ifnum#2=140
$P_{b}ba2$\else
\ifnum#2=141
$P_{C}ba2$\else
\ifnum#2=142
$P_{A}ba2$\else
\ifnum#2=143
$P_{I}ba2$\else
{\color{red}{Invalid MSG number}}
\fi
\fi
\fi
\fi
\fi
\fi
\fi
\fi
\fi
\else
\ifnum#1=33
\ifnum#2=144
$Pna2_{1}$\else
\ifnum#2=145
$Pna2_{1}1'$\else
\ifnum#2=146
$Pn'a2_{1}'$\else
\ifnum#2=147
$Pna'2_{1}'$\else
\ifnum#2=148
$Pn'a'2_{1}$\else
\ifnum#2=149
$P_{a}na2_{1}$\else
\ifnum#2=150
$P_{b}na2_{1}$\else
\ifnum#2=151
$P_{c}na2_{1}$\else
\ifnum#2=152
$P_{A}na2_{1}$\else
\ifnum#2=153
$P_{B}na2_{1}$\else
\ifnum#2=154
$P_{C}na2_{1}$\else
\ifnum#2=155
$P_{I}na2_{1}$\else
{\color{red}{Invalid MSG number}}
\fi
\fi
\fi
\fi
\fi
\fi
\fi
\fi
\fi
\fi
\fi
\fi
\else
\ifnum#1=34
\ifnum#2=156
$Pnn2$\else
\ifnum#2=157
$Pnn21'$\else
\ifnum#2=158
$Pn'n2'$\else
\ifnum#2=159
$Pn'n'2$\else
\ifnum#2=160
$P_{a}nn2$\else
\ifnum#2=161
$P_{c}nn2$\else
\ifnum#2=162
$P_{A}nn2$\else
\ifnum#2=163
$P_{C}nn2$\else
\ifnum#2=164
$P_{I}nn2$\else
{\color{red}{Invalid MSG number}}
\fi
\fi
\fi
\fi
\fi
\fi
\fi
\fi
\fi
\else
\ifnum#1=35
\ifnum#2=165
$Cmm2$\else
\ifnum#2=166
$Cmm21'$\else
\ifnum#2=167
$Cm'm2'$\else
\ifnum#2=168
$Cm'm'2$\else
\ifnum#2=169
$C_{c}mm2$\else
\ifnum#2=170
$C_{a}mm2$\else
\ifnum#2=171
$C_{A}mm2$\else
{\color{red}{Invalid MSG number}}
\fi
\fi
\fi
\fi
\fi
\fi
\fi
\else
\ifnum#1=36
\ifnum#2=172
$Cmc2_{1}$\else
\ifnum#2=173
$Cmc2_{1}1'$\else
\ifnum#2=174
$Cm'c2_{1}'$\else
\ifnum#2=175
$Cmc'2_{1}'$\else
\ifnum#2=176
$Cm'c'2_{1}$\else
\ifnum#2=177
$C_{c}mc2_{1}$\else
\ifnum#2=178
$C_{a}mc2_{1}$\else
\ifnum#2=179
$C_{A}mc2_{1}$\else
{\color{red}{Invalid MSG number}}
\fi
\fi
\fi
\fi
\fi
\fi
\fi
\fi
\else
\ifnum#1=37
\ifnum#2=180
$Ccc2$\else
\ifnum#2=181
$Ccc21'$\else
\ifnum#2=182
$Cc'c2'$\else
\ifnum#2=183
$Cc'c'2$\else
\ifnum#2=184
$C_{c}cc2$\else
\ifnum#2=185
$C_{a}cc2$\else
\ifnum#2=186
$C_{A}cc2$\else
{\color{red}{Invalid MSG number}}
\fi
\fi
\fi
\fi
\fi
\fi
\fi
\else
\ifnum#1=38
\ifnum#2=187
$Amm2$\else
\ifnum#2=188
$Amm21'$\else
\ifnum#2=189
$Am'm2'$\else
\ifnum#2=190
$Amm'2'$\else
\ifnum#2=191
$Am'm'2$\else
\ifnum#2=192
$A_{a}mm2$\else
\ifnum#2=193
$A_{b}mm2$\else
\ifnum#2=194
$A_{B}mm2$\else
{\color{red}{Invalid MSG number}}
\fi
\fi
\fi
\fi
\fi
\fi
\fi
\fi
\else
\ifnum#1=39
\ifnum#2=195
$Abm2$\else
\ifnum#2=196
$Abm21'$\else
\ifnum#2=197
$Ab'm2'$\else
\ifnum#2=198
$Abm'2'$\else
\ifnum#2=199
$Ab'm'2$\else
\ifnum#2=200
$A_{a}bm2$\else
\ifnum#2=201
$A_{b}bm2$\else
\ifnum#2=202
$A_{B}bm2$\else
{\color{red}{Invalid MSG number}}
\fi
\fi
\fi
\fi
\fi
\fi
\fi
\fi
\else
\ifnum#1=40
\ifnum#2=203
$Ama2$\else
\ifnum#2=204
$Ama21'$\else
\ifnum#2=205
$Am'a2'$\else
\ifnum#2=206
$Ama'2'$\else
\ifnum#2=207
$Am'a'2$\else
\ifnum#2=208
$A_{a}ma2$\else
\ifnum#2=209
$A_{b}ma2$\else
\ifnum#2=210
$A_{B}ma2$\else
{\color{red}{Invalid MSG number}}
\fi
\fi
\fi
\fi
\fi
\fi
\fi
\fi
\else
\ifnum#1=41
\ifnum#2=211
$Aba2$\else
\ifnum#2=212
$Aba21'$\else
\ifnum#2=213
$Ab'a2'$\else
\ifnum#2=214
$Aba'2'$\else
\ifnum#2=215
$Ab'a'2$\else
\ifnum#2=216
$A_{a}ba2$\else
\ifnum#2=217
$A_{b}ba2$\else
\ifnum#2=218
$A_{B}ba2$\else
{\color{red}{Invalid MSG number}}
\fi
\fi
\fi
\fi
\fi
\fi
\fi
\fi
\else
\ifnum#1=42
\ifnum#2=219
$Fmm2$\else
\ifnum#2=220
$Fmm21'$\else
\ifnum#2=221
$Fm'm2'$\else
\ifnum#2=222
$Fm'm'2$\else
\ifnum#2=223
$F_{S}mm2$\else
{\color{red}{Invalid MSG number}}
\fi
\fi
\fi
\fi
\fi
\else
\ifnum#1=43
\ifnum#2=224
$Fdd2$\else
\ifnum#2=225
$Fdd21'$\else
\ifnum#2=226
$Fd'd2'$\else
\ifnum#2=227
$Fd'd'2$\else
\ifnum#2=228
$F_{S}dd2$\else
{\color{red}{Invalid MSG number}}
\fi
\fi
\fi
\fi
\fi
\else
\ifnum#1=44
\ifnum#2=229
$Imm2$\else
\ifnum#2=230
$Imm21'$\else
\ifnum#2=231
$Im'm2'$\else
\ifnum#2=232
$Im'm'2$\else
\ifnum#2=233
$I_{c}mm2$\else
\ifnum#2=234
$I_{a}mm2$\else
{\color{red}{Invalid MSG number}}
\fi
\fi
\fi
\fi
\fi
\fi
\else
\ifnum#1=45
\ifnum#2=235
$Iba2$\else
\ifnum#2=236
$Iba21'$\else
\ifnum#2=237
$Ib'a2'$\else
\ifnum#2=238
$Ib'a'2$\else
\ifnum#2=239
$I_{c}ba2$\else
\ifnum#2=240
$I_{a}ba2$\else
{\color{red}{Invalid MSG number}}
\fi
\fi
\fi
\fi
\fi
\fi
\else
\ifnum#1=46
\ifnum#2=241
$Ima2$\else
\ifnum#2=242
$Ima21'$\else
\ifnum#2=243
$Im'a2'$\else
\ifnum#2=244
$Ima'2'$\else
\ifnum#2=245
$Im'a'2$\else
\ifnum#2=246
$I_{c}ma2$\else
\ifnum#2=247
$I_{a}ma2$\else
\ifnum#2=248
$I_{b}ma2$\else
{\color{red}{Invalid MSG number}}
\fi
\fi
\fi
\fi
\fi
\fi
\fi
\fi
\else
\ifnum#1=47
\ifnum#2=249
$Pmmm$\else
\ifnum#2=250
$Pmmm1'$\else
\ifnum#2=251
$Pm'mm$\else
\ifnum#2=252
$Pm'm'm$\else
\ifnum#2=253
$Pm'm'm'$\else
\ifnum#2=254
$P_{a}mmm$\else
\ifnum#2=255
$P_{C}mmm$\else
\ifnum#2=256
$P_{I}mmm$\else
{\color{red}{Invalid MSG number}}
\fi
\fi
\fi
\fi
\fi
\fi
\fi
\fi
\else
\ifnum#1=48
\ifnum#2=257
$Pnnn$\else
\ifnum#2=258
$Pnnn1'$\else
\ifnum#2=259
$Pn'nn$\else
\ifnum#2=260
$Pn'n'n$\else
\ifnum#2=261
$Pn'n'n'$\else
\ifnum#2=262
$P_{c}nnn$\else
\ifnum#2=263
$P_{C}nnn$\else
\ifnum#2=264
$P_{I}nnn$\else
{\color{red}{Invalid MSG number}}
\fi
\fi
\fi
\fi
\fi
\fi
\fi
\fi
\else
\ifnum#1=49
\ifnum#2=265
$Pccm$\else
\ifnum#2=266
$Pccm1'$\else
\ifnum#2=267
$Pc'cm$\else
\ifnum#2=268
$Pccm'$\else
\ifnum#2=269
$Pc'c'm$\else
\ifnum#2=270
$Pc'cm'$\else
\ifnum#2=271
$Pc'c'm'$\else
\ifnum#2=272
$P_{a}ccm$\else
\ifnum#2=273
$P_{c}ccm$\else
\ifnum#2=274
$P_{B}ccm$\else
\ifnum#2=275
$P_{C}ccm$\else
\ifnum#2=276
$P_{I}ccm$\else
{\color{red}{Invalid MSG number}}
\fi
\fi
\fi
\fi
\fi
\fi
\fi
\fi
\fi
\fi
\fi
\fi
\else
\ifnum#1=50
\ifnum#2=277
$Pban$\else
\ifnum#2=278
$Pban1'$\else
\ifnum#2=279
$Pb'an$\else
\ifnum#2=280
$Pban'$\else
\ifnum#2=281
$Pb'a'n$\else
\ifnum#2=282
$Pb'an'$\else
\ifnum#2=283
$Pb'a'n'$\else
\ifnum#2=284
$P_{a}ban$\else
\ifnum#2=285
$P_{c}ban$\else
\ifnum#2=286
$P_{A}ban$\else
\ifnum#2=287
$P_{C}ban$\else
\ifnum#2=288
$P_{I}ban$\else
{\color{red}{Invalid MSG number}}
\fi
\fi
\fi
\fi
\fi
\fi
\fi
\fi
\fi
\fi
\fi
\fi
\else
\ifnum#1=51
\ifnum#2=289
$Pmma$\else
\ifnum#2=290
$Pmma1'$\else
\ifnum#2=291
$Pm'ma$\else
\ifnum#2=292
$Pmm'a$\else
\ifnum#2=293
$Pmma'$\else
\ifnum#2=294
$Pm'm'a$\else
\ifnum#2=295
$Pmm'a'$\else
\ifnum#2=296
$Pm'ma'$\else
\ifnum#2=297
$Pm'm'a'$\else
\ifnum#2=298
$P_{a}mma$\else
\ifnum#2=299
$P_{b}mma$\else
\ifnum#2=300
$P_{c}mma$\else
\ifnum#2=301
$P_{A}mma$\else
\ifnum#2=302
$P_{B}mma$\else
\ifnum#2=303
$P_{C}mma$\else
\ifnum#2=304
$P_{I}mma$\else
{\color{red}{Invalid MSG number}}
\fi
\fi
\fi
\fi
\fi
\fi
\fi
\fi
\fi
\fi
\fi
\fi
\fi
\fi
\fi
\fi
\else
\ifnum#1=52
\ifnum#2=305
$Pnna$\else
\ifnum#2=306
$Pnna1'$\else
\ifnum#2=307
$Pn'na$\else
\ifnum#2=308
$Pnn'a$\else
\ifnum#2=309
$Pnna'$\else
\ifnum#2=310
$Pn'n'a$\else
\ifnum#2=311
$Pnn'a'$\else
\ifnum#2=312
$Pn'na'$\else
\ifnum#2=313
$Pn'n'a'$\else
\ifnum#2=314
$P_{a}nna$\else
\ifnum#2=315
$P_{b}nna$\else
\ifnum#2=316
$P_{c}nna$\else
\ifnum#2=317
$P_{A}nna$\else
\ifnum#2=318
$P_{B}nna$\else
\ifnum#2=319
$P_{C}nna$\else
\ifnum#2=320
$P_{I}nna$\else
{\color{red}{Invalid MSG number}}
\fi
\fi
\fi
\fi
\fi
\fi
\fi
\fi
\fi
\fi
\fi
\fi
\fi
\fi
\fi
\fi
\else
\ifnum#1=53
\ifnum#2=321
$Pmna$\else
\ifnum#2=322
$Pmna1'$\else
\ifnum#2=323
$Pm'na$\else
\ifnum#2=324
$Pmn'a$\else
\ifnum#2=325
$Pmna'$\else
\ifnum#2=326
$Pm'n'a$\else
\ifnum#2=327
$Pmn'a'$\else
\ifnum#2=328
$Pm'na'$\else
\ifnum#2=329
$Pm'n'a'$\else
\ifnum#2=330
$P_{a}mna$\else
\ifnum#2=331
$P_{b}mna$\else
\ifnum#2=332
$P_{c}mna$\else
\ifnum#2=333
$P_{A}mna$\else
\ifnum#2=334
$P_{B}mna$\else
\ifnum#2=335
$P_{C}mna$\else
\ifnum#2=336
$P_{I}mna$\else
{\color{red}{Invalid MSG number}}
\fi
\fi
\fi
\fi
\fi
\fi
\fi
\fi
\fi
\fi
\fi
\fi
\fi
\fi
\fi
\fi
\else
\ifnum#1=54
\ifnum#2=337
$Pcca$\else
\ifnum#2=338
$Pcca1'$\else
\ifnum#2=339
$Pc'ca$\else
\ifnum#2=340
$Pcc'a$\else
\ifnum#2=341
$Pcca'$\else
\ifnum#2=342
$Pc'c'a$\else
\ifnum#2=343
$Pcc'a'$\else
\ifnum#2=344
$Pc'ca'$\else
\ifnum#2=345
$Pc'c'a'$\else
\ifnum#2=346
$P_{a}cca$\else
\ifnum#2=347
$P_{b}cca$\else
\ifnum#2=348
$P_{c}cca$\else
\ifnum#2=349
$P_{A}cca$\else
\ifnum#2=350
$P_{B}cca$\else
\ifnum#2=351
$P_{C}cca$\else
\ifnum#2=352
$P_{I}cca$\else
{\color{red}{Invalid MSG number}}
\fi
\fi
\fi
\fi
\fi
\fi
\fi
\fi
\fi
\fi
\fi
\fi
\fi
\fi
\fi
\fi
\else
\ifnum#1=55
\ifnum#2=353
$Pbam$\else
\ifnum#2=354
$Pbam1'$\else
\ifnum#2=355
$Pb'am$\else
\ifnum#2=356
$Pbam'$\else
\ifnum#2=357
$Pb'a'm$\else
\ifnum#2=358
$Pb'am'$\else
\ifnum#2=359
$Pb'a'm'$\else
\ifnum#2=360
$P_{a}bam$\else
\ifnum#2=361
$P_{c}bam$\else
\ifnum#2=362
$P_{A}bam$\else
\ifnum#2=363
$P_{C}bam$\else
\ifnum#2=364
$P_{I}bam$\else
{\color{red}{Invalid MSG number}}
\fi
\fi
\fi
\fi
\fi
\fi
\fi
\fi
\fi
\fi
\fi
\fi
\else
\ifnum#1=56
\ifnum#2=365
$Pccn$\else
\ifnum#2=366
$Pccn1'$\else
\ifnum#2=367
$Pc'cn$\else
\ifnum#2=368
$Pccn'$\else
\ifnum#2=369
$Pc'c'n$\else
\ifnum#2=370
$Pc'cn'$\else
\ifnum#2=371
$Pc'c'n'$\else
\ifnum#2=372
$P_{b}ccn$\else
\ifnum#2=373
$P_{c}ccn$\else
\ifnum#2=374
$P_{A}ccn$\else
\ifnum#2=375
$P_{C}ccn$\else
\ifnum#2=376
$P_{I}ccn$\else
{\color{red}{Invalid MSG number}}
\fi
\fi
\fi
\fi
\fi
\fi
\fi
\fi
\fi
\fi
\fi
\fi
\else
\ifnum#1=57
\ifnum#2=377
$Pbcm$\else
\ifnum#2=378
$Pbcm1'$\else
\ifnum#2=379
$Pb'cm$\else
\ifnum#2=380
$Pbc'm$\else
\ifnum#2=381
$Pbcm'$\else
\ifnum#2=382
$Pb'c'm$\else
\ifnum#2=383
$Pbc'm'$\else
\ifnum#2=384
$Pb'cm'$\else
\ifnum#2=385
$Pb'c'm'$\else
\ifnum#2=386
$P_{a}bcm$\else
\ifnum#2=387
$P_{b}bcm$\else
\ifnum#2=388
$P_{c}bcm$\else
\ifnum#2=389
$P_{A}bcm$\else
\ifnum#2=390
$P_{B}bcm$\else
\ifnum#2=391
$P_{C}bcm$\else
\ifnum#2=392
$P_{I}bcm$\else
{\color{red}{Invalid MSG number}}
\fi
\fi
\fi
\fi
\fi
\fi
\fi
\fi
\fi
\fi
\fi
\fi
\fi
\fi
\fi
\fi
\else
\ifnum#1=58
\ifnum#2=393
$Pnnm$\else
\ifnum#2=394
$Pnnm1'$\else
\ifnum#2=395
$Pn'nm$\else
\ifnum#2=396
$Pnnm'$\else
\ifnum#2=397
$Pn'n'm$\else
\ifnum#2=398
$Pnn'm'$\else
\ifnum#2=399
$Pn'n'm'$\else
\ifnum#2=400
$P_{a}nnm$\else
\ifnum#2=401
$P_{c}nnm$\else
\ifnum#2=402
$P_{B}nnm$\else
\ifnum#2=403
$P_{C}nnm$\else
\ifnum#2=404
$P_{I}nnm$\else
{\color{red}{Invalid MSG number}}
\fi
\fi
\fi
\fi
\fi
\fi
\fi
\fi
\fi
\fi
\fi
\fi
\else
\ifnum#1=59
\ifnum#2=405
$Pmmn$\else
\ifnum#2=406
$Pmmn1'$\else
\ifnum#2=407
$Pm'mn$\else
\ifnum#2=408
$Pmmn'$\else
\ifnum#2=409
$Pm'm'n$\else
\ifnum#2=410
$Pmm'n'$\else
\ifnum#2=411
$Pm'm'n'$\else
\ifnum#2=412
$P_{b}mmn$\else
\ifnum#2=413
$P_{c}mmn$\else
\ifnum#2=414
$P_{B}mmn$\else
\ifnum#2=415
$P_{C}mmn$\else
\ifnum#2=416
$P_{I}mmn$\else
{\color{red}{Invalid MSG number}}
\fi
\fi
\fi
\fi
\fi
\fi
\fi
\fi
\fi
\fi
\fi
\fi
\else
\ifnum#1=60
\ifnum#2=417
$Pbcn$\else
\ifnum#2=418
$Pbcn1'$\else
\ifnum#2=419
$Pb'cn$\else
\ifnum#2=420
$Pbc'n$\else
\ifnum#2=421
$Pbcn'$\else
\ifnum#2=422
$Pb'c'n$\else
\ifnum#2=423
$Pbc'n'$\else
\ifnum#2=424
$Pb'cn'$\else
\ifnum#2=425
$Pb'c'n'$\else
\ifnum#2=426
$P_{a}bcn$\else
\ifnum#2=427
$P_{b}bcn$\else
\ifnum#2=428
$P_{c}bcn$\else
\ifnum#2=429
$P_{A}bcn$\else
\ifnum#2=430
$P_{B}bcn$\else
\ifnum#2=431
$P_{C}bcn$\else
\ifnum#2=432
$P_{I}bcn$\else
{\color{red}{Invalid MSG number}}
\fi
\fi
\fi
\fi
\fi
\fi
\fi
\fi
\fi
\fi
\fi
\fi
\fi
\fi
\fi
\fi
\else
\ifnum#1=61
\ifnum#2=433
$Pbca$\else
\ifnum#2=434
$Pbca1'$\else
\ifnum#2=435
$Pb'ca$\else
\ifnum#2=436
$Pb'c'a$\else
\ifnum#2=437
$Pb'c'a'$\else
\ifnum#2=438
$P_{a}bca$\else
\ifnum#2=439
$P_{C}bca$\else
\ifnum#2=440
$P_{I}bca$\else
{\color{red}{Invalid MSG number}}
\fi
\fi
\fi
\fi
\fi
\fi
\fi
\fi
\else
\ifnum#1=62
\ifnum#2=441
$Pnma$\else
\ifnum#2=442
$Pnma1'$\else
\ifnum#2=443
$Pn'ma$\else
\ifnum#2=444
$Pnm'a$\else
\ifnum#2=445
$Pnma'$\else
\ifnum#2=446
$Pn'm'a$\else
\ifnum#2=447
$Pnm'a'$\else
\ifnum#2=448
$Pn'ma'$\else
\ifnum#2=449
$Pn'm'a'$\else
\ifnum#2=450
$P_{a}nma$\else
\ifnum#2=451
$P_{b}nma$\else
\ifnum#2=452
$P_{c}nma$\else
\ifnum#2=453
$P_{A}nma$\else
\ifnum#2=454
$P_{B}nma$\else
\ifnum#2=455
$P_{C}nma$\else
\ifnum#2=456
$P_{I}nma$\else
{\color{red}{Invalid MSG number}}
\fi
\fi
\fi
\fi
\fi
\fi
\fi
\fi
\fi
\fi
\fi
\fi
\fi
\fi
\fi
\fi
\else
\ifnum#1=63
\ifnum#2=457
$Cmcm$\else
\ifnum#2=458
$Cmcm1'$\else
\ifnum#2=459
$Cm'cm$\else
\ifnum#2=460
$Cmc'm$\else
\ifnum#2=461
$Cmcm'$\else
\ifnum#2=462
$Cm'c'm$\else
\ifnum#2=463
$Cmc'm'$\else
\ifnum#2=464
$Cm'cm'$\else
\ifnum#2=465
$Cm'c'm'$\else
\ifnum#2=466
$C_{c}mcm$\else
\ifnum#2=467
$C_{a}mcm$\else
\ifnum#2=468
$C_{A}mcm$\else
{\color{red}{Invalid MSG number}}
\fi
\fi
\fi
\fi
\fi
\fi
\fi
\fi
\fi
\fi
\fi
\fi
\else
\ifnum#1=64
\ifnum#2=469
$Cmca$\else
\ifnum#2=470
$Cmca1'$\else
\ifnum#2=471
$Cm'ca$\else
\ifnum#2=472
$Cmc'a$\else
\ifnum#2=473
$Cmca'$\else
\ifnum#2=474
$Cm'c'a$\else
\ifnum#2=475
$Cmc'a'$\else
\ifnum#2=476
$Cm'ca'$\else
\ifnum#2=477
$Cm'c'a'$\else
\ifnum#2=478
$C_{c}mca$\else
\ifnum#2=479
$C_{a}mca$\else
\ifnum#2=480
$C_{A}mca$\else
{\color{red}{Invalid MSG number}}
\fi
\fi
\fi
\fi
\fi
\fi
\fi
\fi
\fi
\fi
\fi
\fi
\else
\ifnum#1=65
\ifnum#2=481
$Cmmm$\else
\ifnum#2=482
$Cmmm1'$\else
\ifnum#2=483
$Cm'mm$\else
\ifnum#2=484
$Cmmm'$\else
\ifnum#2=485
$Cm'm'm$\else
\ifnum#2=486
$Cmm'm'$\else
\ifnum#2=487
$Cm'm'm'$\else
\ifnum#2=488
$C_{c}mmm$\else
\ifnum#2=489
$C_{a}mmm$\else
\ifnum#2=490
$C_{A}mmm$\else
{\color{red}{Invalid MSG number}}
\fi
\fi
\fi
\fi
\fi
\fi
\fi
\fi
\fi
\fi
\else
\ifnum#1=66
\ifnum#2=491
$Cccm$\else
\ifnum#2=492
$Cccm1'$\else
\ifnum#2=493
$Cc'cm$\else
\ifnum#2=494
$Cccm'$\else
\ifnum#2=495
$Cc'c'm$\else
\ifnum#2=496
$Ccc'm'$\else
\ifnum#2=497
$Cc'c'm'$\else
\ifnum#2=498
$C_{c}ccm$\else
\ifnum#2=499
$C_{a}ccm$\else
\ifnum#2=500
$C_{A}ccm$\else
{\color{red}{Invalid MSG number}}
\fi
\fi
\fi
\fi
\fi
\fi
\fi
\fi
\fi
\fi
\else
\ifnum#1=67
\ifnum#2=501
$Cmma$\else
\ifnum#2=502
$Cmma1'$\else
\ifnum#2=503
$Cm'ma$\else
\ifnum#2=504
$Cmma'$\else
\ifnum#2=505
$Cm'm'a$\else
\ifnum#2=506
$Cmm'a'$\else
\ifnum#2=507
$Cm'm'a'$\else
\ifnum#2=508
$C_{c}mma$\else
\ifnum#2=509
$C_{a}mma$\else
\ifnum#2=510
$C_{A}mma$\else
{\color{red}{Invalid MSG number}}
\fi
\fi
\fi
\fi
\fi
\fi
\fi
\fi
\fi
\fi
\else
\ifnum#1=68
\ifnum#2=511
$Ccca$\else
\ifnum#2=512
$Ccca1'$\else
\ifnum#2=513
$Cc'ca$\else
\ifnum#2=514
$Ccca'$\else
\ifnum#2=515
$Cc'c'a$\else
\ifnum#2=516
$Ccc'a'$\else
\ifnum#2=517
$Cc'c'a'$\else
\ifnum#2=518
$C_{c}cca$\else
\ifnum#2=519
$C_{a}cca$\else
\ifnum#2=520
$C_{A}cca$\else
{\color{red}{Invalid MSG number}}
\fi
\fi
\fi
\fi
\fi
\fi
\fi
\fi
\fi
\fi
\else
\ifnum#1=69
\ifnum#2=521
$Fmmm$\else
\ifnum#2=522
$Fmmm1'$\else
\ifnum#2=523
$Fm'mm$\else
\ifnum#2=524
$Fm'm'm$\else
\ifnum#2=525
$Fm'm'm'$\else
\ifnum#2=526
$F_{S}mmm$\else
{\color{red}{Invalid MSG number}}
\fi
\fi
\fi
\fi
\fi
\fi
\else
\ifnum#1=70
\ifnum#2=527
$Fddd$\else
\ifnum#2=528
$Fddd1'$\else
\ifnum#2=529
$Fd'dd$\else
\ifnum#2=530
$Fd'd'd$\else
\ifnum#2=531
$Fd'd'd'$\else
\ifnum#2=532
$F_{S}ddd$\else
{\color{red}{Invalid MSG number}}
\fi
\fi
\fi
\fi
\fi
\fi
\else
\ifnum#1=71
\ifnum#2=533
$Immm$\else
\ifnum#2=534
$Immm1'$\else
\ifnum#2=535
$Im'mm$\else
\ifnum#2=536
$Im'm'm$\else
\ifnum#2=537
$Im'm'm'$\else
\ifnum#2=538
$I_{c}mmm$\else
{\color{red}{Invalid MSG number}}
\fi
\fi
\fi
\fi
\fi
\fi
\else
\ifnum#1=72
\ifnum#2=539
$Ibam$\else
\ifnum#2=540
$Ibam1'$\else
\ifnum#2=541
$Ib'am$\else
\ifnum#2=542
$Ibam'$\else
\ifnum#2=543
$Ib'a'm$\else
\ifnum#2=544
$Iba'm'$\else
\ifnum#2=545
$Ib'a'm'$\else
\ifnum#2=546
$I_{c}bam$\else
\ifnum#2=547
$I_{b}bam$\else
{\color{red}{Invalid MSG number}}
\fi
\fi
\fi
\fi
\fi
\fi
\fi
\fi
\fi
\else
\ifnum#1=73
\ifnum#2=548
$Ibca$\else
\ifnum#2=549
$Ibca1'$\else
\ifnum#2=550
$Ib'ca$\else
\ifnum#2=551
$Ib'c'a$\else
\ifnum#2=552
$Ib'c'a'$\else
\ifnum#2=553
$I_{c}bca$\else
{\color{red}{Invalid MSG number}}
\fi
\fi
\fi
\fi
\fi
\fi
\else
\ifnum#1=74
\ifnum#2=554
$Imma$\else
\ifnum#2=555
$Imma1'$\else
\ifnum#2=556
$Im'ma$\else
\ifnum#2=557
$Imma'$\else
\ifnum#2=558
$Im'm'a$\else
\ifnum#2=559
$Imm'a'$\else
\ifnum#2=560
$Im'm'a'$\else
\ifnum#2=561
$I_{c}mma$\else
\ifnum#2=562
$I_{b}mma$\else
{\color{red}{Invalid MSG number}}
\fi
\fi
\fi
\fi
\fi
\fi
\fi
\fi
\fi
\else
\ifnum#1=75
\ifnum#2=1
$P4$\else
\ifnum#2=2
$P41'$\else
\ifnum#2=3
$P4'$\else
\ifnum#2=4
$P_{c}4$\else
\ifnum#2=5
$P_{C}4$\else
\ifnum#2=6
$P_{I}4$\else
{\color{red}{Invalid MSG number}}
\fi
\fi
\fi
\fi
\fi
\fi
\else
\ifnum#1=76
\ifnum#2=7
$P4_{1}$\else
\ifnum#2=8
$P4_{1}1'$\else
\ifnum#2=9
$P4_{1}'$\else
\ifnum#2=10
$P_{c}4_{1}$\else
\ifnum#2=11
$P_{C}4_{1}$\else
\ifnum#2=12
$P_{I}4_{1}$\else
{\color{red}{Invalid MSG number}}
\fi
\fi
\fi
\fi
\fi
\fi
\else
\ifnum#1=77
\ifnum#2=13
$P4_{2}$\else
\ifnum#2=14
$P4_{2}1'$\else
\ifnum#2=15
$P4_{2}'$\else
\ifnum#2=16
$P_{c}4_{2}$\else
\ifnum#2=17
$P_{C}4_{2}$\else
\ifnum#2=18
$P_{I}4_{2}$\else
{\color{red}{Invalid MSG number}}
\fi
\fi
\fi
\fi
\fi
\fi
\else
\ifnum#1=78
\ifnum#2=19
$P4_{3}$\else
\ifnum#2=20
$P4_{3}1'$\else
\ifnum#2=21
$P4_{3}'$\else
\ifnum#2=22
$P_{c}4_{3}$\else
\ifnum#2=23
$P_{C}4_{3}$\else
\ifnum#2=24
$P_{I}4_{3}$\else
{\color{red}{Invalid MSG number}}
\fi
\fi
\fi
\fi
\fi
\fi
\else
\ifnum#1=79
\ifnum#2=25
$I4$\else
\ifnum#2=26
$I41'$\else
\ifnum#2=27
$I4'$\else
\ifnum#2=28
$I_{c}4$\else
{\color{red}{Invalid MSG number}}
\fi
\fi
\fi
\fi
\else
\ifnum#1=80
\ifnum#2=29
$I4_{1}$\else
\ifnum#2=30
$I4_{1}1'$\else
\ifnum#2=31
$I4_{1}'$\else
\ifnum#2=32
$I_{c}4_{1}$\else
{\color{red}{Invalid MSG number}}
\fi
\fi
\fi
\fi
\else
\ifnum#1=81
\ifnum#2=33
$P\bar{4}$\else
\ifnum#2=34
$P\bar{4}1'$\else
\ifnum#2=35
$P\bar{4}'$\else
\ifnum#2=36
$P_{c}\bar{4}$\else
\ifnum#2=37
$P_{C}\bar{4}$\else
\ifnum#2=38
$P_{I}\bar{4}$\else
{\color{red}{Invalid MSG number}}
\fi
\fi
\fi
\fi
\fi
\fi
\else
\ifnum#1=82
\ifnum#2=39
$I\bar{4}$\else
\ifnum#2=40
$I\bar{4}1'$\else
\ifnum#2=41
$I\bar{4}'$\else
\ifnum#2=42
$I_{c}\bar{4}$\else
{\color{red}{Invalid MSG number}}
\fi
\fi
\fi
\fi
\else
\ifnum#1=83
\ifnum#2=43
$P4/m$\else
\ifnum#2=44
$P4/m1'$\else
\ifnum#2=45
$P4'/m$\else
\ifnum#2=46
$P4/m'$\else
\ifnum#2=47
$P4'/m'$\else
\ifnum#2=48
$P_{c}4/m$\else
\ifnum#2=49
$P_{C}4/m$\else
\ifnum#2=50
$P_{I}4/m$\else
{\color{red}{Invalid MSG number}}
\fi
\fi
\fi
\fi
\fi
\fi
\fi
\fi
\else
\ifnum#1=84
\ifnum#2=51
$P4_{2}/m$\else
\ifnum#2=52
$P4_{2}/m1'$\else
\ifnum#2=53
$P4_{2}'/m$\else
\ifnum#2=54
$P4_{2}/m'$\else
\ifnum#2=55
$P4_{2}'/m'$\else
\ifnum#2=56
$P_{c}4_{2}/m$\else
\ifnum#2=57
$P_{C}4_{2}/m$\else
\ifnum#2=58
$P_{I}4_{2}/m$\else
{\color{red}{Invalid MSG number}}
\fi
\fi
\fi
\fi
\fi
\fi
\fi
\fi
\else
\ifnum#1=85
\ifnum#2=59
$P4/n$\else
\ifnum#2=60
$P4/n1'$\else
\ifnum#2=61
$P4'/n$\else
\ifnum#2=62
$P4/n'$\else
\ifnum#2=63
$P4'/n'$\else
\ifnum#2=64
$P_{c}4/n$\else
\ifnum#2=65
$P_{C}4/n$\else
\ifnum#2=66
$P_{I}4/n$\else
{\color{red}{Invalid MSG number}}
\fi
\fi
\fi
\fi
\fi
\fi
\fi
\fi
\else
\ifnum#1=86
\ifnum#2=67
$P4_{2}/n$\else
\ifnum#2=68
$P4_{2}/n1'$\else
\ifnum#2=69
$P4_{2}'/n$\else
\ifnum#2=70
$P4_{2}/n'$\else
\ifnum#2=71
$P4_{2}'/n'$\else
\ifnum#2=72
$P_{c}4_{2}/n$\else
\ifnum#2=73
$P_{C}4_{2}/n$\else
\ifnum#2=74
$P_{I}4_{2}/n$\else
{\color{red}{Invalid MSG number}}
\fi
\fi
\fi
\fi
\fi
\fi
\fi
\fi
\else
\ifnum#1=87
\ifnum#2=75
$I4/m$\else
\ifnum#2=76
$I4/m1'$\else
\ifnum#2=77
$I4'/m$\else
\ifnum#2=78
$I4/m'$\else
\ifnum#2=79
$I4'/m'$\else
\ifnum#2=80
$I_{c}4/m$\else
{\color{red}{Invalid MSG number}}
\fi
\fi
\fi
\fi
\fi
\fi
\else
\ifnum#1=88
\ifnum#2=81
$I4_{1}/a$\else
\ifnum#2=82
$I4_{1}/a1'$\else
\ifnum#2=83
$I4_{1}'/a$\else
\ifnum#2=84
$I4_{1}/a'$\else
\ifnum#2=85
$I4_{1}'/a'$\else
\ifnum#2=86
$I_{c}4_{1}/a$\else
{\color{red}{Invalid MSG number}}
\fi
\fi
\fi
\fi
\fi
\fi
\else
\ifnum#1=89
\ifnum#2=87
$P422$\else
\ifnum#2=88
$P4221'$\else
\ifnum#2=89
$P4'22'$\else
\ifnum#2=90
$P42'2'$\else
\ifnum#2=91
$P4'2'2$\else
\ifnum#2=92
$P_{c}422$\else
\ifnum#2=93
$P_{C}422$\else
\ifnum#2=94
$P_{I}422$\else
{\color{red}{Invalid MSG number}}
\fi
\fi
\fi
\fi
\fi
\fi
\fi
\fi
\else
\ifnum#1=90
\ifnum#2=95
$P42_{1}2$\else
\ifnum#2=96
$P42_{1}21'$\else
\ifnum#2=97
$P4'2_{1}2'$\else
\ifnum#2=98
$P42_{1}'2'$\else
\ifnum#2=99
$P4'2_{1}'2$\else
\ifnum#2=100
$P_{c}42_{1}2$\else
\ifnum#2=101
$P_{C}42_{1}2$\else
\ifnum#2=102
$P_{I}42_{1}2$\else
{\color{red}{Invalid MSG number}}
\fi
\fi
\fi
\fi
\fi
\fi
\fi
\fi
\else
\ifnum#1=91
\ifnum#2=103
$P4_{1}22$\else
\ifnum#2=104
$P4_{1}221'$\else
\ifnum#2=105
$P4_{1}'22'$\else
\ifnum#2=106
$P4_{1}2'2'$\else
\ifnum#2=107
$P4_{1}'2'2$\else
\ifnum#2=108
$P_{c}4_{1}22$\else
\ifnum#2=109
$P_{C}4_{1}22$\else
\ifnum#2=110
$P_{I}4_{1}22$\else
{\color{red}{Invalid MSG number}}
\fi
\fi
\fi
\fi
\fi
\fi
\fi
\fi
\else
\ifnum#1=92
\ifnum#2=111
$P4_{1}2_{1}2$\else
\ifnum#2=112
$P4_{1}2_{1}21'$\else
\ifnum#2=113
$P4_{1}'2_{1}2'$\else
\ifnum#2=114
$P4_{1}2_{1}'2'$\else
\ifnum#2=115
$P4_{1}'2_{1}'2$\else
\ifnum#2=116
$P_{c}4_{1}2_{1}2$\else
\ifnum#2=117
$P_{C}4_{1}2_{1}2$\else
\ifnum#2=118
$P_{I}4_{1}2_{1}2$\else
{\color{red}{Invalid MSG number}}
\fi
\fi
\fi
\fi
\fi
\fi
\fi
\fi
\else
\ifnum#1=93
\ifnum#2=119
$P4_{2}22$\else
\ifnum#2=120
$P4_{2}221'$\else
\ifnum#2=121
$P4_{2}'22'$\else
\ifnum#2=122
$P4_{2}2'2'$\else
\ifnum#2=123
$P4_{2}'2'2$\else
\ifnum#2=124
$P_{c}4_{2}22$\else
\ifnum#2=125
$P_{C}4_{2}22$\else
\ifnum#2=126
$P_{I}4_{2}22$\else
{\color{red}{Invalid MSG number}}
\fi
\fi
\fi
\fi
\fi
\fi
\fi
\fi
\else
\ifnum#1=94
\ifnum#2=127
$P4_{2}2_{1}2$\else
\ifnum#2=128
$P4_{2}2_{1}21'$\else
\ifnum#2=129
$P4_{2}'2_{1}2'$\else
\ifnum#2=130
$P4_{2}2_{1}'2'$\else
\ifnum#2=131
$P4_{2}'2_{1}'2$\else
\ifnum#2=132
$P_{c}4_{2}2_{1}2$\else
\ifnum#2=133
$P_{C}4_{2}2_{1}2$\else
\ifnum#2=134
$P_{I}4_{2}2_{1}2$\else
{\color{red}{Invalid MSG number}}
\fi
\fi
\fi
\fi
\fi
\fi
\fi
\fi
\else
\ifnum#1=95
\ifnum#2=135
$P4_{3}22$\else
\ifnum#2=136
$P4_{3}221'$\else
\ifnum#2=137
$P4_{3}'22'$\else
\ifnum#2=138
$P4_{3}2'2'$\else
\ifnum#2=139
$P4_{3}'2'2$\else
\ifnum#2=140
$P_{c}4_{3}22$\else
\ifnum#2=141
$P_{C}4_{3}22$\else
\ifnum#2=142
$P_{I}4_{3}22$\else
{\color{red}{Invalid MSG number}}
\fi
\fi
\fi
\fi
\fi
\fi
\fi
\fi
\else
\ifnum#1=96
\ifnum#2=143
$P4_{3}2_{1}2$\else
\ifnum#2=144
$P4_{3}2_{1}21'$\else
\ifnum#2=145
$P4_{3}'2_{1}2'$\else
\ifnum#2=146
$P4_{3}2_{1}'2'$\else
\ifnum#2=147
$P4_{3}'2_{1}'2$\else
\ifnum#2=148
$P_{c}4_{3}2_{1}2$\else
\ifnum#2=149
$P_{C}4_{3}2_{1}2$\else
\ifnum#2=150
$P_{I}4_{3}2_{1}2$\else
{\color{red}{Invalid MSG number}}
\fi
\fi
\fi
\fi
\fi
\fi
\fi
\fi
\else
\ifnum#1=97
\ifnum#2=151
$I422$\else
\ifnum#2=152
$I4221'$\else
\ifnum#2=153
$I4'22'$\else
\ifnum#2=154
$I42'2'$\else
\ifnum#2=155
$I4'2'2$\else
\ifnum#2=156
$I_{c}422$\else
{\color{red}{Invalid MSG number}}
\fi
\fi
\fi
\fi
\fi
\fi
\else
\ifnum#1=98
\ifnum#2=157
$I4_{1}22$\else
\ifnum#2=158
$I4_{1}221'$\else
\ifnum#2=159
$I4_{1}'22'$\else
\ifnum#2=160
$I4_{1}2'2'$\else
\ifnum#2=161
$I4_{1}'2'2$\else
\ifnum#2=162
$I_{c}4_{1}22$\else
{\color{red}{Invalid MSG number}}
\fi
\fi
\fi
\fi
\fi
\fi
\else
\ifnum#1=99
\ifnum#2=163
$P4mm$\else
\ifnum#2=164
$P4mm1'$\else
\ifnum#2=165
$P4'm'm$\else
\ifnum#2=166
$P4'mm'$\else
\ifnum#2=167
$P4m'm'$\else
\ifnum#2=168
$P_{c}4mm$\else
\ifnum#2=169
$P_{C}4mm$\else
\ifnum#2=170
$P_{I}4mm$\else
{\color{red}{Invalid MSG number}}
\fi
\fi
\fi
\fi
\fi
\fi
\fi
\fi
\else
\ifnum#1=100
\ifnum#2=171
$P4bm$\else
\ifnum#2=172
$P4bm1'$\else
\ifnum#2=173
$P4'b'm$\else
\ifnum#2=174
$P4'bm'$\else
\ifnum#2=175
$P4b'm'$\else
\ifnum#2=176
$P_{c}4bm$\else
\ifnum#2=177
$P_{C}4bm$\else
\ifnum#2=178
$P_{I}4bm$\else
{\color{red}{Invalid MSG number}}
\fi
\fi
\fi
\fi
\fi
\fi
\fi
\fi
\else
\ifnum#1=101
\ifnum#2=179
$P4_{2}cm$\else
\ifnum#2=180
$P4_{2}cm1'$\else
\ifnum#2=181
$P4_{2}'c'm$\else
\ifnum#2=182
$P4_{2}'cm'$\else
\ifnum#2=183
$P4_{2}c'm'$\else
\ifnum#2=184
$P_{c}4_{2}cm$\else
\ifnum#2=185
$P_{C}4_{2}cm$\else
\ifnum#2=186
$P_{I}4_{2}cm$\else
{\color{red}{Invalid MSG number}}
\fi
\fi
\fi
\fi
\fi
\fi
\fi
\fi
\else
\ifnum#1=102
\ifnum#2=187
$P4_{2}nm$\else
\ifnum#2=188
$P4_{2}nm1'$\else
\ifnum#2=189
$P4_{2}'n'm$\else
\ifnum#2=190
$P4_{2}'nm'$\else
\ifnum#2=191
$P4_{2}n'm'$\else
\ifnum#2=192
$P_{c}4_{2}nm$\else
\ifnum#2=193
$P_{C}4_{2}nm$\else
\ifnum#2=194
$P_{I}4_{2}nm$\else
{\color{red}{Invalid MSG number}}
\fi
\fi
\fi
\fi
\fi
\fi
\fi
\fi
\else
\ifnum#1=103
\ifnum#2=195
$P4cc$\else
\ifnum#2=196
$P4cc1'$\else
\ifnum#2=197
$P4'c'c$\else
\ifnum#2=198
$P4'cc'$\else
\ifnum#2=199
$P4c'c'$\else
\ifnum#2=200
$P_{c}4cc$\else
\ifnum#2=201
$P_{C}4cc$\else
\ifnum#2=202
$P_{I}4cc$\else
{\color{red}{Invalid MSG number}}
\fi
\fi
\fi
\fi
\fi
\fi
\fi
\fi
\else
\ifnum#1=104
\ifnum#2=203
$P4nc$\else
\ifnum#2=204
$P4nc1'$\else
\ifnum#2=205
$P4'n'c$\else
\ifnum#2=206
$P4'nc'$\else
\ifnum#2=207
$P4n'c'$\else
\ifnum#2=208
$P_{c}4nc$\else
\ifnum#2=209
$P_{C}4nc$\else
\ifnum#2=210
$P_{I}4nc$\else
{\color{red}{Invalid MSG number}}
\fi
\fi
\fi
\fi
\fi
\fi
\fi
\fi
\else
\ifnum#1=105
\ifnum#2=211
$P4_{2}mc$\else
\ifnum#2=212
$P4_{2}mc1'$\else
\ifnum#2=213
$P4_{2}'m'c$\else
\ifnum#2=214
$P4_{2}'mc'$\else
\ifnum#2=215
$P4_{2}m'c'$\else
\ifnum#2=216
$P_{c}4_{2}mc$\else
\ifnum#2=217
$P_{C}4_{2}mc$\else
\ifnum#2=218
$P_{I}4_{2}mc$\else
{\color{red}{Invalid MSG number}}
\fi
\fi
\fi
\fi
\fi
\fi
\fi
\fi
\else
\ifnum#1=106
\ifnum#2=219
$P4_{2}bc$\else
\ifnum#2=220
$P4_{2}bc1'$\else
\ifnum#2=221
$P4_{2}'b'c$\else
\ifnum#2=222
$P4_{2}'bc'$\else
\ifnum#2=223
$P4_{2}b'c'$\else
\ifnum#2=224
$P_{c}4_{2}bc$\else
\ifnum#2=225
$P_{C}4_{2}bc$\else
\ifnum#2=226
$P_{I}4_{2}bc$\else
{\color{red}{Invalid MSG number}}
\fi
\fi
\fi
\fi
\fi
\fi
\fi
\fi
\else
\ifnum#1=107
\ifnum#2=227
$I4mm$\else
\ifnum#2=228
$I4mm1'$\else
\ifnum#2=229
$I4'm'm$\else
\ifnum#2=230
$I4'mm'$\else
\ifnum#2=231
$I4m'm'$\else
\ifnum#2=232
$I_{c}4mm$\else
{\color{red}{Invalid MSG number}}
\fi
\fi
\fi
\fi
\fi
\fi
\else
\ifnum#1=108
\ifnum#2=233
$I4cm$\else
\ifnum#2=234
$I4cm1'$\else
\ifnum#2=235
$I4'c'm$\else
\ifnum#2=236
$I4'cm'$\else
\ifnum#2=237
$I4c'm'$\else
\ifnum#2=238
$I_{c}4cm$\else
{\color{red}{Invalid MSG number}}
\fi
\fi
\fi
\fi
\fi
\fi
\else
\ifnum#1=109
\ifnum#2=239
$I4_{1}md$\else
\ifnum#2=240
$I4_{1}md1'$\else
\ifnum#2=241
$I4_{1}'m'd$\else
\ifnum#2=242
$I4_{1}'md'$\else
\ifnum#2=243
$I4_{1}m'd'$\else
\ifnum#2=244
$I_{c}4_{1}md$\else
{\color{red}{Invalid MSG number}}
\fi
\fi
\fi
\fi
\fi
\fi
\else
\ifnum#1=110
\ifnum#2=245
$I4_{1}cd$\else
\ifnum#2=246
$I4_{1}cd1'$\else
\ifnum#2=247
$I4_{1}'c'd$\else
\ifnum#2=248
$I4_{1}'cd'$\else
\ifnum#2=249
$I4_{1}c'd'$\else
\ifnum#2=250
$I_{c}4_{1}cd$\else
{\color{red}{Invalid MSG number}}
\fi
\fi
\fi
\fi
\fi
\fi
\else
\ifnum#1=111
\ifnum#2=251
$P\bar{4}2m$\else
\ifnum#2=252
$P\bar{4}2m1'$\else
\ifnum#2=253
$P\bar{4}'2'm$\else
\ifnum#2=254
$P\bar{4}'2m'$\else
\ifnum#2=255
$P\bar{4}2'm'$\else
\ifnum#2=256
$P_{c}\bar{4}2m$\else
\ifnum#2=257
$P_{C}\bar{4}2m$\else
\ifnum#2=258
$P_{I}\bar{4}2m$\else
{\color{red}{Invalid MSG number}}
\fi
\fi
\fi
\fi
\fi
\fi
\fi
\fi
\else
\ifnum#1=112
\ifnum#2=259
$P\bar{4}2c$\else
\ifnum#2=260
$P\bar{4}2c1'$\else
\ifnum#2=261
$P\bar{4}'2'c$\else
\ifnum#2=262
$P\bar{4}'2c'$\else
\ifnum#2=263
$P\bar{4}2'c'$\else
\ifnum#2=264
$P_{c}\bar{4}2c$\else
\ifnum#2=265
$P_{C}\bar{4}2c$\else
\ifnum#2=266
$P_{I}\bar{4}2c$\else
{\color{red}{Invalid MSG number}}
\fi
\fi
\fi
\fi
\fi
\fi
\fi
\fi
\else
\ifnum#1=113
\ifnum#2=267
$P\bar{4}2_{1}m$\else
\ifnum#2=268
$P\bar{4}2_{1}m1'$\else
\ifnum#2=269
$P\bar{4}'2_{1}'m$\else
\ifnum#2=270
$P\bar{4}'2_{1}m'$\else
\ifnum#2=271
$P\bar{4}2_{1}'m'$\else
\ifnum#2=272
$P_{c}\bar{4}2_{1}m$\else
\ifnum#2=273
$P_{C}\bar{4}2_{1}m$\else
\ifnum#2=274
$P_{I}\bar{4}2_{1}m$\else
{\color{red}{Invalid MSG number}}
\fi
\fi
\fi
\fi
\fi
\fi
\fi
\fi
\else
\ifnum#1=114
\ifnum#2=275
$P\bar{4}2_{1}c$\else
\ifnum#2=276
$P\bar{4}2_{1}c1'$\else
\ifnum#2=277
$P\bar{4}'2_{1}'c$\else
\ifnum#2=278
$P\bar{4}'2_{1}c'$\else
\ifnum#2=279
$P\bar{4}2_{1}'c'$\else
\ifnum#2=280
$P_{c}\bar{4}2_{1}c$\else
\ifnum#2=281
$P_{C}\bar{4}2_{1}c$\else
\ifnum#2=282
$P_{I}\bar{4}2_{1}c$\else
{\color{red}{Invalid MSG number}}
\fi
\fi
\fi
\fi
\fi
\fi
\fi
\fi
\else
\ifnum#1=115
\ifnum#2=283
$P\bar{4}m2$\else
\ifnum#2=284
$P\bar{4}m21'$\else
\ifnum#2=285
$P\bar{4}'m'2$\else
\ifnum#2=286
$P\bar{4}'m2'$\else
\ifnum#2=287
$P\bar{4}m'2'$\else
\ifnum#2=288
$P_{c}\bar{4}m2$\else
\ifnum#2=289
$P_{C}\bar{4}m2$\else
\ifnum#2=290
$P_{I}\bar{4}m2$\else
{\color{red}{Invalid MSG number}}
\fi
\fi
\fi
\fi
\fi
\fi
\fi
\fi
\else
\ifnum#1=116
\ifnum#2=291
$P\bar{4}c2$\else
\ifnum#2=292
$P\bar{4}c21'$\else
\ifnum#2=293
$P\bar{4}'c'2$\else
\ifnum#2=294
$P\bar{4}'c2'$\else
\ifnum#2=295
$P\bar{4}c'2'$\else
\ifnum#2=296
$P_{c}\bar{4}c2$\else
\ifnum#2=297
$P_{C}\bar{4}c2$\else
\ifnum#2=298
$P_{I}\bar{4}c2$\else
{\color{red}{Invalid MSG number}}
\fi
\fi
\fi
\fi
\fi
\fi
\fi
\fi
\else
\ifnum#1=117
\ifnum#2=299
$P\bar{4}b2$\else
\ifnum#2=300
$P\bar{4}b21'$\else
\ifnum#2=301
$P\bar{4}'b'2$\else
\ifnum#2=302
$P\bar{4}'b2'$\else
\ifnum#2=303
$P\bar{4}b'2'$\else
\ifnum#2=304
$P_{c}\bar{4}b2$\else
\ifnum#2=305
$P_{C}\bar{4}b2$\else
\ifnum#2=306
$P_{I}\bar{4}b2$\else
{\color{red}{Invalid MSG number}}
\fi
\fi
\fi
\fi
\fi
\fi
\fi
\fi
\else
\ifnum#1=118
\ifnum#2=307
$P\bar{4}n2$\else
\ifnum#2=308
$P\bar{4}n21'$\else
\ifnum#2=309
$P\bar{4}'n'2$\else
\ifnum#2=310
$P\bar{4}'n2'$\else
\ifnum#2=311
$P\bar{4}n'2'$\else
\ifnum#2=312
$P_{c}\bar{4}n2$\else
\ifnum#2=313
$P_{C}\bar{4}n2$\else
\ifnum#2=314
$P_{I}\bar{4}n2$\else
{\color{red}{Invalid MSG number}}
\fi
\fi
\fi
\fi
\fi
\fi
\fi
\fi
\else
\ifnum#1=119
\ifnum#2=315
$I\bar{4}m2$\else
\ifnum#2=316
$I\bar{4}m21'$\else
\ifnum#2=317
$I\bar{4}'m'2$\else
\ifnum#2=318
$I\bar{4}'m2'$\else
\ifnum#2=319
$I\bar{4}m'2'$\else
\ifnum#2=320
$I_{c}\bar{4}m2$\else
{\color{red}{Invalid MSG number}}
\fi
\fi
\fi
\fi
\fi
\fi
\else
\ifnum#1=120
\ifnum#2=321
$I\bar{4}c2$\else
\ifnum#2=322
$I\bar{4}c21'$\else
\ifnum#2=323
$I\bar{4}'c'2$\else
\ifnum#2=324
$I\bar{4}'c2'$\else
\ifnum#2=325
$I\bar{4}c'2'$\else
\ifnum#2=326
$I_{c}\bar{4}c2$\else
{\color{red}{Invalid MSG number}}
\fi
\fi
\fi
\fi
\fi
\fi
\else
\ifnum#1=121
\ifnum#2=327
$I\bar{4}2m$\else
\ifnum#2=328
$I\bar{4}2m1'$\else
\ifnum#2=329
$I\bar{4}'2'm$\else
\ifnum#2=330
$I\bar{4}'2m'$\else
\ifnum#2=331
$I\bar{4}2'm'$\else
\ifnum#2=332
$I_{c}\bar{4}2m$\else
{\color{red}{Invalid MSG number}}
\fi
\fi
\fi
\fi
\fi
\fi
\else
\ifnum#1=122
\ifnum#2=333
$I\bar{4}2d$\else
\ifnum#2=334
$I\bar{4}2d1'$\else
\ifnum#2=335
$I\bar{4}'2'd$\else
\ifnum#2=336
$I\bar{4}'2d'$\else
\ifnum#2=337
$I\bar{4}2'd'$\else
\ifnum#2=338
$I_{c}\bar{4}2d$\else
{\color{red}{Invalid MSG number}}
\fi
\fi
\fi
\fi
\fi
\fi
\else
\ifnum#1=123
\ifnum#2=339
$P4/mmm$\else
\ifnum#2=340
$P4/mmm1'$\else
\ifnum#2=341
$P4/m'mm$\else
\ifnum#2=342
$P4'/mm'm$\else
\ifnum#2=343
$P4'/mmm'$\else
\ifnum#2=344
$P4'/m'm'm$\else
\ifnum#2=345
$P4/mm'm'$\else
\ifnum#2=346
$P4'/m'mm'$\else
\ifnum#2=347
$P4/m'm'm'$\else
\ifnum#2=348
$P_{c}4/mmm$\else
\ifnum#2=349
$P_{C}4/mmm$\else
\ifnum#2=350
$P_{I}4/mmm$\else
{\color{red}{Invalid MSG number}}
\fi
\fi
\fi
\fi
\fi
\fi
\fi
\fi
\fi
\fi
\fi
\fi
\else
\ifnum#1=124
\ifnum#2=351
$P4/mcc$\else
\ifnum#2=352
$P4/mcc1'$\else
\ifnum#2=353
$P4/m'cc$\else
\ifnum#2=354
$P4'/mc'c$\else
\ifnum#2=355
$P4'/mcc'$\else
\ifnum#2=356
$P4'/m'c'c$\else
\ifnum#2=357
$P4/mc'c'$\else
\ifnum#2=358
$P4'/m'cc'$\else
\ifnum#2=359
$P4/m'c'c'$\else
\ifnum#2=360
$P_{c}4/mcc$\else
\ifnum#2=361
$P_{C}4/mcc$\else
\ifnum#2=362
$P_{I}4/mcc$\else
{\color{red}{Invalid MSG number}}
\fi
\fi
\fi
\fi
\fi
\fi
\fi
\fi
\fi
\fi
\fi
\fi
\else
\ifnum#1=125
\ifnum#2=363
$P4/nbm$\else
\ifnum#2=364
$P4/nbm1'$\else
\ifnum#2=365
$P4/n'bm$\else
\ifnum#2=366
$P4'/nb'm$\else
\ifnum#2=367
$P4'/nbm'$\else
\ifnum#2=368
$P4'/n'b'm$\else
\ifnum#2=369
$P4/nb'm'$\else
\ifnum#2=370
$P4'/n'bm'$\else
\ifnum#2=371
$P4/n'b'm'$\else
\ifnum#2=372
$P_{c}4/nbm$\else
\ifnum#2=373
$P_{C}4/nbm$\else
\ifnum#2=374
$P_{I}4/nbm$\else
{\color{red}{Invalid MSG number}}
\fi
\fi
\fi
\fi
\fi
\fi
\fi
\fi
\fi
\fi
\fi
\fi
\else
\ifnum#1=126
\ifnum#2=375
$P4/nnc$\else
\ifnum#2=376
$P4/nnc1'$\else
\ifnum#2=377
$P4/n'nc$\else
\ifnum#2=378
$P4'/nn'c$\else
\ifnum#2=379
$P4'/nnc'$\else
\ifnum#2=380
$P4'/n'n'c$\else
\ifnum#2=381
$P4/nn'c'$\else
\ifnum#2=382
$P4'/n'nc'$\else
\ifnum#2=383
$P4/n'n'c'$\else
\ifnum#2=384
$P_{c}4/nnc$\else
\ifnum#2=385
$P_{C}4/nnc$\else
\ifnum#2=386
$P_{I}4/nnc$\else
{\color{red}{Invalid MSG number}}
\fi
\fi
\fi
\fi
\fi
\fi
\fi
\fi
\fi
\fi
\fi
\fi
\else
\ifnum#1=127
\ifnum#2=387
$P4/mbm$\else
\ifnum#2=388
$P4/mbm1'$\else
\ifnum#2=389
$P4/m'bm$\else
\ifnum#2=390
$P4'/mb'm$\else
\ifnum#2=391
$P4'/mbm'$\else
\ifnum#2=392
$P4'/m'b'm$\else
\ifnum#2=393
$P4/mb'm'$\else
\ifnum#2=394
$P4'/m'bm'$\else
\ifnum#2=395
$P4/m'b'm'$\else
\ifnum#2=396
$P_{c}4/mbm$\else
\ifnum#2=397
$P_{C}4/mbm$\else
\ifnum#2=398
$P_{I}4/mbm$\else
{\color{red}{Invalid MSG number}}
\fi
\fi
\fi
\fi
\fi
\fi
\fi
\fi
\fi
\fi
\fi
\fi
\else
\ifnum#1=128
\ifnum#2=399
$P4/mnc$\else
\ifnum#2=400
$P4/mnc1'$\else
\ifnum#2=401
$P4/m'nc$\else
\ifnum#2=402
$P4'/mn'c$\else
\ifnum#2=403
$P4'/mnc'$\else
\ifnum#2=404
$P4'/m'n'c$\else
\ifnum#2=405
$P4/mn'c'$\else
\ifnum#2=406
$P4'/m'nc'$\else
\ifnum#2=407
$P4/m'n'c'$\else
\ifnum#2=408
$P_{c}4/mnc$\else
\ifnum#2=409
$P_{C}4/mnc$\else
\ifnum#2=410
$P_{I}4/mnc$\else
{\color{red}{Invalid MSG number}}
\fi
\fi
\fi
\fi
\fi
\fi
\fi
\fi
\fi
\fi
\fi
\fi
\else
\ifnum#1=129
\ifnum#2=411
$P4/nmm$\else
\ifnum#2=412
$P4/nmm1'$\else
\ifnum#2=413
$P4/n'mm$\else
\ifnum#2=414
$P4'/nm'm$\else
\ifnum#2=415
$P4'/nmm'$\else
\ifnum#2=416
$P4'/n'm'm$\else
\ifnum#2=417
$P4/nm'm'$\else
\ifnum#2=418
$P4'/n'mm'$\else
\ifnum#2=419
$P4/n'm'm'$\else
\ifnum#2=420
$P_{c}4/nmm$\else
\ifnum#2=421
$P_{C}4/nmm$\else
\ifnum#2=422
$P_{I}4/nmm$\else
{\color{red}{Invalid MSG number}}
\fi
\fi
\fi
\fi
\fi
\fi
\fi
\fi
\fi
\fi
\fi
\fi
\else
\ifnum#1=130
\ifnum#2=423
$P4/ncc$\else
\ifnum#2=424
$P4/ncc1'$\else
\ifnum#2=425
$P4/n'cc$\else
\ifnum#2=426
$P4'/nc'c$\else
\ifnum#2=427
$P4'/ncc'$\else
\ifnum#2=428
$P4'/n'c'c$\else
\ifnum#2=429
$P4/nc'c'$\else
\ifnum#2=430
$P4'/n'cc'$\else
\ifnum#2=431
$P4/n'c'c'$\else
\ifnum#2=432
$P_{c}4/ncc$\else
\ifnum#2=433
$P_{C}4/ncc$\else
\ifnum#2=434
$P_{I}4/ncc$\else
{\color{red}{Invalid MSG number}}
\fi
\fi
\fi
\fi
\fi
\fi
\fi
\fi
\fi
\fi
\fi
\fi
\else
\ifnum#1=131
\ifnum#2=435
$P4_{2}/mmc$\else
\ifnum#2=436
$P4_{2}/mmc1'$\else
\ifnum#2=437
$P4_{2}/m'mc$\else
\ifnum#2=438
$P4_{2}'/mm'c$\else
\ifnum#2=439
$P4_{2}'/mmc'$\else
\ifnum#2=440
$P4_{2}'/m'm'c$\else
\ifnum#2=441
$P4_{2}/mm'c'$\else
\ifnum#2=442
$P4_{2}'/m'mc'$\else
\ifnum#2=443
$P4_{2}/m'm'c'$\else
\ifnum#2=444
$P_{c}4_{2}/mmc$\else
\ifnum#2=445
$P_{C}4_{2}/mmc$\else
\ifnum#2=446
$P_{I}4_{2}/mmc$\else
{\color{red}{Invalid MSG number}}
\fi
\fi
\fi
\fi
\fi
\fi
\fi
\fi
\fi
\fi
\fi
\fi
\else
\ifnum#1=132
\ifnum#2=447
$P4_{2}/mcm$\else
\ifnum#2=448
$P4_{2}/mcm1'$\else
\ifnum#2=449
$P4_{2}/m'cm$\else
\ifnum#2=450
$P4_{2}'/mc'm$\else
\ifnum#2=451
$P4_{2}'/mcm'$\else
\ifnum#2=452
$P4_{2}'/m'c'm$\else
\ifnum#2=453
$P4_{2}/mc'm'$\else
\ifnum#2=454
$P4_{2}'/m'cm'$\else
\ifnum#2=455
$P4_{2}/m'c'm'$\else
\ifnum#2=456
$P_{c}4_{2}/mcm$\else
\ifnum#2=457
$P_{C}4_{2}/mcm$\else
\ifnum#2=458
$P_{I}4_{2}/mcm$\else
{\color{red}{Invalid MSG number}}
\fi
\fi
\fi
\fi
\fi
\fi
\fi
\fi
\fi
\fi
\fi
\fi
\else
\ifnum#1=133
\ifnum#2=459
$P4_{2}/nbc$\else
\ifnum#2=460
$P4_{2}/nbc1'$\else
\ifnum#2=461
$P4_{2}/n'bc$\else
\ifnum#2=462
$P4_{2}'/nb'c$\else
\ifnum#2=463
$P4_{2}'/nbc'$\else
\ifnum#2=464
$P4_{2}'/n'b'c$\else
\ifnum#2=465
$P4_{2}/nb'c'$\else
\ifnum#2=466
$P4_{2}'/n'bc'$\else
\ifnum#2=467
$P4_{2}/n'b'c'$\else
\ifnum#2=468
$P_{c}4_{2}/nbc$\else
\ifnum#2=469
$P_{C}4_{2}/nbc$\else
\ifnum#2=470
$P_{I}4_{2}/nbc$\else
{\color{red}{Invalid MSG number}}
\fi
\fi
\fi
\fi
\fi
\fi
\fi
\fi
\fi
\fi
\fi
\fi
\else
\ifnum#1=134
\ifnum#2=471
$P4_{2}/nnm$\else
\ifnum#2=472
$P4_{2}/nnm1'$\else
\ifnum#2=473
$P4_{2}/n'nm$\else
\ifnum#2=474
$P4_{2}'/nn'm$\else
\ifnum#2=475
$P4_{2}'/nnm'$\else
\ifnum#2=476
$P4_{2}'/n'n'm$\else
\ifnum#2=477
$P4_{2}/nn'm'$\else
\ifnum#2=478
$P4_{2}'/n'nm'$\else
\ifnum#2=479
$P4_{2}/n'n'm'$\else
\ifnum#2=480
$P_{c}4_{2}/nnm$\else
\ifnum#2=481
$P_{C}4_{2}/nnm$\else
\ifnum#2=482
$P_{I}4_{2}/nnm$\else
{\color{red}{Invalid MSG number}}
\fi
\fi
\fi
\fi
\fi
\fi
\fi
\fi
\fi
\fi
\fi
\fi
\else
\ifnum#1=135
\ifnum#2=483
$P4_{2}/mbc$\else
\ifnum#2=484
$P4_{2}/mbc1'$\else
\ifnum#2=485
$P4_{2}/m'bc$\else
\ifnum#2=486
$P4_{2}'/mb'c$\else
\ifnum#2=487
$P4_{2}'/mbc'$\else
\ifnum#2=488
$P4_{2}'/m'b'c$\else
\ifnum#2=489
$P4_{2}/mb'c'$\else
\ifnum#2=490
$P4_{2}'/m'bc'$\else
\ifnum#2=491
$P4_{2}/m'b'c'$\else
\ifnum#2=492
$P_{c}4_{2}/mbc$\else
\ifnum#2=493
$P_{C}4_{2}/mbc$\else
\ifnum#2=494
$P_{I}4_{2}/mbc$\else
{\color{red}{Invalid MSG number}}
\fi
\fi
\fi
\fi
\fi
\fi
\fi
\fi
\fi
\fi
\fi
\fi
\else
\ifnum#1=136
\ifnum#2=495
$P4_{2}/mnm$\else
\ifnum#2=496
$P4_{2}/mnm1'$\else
\ifnum#2=497
$P4_{2}/m'nm$\else
\ifnum#2=498
$P4_{2}'/mn'm$\else
\ifnum#2=499
$P4_{2}'/mnm'$\else
\ifnum#2=500
$P4_{2}'/m'n'm$\else
\ifnum#2=501
$P4_{2}/mn'm'$\else
\ifnum#2=502
$P4_{2}'/m'nm'$\else
\ifnum#2=503
$P4_{2}/m'n'm'$\else
\ifnum#2=504
$P_{c}4_{2}/mnm$\else
\ifnum#2=505
$P_{C}4_{2}/mnm$\else
\ifnum#2=506
$P_{I}4_{2}/mnm$\else
{\color{red}{Invalid MSG number}}
\fi
\fi
\fi
\fi
\fi
\fi
\fi
\fi
\fi
\fi
\fi
\fi
\else
\ifnum#1=137
\ifnum#2=507
$P4_{2}/nmc$\else
\ifnum#2=508
$P4_{2}/nmc1'$\else
\ifnum#2=509
$P4_{2}/n'mc$\else
\ifnum#2=510
$P4_{2}'/nm'c$\else
\ifnum#2=511
$P4_{2}'/nmc'$\else
\ifnum#2=512
$P4_{2}'/n'm'c$\else
\ifnum#2=513
$P4_{2}/nm'c'$\else
\ifnum#2=514
$P4_{2}'/n'mc'$\else
\ifnum#2=515
$P4_{2}/n'm'c'$\else
\ifnum#2=516
$P_{c}4_{2}/nmc$\else
\ifnum#2=517
$P_{C}4_{2}/nmc$\else
\ifnum#2=518
$P_{I}4_{2}/nmc$\else
{\color{red}{Invalid MSG number}}
\fi
\fi
\fi
\fi
\fi
\fi
\fi
\fi
\fi
\fi
\fi
\fi
\else
\ifnum#1=138
\ifnum#2=519
$P4_{2}/ncm$\else
\ifnum#2=520
$P4_{2}/ncm1'$\else
\ifnum#2=521
$P4_{2}/n'cm$\else
\ifnum#2=522
$P4_{2}'/nc'm$\else
\ifnum#2=523
$P4_{2}'/ncm'$\else
\ifnum#2=524
$P4_{2}'/n'c'm$\else
\ifnum#2=525
$P4_{2}/nc'm'$\else
\ifnum#2=526
$P4_{2}'/n'cm'$\else
\ifnum#2=527
$P4_{2}/n'c'm'$\else
\ifnum#2=528
$P_{c}4_{2}/ncm$\else
\ifnum#2=529
$P_{C}4_{2}/ncm$\else
\ifnum#2=530
$P_{I}4_{2}/ncm$\else
{\color{red}{Invalid MSG number}}
\fi
\fi
\fi
\fi
\fi
\fi
\fi
\fi
\fi
\fi
\fi
\fi
\else
\ifnum#1=139
\ifnum#2=531
$I4/mmm$\else
\ifnum#2=532
$I4/mmm1'$\else
\ifnum#2=533
$I4/m'mm$\else
\ifnum#2=534
$I4'/mm'm$\else
\ifnum#2=535
$I4'/mmm'$\else
\ifnum#2=536
$I4'/m'm'm$\else
\ifnum#2=537
$I4/mm'm'$\else
\ifnum#2=538
$I4'/m'mm'$\else
\ifnum#2=539
$I4/m'm'm'$\else
\ifnum#2=540
$I_{c}4/mmm$\else
{\color{red}{Invalid MSG number}}
\fi
\fi
\fi
\fi
\fi
\fi
\fi
\fi
\fi
\fi
\else
\ifnum#1=140
\ifnum#2=541
$I4/mcm$\else
\ifnum#2=542
$I4/mcm1'$\else
\ifnum#2=543
$I4/m'cm$\else
\ifnum#2=544
$I4'/mc'm$\else
\ifnum#2=545
$I4'/mcm'$\else
\ifnum#2=546
$I4'/m'c'm$\else
\ifnum#2=547
$I4/mc'm'$\else
\ifnum#2=548
$I4'/m'cm'$\else
\ifnum#2=549
$I4/m'c'm'$\else
\ifnum#2=550
$I_{c}4/mcm$\else
{\color{red}{Invalid MSG number}}
\fi
\fi
\fi
\fi
\fi
\fi
\fi
\fi
\fi
\fi
\else
\ifnum#1=141
\ifnum#2=551
$I4_{1}/amd$\else
\ifnum#2=552
$I4_{1}/amd1'$\else
\ifnum#2=553
$I4_{1}/a'md$\else
\ifnum#2=554
$I4_{1}'/am'd$\else
\ifnum#2=555
$I4_{1}'/amd'$\else
\ifnum#2=556
$I4_{1}'/a'm'd$\else
\ifnum#2=557
$I4_{1}/am'd'$\else
\ifnum#2=558
$I4_{1}'/a'md'$\else
\ifnum#2=559
$I4_{1}/a'm'd'$\else
\ifnum#2=560
$I_{c}4_{1}/amd$\else
{\color{red}{Invalid MSG number}}
\fi
\fi
\fi
\fi
\fi
\fi
\fi
\fi
\fi
\fi
\else
\ifnum#1=142
\ifnum#2=561
$I4_{1}/acd$\else
\ifnum#2=562
$I4_{1}/acd1'$\else
\ifnum#2=563
$I4_{1}/a'cd$\else
\ifnum#2=564
$I4_{1}'/ac'd$\else
\ifnum#2=565
$I4_{1}'/acd'$\else
\ifnum#2=566
$I4_{1}'/a'c'd$\else
\ifnum#2=567
$I4_{1}/ac'd'$\else
\ifnum#2=568
$I4_{1}'/a'cd'$\else
\ifnum#2=569
$I4_{1}/a'c'd'$\else
\ifnum#2=570
$I_{c}4_{1}/acd$\else
{\color{red}{Invalid MSG number}}
\fi
\fi
\fi
\fi
\fi
\fi
\fi
\fi
\fi
\fi
\else
\ifnum#1=143
\ifnum#2=1
$P3$\else
\ifnum#2=2
$P31'$\else
\ifnum#2=3
$P_{c}3$\else
{\color{red}{Invalid MSG number}}
\fi
\fi
\fi
\else
\ifnum#1=144
\ifnum#2=4
$P3_{1}$\else
\ifnum#2=5
$P3_{1}1'$\else
\ifnum#2=6
$P_{c}3_{1}$\else
{\color{red}{Invalid MSG number}}
\fi
\fi
\fi
\else
\ifnum#1=145
\ifnum#2=7
$P3_{2}$\else
\ifnum#2=8
$P3_{2}1'$\else
\ifnum#2=9
$P_{c}3_{2}$\else
{\color{red}{Invalid MSG number}}
\fi
\fi
\fi
\else
\ifnum#1=146
\ifnum#2=10
$R3$\else
\ifnum#2=11
$R31'$\else
\ifnum#2=12
$R_{I}3$\else
{\color{red}{Invalid MSG number}}
\fi
\fi
\fi
\else
\ifnum#1=147
\ifnum#2=13
$P\bar{3}$\else
\ifnum#2=14
$P\bar{3}1'$\else
\ifnum#2=15
$P\bar{3}'$\else
\ifnum#2=16
$P_{c}\bar{3}$\else
{\color{red}{Invalid MSG number}}
\fi
\fi
\fi
\fi
\else
\ifnum#1=148
\ifnum#2=17
$R\bar{3}$\else
\ifnum#2=18
$R\bar{3}1'$\else
\ifnum#2=19
$R\bar{3}'$\else
\ifnum#2=20
$R_{I}\bar{3}$\else
{\color{red}{Invalid MSG number}}
\fi
\fi
\fi
\fi
\else
\ifnum#1=149
\ifnum#2=21
$P312$\else
\ifnum#2=22
$P3121'$\else
\ifnum#2=23
$P312'$\else
\ifnum#2=24
$P_{c}312$\else
{\color{red}{Invalid MSG number}}
\fi
\fi
\fi
\fi
\else
\ifnum#1=150
\ifnum#2=25
$P321$\else
\ifnum#2=26
$P3211'$\else
\ifnum#2=27
$P32'1$\else
\ifnum#2=28
$P_{c}321$\else
{\color{red}{Invalid MSG number}}
\fi
\fi
\fi
\fi
\else
\ifnum#1=151
\ifnum#2=29
$P3_{1}12$\else
\ifnum#2=30
$P3_{1}121'$\else
\ifnum#2=31
$P3_{1}12'$\else
\ifnum#2=32
$P_{c}3_{1}12$\else
{\color{red}{Invalid MSG number}}
\fi
\fi
\fi
\fi
\else
\ifnum#1=152
\ifnum#2=33
$P3_{1}21$\else
\ifnum#2=34
$P3_{1}211'$\else
\ifnum#2=35
$P3_{1}2'1$\else
\ifnum#2=36
$P_{c}3_{1}21$\else
{\color{red}{Invalid MSG number}}
\fi
\fi
\fi
\fi
\else
\ifnum#1=153
\ifnum#2=37
$P3_{2}12$\else
\ifnum#2=38
$P3_{2}121'$\else
\ifnum#2=39
$P3_{2}12'$\else
\ifnum#2=40
$P_{c}3_{2}12$\else
{\color{red}{Invalid MSG number}}
\fi
\fi
\fi
\fi
\else
\ifnum#1=154
\ifnum#2=41
$P3_{2}21$\else
\ifnum#2=42
$P3_{2}211'$\else
\ifnum#2=43
$P3_{2}2'1$\else
\ifnum#2=44
$P_{c}3_{2}21$\else
{\color{red}{Invalid MSG number}}
\fi
\fi
\fi
\fi
\else
\ifnum#1=155
\ifnum#2=45
$R32$\else
\ifnum#2=46
$R321'$\else
\ifnum#2=47
$R32'$\else
\ifnum#2=48
$R_{I}32$\else
{\color{red}{Invalid MSG number}}
\fi
\fi
\fi
\fi
\else
\ifnum#1=156
\ifnum#2=49
$P3m1$\else
\ifnum#2=50
$P3m11'$\else
\ifnum#2=51
$P3m'1$\else
\ifnum#2=52
$P_{c}3m1$\else
{\color{red}{Invalid MSG number}}
\fi
\fi
\fi
\fi
\else
\ifnum#1=157
\ifnum#2=53
$P31m$\else
\ifnum#2=54
$P31m1'$\else
\ifnum#2=55
$P31m'$\else
\ifnum#2=56
$P_{c}31m$\else
{\color{red}{Invalid MSG number}}
\fi
\fi
\fi
\fi
\else
\ifnum#1=158
\ifnum#2=57
$P3c1$\else
\ifnum#2=58
$P3c11'$\else
\ifnum#2=59
$P3c'1$\else
\ifnum#2=60
$P_{c}3c1$\else
{\color{red}{Invalid MSG number}}
\fi
\fi
\fi
\fi
\else
\ifnum#1=159
\ifnum#2=61
$P31c$\else
\ifnum#2=62
$P31c1'$\else
\ifnum#2=63
$P31c'$\else
\ifnum#2=64
$P_{c}31c$\else
{\color{red}{Invalid MSG number}}
\fi
\fi
\fi
\fi
\else
\ifnum#1=160
\ifnum#2=65
$R3m$\else
\ifnum#2=66
$R3m1'$\else
\ifnum#2=67
$R3m'$\else
\ifnum#2=68
$R_{I}3m$\else
{\color{red}{Invalid MSG number}}
\fi
\fi
\fi
\fi
\else
\ifnum#1=161
\ifnum#2=69
$R3c$\else
\ifnum#2=70
$R3c1'$\else
\ifnum#2=71
$R3c'$\else
\ifnum#2=72
$R_{I}3c$\else
{\color{red}{Invalid MSG number}}
\fi
\fi
\fi
\fi
\else
\ifnum#1=162
\ifnum#2=73
$P\bar{3}1m$\else
\ifnum#2=74
$P\bar{3}1m1'$\else
\ifnum#2=75
$P\bar{3}'1m$\else
\ifnum#2=76
$P\bar{3}'1m'$\else
\ifnum#2=77
$P\bar{3}1m'$\else
\ifnum#2=78
$P_{c}\bar{3}1m$\else
{\color{red}{Invalid MSG number}}
\fi
\fi
\fi
\fi
\fi
\fi
\else
\ifnum#1=163
\ifnum#2=79
$P\bar{3}1c$\else
\ifnum#2=80
$P\bar{3}1c1'$\else
\ifnum#2=81
$P\bar{3}'1c$\else
\ifnum#2=82
$P\bar{3}'1c'$\else
\ifnum#2=83
$P\bar{3}1c'$\else
\ifnum#2=84
$P_{c}\bar{3}1c$\else
{\color{red}{Invalid MSG number}}
\fi
\fi
\fi
\fi
\fi
\fi
\else
\ifnum#1=164
\ifnum#2=85
$P\bar{3}m1$\else
\ifnum#2=86
$P\bar{3}m11'$\else
\ifnum#2=87
$P\bar{3}'m1$\else
\ifnum#2=88
$P\bar{3}'m'1$\else
\ifnum#2=89
$P\bar{3}m'1$\else
\ifnum#2=90
$P_{c}\bar{3}m1$\else
{\color{red}{Invalid MSG number}}
\fi
\fi
\fi
\fi
\fi
\fi
\else
\ifnum#1=165
\ifnum#2=91
$P\bar{3}c1$\else
\ifnum#2=92
$P\bar{3}c11'$\else
\ifnum#2=93
$P\bar{3}'c1$\else
\ifnum#2=94
$P\bar{3}'c'1$\else
\ifnum#2=95
$P\bar{3}c'1$\else
\ifnum#2=96
$P_{c}\bar{3}c1$\else
{\color{red}{Invalid MSG number}}
\fi
\fi
\fi
\fi
\fi
\fi
\else
\ifnum#1=166
\ifnum#2=97
$R\bar{3}m$\else
\ifnum#2=98
$R\bar{3}m1'$\else
\ifnum#2=99
$R\bar{3}'m$\else
\ifnum#2=100
$R\bar{3}'m'$\else
\ifnum#2=101
$R\bar{3}m'$\else
\ifnum#2=102
$R_{I}\bar{3}m$\else
{\color{red}{Invalid MSG number}}
\fi
\fi
\fi
\fi
\fi
\fi
\else
\ifnum#1=167
\ifnum#2=103
$R\bar{3}c$\else
\ifnum#2=104
$R\bar{3}c1'$\else
\ifnum#2=105
$R\bar{3}'c$\else
\ifnum#2=106
$R\bar{3}'c'$\else
\ifnum#2=107
$R\bar{3}c'$\else
\ifnum#2=108
$R_{I}\bar{3}c$\else
{\color{red}{Invalid MSG number}}
\fi
\fi
\fi
\fi
\fi
\fi
\else
\ifnum#1=168
\ifnum#2=109
$P6$\else
\ifnum#2=110
$P61'$\else
\ifnum#2=111
$P6'$\else
\ifnum#2=112
$P_{c}6$\else
{\color{red}{Invalid MSG number}}
\fi
\fi
\fi
\fi
\else
\ifnum#1=169
\ifnum#2=113
$P6_{1}$\else
\ifnum#2=114
$P6_{1}1'$\else
\ifnum#2=115
$P6_{1}'$\else
\ifnum#2=116
$P_{c}6_{1}$\else
{\color{red}{Invalid MSG number}}
\fi
\fi
\fi
\fi
\else
\ifnum#1=170
\ifnum#2=117
$P6_{5}$\else
\ifnum#2=118
$P6_{5}1'$\else
\ifnum#2=119
$P6_{5}'$\else
\ifnum#2=120
$P_{c}6_{5}$\else
{\color{red}{Invalid MSG number}}
\fi
\fi
\fi
\fi
\else
\ifnum#1=171
\ifnum#2=121
$P6_{2}$\else
\ifnum#2=122
$P6_{2}1'$\else
\ifnum#2=123
$P6_{2}'$\else
\ifnum#2=124
$P_{c}6_{2}$\else
{\color{red}{Invalid MSG number}}
\fi
\fi
\fi
\fi
\else
\ifnum#1=172
\ifnum#2=125
$P6_{4}$\else
\ifnum#2=126
$P6_{4}1'$\else
\ifnum#2=127
$P6_{4}'$\else
\ifnum#2=128
$P_{c}6_{4}$\else
{\color{red}{Invalid MSG number}}
\fi
\fi
\fi
\fi
\else
\ifnum#1=173
\ifnum#2=129
$P6_{3}$\else
\ifnum#2=130
$P6_{3}1'$\else
\ifnum#2=131
$P6_{3}'$\else
\ifnum#2=132
$P_{c}6_{3}$\else
{\color{red}{Invalid MSG number}}
\fi
\fi
\fi
\fi
\else
\ifnum#1=174
\ifnum#2=133
$P\bar{6}$\else
\ifnum#2=134
$P\bar{6}1'$\else
\ifnum#2=135
$P\bar{6}'$\else
\ifnum#2=136
$P_{c}\bar{6}$\else
{\color{red}{Invalid MSG number}}
\fi
\fi
\fi
\fi
\else
\ifnum#1=175
\ifnum#2=137
$P6/m$\else
\ifnum#2=138
$P6/m1'$\else
\ifnum#2=139
$P6'/m$\else
\ifnum#2=140
$P6/m'$\else
\ifnum#2=141
$P6'/m'$\else
\ifnum#2=142
$P_{c}6/m$\else
{\color{red}{Invalid MSG number}}
\fi
\fi
\fi
\fi
\fi
\fi
\else
\ifnum#1=176
\ifnum#2=143
$P6_{3}/m$\else
\ifnum#2=144
$P6_{3}/m1'$\else
\ifnum#2=145
$P6_{3}'/m$\else
\ifnum#2=146
$P6_{3}/m'$\else
\ifnum#2=147
$P6_{3}'/m'$\else
\ifnum#2=148
$P_{c}6_{3}/m$\else
{\color{red}{Invalid MSG number}}
\fi
\fi
\fi
\fi
\fi
\fi
\else
\ifnum#1=177
\ifnum#2=149
$P622$\else
\ifnum#2=150
$P6221'$\else
\ifnum#2=151
$P6'2'2$\else
\ifnum#2=152
$P6'22'$\else
\ifnum#2=153
$P62'2'$\else
\ifnum#2=154
$P_{c}622$\else
{\color{red}{Invalid MSG number}}
\fi
\fi
\fi
\fi
\fi
\fi
\else
\ifnum#1=178
\ifnum#2=155
$P6_{1}22$\else
\ifnum#2=156
$P6_{1}221'$\else
\ifnum#2=157
$P6_{1}'2'2$\else
\ifnum#2=158
$P6_{1}'22'$\else
\ifnum#2=159
$P6_{1}2'2'$\else
\ifnum#2=160
$P_{c}6_{1}22$\else
{\color{red}{Invalid MSG number}}
\fi
\fi
\fi
\fi
\fi
\fi
\else
\ifnum#1=179
\ifnum#2=161
$P6_{5}22$\else
\ifnum#2=162
$P6_{5}221'$\else
\ifnum#2=163
$P6_{5}'2'2$\else
\ifnum#2=164
$P6_{5}'22'$\else
\ifnum#2=165
$P6_{5}2'2'$\else
\ifnum#2=166
$P_{c}6_{5}22$\else
{\color{red}{Invalid MSG number}}
\fi
\fi
\fi
\fi
\fi
\fi
\else
\ifnum#1=180
\ifnum#2=167
$P6_{2}22$\else
\ifnum#2=168
$P6_{2}221'$\else
\ifnum#2=169
$P6_{2}'2'2$\else
\ifnum#2=170
$P6_{2}'22'$\else
\ifnum#2=171
$P6_{2}2'2'$\else
\ifnum#2=172
$P_{c}6_{2}22$\else
{\color{red}{Invalid MSG number}}
\fi
\fi
\fi
\fi
\fi
\fi
\else
\ifnum#1=181
\ifnum#2=173
$P6_{4}22$\else
\ifnum#2=174
$P6_{4}221'$\else
\ifnum#2=175
$P6_{4}'2'2$\else
\ifnum#2=176
$P6_{4}'22'$\else
\ifnum#2=177
$P6_{4}2'2'$\else
\ifnum#2=178
$P_{c}6_{4}22$\else
{\color{red}{Invalid MSG number}}
\fi
\fi
\fi
\fi
\fi
\fi
\else
\ifnum#1=182
\ifnum#2=179
$P6_{3}22$\else
\ifnum#2=180
$P6_{3}221'$\else
\ifnum#2=181
$P6_{3}'2'2$\else
\ifnum#2=182
$P6_{3}'22'$\else
\ifnum#2=183
$P6_{3}2'2'$\else
\ifnum#2=184
$P_{c}6_{3}22$\else
{\color{red}{Invalid MSG number}}
\fi
\fi
\fi
\fi
\fi
\fi
\else
\ifnum#1=183
\ifnum#2=185
$P6mm$\else
\ifnum#2=186
$P6mm1'$\else
\ifnum#2=187
$P6'm'm$\else
\ifnum#2=188
$P6'mm'$\else
\ifnum#2=189
$P6m'm'$\else
\ifnum#2=190
$P_{c}6mm$\else
{\color{red}{Invalid MSG number}}
\fi
\fi
\fi
\fi
\fi
\fi
\else
\ifnum#1=184
\ifnum#2=191
$P6cc$\else
\ifnum#2=192
$P6cc1'$\else
\ifnum#2=193
$P6'c'c$\else
\ifnum#2=194
$P6'cc'$\else
\ifnum#2=195
$P6c'c'$\else
\ifnum#2=196
$P_{c}6cc$\else
{\color{red}{Invalid MSG number}}
\fi
\fi
\fi
\fi
\fi
\fi
\else
\ifnum#1=185
\ifnum#2=197
$P6_{3}cm$\else
\ifnum#2=198
$P6_{3}cm1'$\else
\ifnum#2=199
$P6_{3}'c'm$\else
\ifnum#2=200
$P6_{3}'cm'$\else
\ifnum#2=201
$P6_{3}c'm'$\else
\ifnum#2=202
$P_{c}6_{3}cm$\else
{\color{red}{Invalid MSG number}}
\fi
\fi
\fi
\fi
\fi
\fi
\else
\ifnum#1=186
\ifnum#2=203
$P6_{3}mc$\else
\ifnum#2=204
$P6_{3}mc1'$\else
\ifnum#2=205
$P6_{3}'m'c$\else
\ifnum#2=206
$P6_{3}'mc'$\else
\ifnum#2=207
$P6_{3}m'c'$\else
\ifnum#2=208
$P_{c}6_{3}mc$\else
{\color{red}{Invalid MSG number}}
\fi
\fi
\fi
\fi
\fi
\fi
\else
\ifnum#1=187
\ifnum#2=209
$P\bar{6}m2$\else
\ifnum#2=210
$P\bar{6}m21'$\else
\ifnum#2=211
$P\bar{6}'m'2$\else
\ifnum#2=212
$P\bar{6}'m2'$\else
\ifnum#2=213
$P\bar{6}m'2'$\else
\ifnum#2=214
$P_{c}\bar{6}m2$\else
{\color{red}{Invalid MSG number}}
\fi
\fi
\fi
\fi
\fi
\fi
\else
\ifnum#1=188
\ifnum#2=215
$P\bar{6}c2$\else
\ifnum#2=216
$P\bar{6}c21'$\else
\ifnum#2=217
$P\bar{6}'c'2$\else
\ifnum#2=218
$P\bar{6}'c2'$\else
\ifnum#2=219
$P\bar{6}c'2'$\else
\ifnum#2=220
$P_{c}\bar{6}c2$\else
{\color{red}{Invalid MSG number}}
\fi
\fi
\fi
\fi
\fi
\fi
\else
\ifnum#1=189
\ifnum#2=221
$P\bar{6}2m$\else
\ifnum#2=222
$P\bar{6}2m1'$\else
\ifnum#2=223
$P\bar{6}'2'm$\else
\ifnum#2=224
$P\bar{6}'2m'$\else
\ifnum#2=225
$P\bar{6}2'm'$\else
\ifnum#2=226
$P_{c}\bar{6}2m$\else
{\color{red}{Invalid MSG number}}
\fi
\fi
\fi
\fi
\fi
\fi
\else
\ifnum#1=190
\ifnum#2=227
$P\bar{6}2c$\else
\ifnum#2=228
$P\bar{6}2c1'$\else
\ifnum#2=229
$P\bar{6}'2'c$\else
\ifnum#2=230
$P\bar{6}'2c'$\else
\ifnum#2=231
$P\bar{6}2'c'$\else
\ifnum#2=232
$P_{c}\bar{6}2c$\else
{\color{red}{Invalid MSG number}}
\fi
\fi
\fi
\fi
\fi
\fi
\else
\ifnum#1=191
\ifnum#2=233
$P6/mmm$\else
\ifnum#2=234
$P6/mmm1'$\else
\ifnum#2=235
$P6/m'mm$\else
\ifnum#2=236
$P6'/mm'm$\else
\ifnum#2=237
$P6'/mmm'$\else
\ifnum#2=238
$P6'/m'm'm$\else
\ifnum#2=239
$P6'/m'mm'$\else
\ifnum#2=240
$P6/mm'm'$\else
\ifnum#2=241
$P6/m'm'm'$\else
\ifnum#2=242
$P_{c}6/mmm$\else
{\color{red}{Invalid MSG number}}
\fi
\fi
\fi
\fi
\fi
\fi
\fi
\fi
\fi
\fi
\else
\ifnum#1=192
\ifnum#2=243
$P6/mcc$\else
\ifnum#2=244
$P6/mcc1'$\else
\ifnum#2=245
$P6/m'cc$\else
\ifnum#2=246
$P6'/mc'c$\else
\ifnum#2=247
$P6'/mcc'$\else
\ifnum#2=248
$P6'/m'c'c$\else
\ifnum#2=249
$P6'/m'cc'$\else
\ifnum#2=250
$P6/mc'c'$\else
\ifnum#2=251
$P6/m'c'c'$\else
\ifnum#2=252
$P_{c}6/mcc$\else
{\color{red}{Invalid MSG number}}
\fi
\fi
\fi
\fi
\fi
\fi
\fi
\fi
\fi
\fi
\else
\ifnum#1=193
\ifnum#2=253
$P6_{3}/mcm$\else
\ifnum#2=254
$P6_{3}/mcm1'$\else
\ifnum#2=255
$P6_{3}/m'cm$\else
\ifnum#2=256
$P6_{3}'/mc'm$\else
\ifnum#2=257
$P6_{3}'/mcm'$\else
\ifnum#2=258
$P6_{3}'/m'c'm$\else
\ifnum#2=259
$P6_{3}'/m'cm'$\else
\ifnum#2=260
$P6_{3}/mc'm'$\else
\ifnum#2=261
$P6_{3}/m'c'm'$\else
\ifnum#2=262
$P_{c}6_{3}/mcm$\else
{\color{red}{Invalid MSG number}}
\fi
\fi
\fi
\fi
\fi
\fi
\fi
\fi
\fi
\fi
\else
\ifnum#1=194
\ifnum#2=263
$P6_{3}/mmc$\else
\ifnum#2=264
$P6_{3}/mmc1'$\else
\ifnum#2=265
$P6_{3}/m'mc$\else
\ifnum#2=266
$P6_{3}'/mm'c$\else
\ifnum#2=267
$P6_{3}'/mmc'$\else
\ifnum#2=268
$P6_{3}'/m'm'c$\else
\ifnum#2=269
$P6_{3}'/m'mc'$\else
\ifnum#2=270
$P6_{3}/mm'c'$\else
\ifnum#2=271
$P6_{3}/m'm'c'$\else
\ifnum#2=272
$P_{c}6_{3}/mmc$\else
{\color{red}{Invalid MSG number}}
\fi
\fi
\fi
\fi
\fi
\fi
\fi
\fi
\fi
\fi
\else
\ifnum#1=195
\ifnum#2=1
$P23$\else
\ifnum#2=2
$P231'$\else
\ifnum#2=3
$P_{I}23$\else
{\color{red}{Invalid MSG number}}
\fi
\fi
\fi
\else
\ifnum#1=196
\ifnum#2=4
$F23$\else
\ifnum#2=5
$F231'$\else
\ifnum#2=6
$F_{S}23$\else
{\color{red}{Invalid MSG number}}
\fi
\fi
\fi
\else
\ifnum#1=197
\ifnum#2=7
$I23$\else
\ifnum#2=8
$I231'$\else
{\color{red}{Invalid MSG number}}
\fi
\fi
\else
\ifnum#1=198
\ifnum#2=9
$P2_{1}3$\else
\ifnum#2=10
$P2_{1}31'$\else
\ifnum#2=11
$P_{I}2_{1}3$\else
{\color{red}{Invalid MSG number}}
\fi
\fi
\fi
\else
\ifnum#1=199
\ifnum#2=12
$I2_{1}3$\else
\ifnum#2=13
$I2_{1}31'$\else
{\color{red}{Invalid MSG number}}
\fi
\fi
\else
\ifnum#1=200
\ifnum#2=14
$Pm\bar{3}$\else
\ifnum#2=15
$Pm\bar{3}1'$\else
\ifnum#2=16
$Pm'\bar{3}'$\else
\ifnum#2=17
$P_{I}m\bar{3}$\else
{\color{red}{Invalid MSG number}}
\fi
\fi
\fi
\fi
\else
\ifnum#1=201
\ifnum#2=18
$Pn\bar{3}$\else
\ifnum#2=19
$Pn\bar{3}1'$\else
\ifnum#2=20
$Pn'\bar{3}'$\else
\ifnum#2=21
$P_{I}n\bar{3}$\else
{\color{red}{Invalid MSG number}}
\fi
\fi
\fi
\fi
\else
\ifnum#1=202
\ifnum#2=22
$Fm\bar{3}$\else
\ifnum#2=23
$Fm\bar{3}1'$\else
\ifnum#2=24
$Fm'\bar{3}'$\else
\ifnum#2=25
$F_{S}m\bar{3}$\else
{\color{red}{Invalid MSG number}}
\fi
\fi
\fi
\fi
\else
\ifnum#1=203
\ifnum#2=26
$Fd\bar{3}$\else
\ifnum#2=27
$Fd\bar{3}1'$\else
\ifnum#2=28
$Fd'\bar{3}'$\else
\ifnum#2=29
$F_{S}d\bar{3}$\else
{\color{red}{Invalid MSG number}}
\fi
\fi
\fi
\fi
\else
\ifnum#1=204
\ifnum#2=30
$Im\bar{3}$\else
\ifnum#2=31
$Im\bar{3}1'$\else
\ifnum#2=32
$Im'\bar{3}'$\else
{\color{red}{Invalid MSG number}}
\fi
\fi
\fi
\else
\ifnum#1=205
\ifnum#2=33
$Pa\bar{3}$\else
\ifnum#2=34
$Pa\bar{3}1'$\else
\ifnum#2=35
$Pa'\bar{3}'$\else
\ifnum#2=36
$P_{I}a\bar{3}$\else
{\color{red}{Invalid MSG number}}
\fi
\fi
\fi
\fi
\else
\ifnum#1=206
\ifnum#2=37
$Ia\bar{3}$\else
\ifnum#2=38
$Ia\bar{3}1'$\else
\ifnum#2=39
$Ia'\bar{3}'$\else
{\color{red}{Invalid MSG number}}
\fi
\fi
\fi
\else
\ifnum#1=207
\ifnum#2=40
$P432$\else
\ifnum#2=41
$P4321'$\else
\ifnum#2=42
$P4'32'$\else
\ifnum#2=43
$P_{I}432$\else
{\color{red}{Invalid MSG number}}
\fi
\fi
\fi
\fi
\else
\ifnum#1=208
\ifnum#2=44
$P4_{2}32$\else
\ifnum#2=45
$P4_{2}321'$\else
\ifnum#2=46
$P4_{2}'32'$\else
\ifnum#2=47
$P_{I}4_{2}32$\else
{\color{red}{Invalid MSG number}}
\fi
\fi
\fi
\fi
\else
\ifnum#1=209
\ifnum#2=48
$F432$\else
\ifnum#2=49
$F4321'$\else
\ifnum#2=50
$F4'32'$\else
\ifnum#2=51
$F_{S}432$\else
{\color{red}{Invalid MSG number}}
\fi
\fi
\fi
\fi
\else
\ifnum#1=210
\ifnum#2=52
$F4_{1}32$\else
\ifnum#2=53
$F4_{1}321'$\else
\ifnum#2=54
$F4_{1}'32'$\else
\ifnum#2=55
$F_{S}4_{1}32$\else
{\color{red}{Invalid MSG number}}
\fi
\fi
\fi
\fi
\else
\ifnum#1=211
\ifnum#2=56
$I432$\else
\ifnum#2=57
$I4321'$\else
\ifnum#2=58
$I4'32'$\else
{\color{red}{Invalid MSG number}}
\fi
\fi
\fi
\else
\ifnum#1=212
\ifnum#2=59
$P4_{3}32$\else
\ifnum#2=60
$P4_{3}321'$\else
\ifnum#2=61
$P4_{3}'32'$\else
\ifnum#2=62
$P_{I}4_{3}32$\else
{\color{red}{Invalid MSG number}}
\fi
\fi
\fi
\fi
\else
\ifnum#1=213
\ifnum#2=63
$P4_{1}32$\else
\ifnum#2=64
$P4_{1}321'$\else
\ifnum#2=65
$P4_{1}'32'$\else
\ifnum#2=66
$P_{I}4_{1}32$\else
{\color{red}{Invalid MSG number}}
\fi
\fi
\fi
\fi
\else
\ifnum#1=214
\ifnum#2=67
$I4_{1}32$\else
\ifnum#2=68
$I4_{1}321'$\else
\ifnum#2=69
$I4_{1}'32'$\else
{\color{red}{Invalid MSG number}}
\fi
\fi
\fi
\else
\ifnum#1=215
\ifnum#2=70
$P\bar{4}3m$\else
\ifnum#2=71
$P\bar{4}3m1'$\else
\ifnum#2=72
$P\bar{4}'3m'$\else
\ifnum#2=73
$P_{I}\bar{4}3m$\else
{\color{red}{Invalid MSG number}}
\fi
\fi
\fi
\fi
\else
\ifnum#1=216
\ifnum#2=74
$F\bar{4}3m$\else
\ifnum#2=75
$F\bar{4}3m1'$\else
\ifnum#2=76
$F\bar{4}'3m'$\else
\ifnum#2=77
$F_{S}\bar{4}3m$\else
{\color{red}{Invalid MSG number}}
\fi
\fi
\fi
\fi
\else
\ifnum#1=217
\ifnum#2=78
$I\bar{4}3m$\else
\ifnum#2=79
$I\bar{4}3m1'$\else
\ifnum#2=80
$I\bar{4}'3m'$\else
{\color{red}{Invalid MSG number}}
\fi
\fi
\fi
\else
\ifnum#1=218
\ifnum#2=81
$P\bar{4}3n$\else
\ifnum#2=82
$P\bar{4}3n1'$\else
\ifnum#2=83
$P\bar{4}'3n'$\else
\ifnum#2=84
$P_{I}\bar{4}3n$\else
{\color{red}{Invalid MSG number}}
\fi
\fi
\fi
\fi
\else
\ifnum#1=219
\ifnum#2=85
$F\bar{4}3c$\else
\ifnum#2=86
$F\bar{4}3c1'$\else
\ifnum#2=87
$F\bar{4}'3c'$\else
\ifnum#2=88
$F_{S}\bar{4}3c$\else
{\color{red}{Invalid MSG number}}
\fi
\fi
\fi
\fi
\else
\ifnum#1=220
\ifnum#2=89
$I\bar{4}3d$\else
\ifnum#2=90
$I\bar{4}3d1'$\else
\ifnum#2=91
$I\bar{4}'3d'$\else
{\color{red}{Invalid MSG number}}
\fi
\fi
\fi
\else
\ifnum#1=221
\ifnum#2=92
$Pm\bar{3}m$\else
\ifnum#2=93
$Pm\bar{3}m1'$\else
\ifnum#2=94
$Pm'\bar{3}'m$\else
\ifnum#2=95
$Pm\bar{3}m'$\else
\ifnum#2=96
$Pm'\bar{3}'m'$\else
\ifnum#2=97
$P_{I}m\bar{3}m$\else
{\color{red}{Invalid MSG number}}
\fi
\fi
\fi
\fi
\fi
\fi
\else
\ifnum#1=222
\ifnum#2=98
$Pn\bar{3}n$\else
\ifnum#2=99
$Pn\bar{3}n1'$\else
\ifnum#2=100
$Pn'\bar{3}'n$\else
\ifnum#2=101
$Pn\bar{3}n'$\else
\ifnum#2=102
$Pn'\bar{3}'n'$\else
\ifnum#2=103
$P_{I}n\bar{3}n$\else
{\color{red}{Invalid MSG number}}
\fi
\fi
\fi
\fi
\fi
\fi
\else
\ifnum#1=223
\ifnum#2=104
$Pm\bar{3}n$\else
\ifnum#2=105
$Pm\bar{3}n1'$\else
\ifnum#2=106
$Pm'\bar{3}'n$\else
\ifnum#2=107
$Pm\bar{3}n'$\else
\ifnum#2=108
$Pm'\bar{3}'n'$\else
\ifnum#2=109
$P_{I}m\bar{3}n$\else
{\color{red}{Invalid MSG number}}
\fi
\fi
\fi
\fi
\fi
\fi
\else
\ifnum#1=224
\ifnum#2=110
$Pn\bar{3}m$\else
\ifnum#2=111
$Pn\bar{3}m1'$\else
\ifnum#2=112
$Pn'\bar{3}'m$\else
\ifnum#2=113
$Pn\bar{3}m'$\else
\ifnum#2=114
$Pn'\bar{3}'m'$\else
\ifnum#2=115
$P_{I}n\bar{3}m$\else
{\color{red}{Invalid MSG number}}
\fi
\fi
\fi
\fi
\fi
\fi
\else
\ifnum#1=225
\ifnum#2=116
$Fm\bar{3}m$\else
\ifnum#2=117
$Fm\bar{3}m1'$\else
\ifnum#2=118
$Fm'\bar{3}'m$\else
\ifnum#2=119
$Fm\bar{3}m'$\else
\ifnum#2=120
$Fm'\bar{3}'m'$\else
\ifnum#2=121
$F_{S}m\bar{3}m$\else
{\color{red}{Invalid MSG number}}
\fi
\fi
\fi
\fi
\fi
\fi
\else
\ifnum#1=226
\ifnum#2=122
$Fm\bar{3}c$\else
\ifnum#2=123
$Fm\bar{3}c1'$\else
\ifnum#2=124
$Fm'\bar{3}'c$\else
\ifnum#2=125
$Fm\bar{3}c'$\else
\ifnum#2=126
$Fm'\bar{3}'c'$\else
\ifnum#2=127
$F_{S}m\bar{3}c$\else
{\color{red}{Invalid MSG number}}
\fi
\fi
\fi
\fi
\fi
\fi
\else
\ifnum#1=227
\ifnum#2=128
$Fd\bar{3}m$\else
\ifnum#2=129
$Fd\bar{3}m1'$\else
\ifnum#2=130
$Fd'\bar{3}'m$\else
\ifnum#2=131
$Fd\bar{3}m'$\else
\ifnum#2=132
$Fd'\bar{3}'m'$\else
\ifnum#2=133
$F_{S}d\bar{3}m$\else
{\color{red}{Invalid MSG number}}
\fi
\fi
\fi
\fi
\fi
\fi
\else
\ifnum#1=228
\ifnum#2=134
$Fd\bar{3}c$\else
\ifnum#2=135
$Fd\bar{3}c1'$\else
\ifnum#2=136
$Fd'\bar{3}'c$\else
\ifnum#2=137
$Fd\bar{3}c'$\else
\ifnum#2=138
$Fd'\bar{3}'c'$\else
\ifnum#2=139
$F_{S}d\bar{3}c$\else
{\color{red}{Invalid MSG number}}
\fi
\fi
\fi
\fi
\fi
\fi
\else
\ifnum#1=229
\ifnum#2=140
$Im\bar{3}m$\else
\ifnum#2=141
$Im\bar{3}m1'$\else
\ifnum#2=142
$Im'\bar{3}'m$\else
\ifnum#2=143
$Im\bar{3}m'$\else
\ifnum#2=144
$Im'\bar{3}'m'$\else
{\color{red}{Invalid MSG number}}
\fi
\fi
\fi
\fi
\fi
\else
\ifnum#1=230
\ifnum#2=145
$Ia\bar{3}d$\else
\ifnum#2=146
$Ia\bar{3}d1'$\else
\ifnum#2=147
$Ia'\bar{3}'d$\else
\ifnum#2=148
$Ia\bar{3}d'$\else
\ifnum#2=149
$Ia'\bar{3}'d'$\else
{\color{red}{Invalid MSG number}}
\fi
\fi
\fi
\fi
\fi
\fi
\fi
\fi
\fi
\fi
\fi
\fi
\fi
\fi
\fi
\fi
\fi
\fi
\fi
\fi
\fi
\fi
\fi
\fi
\fi
\fi
\fi
\fi
\fi
\fi
\fi
\fi
\fi
\fi
\fi
\fi
\fi
\fi
\fi
\fi
\fi
\fi
\fi
\fi
\fi
\fi
\fi
\fi
\fi
\fi
\fi
\fi
\fi
\fi
\fi
\fi
\fi
\fi
\fi
\fi
\fi
\fi
\fi
\fi
\fi
\fi
\fi
\fi
\fi
\fi
\fi
\fi
\fi
\fi
\fi
\fi
\fi
\fi
\fi
\fi
\fi
\fi
\fi
\fi
\fi
\fi
\fi
\fi
\fi
\fi
\fi
\fi
\fi
\fi
\fi
\fi
\fi
\fi
\fi
\fi
\fi
\fi
\fi
\fi
\fi
\fi
\fi
\fi
\fi
\fi
\fi
\fi
\fi
\fi
\fi
\fi
\fi
\fi
\fi
\fi
\fi
\fi
\fi
\fi
\fi
\fi
\fi
\fi
\fi
\fi
\fi
\fi
\fi
\fi
\fi
\fi
\fi
\fi
\fi
\fi
\fi
\fi
\fi
\fi
\fi
\fi
\fi
\fi
\fi
\fi
\fi
\fi
\fi
\fi
\fi
\fi
\fi
\fi
\fi
\fi
\fi
\fi
\fi
\fi
\fi
\fi
\fi
\fi
\fi
\fi
\fi
\fi
\fi
\fi
\fi
\fi
\fi
\fi
\fi
\fi
\fi
\fi
\fi
\fi
\fi
\fi
\fi
\fi
\fi
\fi
\fi
\fi
\fi
\fi
\fi
\fi
\fi
\fi
\fi
\fi
\fi
\fi
\fi
\fi
\fi
\fi
\fi
\fi
\fi
\fi
\fi
\fi
\fi
\fi
\fi
\fi
\fi
\fi
\fi
\fi
\fi
\fi
\fi
\fi
\fi
\fi
\fi
\fi
\fi
\fi
\fi
\fi
\fi
\fi
\fi
}

\newcommand{\msgsymbnum}[2]{MSG #1.#2 (\msgsymb{#1}{#2})}


\newcommand{\sgsymb}[1]{\ifnum#1=1
$P1$\else
\ifnum#1=2
$P\bar{1}$\else
\ifnum#1=3
$P2$\else
\ifnum#1=4
$P2_1$\else
\ifnum#1=5
$C2$\else
\ifnum#1=6
$Pm$\else
\ifnum#1=7
$Pc$\else
\ifnum#1=8
$Cm$\else
\ifnum#1=9
$Cc$\else
\ifnum#1=10
$P2/m$\else
\ifnum#1=11
$P2_1/m$\else
\ifnum#1=12
$C2/m$\else
\ifnum#1=13
$P2/c$\else
\ifnum#1=14
$P2_1/c$\else
\ifnum#1=15
$C2/c$\else
\ifnum#1=16
$P222$\else
\ifnum#1=17
$P222_1$\else
\ifnum#1=18
$P2_12_12$\else
\ifnum#1=19
$P2_12_12_1$\else
\ifnum#1=20
$C222_1$\else
\ifnum#1=21
$C222$\else
\ifnum#1=22
$F222$\else
\ifnum#1=23
$I222$\else
\ifnum#1=24
$I2_12_12_1$\else
\ifnum#1=25
$Pmm2$\else
\ifnum#1=26
$Pmc2_1$\else
\ifnum#1=27
$Pcc2$\else
\ifnum#1=28
$Pma2$\else
\ifnum#1=29
$Pca2_1$\else
\ifnum#1=30
$Pnc2$\else
\ifnum#1=31
$Pmn2_1$\else
\ifnum#1=32
$Pba2$\else
\ifnum#1=33
$Pna2_1$\else
\ifnum#1=34
$Pnn2$\else
\ifnum#1=35
$Cmm2$\else
\ifnum#1=36
$Cmc2_1$\else
\ifnum#1=37
$Ccc2$\else
\ifnum#1=38
$Amm2$\else
\ifnum#1=39
$Aem2$\else
\ifnum#1=40
$Ama2$\else
\ifnum#1=41
$Aea2$\else
\ifnum#1=42
$Fmm2$\else
\ifnum#1=43
$Fdd2$\else
\ifnum#1=44
$Imm2$\else
\ifnum#1=45
$Iba2$\else
\ifnum#1=46
$Ima2$\else
\ifnum#1=47
$Pmmm$\else
\ifnum#1=48
$Pnnn$\else
\ifnum#1=49
$Pccm$\else
\ifnum#1=50
$Pban$\else
\ifnum#1=51
$Pmma$\else
\ifnum#1=52
$Pnna$\else
\ifnum#1=53
$Pmna$\else
\ifnum#1=54
$Pcca$\else
\ifnum#1=55
$Pbam$\else
\ifnum#1=56
$Pccn$\else
\ifnum#1=57
$Pbcm$\else
\ifnum#1=58
$Pnnm$\else
\ifnum#1=59
$Pmmn$\else
\ifnum#1=60
$Pbcn$\else
\ifnum#1=61
$Pbca$\else
\ifnum#1=62
$Pnma$\else
\ifnum#1=63
$Cmcm$\else
\ifnum#1=64
$Cmce$\else
\ifnum#1=65
$Cmmm$\else
\ifnum#1=66
$Cccm$\else
\ifnum#1=67
$Cmme$\else
\ifnum#1=68
$Ccce$\else
\ifnum#1=69
$Fmmm$\else
\ifnum#1=70
$Fddd$\else
\ifnum#1=71
$Immm$\else
\ifnum#1=72
$Ibam$\else
\ifnum#1=73
$Ibca$\else
\ifnum#1=74
$Imma$\else
\ifnum#1=75
$P4$\else
\ifnum#1=76
$P4_1$\else
\ifnum#1=77
$P4_2$\else
\ifnum#1=78
$P4_3$\else
\ifnum#1=79
$I4$\else
\ifnum#1=80
$I4_1$\else
\ifnum#1=81
$P\bar{4}$\else
\ifnum#1=82
$I\bar{4}$\else
\ifnum#1=83
$P4/m$\else
\ifnum#1=84
$P4_2/m$\else
\ifnum#1=85
$P4/n$\else
\ifnum#1=86
$P4_2/n$\else
\ifnum#1=87
$I4/m$\else
\ifnum#1=88
$I4_1/a$\else
\ifnum#1=89
$P422$\else
\ifnum#1=90
$P42_12$\else
\ifnum#1=91
$P4_122$\else
\ifnum#1=92
$P4_12_12$\else
\ifnum#1=93
$P4_222$\else
\ifnum#1=94
$P4_22_12$\else
\ifnum#1=95
$P4_322$\else
\ifnum#1=96
$P4_32_12$\else
\ifnum#1=97
$I422$\else
\ifnum#1=98
$I4_122$\else
\ifnum#1=99
$P4mm$\else
\ifnum#1=100
$P4bm$\else
\ifnum#1=101
$P4_2cm$\else
\ifnum#1=102
$P4_2nm$\else
\ifnum#1=103
$P4cc$\else
\ifnum#1=104
$P4nc$\else
\ifnum#1=105
$P4_2mc$\else
\ifnum#1=106
$P4_2bc$\else
\ifnum#1=107
$I4mm$\else
\ifnum#1=108
$I4cm$\else
\ifnum#1=109
$I4_1md$\else
\ifnum#1=110
$I4_1cd$\else
\ifnum#1=111
$P\bar{4}2m$\else
\ifnum#1=112
$P\bar{4}2c$\else
\ifnum#1=113
$P\bar{4}2_1m$\else
\ifnum#1=114
$P\bar{4}2_1c$\else
\ifnum#1=115
$P\bar{4}m2$\else
\ifnum#1=116
$P\bar{4}c2$\else
\ifnum#1=117
$P\bar{4}b2$\else
\ifnum#1=118
$P\bar{4}n2$\else
\ifnum#1=119
$I\bar{4}m2$\else
\ifnum#1=120
$I\bar{4}c2$\else
\ifnum#1=121
$I\bar{4}2m$\else
\ifnum#1=122
$I\bar{4}2d$\else
\ifnum#1=123
$P4/mmm$\else
\ifnum#1=124
$P4/mcc$\else
\ifnum#1=125
$P4/nbm$\else
\ifnum#1=126
$P4/nnc$\else
\ifnum#1=127
$P4/mbm$\else
\ifnum#1=128
$P4/mnc$\else
\ifnum#1=129
$P4/nmm$\else
\ifnum#1=130
$P4/ncc$\else
\ifnum#1=131
$P4_2/mmc$\else
\ifnum#1=132
$P4_2/mcm$\else
\ifnum#1=133
$P4_2/nbc$\else
\ifnum#1=134
$P4_2/nnm$\else
\ifnum#1=135
$P4_2/mbc$\else
\ifnum#1=136
$P4_2/mnm$\else
\ifnum#1=137
$P4_2/nmc$\else
\ifnum#1=138
$P4_2/ncm$\else
\ifnum#1=139
$I4/mmm$\else
\ifnum#1=140
$I4/mcm$\else
\ifnum#1=141
$I4_1/amd$\else
\ifnum#1=142
$I4_1/acd$\else
\ifnum#1=143
$P3$\else
\ifnum#1=144
$P3_1$\else
\ifnum#1=145
$P3_2$\else
\ifnum#1=146
$R3$\else
\ifnum#1=147
$P\bar{3}$\else
\ifnum#1=148
$R\bar{3}$\else
\ifnum#1=149
$P312$\else
\ifnum#1=150
$P321$\else
\ifnum#1=151
$P3_112$\else
\ifnum#1=152
$P3_121$\else
\ifnum#1=153
$P3_212$\else
\ifnum#1=154
$P3_221$\else
\ifnum#1=155
$R32$\else
\ifnum#1=156
$P3m1$\else
\ifnum#1=157
$P31m$\else
\ifnum#1=158
$P3c1$\else
\ifnum#1=159
$P31c$\else
\ifnum#1=160
$R3m$\else
\ifnum#1=161
$R3c$\else
\ifnum#1=162
$P\bar{3}1m$\else
\ifnum#1=163
$P\bar{3}1c$\else
\ifnum#1=164
$P\bar{3}m1$\else
\ifnum#1=165
$P\bar{3}c1$\else
\ifnum#1=166
$R\bar{3}m$\else
\ifnum#1=167
$R\bar{3}c$\else
\ifnum#1=168
$P6$\else
\ifnum#1=169
$P6_1$\else
\ifnum#1=170
$P6_5$\else
\ifnum#1=171
$P6_2$\else
\ifnum#1=172
$P6_4$\else
\ifnum#1=173
$P6_3$\else
\ifnum#1=174
$P\bar{6}$\else
\ifnum#1=175
$P6/m$\else
\ifnum#1=176
$P6_3/m$\else
\ifnum#1=177
$P622$\else
\ifnum#1=178
$P6_122$\else
\ifnum#1=179
$P6_522$\else
\ifnum#1=180
$P6_222$\else
\ifnum#1=181
$P6_422$\else
\ifnum#1=182
$P6_322$\else
\ifnum#1=183
$P6mm$\else
\ifnum#1=184
$P6cc$\else
\ifnum#1=185
$P6_3cm$\else
\ifnum#1=186
$P6_3mc$\else
\ifnum#1=187
$P\bar{6}m2$\else
\ifnum#1=188
$P\bar{6}c2$\else
\ifnum#1=189
$P\bar{6}2m$\else
\ifnum#1=190
$P\bar{6}2c$\else
\ifnum#1=191
$P6/mmm$\else
\ifnum#1=192
$P6/mcc$\else
\ifnum#1=193
$P6_3/mcm$\else
\ifnum#1=194
$P6_3/mmc$\else
\ifnum#1=195
$P23$\else
\ifnum#1=196
$F23$\else
\ifnum#1=197
$I23$\else
\ifnum#1=198
$P2_13$\else
\ifnum#1=199
$I2_13$\else
\ifnum#1=200
$Pm\bar{3}$\else
\ifnum#1=201
$Pn\bar{3}$\else
\ifnum#1=202
$Fm\bar{3}$\else
\ifnum#1=203
$Fd\bar{3}$\else
\ifnum#1=204
$Im\bar{3}$\else
\ifnum#1=205
$Pa\bar{3}$\else
\ifnum#1=206
$Ia\bar{3}$\else
\ifnum#1=207
$P432$\else
\ifnum#1=208
$P4_232$\else
\ifnum#1=209
$F432$\else
\ifnum#1=210
$F4_132$\else
\ifnum#1=211
$I432$\else
\ifnum#1=212
$P4_332$\else
\ifnum#1=213
$P4_132$\else
\ifnum#1=214
$I4_132$\else
\ifnum#1=215
$P\bar{4}3m$\else
\ifnum#1=216
$F\bar{4}3m$\else
\ifnum#1=217
$I\bar{4}3m$\else
\ifnum#1=218
$P\bar{4}3n$\else
\ifnum#1=219
$F\bar{4}3c$\else
\ifnum#1=220
$I\bar{4}3d$\else
\ifnum#1=221
$Pm\bar{3}m$\else
\ifnum#1=222
$Pn\bar{3}n$\else
\ifnum#1=223
$Pm\bar{3}n$\else
\ifnum#1=224
$Pn\bar{3}m$\else
\ifnum#1=225
$Fm\bar{3}m$\else
\ifnum#1=226
$Fm\bar{3}c$\else
\ifnum#1=227
$Fd\bar{3}m$\else
\ifnum#1=228
$Fd\bar{3}c$\else
\ifnum#1=229
$Im\bar{3}m$\else
\ifnum#1=230
$Ia\bar{3}d$\else
{\color{red}{Invalid SG number}}
\fi
\fi
\fi
\fi
\fi
\fi
\fi
\fi
\fi
\fi
\fi
\fi
\fi
\fi
\fi
\fi
\fi
\fi
\fi
\fi
\fi
\fi
\fi
\fi
\fi
\fi
\fi
\fi
\fi
\fi
\fi
\fi
\fi
\fi
\fi
\fi
\fi
\fi
\fi
\fi
\fi
\fi
\fi
\fi
\fi
\fi
\fi
\fi
\fi
\fi
\fi
\fi
\fi
\fi
\fi
\fi
\fi
\fi
\fi
\fi
\fi
\fi
\fi
\fi
\fi
\fi
\fi
\fi
\fi
\fi
\fi
\fi
\fi
\fi
\fi
\fi
\fi
\fi
\fi
\fi
\fi
\fi
\fi
\fi
\fi
\fi
\fi
\fi
\fi
\fi
\fi
\fi
\fi
\fi
\fi
\fi
\fi
\fi
\fi
\fi
\fi
\fi
\fi
\fi
\fi
\fi
\fi
\fi
\fi
\fi
\fi
\fi
\fi
\fi
\fi
\fi
\fi
\fi
\fi
\fi
\fi
\fi
\fi
\fi
\fi
\fi
\fi
\fi
\fi
\fi
\fi
\fi
\fi
\fi
\fi
\fi
\fi
\fi
\fi
\fi
\fi
\fi
\fi
\fi
\fi
\fi
\fi
\fi
\fi
\fi
\fi
\fi
\fi
\fi
\fi
\fi
\fi
\fi
\fi
\fi
\fi
\fi
\fi
\fi
\fi
\fi
\fi
\fi
\fi
\fi
\fi
\fi
\fi
\fi
\fi
\fi
\fi
\fi
\fi
\fi
\fi
\fi
\fi
\fi
\fi
\fi
\fi
\fi
\fi
\fi
\fi
\fi
\fi
\fi
\fi
\fi
\fi
\fi
\fi
\fi
\fi
\fi
\fi
\fi
\fi
\fi
\fi
\fi
\fi
\fi
\fi
\fi
\fi
\fi
\fi
\fi
\fi
\fi
\fi
\fi
\fi
\fi
\fi
\fi
\fi
\fi
\fi
\fi
\fi
\fi}

\newcommand{\sgsymbnum}[1]{SG #1 (\sgsymb{#1})}

\newcommand{\webBCSMAG}{\href{http://webbdcrista1.ehu.es/magndata/
}{MAGDATA}}
\newcommand{\webBCS}{\href{https://www.cryst.ehu.es/}{Bilbao Crystallographic Server}}

\newcommand{\icsdwebshort}[1]{\href{https://www.topologicalquantumchemistry.fr/\#/detail/#1}{#1}}
\newcommand{\icsdweb}[1]{\href{https://www.topologicalquantumchemistry.fr/\#/detail/#1}{ICSD #1}}
\newcommand{\icsdwebdirectlink}[2]{\href{https://www.topologicalquantumchemistry.fr/\#/detail/#1}{#2}}
\newcommand{\webNoICSD}{\href{https://www.topologicalquantumchemistry.fr/}}
\newcommand{\webTQC}{\href{https://www.topologicalquantumchemistry.org/}{Topological Quantum Chemistry website}}
\newcommand{\webMTQC}{\href{https://www.topologicalquantumchemistry.fr/magnetic}{Topological Magnetic Materials website}}
\newcommand{\webflatband}{\href{https://www.topologicalquantumchemistry.fr/flatbands/}{Materials Flatband Database website}}

\newcommand{\bcsidwebshort}[1]{\href{https://www.topologicalquantumchemistry.fr/magnetic/index.html?BCSID=#1}{#1}}
\newcommand{\bcsidweblong}[1]{\href{https://www.topologicalquantumchemistry.fr/magnetic/index.html?BCSID=#1}{BCSID #1}}
\newcommand{\flatwebdirectlink}[2]{\href{https://www.topologicalquantumchemistry.fr/flatbands/index.html?ICSD=#1}{#2}}

\newcommand{\TQCDBNbrICSDs}{73,234}
\newcommand{\TQCDBNbrUniqueMaterials}{38,298}

\newcommand{\TQCDBNbrICSDsTrivial}{34,013}
\newcommand{\TQCDBNbrICSDsTrivialPercent}{46.44\%}

\newcommand{\TQCDBNbrMaterialsTrivial}{18,133}
\newcommand{\TQCDBNbrMaterialsTrivialPercent}{47.35\%}

\newcommand{\TQCDBNbrICSDsFeOAI}{957}
\newcommand{\TQCDBNbrICSDsFeOAIPercent}{2.81\%}

\newcommand{\TQCDBNbrMaterialsFeOAI}{638}
\newcommand{\TQCDBNbrMaterialsFeOAIPercent}{3.52\%}

\newcommand{\TQCDBNbrICSDsFeOAIIndirectGap}{738}
\newcommand{\TQCDBNbrICSDsFeOAIIndirectGapPercent}{2.17\%}

\newcommand{\TQCDBNbrMaterialsFeOAIIndirectGap}{475}
\newcommand{\TQCDBNbrMaterialsFeOAIIndirectGapPercent}{2.62\%}

\newcommand{\TQCDBNbrICSDsOAI}{3,383}
\newcommand{\TQCDBNbrICSDsOAIPercent}{9.95\%}

\newcommand{\TQCDBNbrMaterialsOAI}{1,788}
\newcommand{\TQCDBNbrMaterialsOAIPercent}{9.86\%}

\newcommand{\TQCDBNbrICSDsOAIIndirectGap}{2,061}
\newcommand{\TQCDBNbrICSDsOAIIndirectGapPercent}{60.92\%}
\newcommand{\TQCDBNbrICSDsOAIIndirectGapLarge}{1,545}
\newcommand{\TQCDBNbrICSDsOAIIndirectGapLargePercent}{45.67\%}
\newcommand{\TQCDBNbrICSDsOAIOnlyDirectGap}{1,322}
\newcommand{\TQCDBNbrICSDsOAIOnlyDirectGapPercent}{39.08\%}

\newcommand{\TQCDBNbrMaterialsOAIIndirectGap}{1,096}
\newcommand{\TQCDBNbrMaterialsOAIIndirectGapPercent}{61.30\%}
\newcommand{\TQCDBNbrMaterialsOAIIndirectGapLarge}{818}
\newcommand{\TQCDBNbrMaterialsOAIIndirectGapLargePercent}{45.75\%}
\newcommand{\TQCDBNbrMaterialsOAIIndirectGapOneAtomType}{48}
\newcommand{\TQCDBNbrMaterialsOAIIndirectGapOneAtomTypePercent}{4.38\%}
\newcommand{\TQCDBNbrMaterialsOAIIndirectGapTwoAtomTypes}{281}
\newcommand{\TQCDBNbrMaterialsOAIIndirectGapTwoAtomTypesPercent}{25.64\%}

\newcommand{\TQCDBNbrMaterialsOAIOnlyDirectGap}{692}
\newcommand{\TQCDBNbrMaterialsOAIOnlyDirectGapPercent}{38.70\%}
\newcommand{\TQCDBNbrMaterialsOAIOnlyDirectGapOneAtomType}{13}
\newcommand{\TQCDBNbrMaterialsOAIOnlyDirectGapOneAtomTypePercent}{1.88\%}
\newcommand{\TQCDBNbrMaterialsOAIOnlyDirectGapTwoAtomTypes}{249}
\newcommand{\TQCDBNbrMaterialsOAIOnlyDirectGapTwoAtomTypesPercent}{35.98\%}

\newcommand{\TQCDBNbrICSDsOOAI}{121}
\newcommand{\TQCDBNbrICSDsOOAIPercent}{9.15\%}

\newcommand{\TQCDBNbrICSDsOOAIandOAI}{17}
\newcommand{\TQCDBNbrICSDsOOAIandOAIPercent}{1.29\%}

\newcommand{\TQCDBNbrMaterialsOOAI}{62}
\newcommand{\TQCDBNbrMaterialsOOAIPercent}{0.34\%}

\newcommand{\TQCDBNbrMaterialsOOAIandOAI}{8}
\newcommand{\TQCDBNbrMaterialsOOAIandOAIPercent}{0.04\%}

\newcommand{\TQCDBNbrICSDsOOAIIndirectGap}{70}
\newcommand{\TQCDBNbrICSDsOOAIIndirectGapPercent}{100.00\%}

\newcommand{\TQCDBNbrMaterialsOOAIIndirectGap}{33}
\newcommand{\TQCDBNbrMaterialsOOAIIndirectGapPercent}{0.18\%}

\newcommand{\MTQCDBNbrBCSIDs}{556}

\newcommand{\MTQCDBNbrBCSIDsWithPhaseDiagram}{372}

\newcommand{\MTQCDBNbrBCSIDsTI}{52}
\newcommand{\MTQCDBNbrBCSIDsTIPercent}{13.98\%}
\newcommand{\MTQCDBNbrBCSIDsSM}{89}
\newcommand{\MTQCDBNbrBCSIDsSMPercent}{23.92\%}
\newcommand{\MTQCDBNbrBCSIDsTrivial}{296}
\newcommand{\MTQCDBNbrBCSIDsTrivialPercent}{79.57\%}

\newcommand{\MTQCDBNbrBCSIDsMOAI}{30}
\newcommand{\MTQCDBNbrBCSIDsMOAIPercent}{8.06\%}

\title{Three-Dimensional Real Space Invariants, Obstructed Atomic Insulators and A New Principle for Active Catalytic Sites }

\author{Yuanfeng Xu}
\thanks{These authors contributed equally}
\affiliation{Max Planck Institute of Microstructure Physics, 06120 Halle, Germany}
\affiliation{Department of Physics, Princeton University, Princeton, New Jersey 08544, USA}

\author{Luis Elcoro}
\thanks{These authors contributed equally}
\affiliation{Department of Physics, University of the Basque Country UPV/EHU, Apartado 644, 48080 Bilbao, Spain}

\author{Guowei Li}
\thanks{These authors contributed equally}
\affiliation{Max Planck Institute for Chemical Physics of Solids, Dresden D-01187, Germany}
\affiliation{CAS Key Laboratory of Magnetic Materials and Devices, and Zhejiang Province Key Laboratory of Magnetic Materials and Application Technology, Ningbo Institute of Materials Technology and Engineering, Chinese Academy of Sciences, Ningbo 315201, China.}

\author{Zhi-Da Song}
\email{zhidas@princeton.edu}
\affiliation{Department of Physics, Princeton University, Princeton, New Jersey 08544, USA}

\author{Nicolas Regnault}
\affiliation{Department of Physics, Princeton University, Princeton, New Jersey 08544, USA}
\affiliation{Laboratoire de Physique de l'Ecole normale sup\'{e}rieure, ENS, Universit\'{e} PSL, CNRS, Sorbonne Universit\'{e}, Universit\'{e} Paris-Diderot, Sorbonne Paris Cit\'{e}, 75005 Paris, France}

\author{Qun Yang}
\affiliation{Max Planck Institute for Chemical Physics of Solids, Dresden D-01187, Germany}

\author{Yan Sun}
\affiliation{Max Planck Institute for Chemical Physics of Solids, Dresden D-01187, Germany}

\author{Stuart Parkin}
\affiliation{Max Planck Institute of Microstructure Physics, 06120 Halle, Germany}

\author{Claudia Felser}
\email{Claudia.Felser@cpfs.mpg.de}
\affiliation{Max Planck Institute for Chemical Physics of Solids, Dresden D-01187, Germany}

\author{B. Andrei Bernevig }
\email{bernevig@princeton.edu}
\affiliation{Department of Physics, Princeton University, Princeton, New Jersey 08544, USA}
\affiliation{Donostia International Physics Center, P. Manuel de Lardizabal 4, 20018 Donostia-San Sebastian, Spain}
\affiliation{IKERBASQUE, Basque Foundation for Science, Bilbao, Spain}

\begin{abstract}
Topologically trivial insulators come in two kinds: atomic, where the Wannier charge centers (WCCs) are localized on the atoms, and obstructed atomic, where the WCCs are located away from the atoms. The latter, which can exhibit interesting surface states and possibly have much larger band gaps than the topological insulators, have so far not been classified in three-dimensional (3D) crystalline materials. In this paper, we developed the 3D real space invariants (RSIs) for the 1,651 Shubnikov space groups with the spin-orbit coupling and provide the full classification of 3D obstructed atomic insulators (OAIs) by the RSIs. We then apply the theory to the entire database of materials on the \webTQC\ and \webMTQC, obtaining all the paramagnetic and magnetic OAIs so far existing in nature. We find that, out of the \TQCDBNbrICSDsTrivial\ paramagnetic and \MTQCDBNbrBCSIDsTrivial\ magnetic  topologically trivial insulators, there are \TQCDBNbrICSDsOAI\ paramagnetic and \MTQCDBNbrBCSIDsMOAI\ magnetic OAIs. All of them present a filling anomaly under certain open-boundary conditions and exhibit obstructed surface states (OSSs). We then derive the Miller indices of the cleavage planes which show the OSSs for all the OAIs and pick some of the best examples with large band gap to showcase their OSSs. We further refine the atomic insulator concept to obtain the orbital-selected OAIs (OOAIs), where the WCC of the system is located at a Wyckoff position occupied by an atom but forms a symmetric representation (orbital) that does not belong to the outer-shell electrons of the given atom. In such a way, we obtain a further \TQCDBNbrICSDsOOAI\ OOAIs. Furthermore, we analyze the catalytic properties of one of the OAIs in a ``proof of principle'' experiment. The surface of a high-performance heterogeneous catalyst is characterized by high stability, good electrical conductivity, and high charge carrier density near the Fermi level. These are also characteristic properties of the OSSs. By using a high-quality, single crystal of 2H-\ce{MoS2}, that has well-defined surfaces, as a hydrogen evolution catalyst, we directly proved that the catalytic activities arise from the surfaces with OSSs, which is consistent with previous results based on electronic-structure calculations. Additional potential applications of the 3D RSIs and OAIs in, for example, electrochemistry, asymmetric catalysis, superconductivity and Josephson diode will be discussed.
\end{abstract}

\maketitle

\section{Introduction}
Two-dimensional (2D) metallic states on the surface or at the interface of solid state materials are key to the physical and chemical properties of many materials that enable important technologies including superconductivity \cite{ohtomo2004high,reyren2007superconducting}, catalysis and electrochemistry \cite{sivula2016semiconducting}. As the nano-structuring of materials has evolved over the recent years, the surface to volume ratio has increased, making the surfaces even more critical to the properties of nano-systems. 
The symmetry-protected topological materials are known to have gapless states on the boundary of lower dimension, which are evolving from the non-trivial topology of the bulk states and hence are robust when special symmetries persist \cite{TKNN1982,bernevig2006quantum,PhysRevLett.95.146802}.  
For example, in a 3D topological insulator which is protected by the time reversal symmetry (TRS), the topological surfaces states residing in the bulk insulating gap arise at all crystal facets and are robust in the presence of TRS \cite{zhang2009topological,chen2009experimental,xia2009observation}. Another example is the 3D Weyl semi-metal, which has Fermi arc states on the surface if the translation symmetry prohibiting hybridization between Weyl nodes is respected \cite{wang2012dirac,wang2013three,weng2015weyl,huang2015weyl}. Whether a material is topologically nontrivial can be efficiently diagnosed via the general theory of Topological
quantum chemistry (TQC) \cite{bradlyn_topological_2017, MTQC} or the equivalent symmetry indicators \cite{po_symmetry-based_2017,watanabe2018structure,song_quantitative_2018,SlagerSymmetry} and the complete catalogues of topological materials were obtained in several works \cite{vergniory_complete_2019,zhang2019catalogue,tang2019comprehensive,xu2020high,Vergniory2021}. 
However, as the bulk band gaps of the topological materials are always small (in general, smaller than 0.3 eV), the surface states of topological materials have not been widely used in practical applications. In the present work, using TQC theory,
we explore the intrinsic electronic surface states of all the topologically trivial inorganic insulating compounds in the Inorganic Crystal Structure Database (ICSD) \cite{ICSD}, whose bulk band gaps are likely to be much larger than those of topological insulators. 

In the TQC theory \cite{bradlyn_topological_2017,MTQC}, as briefly reviewed in Appendix~\ref{app:concepts}, the symmetry eigenvalues at high symmetry momenta \cite{zak1980symmetry,zak1981band} of topologically trivial insulators can be expressed as an integer linear combination of elementary band representations (EBRs) (\ie the basis of atomic limits) \cite{zak1982band} with non-negative coefficients in terms of symmetry eigenvalues. The EBRs are the induced representations into the space or magnetic group of the crystal from the irreducible representations (irreps) of the atomic orbitals sitting at the (maximal) high symmetry Wyckoff positions. We refer to the topology of this kind of symmetry-indicated topologically trivial insulator to as linear combination of EBRs (LCEBR) \cite{vergniory_complete_2019,xu2020high,Vergniory2021}. We emphasize that an insulator identified as LCEBR can also be a topological insulator or semimetal that \emph{cannot} be diagnosed through symmetry eigenvalues at the high-symmetry momenta. For example, the TRS-protected topological insulators of a trivial space group $\sgsymb{1}$ can not be identified by symmetry eigenvalues. For a LCEBR-type insulator, its symmetry eigenvalues of the occupied bands at high symmetry momenta are consistant with a band representation (BR) induced from irreps at either occupied or empty sites. If the BR cannot be induced \emph{only} from orbitals located at the atom occupied sites, there must be a subset of Wannier functions exponentially localized at some empty Wyckoff positions. In this case, we refer to the BR as an obstructed atomic insulator (OAI) \cite{bradlyn_topological_2017,MTQC,song2020,xu2021filling} and an empty position where the Wannier function localize to as an obstructed Wannier charge center (OWCC). 
When all the local orbitals in the decomposition can be located at the occupied Wyckoff positions, the BR can still be further diagnosed as the most trivial atomic insulator (AI) (altough it can be of further types, such as non-Compact atomic insulator \cite{schindler2021non}). However, it may happen that, \emph{at least} one irrep at an occupied Wyckoff position does not correspond to the valence orbitals of the specific atom that occupies the site. We refer to these materials as orbital-selected obstructed atomic insulator (OOAI). As the OWCC of OAIs is away from the atom sites, it is possible to have a finite-size crystal with the OWCCs on the boundary and preserving the crystal symmetry, which exhibits the filling anomaly \cite{song_d-2-dimensional_2017,PhysRevB.95.035421,benalcazar2019quantization,schindler2019fractional} and hence gives rise to metallic obstructed surface states (OSSs) or hinge states. Unlike OAIs, the obstructed orbitals of OOAIs are centered on the top of atoms and there has no filling anomaly.
(See Appendix~\ref{app:concepts} for more details about the concepts of TQC, OAIs, OOAIs and Appendix~\ref{app:filling_anomaly} for filling anomaly of OAIs.)

Due to the gauge choice of Wannierization, for topologically trivial bands the symmetric Wannier functions are generally not unique.
For example, in a 1D system with inversion symmetry if the occupied bands form a pair of Wannier functions with even ($|+\rangle$) and odd ($|-\rangle$) parities at $x=0$, then one can recombine them as two Wannier functions at $\pm x$ positions as $|x\rangle = ( |+\rangle +  |-\rangle)/\sqrt2$ and $|-x\rangle = ( |+\rangle - |-\rangle)/\sqrt2$, respectively. 
In order to diagnose the OAI states one in principle needs to enumerate all the possible Wannier function realizations (BR decompositions) to guarantee that in every realization some empty sites are occupied by Wannier functions.
In this work we make use of an efficient theoretical tool to simplify this analysis: the real space invariants (RSIs) which were initially developed for 2D wallpaper groups in Ref.~\cite{song2020}.
RSIs are local invariants defined at Wyckoff positions with nontrivial site-symmetry group whose nonzero values imply unavoidable Wannier functions at the corresponding positions.
In the 1D example we can define the RSI at $x=0$ as $\delta=m_+ - m_-$ with $m_\pm$ being the number of Wannier functions with parity $\pm1$ at $x=0$.
In the above example with two Wannier functions $|+\rangle$ and $|-\rangle$, the RSI index $\delta=0$ and the Wannier functions at $x=0$ can be moved away from this site. 
However, if $\delta\neq 0$ (for example, there is only a single even Wannier function at $x=0$), one cannot remove all the Wannier functions from this site without breaking the inversion symmetry. 
For generic space groups the RSIs can be obtained by the induction-subduction process between irreps in group-subgroup pairs (detailed in Appendix~\ref{app:rsi3D}). 
Another advantage of RSIs is that they can be directly calculated from the momentum space irreps formed by the band structure.
This allows us to diagnose OAIs by applying RSI formulae to the irreps of the bands without explicitly decomposing the bands into BRs.

In the present work, we first develop a \emph{general} framework to obtain the 3D RSIs in all the 1,651 Shubnikov space groups (SSGs) with spin-orbit coupling (SOC), including the 230 paramagnetic space groups (SGs) and the 1,421 magnetic space groups (MSGs), and provide the expressions of all the RSIs in the \webBCS\ (BCS). 
Compared with the RSIs in 2D wallpaper groups \cite{song2020}, the RSIs in 3D SSGs could be of $Z_4$-type and composite, which are defined at multiple Wyckoff positions; moreover, the 3D RSIs are applicable to the 3D stoichiometric materials in the topological (magnetic) materials databases~\cite{vergniory_complete_2019,zhang2019catalogue,tang2019comprehensive,xu2020high,Vergniory2021}.
By applying the RSIs to all the (paramagnetic and magnetic) topologically trivial insulators on the respective \webTQC\ (TQCDB) and \webMTQC\ (MTQCDB), we perform a high-throughput search for paramagnetic OAIs, OOAIs and the magnetic OAIs (mOAIs). For each OAI or mOAI, we also provide its cleavage planes that exhibit filling-anomaly OSSs. All the results obtained from the high-throughput calculations can be found on the TQCDB (or \webflatband) for OAIs and OOAIs and MTQCDB for mOAIs. The surface states on OAIs and mOAIs are calculated and analyzed in several material examples. 
Finally, we present a ``proof of principle'' experimental explaination of the application of OAIs and their OSSs in catalysis. When used as heterogeneous catalysts, we find that the OSSs of OAIs fulfill the essential factors of catalytic active sites, and could explain their activity origins. We confirm this methodology with 2H-\ce{MoS2}, one of the most studied and promising catalysts for catalytic reactions \cite{Jaramillo100,kibsgaard2012engineering}. As a hydrogen evolution reaction catalyst, we directly observe the hydrogen production process at the edge surfaces, consistent with our theoretical prediction. In Ref.~\cite{Li2021catalysts} we provide a full analysis of the catalytic properties of many new compounds predicted in this way.

\section{RSI theory in 3D}

The RSI indices \cite{song2020} are sets of indicators that show the possibility (or not) of moving charge centers (atomic orbitals) along different Wyckoff positions in a crystal system in an adiabatic process, in which the symmetry group does not change. In the first step, it is possible to assign a set of \emph{local} RSIs to every Wyckoff position in a SSG. The calculation relies on the group-subgroup relations (see Appendix~\ref{RSI:theory}) between the site-symmetry groups of the given Wyckoff position and each of the Wyckoff positions of lower symmetry connected to it. These local RSIs are thus defined in the real space (\ie direct space). However, in a given material, it is not easy to obtain the actual positions of the charge centers and their symmetry properties (orbitals). Usually, the {\it ab-initio} calculations give access to the extended states in the reciprocal space which allows the determination of the symmetry properties (irreps) of Bloch states at every {\it k}-point in the Brillouin zone. Therefore, it is necessary to \emph{translate} the definitions of the local RSIs in real space to the reciprocal space in order to make possible the calculation of the RSIs through {\it ab initio} calculations. 

In this section we extend the definition of RSIs from the 2D wallpaper groups \cite{song2020} to the 1,421 MSGs and the 230 non-magnetic SGs in 3D, for materials where the spin-orbit coupling is relevant. As the procedure to determine the indices is exactly the same in magnetic and non-magnetic groups, in this work we will refer, in general, to the 1,651 SSGs, being the indices of non-magnetic groups the calculated indices of the Type-II SSGs.
We will first develop the general procedure to calculate the local RSIs at a Wyckoff position which are given as combinations of irreps in direct space (irreps of the site-symmetry group). 
Then, we define the RSIs of a SSG which are, in general,
linear combinations of the local RSIs defined at all Wyckoff 
positions of the SSG, but expressed in terms of irreps in the momentum space. More details about the method are given in Appendix~\ref{app:rsi3D} which also includes a detailed calculation of the RSIs of SSG $C_cmma$ (N. 67.508) both in direct and momentum space.

\subsection{Local RSI indices}

In a SSG $G$ we consider a maximal Wyckoff position $Q$ of site-symmetry group $G^Q\subset G$ and another Wyckoff position $q$ of lower site-symmetry group $H^q$ connected to $Q$. Let us consider that $H^q$ is a maximal subgroup of $G^Q$. An irrep $\rho^i_{H^q}$ of $H^q$ induces a representation into $G^Q$ which is, in general, reducible, \ie
\begin{equation}
\label{RSI:induction}
\rho_{H^q}^i\uparrow G^Q=\bigoplus_{j=1}^{n_{G^Q}}m_{ji}\rho_{G^Q}^j
\end{equation}
where \{\rhQ[j], $j=1,2,...,n_{G^Q}$\} are the irreps of $G^Q$, $n_{G^Q}$ is the number of irreps of $G^Q$ and $m_{ji}$ are non-negative-integer coefficients. These numbers give the multiplicity of the irreps \rhQ[j] in the decomposition in Eq.~(\ref{RSI:induction}). This induction relation means that a single orbital that is located at $q$ and that transforms under the irrep $\rho_{H^q}^i$ can be adiabatically moved to $Q$, where the formed set of charge centers transform under $\rho_{H^q}^i\uparrow G^Q$. 
In the opposite process, the set of irreps on the right side of Eq.~(\ref{RSI:induction}) centered at $Q$ can be moved away from $Q$ along the Wyckoff position $q$ giving a center of charges that transform under $\rho_{H^q}^i$. 
Using the group-subgroup relations for each Wyckoff position $Q$, we can obtain all the induction relations in Eq.~(\ref{RSI:induction}) from all the irreps at the Wyckoff positions of lower symmetry connected to $Q$. By collecting all the induction processes at $Q$, we construct a $n_{G^Q}\times N_H$-dimentional \emph{induction matrix} $C_{G^Q}$ where the integer $m_{ji}$ represents the multiplicity of $\rho_{G^Q}^j$ in Eq.~(\ref{RSI:induction}), being $i$ the label of one of the irreps of one of the $H^q$ groups. $N_H$  is the total number of irreps in all these $H^q$ groups.

Once the $C_{G^Q}$ matrix has been constructed, we now analyze the possibility of moving a set of orbitals located initially at $Q$ along the Wyckoff positions of lower symmetry $q$. Let us thus consider a set of orbitals at $Q$ that transform under a representation \rhQ of $G^Q$. In general this representation can be expressed as a direct sum of irreps of $G^Q$,
\begin{equation}
\rho_{G^Q}=\bigoplus_{i=1}^{n_{G^Q}}m({\rho_{G^Q}^i})\rho_{G^Q}^i
\end{equation}
with non-negative integer multiplicities $m({\rho_{G^Q}^i})$. The set of orbitals \rhQ can thus be represented by a $n_{G^Q}$-dimensional vector $p$, known as the symmetry-data-vector \cite{bradlyn_topological_2017,MTQC,po_symmetry-based_2017,SlagerSymmetry} , with components $m({\rho_{G^Q}^i})$. The set of orbitals represented by $p$ can all be moved away from the Wyckoff position $Q$ to the set of Wyckoff positions $\{q_k\}$ connected to it if there exists a $N_H$-dimensional vector $X$ of integer components ($x^i$) such that the following equation is fulfilled,
\begin{equation}
\label{RSI:maineq}
C_{G^Q}\cdot X=p
\end{equation}
which is equivalent to 
\begin{equation}
\rho_{G^Q}=\sum_{i=1}^{N_H} x^i\rho^{i}_{H_q}\uparrow G^Q.
\end{equation}
If a solution of Eq.~(\ref{RSI:maineq}) exists, in general it is not unique. Different solutions give different ways to distribute the original charge centers at $Q$ into charge centers at $\{q_k\}$. If there is no solution to Eq.~(\ref{RSI:maineq}), not all the orbitals in $Q$ can be moved away. Therefore we can state the following result: given a set of orbitals at a maximal Wyckoff position $Q$ expressed by a symmetry-data-vector $p$, if Eq.~(\ref{RSI:maineq}) has no solution (\ie no vector $X$ with integer components exists) and the Wyckoff position $Q$ is empty, the set of orbitals are obstructed and at least one of them is pinned at $Q$.

It is important to point out that the solution of the Eq.~(\ref{RSI:maineq}) can contain negative components of $X$. If all solutions to Eq.~(\ref{RSI:maineq}) have \emph{at least} one negative-integer component, the corresponding BR is not inducible at the related Wyckoff position either. We leave the identification of this case for future works \cite{fragileOAI}.
In Appendix~\ref{sub:exchange}, we also provide the derivation of the induction matrices of non-maximal Wyckoff positions, which is slightly different from the maximal Wyckoff positions.

An easy way to analyze under which conditions has Eq.~(\ref{RSI:maineq}) a solution is by performing the Smith decomposition of $C_{G^Q}$ \cite{song2020}.
As detailed in the Appendix~\ref{calcRSI}, using this method we have calculated all the local RSI indices in all the 1,651 SSGs. We find that there are only $\mathbb{Z}$ and $\mathbb{Z}_2$ type RSI indices at any Wyckoff position in any of the 1,651 SSGs. The complete lists of RSI indices in the SSGs have been added to the BCS tool  \href{http://www.cryst.ehu.es/cryst/MagRSI}{RSImag}. The RSI indices of the 230 space groups have been included in the tool \href{http://www.cryst.ehu.es/cryst/RSI}{RSIsg}.

\subsection{RSI indices of SSGs}
The specific orbitals located at different Wyckoff positions in a given material band structure are usually not easily accessible. Therefore, it is necessary to determine a set of RSIs of a SSG based on the local RSIs. The set of RSI indices of a SSG are defined in momentum space, where information about the symmetry of the Bloch states (in terms of irreps of the little group of every k-vector) is accessible by \emph{ab initio} calculations.

In the (M)TQC theory \cite{zak1982band,Bacry1988,bradlyn_topological_2017,vergniory_graph_2017,elcoro_double_2017,SlagerSymmetry,song_quantitative_2018,cano_building_2018,MTQC}  and other alternative methods \cite{watanabe2018structure,peng2021}, each orbital at a given Wyckoff position induces an extended state in the momentum space: an electronic band defined in the whole Brillouin zone. If the local orbital transforms as an irrep $\rho_{G^Q}$ of the site-symmetry group $G^Q$ of a Wyckoff position $Q$, the electronic band transforms as the representation induced by $\rho_{G^Q}$ into the whole SSG, $\rho_{G^Q}\uparrow G$, known as band representation (BR) characterized in the momentum space by a symmetry-data-vector $B$ whose components are the identified irreps through \emph{ab initio} calculations at each maximal $k$-vec. (See Appendix~\ref{app:momentum} for more details).

As stated in the introduction, a compound tagged as LCEBR is compatible with a topologically trivial insulator and its symmetry-data-vector can be written as a linear combination of BRs induced from located orbitals, although the combination is not unique. To determine all possible ways to decompose the symmetry-data-vector into linear combinations of BRs, it can be assumed that there are sets of orbitals pinned at Wyckoff positions that represent points, lines and planes, and a set of orbitals at the general position (GP) (where at this last position no local RSI is defined). The orbitals at the general position transform under the unique irrep $\rho_1^{GP}$. If $\delta_i^{q_j}$ represents the $i^{\textrm{th}}$ RSI index at the $q_j$ Wyckoff position, we can define a basis of orbitals (indexed by $k$) at $q_j$ as,
\begin{equation}
	\label{comb:orb}
	\Tilde{\rho}_{kj}=\sum_n m_k(\rho_n^{q_j})\rho_n^{q_j}
\end{equation}
with $m_k(\rho_n^{q_j})$ any integer numbers such that the values of local RSIs at $q_j$ are,
\begin{equation}
	\delta_i^{q_j}(\Tilde{\rho}_{kj})=\left\{\begin{array}{ll}
			1&k=i\\
			0&k\neq i	
		\end{array}\right.
	\end{equation}
The set of orbitals in Eq.~(\ref{comb:orb}) can be considered as a basis of pinned orbitals at $q_j$, once all the orbitals that can be moved away from $q_j$ have been removed from that position. We then construct a $n_{\rho}\times n_d$-dimensional $BR$ matrix whose first $n_d-1$ columns are the symmetry-data-vectors in momentum space of the basis orbitals $\rho_{kj}$. The last column is the symmetry-data-vector of the irrep of the general position $\rho_1^{GP}$. The number of rows $n_{\rho}$ of $BR$ is the number of irreps at the maximal k-vectors in reciprocal space. Any symmetry-data-vector $B$ of a LCEBR compound can thus be written as,
\begin{equation}
	\label{eq:BR}
	BR\cdot p=B
\end{equation}
where the components of $p$ are the multiplicities (integers) of the orbitals in Eq.~(\ref{comb:orb}) and of $\rho_1^{GP}$. In general, the solution $p$ to Eq.~(\ref{eq:BR}) is not unique. As detailed in the Appendix~\ref{app:momentum}, using again the Smith decomposition of $BR$, we can obtain the solutions $p$ of Eq.~(\ref{eq:BR}) whose element $p_i$ corresponds to the $i^{th}$ column of the $BR$ matrix and thus to a local RSI index of a Wyckoff position.

In general, some components of the solution $p$ might depend on arbitrary parameters and the corresponding local RSI indices are not well defined. However, it is also possible to define linear combinations of these $p_i$ indices such that the combination is independent on the arbitrary parameters. In this case, the corresponding combined indices are well defined and are referred to as the composite RSI indices of a SSG (See Appendix~\ref{app:momentum} for details about composite RSIs).
A non-zero value of a composite RSI index for a given symmetry-data-vector $B$ means that, at least, one orbital must be pinned at one of the involved Wyckoff positions in the definition of the composite RSI index, but not at an specific position. 

We have calculated all the RSI indices for the 1,651 SSGs and checked that there are only indices of the types $\mathbb{Z}$, $\mathbb{Z}_2$ and $\mathbb{Z}_4$. In the calculations we have removed indices that are redundant and restrict the final list of RSI indices to those that are independent. The results have been implemented in the BCS tools  \href{http://www.cryst.ehu.es/cryst/RSI}{RSIsg} and \href{http://www.cryst.ehu.es/cryst/MagRSI}{RSImag} for SGs and MSGs, respectively. We have denoted the $\mathbb{Z}$, $\mathbb{Z}_2$ and $\mathbb{Z}_4$ types of indices as $\delta_i(WP)$, $\eta_i(WP)$ and $\zeta_i(WP)$, respectively, where $WP$ is the list of Wyckoff positions involved in the definition of the (in general composite) RSI index.

\section{Implementation of RSIs to OAIs and OOAIs}
In the above section, we have generalized the RSI indices to 3D crystalline structures of the 1,651 SSGs with SOC and derived and tabulated their expression in terms of the multiplicities of the momentum-space irreps. For a given band structure, its RSIs can be obtained by substituting its symmetry-data-vector into the RSIs' formula.
Using the RSI indices, we are able to diagnose the OAIs and OOAIs from the LCEBR-type topologically trivial insulators.
For a given topologically trivial insulator, if its BR has a non-zero-integer (in general, composite) RSI which is defined at a set of Wyckoff positions $\{\alpha\}$, the BR is induced by a set of orbitals that include at least one orbital in one of the positions $\{\alpha\}$.
If all the positions $\{\alpha\}$ and all the positions of higher symmetry connected to them are empty, we refer the topologically trivial insulator to as an OAI and the Wyckoff positions $\{\alpha\}$ to as OWCCs. 
Although the band topology in terms of EBRs of these systems are trivial, some Wannier charge centers of the occupied bands are out of the atoms. 
However, even when some Wyckoff positions in the subset $\{\alpha\}$ are occupied, we can distinguish two cases: if the irreps that enter into the definition of a non-zero-integer RSI do not correspond to orbitals in the outer-shell of the specific atoms  sitting at $\{\alpha\}$, this RSI indicates an orbital-selected OAI (OOAI); otherwise, this RSI indicates the most trivial atomic insulator (AI). (See Appendix~\ref{app:highthroughput} for step-by-step methods of diagnosing OAIs and OOAIs.) 

\begin{figure}
\centering\includegraphics[width=3.4in]{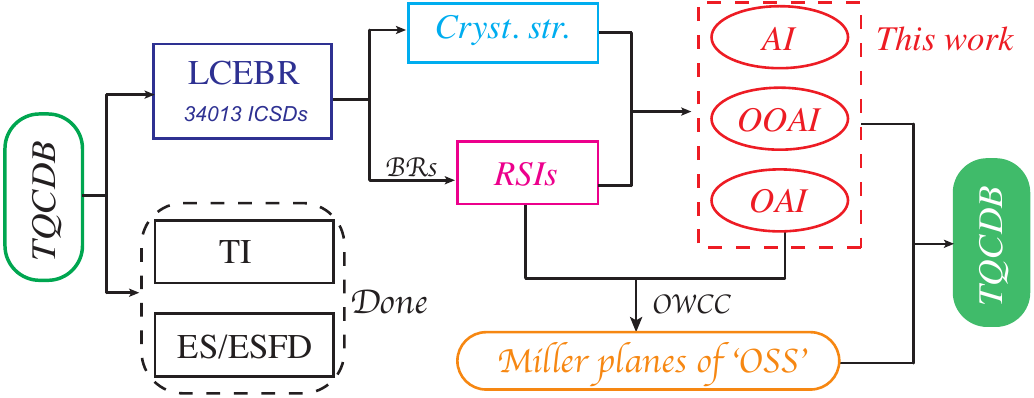}
\caption{Workflow of the high-throughput search for OAIs and OOAIs from the TQCDB. On the TQCDB, the catalogues of topological materials, including topological insulators (TI) and enforced semimetals (ES/ESFD), were completed in Refs.~\cite{vergniory_complete_2019,Vergniory2021}. In this work, we use as input the \TQCDBNbrICSDsTrivial\ (\TQCDBNbrMaterialsTrivial) ICSD entries (unique materials) of topologically trivial insulators, namely the LCEBRs, in the high-throughput search. By calculating the RSIs and the occupied Wyckoff positions of each ICSD entry, the material is diagnosed as an OAI or OOAI.  Finally, Miller indices of the cleavage planes which have OSSs are identified for each OAI.}\label{fig:fig1}
\end{figure}

Using the above methods, based on the TQCDB \cite{vergniory_complete_2019,Vergniory2021}, we have performed a high-throughput calculations for stoichiometric OAIs and OOAIs. 
As schematically shown in Fig.~\ref{fig:fig1}, in the first step, we filter out \TQCDBNbrICSDsTrivial\ (\TQCDBNbrMaterialsTrivial) ICSD entries (unique materials) of topologically trivial insulators from the TQCDB. Unique materials are defined as ICSD entries having the same chemical formula, space group and topology at the Fermi energy.
In the second step, using the \textbf{$Phonopy~package$} \cite{phonopy}, we have obtained the occupied Wyckoff positions and the corresponding atoms of each material. 
In the third step, by substituting  the symmetry-data-vector, as provided on the \webTQC, into the formula of RSIs, we calculate all the 3D RSIs for each material. 
In the fourth step, for each material with non-zero-integer RSIs, we check if the RSIs indicate an OAI or OOAI. Finally, if a material is diagnosed as an OAI, we identify its Miller indices of cleavage planes that exhibit filling anomaly and metallic OSSs. (See Appendix~\ref{app:filling_anomaly} for more details about the cleavage planes of the OSSs.)

As detailed in Appendix~\ref{app:subhighthroughput}, \TQCDBNbrICSDsOAI\ (\TQCDBNbrMaterialsOAI) and \TQCDBNbrICSDsOOAI\ (\TQCDBNbrMaterialsOOAI) ICSD entries (unique materials) are diagnosed as OAIs and OOAIs, respectively.
Using the same method of diagnosing OAIs, we have performed the high-throughput search for mOAIs from the MTQCDB\cite{xu2020high}. By scanning the topological phase diagram as a function of Hubbard-$U$ of each magnetic material, we find \MTQCDBNbrBCSIDsTrivial\ magnetic materials that were classified as LCEBR-type topologically trivial insulators in their phase diagram, among which \MTQCDBNbrBCSIDsMOAI\ materials are diagnosed as mOAIs. The detailed statistics of (m)OAI materials in each (M)SG are provided in the Appendix~\ref{app:highthroughput}. 

In Appendix~\ref{app:3DOAI}, we have tabulated all the OAIs, OOAIs and mOAIs found in the high-throughput searches. 
For each material, we provide its crystal structure, electronic band gaps, the RSIs indicating an OAI, OOAI or mOAI and the Miller indices of cleavage planes that have OSSs. Moreover, to facilitate the potential application of OAIs to asymmetric catalysis \cite{noyori2002asymmetric,list2000proline,ahrendt2000new}, we also indicate the OAIs in chiral space groups and the cleavage planes with 2D chiral plane groups.
All the results can also be found on the \webTQC\ and \webflatband\ (for the OAIs and OOAIs) and \webMTQC\ (for the mOAIs).

\section{Material examples}
Among the \TQCDBNbrICSDsOAI\ OAIs found in the high-throughput search, \TQCDBNbrICSDsOAIIndirectGap\ have a non-zero indirect band gap and \TQCDBNbrICSDsOAIIndirectGapLarge\ have an indirect band gap larger than $0.5eV$ along all the high-symmetry lines. Thus, these OAIs are ideal candidates for Angle-resolved photoemission spectroscopy (ARPES) experiments to prove the OSSs.
In this section, we present the prototypical material with an indirect gap for OAI, OOAI and mOAI. 
The three large-gap materials are, \ce{InS} [\icsdweb{15931}, SG 58 (\sgsymb{58})] for the OAI, \ce{La2Ti2O7} [\icsdweb{164027}, SG 227 (\sgsymb{227})] for the OOAI and \ce{Mn3Si2Te6} [\bcsidweblong{0.176}, MSG 15.89 (\msgsymb{15}{89})] for the mOAI. For each case, we compute and analyze their RSIs. For the OAI material, \ce{InS}, and the mOAI material, \ce{Mn3Si2Te6}, we also perform the surface state calculations and showcase their OSSs. We refer to Appendix~\ref{app:filling_anomaly} for the RSI and OSSs calculations of another eight OAIs including seven paramagnetic OAIs, \ce{NbBr2O} [\icsdweb{416669}), SG 5 (\sgsymb{5})], \ce{CaIn2P2} [\icsdweb{260562}, SG 194 (\sgsymb{194})], \ce{InSe} [\icsdweb{185172}, SG 194 (\sgsymb{194})], \ce{Hg2IO} [\icsdweb{33275}, SG 15 (\sgsymb{15})], \ce{PtSbSi} [\icsdweb{413194}, SG 61 (\sgsymb{61})], \ce{Nb3Br8} [\icsdweb{25766}, SG 166 (\sgsymb{166})], \ce{B12} [\icsdweb{431636}, SG 166 (\sgsymb{166})] and a magnetic OAI \ce{CsFe2Se3} [\bcsidweblong{1.26}, MSG 14.82 (\msgsymb{14}{82})].

\subsection{OAI material: InS}
\begin{figure*}
\centering\includegraphics[width=6.8in]{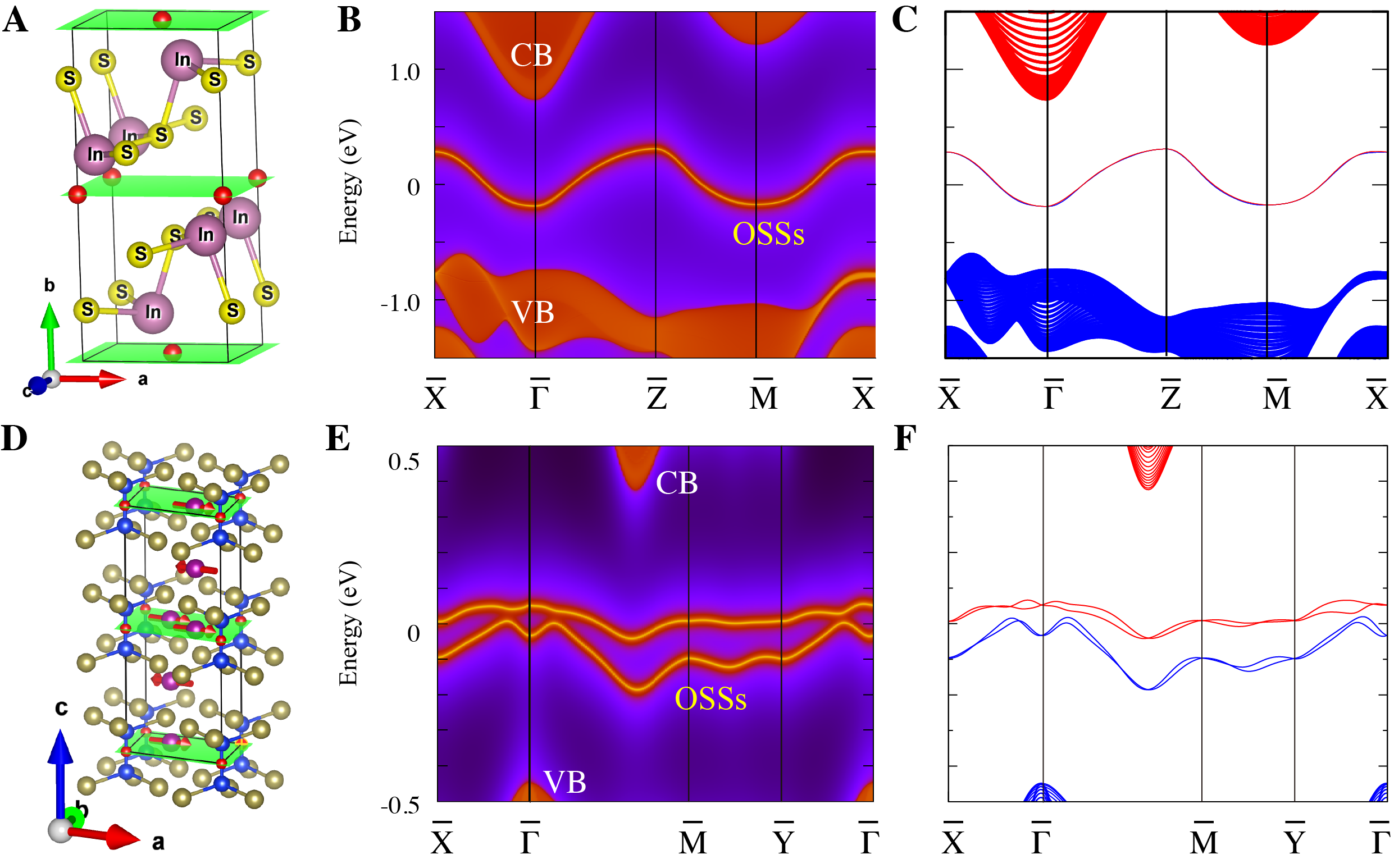}
\caption{Surface states calculations of the OAI InS and the mOAI \ce{Mn3Si2Te6}. (A) and (D) are the crystal structures of InS and \ce{Mn3Si2Te6}, where the red balls are the positions of OWCCs. The green planes cutting through the OWCCs are the cleavage planes hosting surface states. (B) Surface states of InS with a semi-infinite crystal structure along the (010) direction and with the cleavage plane as defined in (A). The surface states are highlighted in between the gap of bulk states. (C) Surface states of InS with a finite slab structure along the (010) direction. As the top and bottom cleavage planes are related by inversion symmetry, as defined in (A), each band of the surface states is two-fold degenerate but localized on different planes. Due to the time reversal symmetry, the four bands of surface states are degenerate at the four time reversal invariant momentum points. (E) and (F) are the same with (B) and (C) but for the (001) surface of \ce{Mn3Si2Te6}.}\label{fig:fig2}
\end{figure*}

As shown in Fig.~\ref{fig:fig2}(A), in the crystal lattice of \ce{InS}, both In and S atoms occupy the Wyckoff position $4g$ of SG 58 (\sgsymb{58}). The band structure of \ce{InS} is diagnosed as a topologically trivial insulator of an indirect band gap (0.675 eV) on the TQCDB \cite{vergniory_complete_2019,Vergniory2021}.
Using the symmetry-data-vector of its valence bands, we calculate all the RSI indices of SG 58 (See Appendix~\ref{app:rsical}). We find that there is a non-zero-integer $Z$-type RSI index, namely $\delta_4(d)=-1$, which is defined at the empty sites of Wyckoff position $2d$. Hence \ce{InS} is an OAI.
As the $2d$ positions (red spheres) are empty and non-coplanar with any atom on the (010) plane (Fig.~\ref{fig:fig2}(A)), the cleavage plane of Miller index (010) could cut through the OWCCs and the finite-size slab structure preserves the bulk crystal symmetry. In Fig.~\ref{fig:fig2}(B) and (c), using the WannierTools package \cite{WU2017}, we have calculated the (010) surface states of a semi-infinite and finite slab structures, where the filling-anomaly OSSs are localized in the gap between the valence and conduction bands.

\subsection{mOAI material: Mn$_3$Si$_2$Te$_6$}

As shown in Fig.~\ref{fig:fig2}(D), \ce{Mn3Si2Te6} with~\bcsidweblong{0.176} and MSG 15.89 (\msgsymb{15}{89}) is an anti-ferromagnetic material of N\'{e}el temperature $T_N=78K$. The magnetic Mn atoms occupy the two Wyckoff positions $4e$ and $8f$. Si atoms occupy the Wyckoff position $8f$. Fe atoms occupy three non-equivalent $4f$ positions. From the \emph{ab initio} calculations in Ref.~\cite{xu2020high} and on the \webMTQC\, \ce{Mn3Si2Te6} is a magnetic topologically trivial insulator and the topology does not change even with different Hubbard-$U$ values included. 
By substituting the symmetry-data-vector into the formula of RSIs of MSG 15.89, we found \ce{Mn3Si2Te6} is a mOAI indicated by a non-zero RSI $\delta_1(a)=2$ at the empty Wyckoff position $4a$ (See Appendix~\ref{app:rsical}).

The OWCCs at $4a$ are indicated by the red spheres in Fig.~\ref{fig:fig2}(D). As there are no atoms co-planar with the $4a$ position on the $(001)$ plane, it is possible to have a cleavage plane of Miller index $(001)$ cutting through the OWCCs and exhibit the OSSs. In Fig.~\ref{fig:fig2}(E) and (F), we have calculated the $(001)$ surface states of a semi-infinite and finite slab structures, where the OSSs are localized in the gap between the valence and conduction bands. Compared with the paramagnetic OAIs, the surface bands in the mOAI of \ce{Mn3Si2Te6} are spin-polarized with in-plane ferromagnetism.

\subsection{OOAI material: La$_2$Ti$_2$O$_7$}

The band structure of the pyrochlore-structure \ce{La2Ti2O7} [\icsdweb{164027}, SG 227 (\sgsymb{227})] is diagnosed as a topologically trivial insulator of a direct band gap 2.561 eV on the TQCDB \cite{vergniory_complete_2019,Vergniory2021}. In the crystal lattice of \ce{La2Ti2O7}, La and Ti atoms occupy the respective Wyckoff positions $16d$ of coordinate $(0.5,0.5,0.5)$ and $16c$ of coordinate $(0,0,0)$, both of which are of point group $\bar 3m$. There are two non-equivalent \ce{O} atoms occupying the Wyckoff positions $8b$ and $48f$. 
Using the symmetry-data-vector of \ce{La2Ti2O7}, we calculate all the RSI indices defined in SG 227 and found that three $Z$-type RSIs have non-zero-integer values: $\delta_3(c)=1$, $\delta_4(c)=1$ and $\delta_7(e)=2$ (See Appendix~\ref{app:rsical}). 
As the maximal Wyckoff position $16c$ is occupied by Ti and the non-maximal Wyckoff position $32e$ of coordinate $(x,x,x)$ is connected with the Ti atoms at $16c$ and La atoms at $16d$,  none of the above three RSIs indicate an OAI. 

As derived above and tabulated in the BCS tool  \href{http://www.cryst.ehu.es/cryst/RSI}{RSIsg}, the local RSI defined in real space at $16c$ are $\delta_3(c)=m(^1\bar {E}_{u}^2\bar {E}_{u})-m(^1\bar {E}_{g}^2\bar {E}_{g})$ and $\delta_4(c)=m(\bar{E}_{1u})-m(\bar{E}_{1g})$ (where $m(\rho)$ is the multiplicy of irrep $\rho$ at $16c$). The RSIs $\delta_3(c)=1$ and $\delta_4(c)=1$ imply that there are at least two odd-parity irreps, 
$^1\bar E_{u}^2\bar E_{u}$ and $\bar E_{1u}$, pinned at the $16c$ position which is occupied by Ti atoms. Similarly, the RSI index $\delta_7(e)=2$, whose definition in real space is $\delta_7(e)=-2m(^1\bar E^2\bar E)+m(\bar E_1)$, implies at least two $\bar E_1$ irreps at the $32e$ position which is connected with La and Ti atoms.
In the \emph{ab initio} calculations of \ce{La2Ti2O7} on the TQCDB \cite{vergniory_complete_2019,Vergniory2021}, the outer-shell electronic configuration in the adopted pseudopotential of Ti and La atoms are $3d^34s^1$ and $5s^2 5p^6 6s^2 5d^1$, respectively. As detailed in Appendix~\ref{app:methodOOAI}, under the site symmetry group $\bar 3m$, $s$ and $d$ orbitals induce the even-parity irreps $\bar E_{1g}$ and $^1\bar {E}_{g}^2\bar {E}_{g}$, and $p$ orbitals induce the odd-parity irreps $^1\bar {E}_{u}^2\bar {E}_{u}$ and $\bar E_{1u}$. 
Thus, the odd-parity irreps $^1\bar {E}_{u}^2\bar {E}_{u}$ and $\bar E_{1u}$ pinned at $16c$ are not coming from the atomic orbitals on Ti and the RSIs $\delta_3(c)=1$ and $\delta_4(c)=1$ indicate an OOAI.

\begin{equation}
\begin{aligned}
    &\bar E_{1g} \downarrow G^e = \bar E_1, &\bar E_{1u} \downarrow G^e = \bar E_1
\end{aligned}
\label{eq:sub227}
\end{equation}
\begin{equation}
    \bar E_1 \uparrow G^d = \bar E_{1g}+\bar E_{1u}
\label{eq:ind227}
\end{equation}

From the subduction and induction relations between irreps at $16d$ (of point group $\bar 3m$) and $32e$ (of point group $3m$) in Eqs.~(\ref{eq:sub227}) and (\ref{eq:ind227}), the RSI $\delta_7(e)=2$ can be contributed by the irreps $\bar E_{1g}$ and $\bar E_{1u}$ at the Wyckoff position $16d$, which are induced from the $s$ and $p$ orbitals on La atom. So, this RSI can be contributed by atomic orbitals and does not indicate either an OAI or OOAI.

\section{Application of OAIs in catalysis}

Now we are going to discuss the potential applications of OAIs and the associated OSSs. Surface chemical reactions such as catalysis attract our attention immediately as they are exactly happened at the surface of crystals and require the surface states for adsorption and electron transfer. However, understanding the origin of catalytic activities is still remains a challenge for the design of high-performance catalysts for reactions including water splitting, nitrogen reduction, fuel cell reactions. Although the descriptors based on the electronic structure calculations are powerful in identifying the actives sites, they require the time-consuming calculation of projected density of states. It is interesting to find that many compounds from our OAI lists are good candidates for various catalysis reactions, and most importantly, the surfaces with OSSs are exactly the active sites, such as the (100) surface of \ce{NiP2} \cite{owens2020crystallographic}, the (110) surface of \ce{FeS2} \cite{wu2019first}, the (100) surface of 2H-\ce{MoS2} and 2H-\ce{MoSe2} \cite{Jaramillo100}, and the (111) surface of \ce{Fe3O4} \cite{huang2006density}.
A further in-depth experimental analysis of many known and, most importantly, new unknown catalysts are presented in Ref. \cite{Li2021catalysts}. Thus, we propose in this work that the OSSs are responsible for the catalytic activities and can be used for the fast determination of active sites.

\begin{figure*}
\centering\includegraphics[width=7.0in]{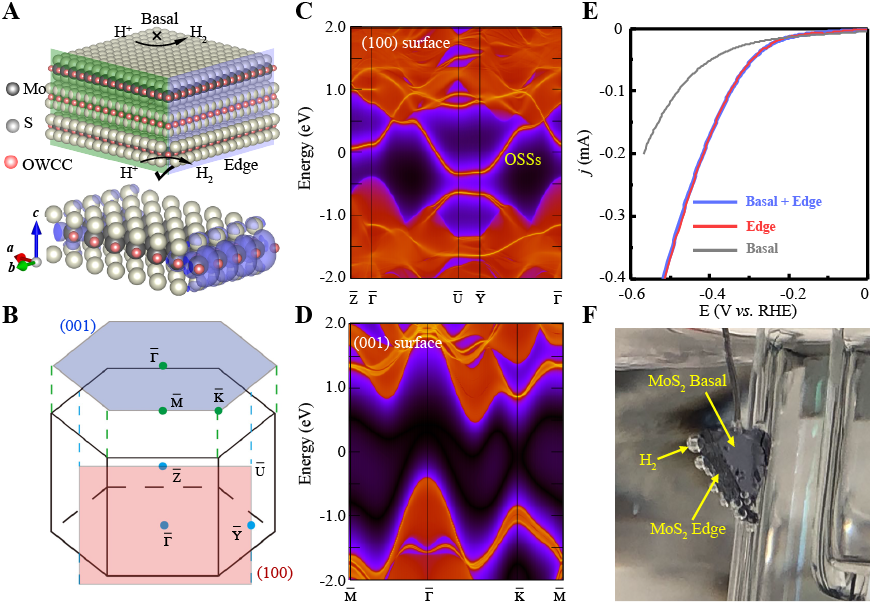}
\caption{The role of OSSs in 2H-\ce{MoS2} for hydrogen evolution. (A) Up panel: The crystal structure of 2H-\ce{MoS2} with the OWCCs represented by red spheres and its three cleavage planes of Miller indices $(001)$, $(100)$ and $(010)$, of which $(100)$ and $(010)$ planes exhibit the metallic OSSs. Bottom panel: The charge density distribution at the edge and basal planes of 2H-\ce{MoS2}. (B) The 3D bulk Brilliouin zone and the projected 2D Brillouin zone on the $(001)$ and $(100)$ surfaces. Surface states calculation of 2H-\ce{MoS2} at the (C) $(100)$ and (D) $(001)$ surfaces, respectively. The OSSs only exist on the $(100)$ but not the $(001)$ surface. (E) LSV curves of the whole crystal, edge surfaces, and $(001)$ basal plane, respectively. (F) Photo of the HER process, which clearly shows the production of hydrogen bubbles on the edge surfaces but not on the $(001)$ basal plane.}\label{fig:MoS2}
\end{figure*}

The Van der Waals compound 2H-\ce{MoS2} [\icsdweb{105091}, SG 194 (\sgsymb{194})] is a desirable objective to validate our prediction as hydrogen evolution reaction (HER) catalyst as a ``proof of principle''. Studies on \ce{MoS2} of monolayer \cite{Jaramillo100} and thin film \cite{kibsgaard2012engineering} structures have confirmed that the HER activity originates from the edge sites, rather than the preferentially exposed $(001)$ basal planes (Fig.~\ref{fig:MoS2}(A)). Here, with \emph{bulk} single crystals, we directly observe the relationship between the calculated OSSs and the measured catalytic activity. From our calculations of \ce{MoS2} (See Appendix~\ref{app:rsical}), there is a non-zero RSI index $\delta_3(b)=1$ which is defined at the Wyckoff position $2b(0,0,\frac{1}{4})$. As $2b$ is not occupied by any atom, \ce{MoS2} is an OAI and the OWCC is pinned at the $2b$ position. 
From the crystal structure in Fig.~\ref{fig:MoS2}(A), the OWCCs are co-planar with the Mo atoms on the $(001)$ plane. Hence, it is impossible to have a cleavage plan of Miller index $(001)$ which cuts through the OWCCs. We further identify that the side cleavage that parallel to $(001)$ direction always cuts through the OWCCs and away from all the atoms.
This suggests the OSSs are absent on the $(001)$ surface but exist on all the surfaces that are perpendicular to the basal $(001)$ plane. This is confirmed by the significantly high charge density around the OWCCs (Fig.~\ref{fig:MoS2}(A)) and the surface state calculations in Fig.~\ref{fig:MoS2}(C) and (D). 
By calculating the contribution of each unit cell in a slab model with a thickness of 50 unit cells, we have analyzed the decay of the OSSs on the $(100)$ surface into bulk and find that all the OSSs on the Fermi level are located within the first unit cell on the surface of the slab (to a depth of about 0.8 nm) (See Appendix~\ref{app:experiments}). 
To directly detect the location of catalytic active surfaces for 2H-\ce{MoS2}, bulk crystals with macroscopic edges $(1.1mm\times5mm)$ and basal planes $(6mm^2)$ are synthesized and tested as an electrochemical HER catalyst. These crystals are confirmed to be the stoichiometric 2H phase by energy dispersive spectroscopy and Raman spectroscopy \cite{LI20191} (See Appendix~\ref{app:experiments}). Linear sweep voltammetry (LSV) curves of the entire crystal, 
the edge surfaces (by covering the basal plane with gel), and basal $(001)$ plane (by covering most of the edge sites with a gel) were recorded in an Argon saturated 0.5 M \ce{H2SO4} electrolyte. The crystal with gel-covered $(001)$ basal plane exhibited almost the same activity as the entire crystal (Fig.~\ref{fig:MoS2}(E)). However, HER activity is significantly depressed by partially covering the crystal edge sites. Photography and video recorded during the chronoamperometric measurements display the active sites directly, with hydrogen bubbles only visible on the edge sites (Fig.~\ref{fig:MoS2}(F)) and the Extended Video). Furthermore, electrochemical impedance spectroscopy demonstrated a much lower charge transfer resistance for the edge sites than for the basal plane, suggesting enhanced electron transfer kinetics on the edge surface due to the highly conducting metallic surface states \cite{li2015charge} (See Appendix~\ref{app:experiments}).

\section{Discussion}
Based on the theory of TQC and MTQC \cite{bradlyn_topological_2017,MTQC}, we have derived the 3D RSIs for all the 1,651 SSGs with SOC. By applying the RSIs to all the symmetry-eigenvalue indicated topologically trivial insulators on the \webTQC\ and \webMTQC, we have found \TQCDBNbrICSDsOAI\ OAIs, \TQCDBNbrICSDsOOAI\ OOAIs and \MTQCDBNbrBCSIDsMOAI\ mOAIs. For the OAIs and mOAIs, there exist symmetric Wannier functions centered and pinned at empty sites. For the OOAIs, there exist Wannier functions pinned at the occupied sites but the Wannier functions are not expressed as states on the atoms sitting at these sites. All of the OAIs and mOAIs are predicted to exhibit OSSs on special cleavage planes. 
The 3D RSI indices can also be applied to metals such as the electride materials \cite{wagner1994electride,dye2003electrons,lee2013dicalcium,PhysRevB.103.205133} whose electron distribution is concentrated at the empty sites. 

The classifications and properties of OAIs presented in this work have not only provided a thorough explanation for the surface states of topologically trivial insulators (for example, the surface states of silicon \cite{PhysRev.106.455}),
but also provide fruitful platforms for the experimental studies and the potential applications such as the 2D superconductivity of 2D electron gas, catalysts and junctions. Although the net electric polarization of OAIs is zero in the bulk, it is non-zero on the surface. This electric dipole on surface provides an intrinsic electric field and could be applied to electronic technologies, such as the field-free diode effect in Josephson junction devices \cite{wu2021realization}.
As the metallic surface states are key to heterogeneous catalysis, this work will also drive the discovery of entirely new catalysts as well as the further exploitation and theoretical explanations of known catalysts \cite{Li2021catalysts}. For example, the catalytic property of 2H-\ce{MoS2} studied in this work and the boron catalysis reported in Ref.~\cite{duan2019origins} could be explained with the OSSs  (See Appendix~\ref{app:B12} for the RSI indices and OSSs of the OAI material \ce{B12} [\icsdweb{431636} SG 166 (\sgsymb{166})]).
Furthermore, this work provides a path to novel transparent conducting oxides and novel two dimensional electron gas that is formed by interfacing a surface with an OAI or an OOAI. A particularly interesting case is the use of a magnetic OAIs such as the example we identify in \ce{Mn3Si2Te6}. Compared with the topological insulators, the band gap of the OAIs could be much larger and hence the OSSs are easier to distinct from the bulk states and detected in the ARPES or transport experiments.

\acknowledgments
We acknowledge the computational resources Cobra/Draco in the Max Planck Computing and Data Facility (MPCDF).
This work is part of a project that has received funding from the European Research Council (ERC) under the European Union's Horizon 2020 research and innovation programme (grant agreement no. 101020833. B.A.B., N.R. and Z-D.S. were also supported by the U.S. Department of Energy (Grant No. DE-SC0016239), and were partially supported by the National Science Foundation (EAGER Grant No. DMR 1643312), a Simons Investigator grant (No. 404513), the Office of Naval Research (ONR Grant No. N00014-20-1-2303), the Packard Foundation, the Schmidt Fund for Innovative Research, the BSF Israel US foundation (Grant No. 2018226), the Gordon and Betty Moore Foundation through Grant No. GBMF8685 towards the Princeton theory program and the NSF-MRSEC (Grant No. DMR-2011750). B.A.B. and N.R. gratefully acknowledge financial support from the Schmidt DataX Fund at Princeton University made possible through a major gift from the Schmidt Futures Foundation. L.E. was supported by the Government of the Basque Country (Project IT1301-19) and the Spanish Ministry of Science and Innovation (PID2019-106644GB-I00). 
C.F. was supported by the European Research Council (ERC) Advanced Grant No.  742068 ``TOP-MAT'', Deutsche Forschungsgemeinschaft (DFG) through SFB 1143, and the Würzburg-Dresden Cluster of Excellence on Complexity and Topology in Quantum Matter-ct.qmat (EXC 2147, Project No. 390858490). S.S.P.P. acknowledges funding by the Deutsche Forschungsgemeinschaft (DFG, German Research Foundation) – Project number 314790414.
L.E. and N.R. were also sponsored by a Global Collaborative Network grant. Y.X. received support from the Max Planck Society.

\textit{Note added}. The implementation of RSIs in searching for 3D obstructed atomic insulators and their applications were originally presented on the APS March Meeting of 2021 \cite{xu2021three}. 
During the preparation of this manuscript, one work appeared on arXiv \cite{gao2021unconventional} where examples of OAI are presented.

%
\clearpage

\onecolumngrid
\renewcommand{\thesection}{Appendix~\arabic{section}}
\renewcommand{\thesubsection}{\arabic{section}.\arabic{subsection}}

\appendix

\clearpage
\begin{center}
{\bf Supplementary materials of "Three-Dimensional Real Space Invariants, Obstructed Atomic Insulators and A New Catalytic Principle"}
\end{center}

\tableofcontents

\clearpage

\section{Introduction of the appendices}\label{app:introduction}

In the present work, we have extended the real space invariant (RSI), as initially introduced in Ref. \cite{song2020}, to the three-dimensional (3D) space, both for the 230 double space groups (DSGs) and for the 1421 double magnetic space groups (MSGs), namely the 1651 Shubnikov space groups (SSGs). By implementing the RSIs to all the symmetry-indicated topologically trivial insulators found in the Ref. \cite{vergniory_complete_2019,zhang2019catalogue,Vergniory2021} and Ref. \cite{xu2020high}, we
presented a catalogue of stoichiometric materials in the obstructed atomic insulator (OAI) phase.
In addition to the main text, we provide below Supplementary Appendices giving an in-depth discussion of our
methodology and several lists of materials. 
In Appendix \ref{app:concepts}, we briefly introduce the theory of (magnetic) topological quantum chemistry ((M)TQC) \cite{bradlyn_topological_2017, MTQC}, the concept of (magnetic) obstructed atomic insulator ((m)OAI) and the orbital-selected OAI (OOAI).
In Appendix \ref{app:rsi3D}, we detail the general methodologies to obtain RSIs of the 1651 SSGs with spin-orbit coupling (SOC), which can be trivially extended also to systems without SOC.
In Appendix \ref{app:TQCdatabase}, we give an overview of the \webTQC~and \webMTQC~on which we relied for our catalogue of (m)OAIs and OOAIs.
In Appendix \ref{app:highthroughput}, we detail the methods to diagnose the (m)OAIs and OOAIs using the RSIs and the crystal structures of stoichiometric materials, and perform the high-throughput searches for (m)OAIs and OOAIs.
In Appendix~\ref{app:rsical}, we provide the calculations of the RSI indices which indicate an OAI or OOAI phase for the materials appeared in the main text. The materials include \ce{InS} [\icsdweb{15931}, SG 58 (\sgsymb{58})], \ce{Mn3Si2Te6} , \ce{La2Ti2O7} [\icsdweb{164027}, SG 227 (\sgsymb{227})] and \ce{MoS2} [\icsdweb{105091}, SG 194 (\sgsymb{194})]. 
In Appendix \ref{app:filling_anomaly}, we introduce the obstructed surface states (OSSs) exhibited on special cleavage planes of OAIs. We also select eight typical OAIs to illustrate the process of identifying an OAI using the RSIs and calculate their OSSs. The eight OAIs include seven paramagnetic OAIs, \ce{NbBr2O} [\icsdweb{416669}), SG 5 (\sgsymb{5})], \ce{CaIn2P2} [\icsdweb{260562}, SG 194 (\sgsymb{194})], \ce{InSe} [\icsdweb{185172}, SG 194 (\sgsymb{194})], \ce{Hg2IO} [\icsdweb{33275}, SG 15 (\sgsymb{15})], \ce{PtSbSi} [\icsdweb{413194}, SG 61 (\sgsymb{61})], \ce{Nb3Br8} [\icsdweb{25766}, SG 166 (\sgsymb{166})], \ce{B12} [\icsdweb{431636}, SG 166 (\sgsymb{166})] and one magnetic OAI \ce{CsFe2Se3} [\bcsidweblong{1.26}, MSG 14.82 (\msgsymb{14}{82})].
In Appendix~\ref{app:experiments}, the methods and extended data for the catalytic experiments on 2H-\ce{MoS2} are provided.
Finally in Appendix \ref{app:3DOAI}, we provide three material lists including the lists of \TQCDBNbrICSDsOAI\ paramagnetic OAIs, \TQCDBNbrICSDsOOAI\ paramagnetic OOAIs and \MTQCDBNbrBCSIDsMOAI\ mOAIs.

\section{Concepts}\label{app:concepts}

In this Appendix, we first briefly review the (magnetic) topological quantum chemistry ((M)TQC) theory \cite{bradlyn_topological_2017, MTQC} and the symmetry indicated topologies as defined in the (M)TQC. 
Then, we introduce the concepts of atomic insulator (AI), obstructed atomic insulator (OAI) \cite{bradlyn_topological_2017} and the orbital-selected obstructed atomic insulator (OOAI). 

\subsection{Topological quantum chemistry}\label{app:TQC}
In band theory, the symmetry properties of a band structure are characterized by the irreducible (co-)representations (irreps) at all the maximal $k$-vectors, whose little groups are maximal subgroups of the (magnetic) space group. Each maximal $k$-vector in the Brillouin zone (BZ) is kept invariant (mod translations of the reciprocal lattice) under a subset of operations of the space group that form the little group $G_{k}$ of $k$, and whose irreducible representations are denoted as $\rho_{k}^i$.

With $N_{e}$ the number of valence electrons of the crystal material in one unit cell, the first $N_{e}$ bands at every $k$ point represents the set of \emph{occupied} bands.
These bands are characterized by the set of multiplicities $m(\rho_{k}^i)$ of every irrep $\rho_{k}^i$ of $G_{k}$ at every maximal $k$-vector, $\{m(\rho_{k_j}^i)|i=1,2,...,N(k_j), j=1,2,...,N_k\}$, where $N(k_j)$ is the number of irreps of $G_{k_j}$ and $N_k$ is the number of maximal $k$-vectors in the BZ of the considered space group.
For convenience, we introduce the symmetry-data-vector \cite{bradlyn_topological_2017,cano_building_2018,elcoro_double_2017,vergniory_graph_2017,SlagerSymmetry},
\begin{equation}
B = (
m(\rho_{k_1}^1), m(\rho_{k_1}^2), ..., m(\rho_{k_1}^{N(k_1)}),
m(\rho_{k_2}^1), m(\rho_{k_2}^2), ..., m(\rho_{k_2}^{N(k_2)}),
...,
m(\rho_{k_{N_k}}^1), m(\rho_{k_{N_k}}^2), ..., m(\rho_{k_{N_k}}^{N(k_{N_k})}) )^T.\label{eq:B-vector}
\end{equation}
with $\sum_{i=1}^{N_k}N(k_i)$ components, that give the multiplicities of the corresponding irrep.

In the analysis of the electronic band structures done in the framework of (M)TQC \cite{bradlyn_topological_2017, MTQC} and the equivalent method of symmetry-based indicators \cite{po_symmetry-based_2017,watanabe2018structure,song_quantitative_2018}, a band structure is gapped if its symmetry-data-vector satisfies all the compatibility relations between any two maximal $k$-vectors, \ie along the lines or planes that connect every pair of maximal $k$-vectors, the multiplicities of the irreps of the litte group of the intermediate line or plane, given by the compatibility relations at both end points, are exactly the same. Otherwise, the band structure is \emph{necessarily} gapless and referred to as \emph{enforced semi-metal}. In the (M)TQC analysis, the gapped band structures have been divided into three different types of topological insulating phases: strong topology, fragile topology and topologically trivial insulator.

Following the terminology of Zak \cite{zak1980symmetry,zak1981band, zak1982band}, the electronic band structure of an atomic insulator is a band representation (BR) and the generators of the BR are elementary BRs (EBRs), which are the induced band representations from all the possible atomic orbitals symmetrically centered at the maximal Wyckoff positions \cite{zak1982band,bradlyn_topological_2017,MTQC}. A BR can always be decomposed into a linear combination of EBRs (LCEBR) with non-negative-integer coefficients. The EBRs are thus the basis of the BRs of topologically trivial insulators that can be induced from the atomic orbitals and are then Wannierizable. 
A gapped band structure can be symmetry-indicated as topologically trivial insulator if its symmetry data-vector (\ref{eq:B-vector}) is a linear combination of the symmetry-data-vectors of the EBRs. If not, it is topologically nontrivial and the material has strong or fragile topology.  

\subsection{Atomic and obstructed atomic insulators}\label{app:OAI}

In real crystalline materials, the BR of an insulator tagged as topologically trivial in the (M)TQC analysis is always induced from local orbitals centered at different WPs that are occupied or not occupied by atoms. If a BR cannot be induced \emph{only} from the atomic orbitals centered at the occupied Wyckoff positions, it is in the obstructed atomic insulating phase.
In this work, topologically trivial insulators in an obstructed atomic insulating phase are referred to as obstructed atomic insulators (OAIs). A small subset of them, identified by only their electron number without the need for further calculations, was presented in Ref. \cite{xu2021filling}.
For the OAIs, there \emph{necessarily} have localized Wannier functions pinned at some empty sites.

Given the symmetry-data-vector $B$ (Eq. \ref{eq:B-vector}) of the BR of a trivial insulator, the material is identified as an OAI if $B$ \emph{does not} satisfy the following condition,
\begin{equation}
B=\sum_{j=1}^{M}  \sum_{i=1}^{N_{rep, \alpha_j}} N_{i,j}(\rho_{\alpha_j}^i\uparrow\mathcal{G}), \;\;\; N_{i,j} \ge 0, \in \mathbf{Z}. \label{eq:OAI}
\end{equation} 
where $\{\alpha_j, j=1,2,...,M\}$ is the set of occupied Wyckoff positions, $\rho_{\alpha_j}^i\uparrow\mathcal{G}$ is the induced BR of an irrep $\rho_{\alpha_j}^i$ at the occupied position $\alpha_j$, $N_{rep, \alpha_j}$ is the number of irreps of the site-symmetry group of $\alpha_j$ and $N_{i,j}$ are non-negative integers. 

If the BR of a topologically trivial insulator satisfies the condition in Eq. \ref{eq:OAI}, it still can be diagnosed as an orbital-selected OAI (OOAI) if Eq. \ref{eq:OAI} \emph{necessarily} contains an irrep $\rho_{\alpha_j}^i$ (with $N_{i,j} > 0$) which \emph{cannot} be contributed by the outer-shell electrons of the atoms occupying ${\alpha_j}$. The topologically trivial insulators that are not diagnosed as either OAIs or OOAIs are referred to as atomic insulators (AIs) in this work, although a further even finer classification is possible.

\section{Real Space Invariant indices in magnetic and non-magnetic space groups}\label{app:rsi3D}

The real space invariants (RSI) in the context of Topological Quantum Chemistry \cite{bradlyn_topological_2017} were introduced in Ref. \onlinecite{song2020}, where the complete lists of RSIs in 2-dimensional (2D) point groups were explicitly given, with and without SOC and with and without time reversal symmetry (TR). In this section, we generalize the calculation of the RSIs to 3-dimensional (3D) non-magnetic and magnetic point and space groups with SOC. 

The symmetry of a magnetic material with lattice-commensurate periodic order is given by one of the 1,421 magnetic magnetic groups, which are divided into Shubnikov groups of Type I, III and IV. The set of symmetry operations of a non-magnetic structure correspond to one of the 230 space groups or Shubnikov groups of Type II, which contain the time-reversal symmetry. Therefore the 1,651 Shubnikov groups characterize both the magnetic and non-magnetic symmetries. As the procedure to calculate the RSI indices described in this section is general for all types of Shubnikov groups, we will not distinguish between magnetic and non-magnetic groups and the general term \emph{Shubnikov space group} (SSG) will be used.

For completeness, we first summarize the main ideas introduced in Ref. \onlinecite{song2020}, although our procedure to determine the RSIs is slightly different. Afterwords, we develop the algorithm used in the present derivation of the RSIs in 3D, stressing the main differences with respect to the calculation in 2D.
Finally as an example we give the detailed calculation of the RSI indices for double irreps (a system where SOC is considered), both in direct and momentum spaces, for the magnetic space group \msgsymb{67}{508} (N. 67.508).

\subsection{Wyckoff positions and site-symmetry groups}
\label{RSI:theory}
In a SSG, the points in real space are classified into different subsets (Wyckoff positions or WPs), according to their site-symmetry group. The site-symmetry group of a point in direct space is the set of symmetry operations of the SSG that keep the point invariant. This subset of operations form a finite group that is isomorphic to one of the 122 Shubnikov point groups. The set of positions whose site-symmetry groups are conjugated under symmetry operations of the SSG form a WP. In particular, all the (infinitely many) symmetry-related points to a given point, i.e., all the points obtained after the application of the symmetry operations of the SSG belong to the same WP. The tables of crystallography usually include the coordinates of the points in the standard unit cell that correspond to every WP. The rest of points that belong to the WP are obtained by translations with linear combinations of the basis vectors of the unit cell. The number of points in a single unit cell is the multiplicity of the WP.

In all the SSGs, the general WP corresponds to the set of points $(x,y,z)$ whose site-symmetry group is the identity. However, for some special values of $x$, $y$ and/or $z$, the site-symmetry group will include additional symmetry operations. These WPs are called special WPs and correspond, in general, to planes, lines and points of symmetry. For instance, in the type-II SSG $P4/m1'$ (N. 83.44), the line $(0,0,z)$ is denoted by the symbol $2g$, where $g$ is known as the Wyckoff letter of this line and 2 is the multiplicity. The multiplicity of the WP is explicitly added to the Wyckoff letter when it is important to stress the number of symmetry-related points that belong to the WP. In this section we will use indistinctly a WP symbol with and without the multiplicity, because the letter identifies unambiguously the WP.
The site-symmetry group of the line $(0,0,z)$ consists of the $\{R|0,0,0\}$ symmetry operations of the SSG whose rotational part $R$ belongs to the point group $41'$. The 4-fold axis is parallel to the $z$ axis and $1'$ represents TR. However, for the special values of the continuous parameter,  $z=0$ and $z=\frac{1}{2}$ in the unit cell, the site-symmetry group contains also an inversion center ($\{I|0,0,0\}$ in the first case and $\{I|0,0,1\}$ in the second one). The site-symmetry groups of these two special points are different and not conjugated under any operation of the SSG. Therefore, they belong to different WPs. Both site-symmetry groups of $z=0$ and $z=\frac{1}{2}$ are isomorphic to the point group $4/m1'$. 

Among all the WPs in a SSG, a special attention is paid to the maximal WPs. We say that a Wyckoff position $Q$ is of maximal symmetry if there does not exist another Wyckoff position $Q'$ and a continuous path that goes from $Q$ to $Q'$ such that the site-symmetry group of $Q$ is a proper subgroup of the site-symmetry group of $Q'$ and a subgroup (proper or not) of the site-symmetry groups of all the points along the path. In our example, $1a$:$(0,0,0)$ and $1b$:$(0,0,1/2)$ points are of maximal symmetry. However, the site-symmetry group of a point of the line $2g$:$(0,0,z)$  with $0<z<1/2$ is not a maximal WP because a point of this line can be continuously connected to point $a$ or point $b$ following the path $(0,0,z)$. The site-symmetry subgroup of all the points in the path are exactly the same and it is a proper subgroup of the site-symmetry group of $a$ and $b$.

The site-symmetry group of a position in the direct space plays an important role in the description and classification of the different physical quantities that characterize a solid material. Any function defined around a specific point (as for example the atomic orbitals located around a point of the crystal) must transform according to a representation of the site-symmetry group of that point. In general, this representation is reducible. As the site-symmetry group of a given point in a crystal of a given SSG is isomorphic to a point group, the tables of irreducible representations (irreps) of the crystallographic point groups usually contain the atomic orbitals or sets of atomic orbitals that transform under every irrep \cite{PointGroupTables,repres}. These sets of orbitals can be located around the positions occupied by the atoms or can be centered out of the atomic positions. These well localized orbitals induce extended states (electronic energy bands) in the reciprocal space, whose connectivity and degeneracy in the different points of the momentum space are governed by the SSG. For a detailed analysis of the relations between local orbitals and the extended states along the momentum space from the point of view of the symmetry see, for instance, the Refs. \onlinecite{Evarestov1997,bradlyn_topological_2017,elcoro_double_2017}.

To raise the problem of the determination of the RSI indices, let us take a gapped band or set of bands in the band structure of a material (\ie there is an energy gap between the states in the chosen set of bands and the states in the bands immediately above and another gap between the states of the chosen set of bands and the states in the band or bands immediately below in the entire first Brillouin zone). Let us consider an adiabatic process that maintains the symmetry of the crystal and that keeps our band(s) gapped along the whole process (the band(s) does not \emph{touch} the bands above and below at any $k$-vector in the momentum space). In principle, the center of charges (or orbitals) can move in the adiabatic process, but the site-symmetry group of every point in the direct space must remain invariant. The fact that the site-symmetry group of every point in space must remain invariant restricts the possible shifts of the orbitals in the adiabatic process. The RSI indices characterize precisely the possible displacements of the atomic orbitals along the high symmetry lines, high symmetry planes or along arbitrary paths in direct space.

For instance, let us consider a WP $Q$ of maximal symmetry (an isolated point of maximal symmetry) and a line $q$ where $Q$ sits, i.e., $G^{q}\subset G^{Q}$, where $G^{Q}$ and $G^{q}$ are the site-symmetry groups of $Q$ and $q$, respectively. Let us also consider that at the beginning of the adiabatic process there is a set of atomic orbitals centered at $Q$. This set of orbitals transforms necessarily under a representation $\rho^Q$ of $G^Q$ which is, in general, reducible. Let us assume that during the adiabatic process the orbitals move, and are now the center of charges located at different points of the $q$ line. We can say that the starting set of orbitals split into different subsets of orbitals now located along the line $q$. Of course, the WP $q$ represents, in general, different symmetry-related lines and the orbitals split symmetrically along all these equivalent lines. This is what is understood in this whole section when the simplified form ``\emph{orbitals move along the line q}" is used.  Every set of orbitals now centered around a given point of $q$ must transform under a representation $\rho^q$, in general reducible, of $G^q$. However, the site-symmetry group $G^{Q}$ of the WP of higher symmetry imposes additional restrictions on this split. In general, the different centers of charges along $q$ must be related through a symmetry element in $G^{Q}$ but not in $G^{q}$, and moreover there must be a relation between the representation $\rho^q$ and the representation $\rho^Q$ of the original situation.

As stated in the main text, it is not always possible to move a set or orbitals from a WP to another WP of lower symmetry, depending on the orbitals at $Q$. In some cases, this could only happen breaking the symmetry (i.e. inducing a phase transition) or closing the gap between the set of bands and the bands above or below. We can say that some orbitals are pinned at $Q$. If the WP $Q$ is empty, i.e., no atom sits at $Q$ and the set of chosen gapped bands are the valence bands below the Fermi level in an insulator, the set of locked orbitals at an empty WP indicates that the band structure is an OAI. In the opposite direction, starting from a set of orbitals in the line $q$, during an adiabatic process it is always possible to move these centers of charges to $Q$. The shift from points of lower symmetry to points of higher symmetry is always possible without closing the gaps and without breaking the space (magnetic) group symmetry.

The identification of those configurations in which the orbitals can move from high symmetry WPs to WPs of lower symmetry (or the configurations in which the complete shift of all orbitals is not possible) can be determined in a group theory analysis based on the induction-subduction relations between the irreps in a group-subgroup pair.
Before the description of the general procedure, we will consider a simple example. Let us assume that in the type-II SSG $P4/m1'$ (N. 83.44) mentioned above, there are centers of charges (orbitals) located at the $2g$;$(0,0,z)$ line at $\pm z_0$ positions. If there is a center of charges centered at $z_0$ there must be another center at $-z_0$ due to the inversion center located at the $1a$:$(0,0,0)$ point. We want to analyze the possibility of moving orbitals between WPs $2g$ and $1a$. For this purpose, we include in Table \ref{rep:30and36} the traces of the irreps of the point groups $4/m1'$ and $41'$, isomorphic to the site-symmetry groups of the WPs $1a$ and $2g$, respectively. The table contains only the double valued irreps, and only half of the elements have been included. The other operations in this point group are obtained from the elements in the table adding the inversion in the spinor space. The traces of the extra operations are the opposite ones of the corresponding operations in the table (see the program {\color{blue} CorepresentationsPG} (\href{http://www.cryst.ehu.es/cryst/corepresentationsPG}{www.cryst.ehu.es/cryst/corepresentationsPG}) in the BCS, for example, for more details). 

\begin{table}
	\caption[Traces of the double valued irreps of the point groups $41'$ and $4/m1'$]{Traces of the double valued irreps of the point groups $41'$ (left) and $4/m1'$ (right). Only half of the symmetry operations have been included. The other symmetry elements are obtained multiplying the operations in the table by the the operation that represents the inversion in the spinor space and the identity in the orbital space. The traces of the resulting operations are the opposite ones of the corresponding elements (see the program {\color{blue} CorepresentationsPG} (\href{http://www.cryst.ehu.es/cryst/corepresentationsPG}{www.cryst.ehu.es/cryst/corepresentationsPG}) in the BCS, for example, for more details).}
	\label{rep:30and36} 
	\begin{tabular}{c|rrrr}
		\hline
		$41'$&1&$2_{001}$&$4_{001}^{+}$&$4_{001}^{-}$\\
		\hline
		$\,^2\overline{E}_{1}\,^1\overline{E}_{1}$&2&0& $\sqrt{2}$& $\sqrt{2}$\\
		$\,^2\overline{E}_{2}\,^1\overline{E}_{2}$&2&0&$-\sqrt{2}$&$-\sqrt{2}$\\
		\hline
	\end{tabular}\hspace{1cm}
	\begin{tabular}{c|rrrrrrrr}
		\hline
		$4/m1'$&1&$2_{001}$&$4_{001}^{+}$&$4_{001}^{-}$&$\bar{1}$&$m_{001}$&$\bar{4}_{001}^{+}$&$\bar{4}_{001}^{-}$\\
		\hline
		$\,^2\overline{E}_{1g}\,^1\overline{E}_{1g}$&2&0& $\sqrt{2}$& $\sqrt{2}$& 2&0& $\sqrt{2}$& $\sqrt{2}$\\
		$\,^2\overline{E}_{1u}\,^1\overline{E}_{1u}$&2&0& $\sqrt{2}$& $\sqrt{2}$&-2&0&$-\sqrt{2}$&$-\sqrt{2}$\\
		$\,^2\overline{E}_{2g}\,^1\overline{E}_{2g}$&2&0&$-\sqrt{2}$&$-\sqrt{2}$& 2&0&$-\sqrt{2}$&$-\sqrt{2}$\\
		$\,^2\overline{E}_{2u}\,^1\overline{E}_{2u}$&2&0&$-\sqrt{2}$&$-\sqrt{2}$&-2&0& $\sqrt{2}$& $\sqrt{2}$\\
		\hline
	\end{tabular}
\end{table}

Using the traces in Table \ref{rep:30and36} we establish the induction and subduction relations between the irreps of this group-subgroup pair. The subduction relations are,
\begin{eqnarray}
\label{eq:exasub1}
\,^2\overline{E}_{1g}\,^1\overline{E}_{1g}\downarrow G^{g}&=&\,^2\overline{E}_{1}\,^1\overline{E}_{1}\\
\label{eq:exasub2}
\,^2\overline{E}_{1u}\,^1\overline{E}_{1u}\downarrow G^{g}&=&\,^2\overline{E}_{1}\,^1\overline{E}_{1}\\
\label{eq:exasub3}
\,^2\overline{E}_{2g}\,^1\overline{E}_{2g}\downarrow G^{g}&=&\,^2\overline{E}_{2}\,^1\overline{E}_{2}\\
\label{eq:exasub4}
\,^2\overline{E}_{2u}\,^1\overline{E}_{2u}\downarrow G^{g}&=&\,^2\overline{E}_{2}\,^1\overline{E}_{2}
\end{eqnarray}
and the induction relations are,
\begin{eqnarray}
\label{eq:exaind1}
\,^2\overline{E}_{1}\,^1\overline{E}_{1}\uparrow G^{a}&=&\,^2\overline{E}_{1g}\,^1\overline{E}_{1g}\oplus\,^2\overline{E}_{1u}\,^1\overline{E}_{1u}\\
\label{eq:exaind2}
\,^2\overline{E}_{2}\,^1\overline{E}_{2}\uparrow G^{a}&=&\,^2\overline{E}_{2g}\,^1\overline{E}_{2g}\oplus\,^2\overline{E}_{2u}\,^1\overline{E}_{2u}
\end{eqnarray}
According to these relations, an irrep $\,^2\overline{E}_{1}\,^1\overline{E}_{1}$ in the point group $41'$, for instance, induces two irreps in the point group $4/m1'$, $\,^2\overline{E}_{1g}\,^1\overline{E}_{1g}$ and $\,^2\overline{E}_{1u}\,^1\overline{E}_{1u}$. This means that we can move a pair of orbitals centered at the points $(0,0,\pm z_0)$ that belong to the WP $2g$:$(0,0,z)$ and that transform under the irrep $\,^2\overline{E}_{1}\,^1\overline{E}_{1}$ following the line $2g$ up to the WP $a$. The pair of centers of charge (orbitals) merge to give a single center of charges that transforms under the direct sum of irreps $\,^2\overline{E}_{1g}\,^1\overline{E}_{1g}\oplus\,^2\overline{E}_{1u}\,^1\overline{E}_{1u}$, i.e. a sum of two orbitals that transform under $\,^2\overline{E}_{1g}\,^1\overline{E}_{1g}$ and $\,^2\overline{E}_{1u}\,^1\overline{E}_{1u}$. We could also consider the opposite process: starting from a pair of orbitals  that transform under the irreps $\,^2\overline{E}_{1g}\,^1\overline{E}_{1g}$ and $\,^2\overline{E}_{1u}\,^1\overline{E}_{1u}$ of $G^{a}$ at the WP $a$, we can move the orbitals along the $2g$ line (one orbital along each each side of the $a$ position). These two orbitals transform under the $\,^2\overline{E}_{1}\,^1\overline{E}_{1}$ irrep of $G^{g}$.

However, if the starting point is a single orbital $\,^2\overline{E}_{1g}\,^1\overline{E}_{1g}$ (or $\,^2\overline{E}_{1u}\,^1\overline{E}_{1u}$) at the $a$ position, this orbital cannot be moved to line $g$ without breaking the symmetry of the system (or without closing a gap above or below the band). This orbital is pinned at $a$. In our example, it is clear that we need an identical number of $\,^2\overline{E}_{1g}\,^1\overline{E}_{1g}$ and $\,^2\overline{E}_{1u}\,^1\overline{E}_{1u}$ orbitals to be possible to move them all away from the WP $a$ to the WP $g$. Every pair  ($\,^2\overline{E}_{1g}\,^1\overline{E}_{1g}$, $\,^2\overline{E}_{1u}\,^1\overline{E}_{1u}$) can be transformed into a single $\,^2\overline{E}_{1}\,^1\overline{E}_{1}$ orbital at each branch of $g$. Therefore, we can define the RSI index $\delta_a$,
\begin{equation}
\delta_a=m(\,^2\overline{E}_{1g}\,^1\overline{E}_{1g})-m(\,^2\overline{E}_{1u}\,^1\overline{E}_{1u})
\label{eq:exatetra}
\end{equation}
that computes the difference of orbitals in $a$ that transform under each irrep. A non-zero value of $\delta_a$ implies that, at least, $|\delta_a|$ orbitals will always remain at $a$ in an adiabatic process (of the $\,^2\overline{E}_{1g}\,^1\overline{E}_{1g}$ type if $\delta_a>0$ and of the $\,^2\overline{E}_{1u}\,^1\overline{E}_{1u}$ type if $\delta_a<0$). When $\delta_a=0$, all the orbitals can be moved away from $a$ to $g$. A non-zero value of $\delta_a$ is specially meaningful when the WP $a$ is atom-empty in the material structure. It means that, at least, $|\delta_a|$ orbitals must be located at an empty position. Therefore, according to the discussion in the main text, if the material is a topologically trivial insulator (no strong and no fragile topological) it is an obstructed atomic insulator (OAI).

In the next section we will explain the general algorithm to calculate all the RSI indices similar to (\ref{eq:exatetra}) in all the SSGs for double valued irreps. 
\subsection{Calculation of the local RSI indices at a Wyckoff position}
\label{calcRSI}
The example described in the previous section shows that the possibility of moving orbitals from a high symmetry WP to a WP of lower symmetry connected to it relies in the induction-subduction relations between the irreps of the site-symmetry groups of both WPs. In principle, the orbitals could move along any of the WPs of lower symmetry where the WP of high symmetry sits. In the example of the previous section of  SSG $P4/m1'$, we have only considered the possibility of moving orbitals from the maximal WP $1a$:$(0,0,0)$ with site-symmetry group isomorphic to PG $4/m1'$ to the line $2g$:$(0,0,z)$ with site-symmetry group isomorphic to $41'$. However, the plane $4j:(x,y,0)$ is also connected to the WP $a$, so we could also consider the possibility of moving the orbitals from the point $a$ to the plane $j$. As the site-symmetry group of this plane is isomorphic to $m1'$, we should calculate the induction-subduction relations equivalent to Eqs. (\ref{eq:exasub1})-(\ref{eq:exasub4}) and (\ref{eq:exaind1})-(\ref{eq:exaind2}) for the $4/m1'$-$m1'$ group-subgroup pair.

In general, to determine whether a set of orbitals at a given WP of high symmetry can be moved away from that position, we need to consider the shift of the orbitals along all possible WPs of lower symmetry connected to it. However, it is enough to consider only the maximal site-symmetry subgroups. If there are three WPs is a SSG such that, $G^{q_2}\subset G^{q_1}\subset G^{Q}$ and it is possible to move a set of orbitals from $Q$ to $q_2$, then it is possible to move the orbitals from $Q$ to $q_1$. Note that it is always possible to move an orbital from a WP to another one of higher symmetry, so once the orbitals have been moved from $Q$ to $q_2$, they can be moved to $q_1$, i.e., $Q\to q_2\to q_1$ implies $Q\to q_1$. Therefore, we need just to consider the induction-subduction relations between a site-symmetry group and its maximal site-symmetry subgroups to determine the RSIs.

It is crucial to recognize at this point that, although all the site-symmetry groups realized in all the 1651 SSGs are isomorphic to one of the 122 magnetic point groups, the maximal site-symmetry subgroups of a given site-symmetry group are not isomorphic to one of the maximal subgroups of the point group. For instance, in our example of the previous section, SSG $P4/m1'$ (N. 83.44), the site-symmetry group of the WP $a$ is isomorphic to the magnetic point group $4/m1'$ whose generators include $\{C_{2z}|(0,0,0)\}$,$\{C_{4z}|(0,0,0)\}$, $\{I|(0,0,0)\}$ and the time reversal symmetry $T$. The maximal subgroups of this point group are $41'$ (of generators $\{C_{2z}|(0,0,0)\}$, $\{C_{4z}|(0,0,0)\}$ and $T$) and $2/m1'$ (of generators $\{C_{2z}|(0,0,0)\}$, $\{I|(0,0,0)\}$ and $T$). The site-symmetry group of the WP $2g$:$(0,0,z)$ is isomorphic to the point group $41'$, but no WP connected to $a$ has a site-symmetry group isomorphic to $2/m1'$. The reason is evident: as $2/m1'$ contains the inversion, the subspace kept invariant by such a point group is a single point and, thus, not connected to $a$. In SSG $P4/m1'$, the maximal site-symmetry groups of $G^{a}$ (isomorphic to point group $4/m1'$) are $G^{g}$ and $G^{j}$, isomorphic to $41'$ and $m1'$ point groups, respectively.

We have determined the maximal site-symmetry groups of all the 122 different site-symmetry groups realized in the 1651 SSGs, which are isomorphic to the 122 magnetic point groups. The list of group-subgroups are included in the first three columns of Table \ref{tab:ci}. The first two columns give the number and the symbol of all the 122 magnetic point groups isomorphic to the 122 different site-symmetry groups and the third column gives the symbols and numbers of the subgroups.

Once the $n_{H}$ maximal subgroups $H^i$, $i=1,\ldots,n_{H}$, of a given site-symmetry group $G^{Q}$ of a WP $Q$ have been determined, we first consider the irreps of $G^{Q}$ in a specific order, $\rho_{G}^1,\ldots,\rho_{G^Q}^{n_{G^Q}}$, being $n_{G^Q}$ the number of irreps of $G^{Q}$. Now we consider the subduction relations between these irreps and the irreps of every maximal subgroup $H^i$. For every irrep $\rho_{G^Q}^j$ of $G^{Q}$ we calculate the subduced representation into the subgroup $H^i$. In general, the resulting representation is reducible and can be expressed as a direct sum of irreps of $H^{i}$,
\begin{equation}
\label{sub}
\rho_{G^Q}^j\downarrow H^i=\bigoplus_{k=1}^{n_{H^i}}s_{jk}^{H^i}\rho_{H^i}^k
\end{equation}
where $\rho_{H_i}^k$ are the irreps of the subgroup $H^i$ and $s_{jk}^{H^i}$ is the integer multiplicity of $\rho_{H^i}^k$ in the subduced representation. In the opposite process, every irrep $\rho_{H^i}^k$ of a given subgroup $H^i$ induces a representation in $G^{Q}$ which is, in general reducible and can be expressed as a direct sum of irreps of $G^{Q}$,
\begin{equation}
\label{ind}
\rho_{H^i}^k\uparrow G^Q=\sum_{j=1}^{n_{G^Q}}c_{jk}^{H^i}\rho_{G^Q}^j
\end{equation}
where $c_{jk}^{H^i}$ is the integer multiplicity of $\rho_{G^Q}^j$ in the induced representation.

When both groups $G^Q$ and $H^i$ are unitary groups (\ie do not contain TR or the combinations of TR and the crystalline symmetries), the relation between the multiplicities $s_{jk}^{H^i}$ in the subduction process (\ref{sub})  and the multiplicities $c_{jk}^{H^i}$ in the induction process (\ref{ind}) are, according to the Frobenius reciprocity theorem,
\begin{equation}
c_{jk}^{H^i}=s_{jk}^{H^i}
\end{equation}

For a group $G$ which contains anti-unitary operations such as TRS, it can be divided into an unitary part $G_U$ and an anti-unitary part $\mathcal{T}\cdot G_U$, where $\mathcal{T}$ is the TRS. The basis functions of irreps of $G_U$, and their TRS-counterparts, if $\mathcal{T}$ transform them to other basis functions, form co-representation of $G$.
The group $G$ has three types of irreps distinguished by the Frobenius-Schur indicator  \cite{frobenius1906reellen,BigBook,song2020}.
\begin{enumerate}
\item[($a$)] $\rho_{G_U}$ is equivalent to its complex conjugate $\rho_{G_U}^*$, \ie $\rho_{G_U} = N \rho_G^* N^{-1} $ for some unitary matrix $N$, and $N N^* = \mathcal{T}^2 $. Then $\rho_{G_U}$ itself form a co-representation of $G$.
\item[($b$)] $\rho_{G_U}$ is equivalent to $\rho_{G_U}^*$ and the transformation matrix satisfies $N N^* = -\mathcal{T}^2 $. Then the corresponding co-representation consists of the direct sum of two $\rho_{G_U}$ irreps, \ie $\rho_{G} = \rho_{G_U} \oplus \rho_{G_U}$ for unitary operations.
\item[($c$)] $\rho_{G_U}$ is not equivalent to $\rho_{G_U}^*$, then the corresponding co-representation consists of $\rho_{G_U}$ and its complex conjugate, \ie  $\rho_{G} = \rho_{G_U} \oplus \rho_{G_U}^*$. 
\end{enumerate}

Hence, when the groups $G^Q$ and/or $H^i$ contain anti-unitary operations, the relation between the multiplicities in Eq. (\ref{sub}) and Eq. (\ref{ind}) are different. The relations between the coefficients in the subduction and the induction processes depend on the type ($a$), ($b$) or ($c$) of the co-representations to which both $\rho_{H^i}^k$ and $\rho_{G^Q}^j$ belong, and whether the group $G^Q$ is anti-unitary or not when $H^i$ is unitary. 

The general relation is (a detailed calculation of the relations between these coefficients can be found in the supplementary material of Ref. \cite{song2020}),
\begin{equation}
\label{frobnounit}
c_{jk}^{H^i}=u(G^Q,H^i)f(t_{\rho_{H^i}^k},t_{\rho_{G^Q}^j})s_{jk}^{H^i}
\end{equation}
where,
\begin{equation}
\label{factoru}
u(G^Q,H^i)=\left\{\begin{array}{lll}
2&\hspace{0.5cm}&H^i\textrm{ is unitary and }G^{Q}\textrm{ is not unitary}\\
1&\hspace{0.5cm}&\textrm{otherwise}
\end{array}\right.
\end{equation}
and $f(t_{\rho_{H^i}^k},t_{\rho_{G^Q}^j})$ is a factor that depends of $t_{\rho}=a,b,c$ the type of co-representations, according to the relations (\ref{nounit}), 

\begin{equation}
\label{nounit}
	\begin{array}{ccc}
		f(a,a)=1&f(a,b)=\frac{1}{2}&f(a,c)=\frac{1}{2}\\
		f(b,a)=1&f(b,b)=\frac{1}{4}&f(b,c)=\frac{1}{2}\\
		f(c,a)=1&f(c,b)=\frac{1}{4}&f(c,c)=\frac{1}{2}\\
			\end{array}
		\end{equation}
Considering the relations (\ref{nounit}), it is straightforward to calculate the induction relations from each $\rho_{H^i}^k$ irrep of each maximal subgroup ${H^i}$ of $G^{Q}$. Using the tables of characters of the magnetic point groups in the database {\color{blue} CorepresentationsPG} (\href{http://www.cryst.ehu.es/cryst/corepresentationsPG}{www.cryst.ehu.es/cryst/corepresentationsPG}) of the BCS, first we calculate the coefficients $s_{jk}^{H^i}$ in the subsuction relations of Eq. (\ref{sub}). Then, using the relations (\ref{frobnounit}), (\ref{factoru}) and (\ref{nounit}) we determine the induction coefficients $c_{jk}^{H^i}$ in Eq. (\ref{ind}). 

For instance, Table (\ref{tab:exampleind}) shows the traces of the symmetry operations in the double irreps of the unitary Shubnikov point group (SPG) $\overline{1}$ (N. 3 in table (\ref{tab:ci})), of its unitary subgroup 1 (N. 1) and of the non-unitary SPG $\overline{1}1'$ (N. 4) and its subgroup $11'$ (N. 2). The labels of the symmetry operations of the double groups follow the notation in Ref. (\onlinecite{elcoro_double_2017}). The action of two operations  $R$ and $^dR$ of the double point group differ in the spin space (one of them acts as the other one followed by the inversion in the spin space).
\begin{table}
	\caption{Traces of the matrices of the symmetry operations for the double valued irreps in the Shubnikov point groups $\overline{1}$ (N. 3 in \ref{tab:ci}), 1 (N. 1), $\overline{1}1'$ (N. 4) and $11'$ (N. 2). In the last two cases, only the unitary operations are given.}
	\label{tab:exampleind} 	\begin{tabular}{l|c|rrrr|}
		SPG&irrep&1&$^d1$&$\overline{1}$&$^d\overline{1}$\\
		\hline
		\multirow{2}{*}{$\overline{1}$}&$\overline{A}_g$&1&-1&1&-1\\
		&$\overline{A}_u$&1&-1&-1&1\\
		\hline
		1&$\overline{A}$&1&-1&&\\
		\hline
		\multirow{2}{*}{$\overline{1}1'$}&$\overline{A}_g\overline{A}_g$&2&-2&2&-2\\
		&$\overline{A}_u\overline{A}_u$&2&-2&-2&2\\
		\hline
		11'&$\overline{A}\overline{A}$&2&-2&&\\
	\end{tabular}
\end{table}
Using the Schur orthogonality the subduction from the irreps of the unitary point group $\overline{1}$ to the irreps of its subgroup 1 is,
\begin{eqnarray}
\overline{A}_g\downarrow1&=&\overline{A}\\
\overline{A}_u\downarrow1&=&\overline{A}
\end{eqnarray}
and the induction relation from the irreps of point group 1 to the irreps of $\overline{1}$ is,
\begin{equation}
\label{inductionfrom1}
\overline{A}\uparrow\overline{1}=\overline{A}_g\oplus\overline{A}_u
\end{equation}
If we consider now the non-unitary point groups, according to (\ref{inductionfrom1}), every irrep $\overline{A}$ in SPG 1 induces a pair of irreps, $\overline{A}_g,\overline{A}_u$. Therefore, the \emph{doubled} irrep $\overline{A}\overline{A}$ in SPG 1 will induce two pairs of irreps $\overline{A}_g,\overline{A}_u$ and then a single $\overline{A}\overline{A}$ in 11' induces the following representation into $\overline{1}1'$,
\begin{equation}
\overline{A}\overline{A}\uparrow\overline{1}1'=\overline{A}_g\overline{A}_g\oplus\overline{A}_u\overline{A}_u
\end{equation}
This result can be obtained using eqs. (\ref{ind}), (\ref{frobnounit}) and (\ref{nounit}), noting that $s_{\overline{A}_g\overline{A}_g,\overline{A}\overline{A}}^{11'}=s_{\overline{A}_u\overline{A}_u,\overline{A}\overline{A}}^{11'}=(2\times2+(-2)\times(-2))/2=4$ in eq. (\ref{nounit}), $u(\overline{1}1',11')=1$ and $f(\overline{A}_g\overline{A}_g,\overline{A}\overline{A})=f(\overline{A}_u\overline{A}_u,\overline{A}\overline{A})=f(b,b)=\frac{1}{4}$.

For every irrep $\rho_{H^i}^k$ of every maximal subgroup ${H^i}$, we define a column vector $(c_{1k}^{H^i},c_{2k}^{H^i},\ldots,c_{n_{G^Q}k}^{H^i})^{T}$ whose coordinates represent the multiplicities of the $\rho_{G^Q}^j$ irreps of ${G^Q}$ for the induced representation $\rho_{H^i}^k\uparrow G^Q$. We can then write the \emph{induction matrix} $C_{G^{Q}}$ introduced in the supplementary material of Ref. (\onlinecite{song2020}):
\begin{equation}
\label{inductionmatrix}
C_{G^{Q}}=\left(\begin{array}{ccccccc}
c_{1,1}^{H^1}&\ldots&c_{1,n_{H^1}}^{H^1}&c_{1,1}^{H^2}&\ldots&c_{1,n_{H^2}}^{H^2}&\ldots\\
c_{2,1}^{H^1}&\ldots&c_{1,n_{H^1}}^{H^1}&c_{2,1}^{H^2}&\ldots&c_{2,n_{H^2}}^{H^2}&\ldots\\
\ldots&\ldots&\ldots&\ldots&\ldots&\ldots&\\
c_{n_{G^Q},1}^{H^1}&\ldots&c_{n_{G^Q},n_{H^1}}^{H^1}&c_{n_{G^Q},1}^{H^2}&\ldots&c_{n_{G^Q},n_{H^2}}^{H^2}&\ldots
\end{array}\right)
\end{equation}
It is a $n_{G^Q}\times N_{H}$ matrix whose number of rows $n_{G^Q}$ is the number of irreps of $G^{Q}$, and the number of columns $N_{H}$ is the total number of irreps of the maximal subgroups,
\begin{equation}
N_{H}=\sum_{i=1}^{n_{H}}n_{H^i}
\end{equation}
($n_{H}$ is the number of maximal subgroups of $G^{Q}$ and $n_{H^i}$ is the number of irreps of subgroup $H^i$).

We have calculated the induction matrices of all the 122 site-symmetry groups realized in the 1651 magnetic groups. Table \ref{tab:ci} includes the result of the calculation. The first two columns show the number and symbol of the magnetic point group isomorphic to the site-symmetry group $G^{Q}$. The third column gives the list of symbols(numbers) of the magnetic point groups isomorphic to the maximal site-symmetry subgroups of $G^{Q}$. Subgroups with the
same number are non-conjugated subgroups with respect to the operations in the supergroup. In general, they can give different induction relations (note that although, in general, they can give different induction relations, for double valued irreps the induction relations are exactly the same.
This result is different from the result obtained for single valued irreps, not considered in this work.). 
For instance,  the point group $4/mmm$ (number 53 in Table (\ref{tab:ci})) has two non-conjugated subgroups with standard notation $mm2$. In one case the 2-fold axis is parallel to the (1,0,0) or (0,1,0) direction (both options give conjugated groups with respect to the 4-fold axis of $4/mmm$) and in the other case the 2-fold axis is parallel to the direction (1,1,0) or (1,-1,0) (both options give also conjugated groups). The fourth column gives the list of irreps of the site-symmetry group and the last column the induction matrix (\ref{inductionmatrix}).

The $C$ matrix contains all the required information to calculate the RSI indices at the maximal WPs. However, the calculation of the
RSI indices at non-maximal WPs requires extra terms in the $C$ matrix (\ref{inductionmatrix}). In the rest of this section, we restrict
ourselves to the calculation of the RSI indices at maximal WPs and leave the complete analysis at non-maximal WPs
for the next section.

With the help of Table \ref{tab:ci}, we can determine whether a set of orbitals in a given maximal WP $Q$ can be moved away from $Q$ along the WPs of lower symmetry connected to it. First we identify  the site-symmetry group $G^Q$ of $Q$, the number of the point group isomorphic to $G^Q$ and the list of irreps of $G^Q$. Next we consider a list of orbitals located at $Q$ that transform as a representation $\rho$ that, in general, is the direct sum of the irreps of $G^Q$. We can represent the set of orbitals through a symmetry data vector $p$, whose components are the multiplicities of the irreps of  $G^Q$ in the direct sum of $\rho$,
\begin{equation}
\label{simdata}
p=(m(\rho_{G^Q}^1),m(\rho_{G^Q}^2),\ldots,m(\rho_{G^Q}^{n_Q}))^T
\end{equation}
Every column $(c_{1k}^{H^i},c_{2k}^{H^i},\ldots,c_{n_{G^Q}k}^{H^i})^{T}$ in Eq. (\ref{inductionmatrix}) represents the multiplicities of the irreps in $G^Q$ of the induced representation $\rho_{H^i}^k\uparrow G^{Q}$. Physically this means that $c_{1k}^{H^i}$ orbitals that transform as $\rho_{G{^Q}}^1$ and $c_{2k}^{H^i}$ orbitals that transform as $\rho_{G{^Q}}^2$, \ldots and $c_{n_{G^Q}k}^{H^i}$ orbitals that transform under $\rho_{G{^Q}}^{n_{G^Q}}$ can be moved away from $Q$ to the WP with site-symmetry group isomorphic to $H^i$ to give an orbital or sum of orbitals that transform under $\rho_{H^i}^k$.
Therefore, a set of orbitals at $Q$ given by the symmetry data vector $p$ in Eq. (\ref{simdata}) can be moved away from $Q$ iff it is a linear combination of the columns in (\ref{inductionmatrix}), i.e., if there exists a vector $X=(x_1,\ldots,x_{n_Q})^{T}$ with integer components such that the relation,
\begin{equation}
\label{eq:cmat}
C\cdot X=p
\end{equation}
is fulfilled.

Note that we are looking for an integer vector $X$ which represents the multiplicities of the irreps at the non-maximal WPs connected to $Q$, where the orbitals can be moved. A solution of Eq. (\ref{eq:cmat}) could, in principle, contain negative components which means that the resulting set of orbitals transform as a difference of irreducible representations. Such a case (if no other solution to Eq. (\ref{eq:cmat}) exists with all the components being non-negative integers) corresponds to a fragile obstructed atomic insulator which will be discussed in Ref. \cite{fragileOAI}. Our present analysis thus,  enables the identification of this type of materials. In such a case, we will find that the RSI indices at the maximal WP $Q$ are all 0 and that the structure is compatible with no orbitals hosted at $Q$. We leave the analysis of these fragile OAIs for future works \cite{fragileOAI}.

The determination of the conditions to be fulfilled by the symmetry data vector $p$ to be, at least, a solution of Eq. (\ref{eq:cmat}) can be simplified if we perform the Smith decomposition of the (integer) $C$ matrix of dimension $n_{G^Q}\times N_{H}$,
\begin{equation}
\label{eq:cdesc}
C=L_C\cdot\Delta_C\cdot R_C
\end{equation}
$L_C$ and $R_C$ are unimodular matrices of dimension $n_{G^Q}\times n_{G^Q}$ and $N_{H}\times N_{H}$, respectively, and the Smith normal form $\Delta_C$ is an integer diagonal matrix ($\Delta_{ij}=0$ for $i\ne j$) of the same dimension and rank of $C$. The number of non-zero elements in the diagonal is the rank $r$ of $C$, and $L_C$ and $R_C$ can be chosen such that $1\le\Delta_{11}\le\Delta_{22}\le\ldots\le\Delta_{rr}$. In the following we will assume this choice. 

Being that $L_C$ and $R_C$ in Eq. (\ref{eq:cdesc}) are unimodular matrices, we can define the vector with integer components $Y=R_CX$. Substituting the decomposition (\ref{eq:cdesc}) into Eq. (\ref{eq:cmat}), using the definition of $Y$ and multiplying both sides by $L_C^{-1}$ , Eq. (\ref{eq:cmat}) takes the following form where we must solve for $Y$,
\begin{equation}
\label{eq:cmat2}
\Delta\cdot Y=L_C^{-1}\cdot p
\end{equation}
The left hand side of Eq. (\ref{eq:cmat2}) is,
\begin{equation}
\label{eq:ycomp}
(\Delta_{11}Y_1,\Delta_{22}Y_2,\ldots,\Delta_{rr}Y_r,0,\ldots,0)^T
\end{equation}
being the number of 0 the difference between the number of irreps of $G^Q$ and the rank of the $C$ matrix, $n_{G^Q}-r$. The symmetry data vector $p$ must fulfill several constrains such that Eq. (\ref{eq:cmat}) or (\ref{eq:cmat2}) has a solution for $Y$. On the one hand, if $n_{G^Q}>r$, Eq. (\ref{eq:cmat2}) implies,
\begin{equation}
\label{eq:delta}
\left(L_C^{-1}\cdot p\right)_i=0\hspace{0.5cm}\textrm{ for }i=r+1,\ldots,n_{G^Q}
\end{equation}
On the other hand, as the $Y_i$ components are integer numbers, for every diagonal component of the Smith form $\Delta_{jj}>1$, the condition,
\begin{equation}
\label{eq:zeta}
\left(L_C^{-1}\cdot p\right)_j\mod\Delta_{jj}=0
\end{equation}
has to be fulfilled.

Following the analysis of 2D groups done in the supplemental material of Ref. \cite{song2020}, we can define two types of indices in every site-symmetry group,
\begin{equation}
\label{listindices}
\delta_i(p)=\left\{\begin{array}{lll}
\left(L_C^{-1}\cdot p\right)_i\mod\Delta_{ii}&\hspace{0.5cm}&i\le r\\
\left(L_C^{-1}\cdot p\right)_i&\hspace{0.5cm}&i> r
\end{array}\right.
\end{equation}
which correspond to conditions (\ref{eq:zeta}) and (\ref{eq:delta}), respectively. Obviously, $\Delta_{ii}=1$ components do not impose any restriction and only $\Delta_{ii}>1$ components give meaningful indices. If any index in Eq. (\ref{listindices}) $\delta(p)\neq0$, it means that the orbitals at the corresponding Wyckoff position $Q$ cannot be moved away from $Q$.

We have checked that, among the 122 $C$ matrices listed in Table \ref{tab:ci}, the largest $\Delta_{ii}$ component is 2, so the first kind of indices in Eq. (\ref{listindices}), if any, are $Z_2$-type indices. The second kind of indices in Eq. (\ref{listindices}), if any, are referred to as $Z$-type indices.

\subsubsection{RSI indices in non-maximal Wyckoff positions: the exchange term}
\label{sub:exchange}
The algorithm developed in section \ref{calcRSI} allows  the determination of the RSI indices in the maximal WPs given by Eq. (\ref{listindices}) and, using these indices, to check whether a set of orbitals can be moved (or not) away from the maximal WPs. However, the $C$ matrix (\ref{inductionmatrix}) obtained by the subduction-induction relations between a site-symmetry group and its maximal subgroups is not enough to determine the RSI indices at non-maximal WPs, precisely because they are connected to maximal WPs. This connection gives an extra possibility to interchange some orbitals with these maximal WPs.

Let us take a pair of WPs, $q$ and $Q$ such that $G^{q}\subset G^{Q}$, \ie the WP $Q$ sits on $q$ ($Q$ can represent an isolated point and $q$ a line or a plane which contains $Q$ or $Q$ can represent a line contained in a plane $q$). Let us also assume that there exists a symmetry operation $g\in G^{Q}$ but $g\notin G^{q}$ such that it keeps $G^{q}$ invariant under conjugation, \ie
\begin{equation}
\label{conjugation}
\exists g\in G^{Q}\,and\, g\notin G^{q},~s.t.,~\forall h\in G^{q}\,|\,g^{-1}hg\in G^{q}
\end{equation}
In a geometrical picture, if $Q$ is an isolated point and $q$ a line (plane) where $Q$ sits, the operation $g$ transforms any point $q_i-Q$ in the line(plane) $q$ into a different point $-q_i-Q$, also in the same line(plane) $q$. For instance, in SSG $P2$ the high symmetry point $Q$ can be the origin and $q$ the line $(x,0,0)$. The two-fold axis $\{2_y|0,0,0\}$ parallel to $y$ transforms the point $(x_0,0,0)$ into $(-x_0,0,0)$, in the same line $q$. 
The existence of such operations implies relations of conjugation between the irreps of $G^{q}$. In fact, the relations between the irreps in a subgroup under the conjuation by elements of a supergroup are the central point of the determination of the irreps of a group from the irreps of one of its subgroups in the induction process (see for example Ref. \cite{aroyo2006}).

Let $D(h)$ the matrix of the symmetry operation $h$ of an irrep $\rho_{G^{q}}$ of the site-symmetry group $G^{q}$ and $g$ an operation of the site-symmetry group of $G^{Q}$ such that $G^{q}\subset G^{Q}$, $g\in G^{Q}$ but $g\notin G^{q}$. Let us also assume that the relation in Eq. (\ref{conjugation}) is fulfilled. 
The set of matrices $D(g^{-1}hg)$ are also the matrices of an irrep of $G^{q}$. There are two possibilities: (1) the two sets of matrices are equivalent, \ie they correspond to the same irrep of $G^{q}$. In such a case, it is said that the irrep $\rho_{G^{q}}$ is self-conjugated under $g$ and the induced representation $\rho_{G^{q}}\uparrow G^{Q}$ is a direct sum of irreps of $G^{Q}$. (2) the set of matrices $D(h)$ and $D(g^{-1}hg)$ are not equivalent, but the two irreps given by $D(h)$, $\rho_{G^{q}}$ and by $D(g^{-1}hg)$, $\rho_{G^{q}}'\neq \rho_{G^{q}}$ are mutually conjugated. Both irreps in $G^{q}$ yield a single irrep in the supergroup $G^{Q}$. For instance, the group 2 (N. 6 in (\ref{tab:ci})), with the 2-fold axis paralell to $y$, has two double valued irreps whose traces are shown in Table (\ref{example:exchange}).
\begin{table}
	\caption{Traces of the matrices of the symmetry operations (given in the first row) for the double valued irreps in the Shubnikov point group $2$ (N. 6 in (\ref{tab:ci})). The first column shows the symbol of the two irreps in the group.}
	\label{example:exchange} 	\begin{tabular}{c|rrrr|}
		&1&$^d1$&$2_y$&$^d2_y$\\
		\hline
		$^1\overline{E}$&1&-1&$i$&$-i$\\
		$^2\overline{E}$&1&-1&$-i$&$i$
	\end{tabular}
\end{table}
 When this group is considered as a subgroup of 222 (N. 17 in Table (\ref{tab:ci})), the two irreps are conjugated with respect to the $2_x$ operation. Taking into consideration the following relations,
 \begin{eqnarray}
 \label{example:conjugation}
 2_x^{-1}\cdot1\cdot2_x&=&1\nonumber\\
 2_x^{-1}\cdot\,^d1\cdot2_x&=&\,^d1\nonumber\\
 2_x^{-1}\cdot2_y\cdot2_x&=&\,^d2_y\nonumber\\
 2_x^{-1}\cdot\,^d2_y\cdot2_x&=&2_y
 \end{eqnarray}
the matrices of the irrep $^1\overline{E}$ in Table (\ref{example:exchange}), $D(1)=1$, $D(\,^d1)=-1$, $D(2_y)=i$ and $D(\,^d2_y)=-i$ are transformed under conjugation into the matrices $D(2_x^{-1}\cdot1\cdot2_x)=D(1)=1$, $D(2_x^{-1}\cdot\,^d1\cdot2_x)=D(\,^d1)=-1$, $D(2_x^{-1}\cdot2_y\cdot2_x)=D(\,^d2_y)=-i$ and $D(2_x^{-1}\cdot\,^d2_y\cdot2_x)=D(2_y)=i$, which correspond to the traces of $^2\overline{E}$ in Table (\ref{example:exchange}). Therefore, the two irreps of the point group 2 are mutuallty conjugated under the operation $2_x$ of the supergroup 222. An identical relation is obtained under conjugation by $^d2_x$, $2_z$ or $^d2_z$.

In our physical analysis of electronic orbitals, the existence of a WP $Q$ of high symmetry connected to a WP of lower symmetry has important consequences in the possibility of moving orbitals away from the WP $q$. If there is an orbital centered at a given point $q_i$ of the WP $q$ at one side of $Q$, there must be another orbital at another $q_i'$ point of the WP $q$ such that $q_i-Q=-(q_i'-Q)$. Depending on $Q$ and $q_i$, the symmetry operation which transforms $q_i-Q$ into $-(q_i'-Q)$ can be a 2-fold axis, the inversion, a mirror plane $m$ or a roto-inversion $\overline{3}$, $\overline{4}$, $\overline{6}$. If the orbital at $q_i$ transforms under an irrep $\rho_{G^Q}$ of $G^{q}$ and this irrep is self-conjugated with respect to a symmetry operation $g\in G^{Q}$ but $g\notin G^{q}$, then the orbital centered at $q_i'$ transforms also under the irrep $\rho_{G^Q}$. In an adiabatic process, both orbitals can move towards the WP $Q$, and induce a combination of electronic orbitals that transform under the induced representation. Next, it would be possible to go back and move the orbitals again to $q$.

\begin{figure}[htbp]
\centering\includegraphics[width=3.5in]{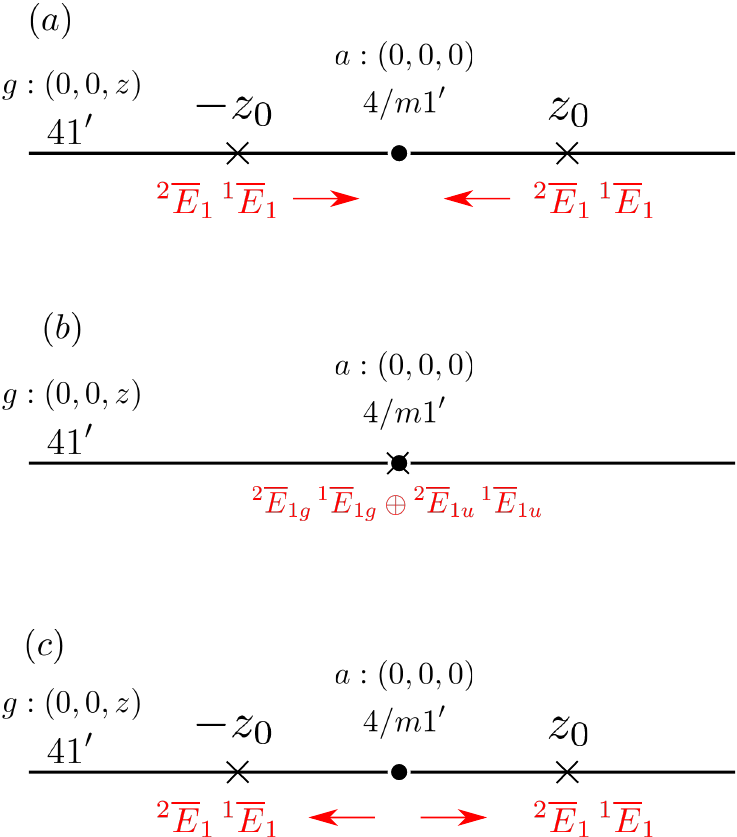}
\caption{Evolution of a set orbitals in an adiabatic process along the high-symmetry line $g:(0,0,z)$ with site-symmetry group $4'$ in the Shubnikov group $4/m1'$ (N. 83.44). (a) Two orbitals that transform under the same irrep $^2\overline{E}_1\,^1\overline{E}_1$ are symmetrically located at $z=\pm z_0$ and move towards the WP $a:(0,0,0)$. The $^2\overline{E}_1\,^1\overline{E}_1$ irrep of the point group $41'$ is self-conjugated under the symmetry operation $\mathcal{M}_z$ of the site-symmetry group of WP $a$ ($4/m1'$), so that the irreps at $\pm z_0$ are forced to be the same. (b) During the adiabatic process both irreps \emph{merge} at $a$ in a set of orbitals that transform under the direct sum of irreps $^2\overline{E}_{1g}\,^1\overline{E}_{1g}\oplus ^2\overline{E}_{1u}\,^1\overline{E}_{1u}$ of the point group $4/m1'$. (c) Following the adiabatic process, the orbitals move away from the $a$ point along the line $g$. The final configuration is the same as the first one and there is no exchange term in this case}
\label{selfconj}
\end{figure}

 An example of self conjugated irreps (see Fig. \ref{selfconj}) are the irreps of the example of type-II SSG $P4/m1'$ (N. 83.44) given in the text surrounding Eqs. (\ref{eq:exaind1}) and (\ref{eq:exaind2}). On the left side of Eq. (\ref{eq:exaind1}), $\,^2\overline{E}_{1}\,^1\overline{E}_{1}$ is an irrep of the site-symmetry group $G^{g}$ of the line $g:(0,0,z)$. The group is isomorphic to the point group $41'$. This line is connected to the WP $a$, which is isomorphic to the point group $4/m1'$. For the mirror symmetry $\mathcal{M}_z \in G^a$ but $\notin G^g$, we have $\mathcal{M}_z(\,^2\overline{E}_{1}\,^1\overline{E}_{1})\mathcal{M}_z^{-1}=\,^2\overline{E}_{1}\,^1\overline{E}_{1}$. Hence, the irrep $\,^2\overline{E}_{1}\,^1\overline{E}_{1}$ is self conjugated with respect to $\mathcal{M}_z$.
 The Eq. (\ref{eq:exaind1}) shows that the induced representation $\,^2\overline{E}_{1}\,^1\overline{E}_{1}\uparrow G^{a}$  is the direct sum of the two irreps, $\,^2\overline{E}_{1g}\,^1\overline{E}_{1g}$ and $\,^2\overline{E}_{1u}\,^1\overline{E}_{1u}$ in $G^{a}$. In our analysis of electronic bands, this means that two orbitals centered at two points $(0,0,\pm z_0)$ of the line $g$ that transform both under the irrep $\,^2\overline{E}_{1}\,^1\overline{E}_{1}$ can be moved to $a$ to form two orbitals that transform under $\,^2\overline{E}_{1g}\,^1\overline{E}_{1g}$ and $\,^2\overline{E}_{1u}\,^1\overline{E}_{1u}$. These two orbitals can be moved back to the $g$ line recovering the initial distribution of charges.
		
\begin{figure}[htbp]
\centering\includegraphics[width=3.5in]{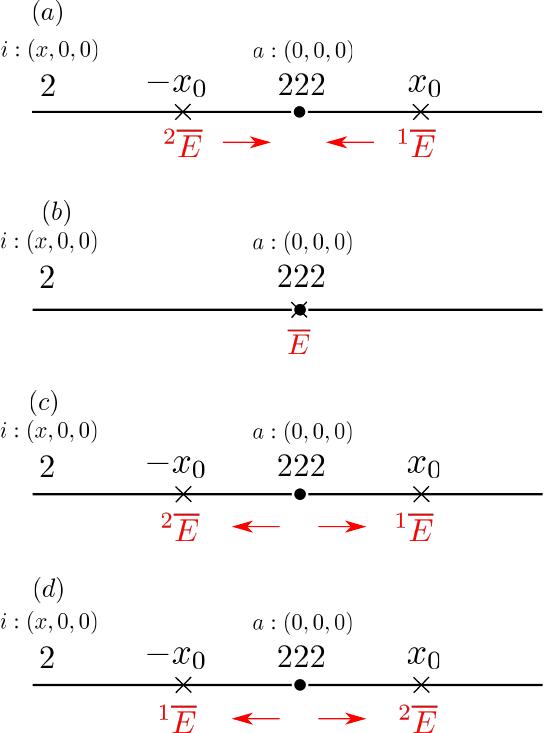}
\caption{Evolution of a set orbitals in an adiabatic process along the high-symmetry line $i:(x,0,0)$ with site-symmetry group $2$ in the Shubnikov group $P222$ (N. 16.1). (a) Two orbitals that transform under the irreps $^1\overline{E}$ and $^2\overline{E}$ are symmetrically located at $x=\pm x_0$ and move towards the WP $a:(0,0,0)$. The irreps $^1\overline{E}$ and $^2\overline{E}$ of the point group $2$ are mutually conjugated under the symmetry operations of the site-symmetry group of WP $a$ ($222$), so that the irreps at $\pm z_0$ are forced to be $^1\overline{E}$ and $^2\overline{E}$. (b) During the adiabatic process both irreps \emph{merge} at $a$ in a set of orbitals that transform under the irrep $\overline{E}$ of the point group $222$. (c) Following the adiabatic process, the orbitals move away from the $a$ point along the line $i$. The final configuration is the same as the first one. (d) Alternatively, in the adiabatic process the orbitals can move away from (a) in a different way. The final configuration is different from the initial one. In this case, it is necessary to consider the exchange term in the induction matrix $C$ (see the text) to take into consideration these interchanges of irreps.}
\label{fig:mutconj}
\end{figure}

However, when the orbitals at the line of lower symmetry transform under mutually conjugated irreps at both sides of the high symmetry point, during the closed adiabatic process, the final configuration can be different from the initial one. For instance, let us consider the simple case of WPs $i$ and $a$ in the type-I double SSG $P222$ (N. 16.1) which does not contain anti-unitary element. The site-symmetry groups $G^{i}$ and $G^{a}$ are isomorphic to point groups $2$ (with element $C_{2x}$) and $222$ (with elements $C_{2x}$ and $C_{2y}$), respectively. $a$:$(0,0,0)$ lies in the line $i:(x,0,0)$ such that $G^{i}\subset G^{a}$. The unique double valued irrep $\overline{E}$ of point group $222$ subduces into irreps of $G^{i}$ according to the relation,
\begin{equation}
\overline{E}\downarrow G^{i}=\,^2\overline{E}\oplus\,^1\overline{E}
\end{equation}
and the induction relations are,
\begin{eqnarray}
^2\overline{E}\uparrow G^{a}=\overline{E}\\
^1\overline{E}\uparrow G^{a}=\overline{E}
\end{eqnarray}
Physically, these relations imply that, if there is an orbital at some point $(x_0,0,0)$ in the line $i$ that transforms under the 1-$d$ irrep $\,^2\overline{E}$, there must be another orbital at point $(-x_0,0,0)$ that transforms under the 1-$d$ irrep $\,^1\overline{E}$, which is the conjugated irrep of $\,^2\overline{E}$ with respect to the operations $C_{2y},C_{2z}\in G^{a}$. This pair of orbitals can be adiabatically moved to WP $a$ to form a pair of orbitals that transform under the 2-$d$ irrep $\overline{E}$. Next, this pair of orbitals can be moved back to the line $i$, but in two different ways. We can recover the starting state with the orbital $\,^2\overline{E}$ at $(0,0,x_0)$ and the orbital $\,^1\overline{E}$ at $(-x_0,0,0)$, or it is also possible to move the orbital $\,^1\overline{E}$ to $(x_0,0,0)$ and $\,^2\overline{E}$ at $(-x_0,0,0)$. In this second case, the two regions of the $i$ line have \emph{exchanged} the irreps through the high symmetry WP $a$ during the adiabatic process (see Fig. \ref{fig:mutconj}). Therefore, we must also consider these possible re-organizations of the orbitals introducing extra columns in the induction matrix $C$ that defined in Eq. (\ref{inductionmatrix}) for all the group-subgroup pairs. 

In general, if the $i^{\textrm{th}}$ and $j^{\textrm{th}}$ irreps of a site-symmetry group $G^{q}$ of a non-maximal Wyckoff position $q$ are mutually conjugated under the operations of the site-symmetry group $G^{Q}$ of the Wyckoff position $Q$, the column added to the induction matrix $C$ is given by,
\begin{equation}
\label{exchangematrix}
C_{ex}=(0,\ldots,1,0,\ldots,-1,0,\ldots,0)^T
\end{equation}
where the integers 1 and -1 are the $i^{\textrm{th}}$ and $j^{\textrm{th}}$ coordinates of the column vector, respectively, or vice-versa. We should add as many exchange columns as pairs of conjugated irreps under the operations of $Q$ or under the operations of any other WP of higher symmetry that lies on $q$.

Table \ref{tab:cex} shows all the possible group-subgroup pairs in the 1651 SSGs that give at least an exchange column. Group-subgroup pairs that give no exchange terms have not been included in the list. This happens when all the irreps of $G^{q}$ are self-conjugated with respect to all operations of the site-symmetry groups of maximal WPs $Q$ connected to it. First and second columns in the table show the number and symbol of the subgroup, third and fourth columns the number and symbol of the supergroup, and the fifth and sixth columns give the irreps of the subgroup and the exchange matrix due to the conjugation relations under the symmetry operations in the supergroup. Note that all SSGs in this table correspond to magnetic groups of type I, III and IV. No SSG of type II (then, no space group of a non-magnetic system) has pairs of WPs that exchange two double valued irreps. For example in the case of Fig.~\ref{fig:mutconj}, adding TRS will make the mutually conjugated irreps $^1\overline{E}$ and $^2\overline{E}$ degenerated and form a 2-$d$ self-conjugated irrep $^1\overline{E} ^2\overline{E}$.

Tables \ref{tab:ci} and \ref{tab:cex} contain all the necessary information to construct the full $C$ matrix of the site-symmetry group of any WP in any SSG and, from this matrix, to calculate the RSI indices of the WP following the procedure of section \ref{calcRSI}. It must be stressed that the $C$ matrix is not unique for every point group, because it depends on the existence of other point groups isomorphic to site-symmetry groups of higher symmetry in the SSG. For instance, the $C$ matrix of the WP $a$ in the SSG $P2$ (N. 3.1), whose site-symmetry group is isomorphic to point group 2 is just the matrix in the 6$^{\textrm{th}}$ entry of Table \ref{tab:ci}, because $a$ is not connected to any other WP with higher symmetry,
\begin{equation}
\label{eqexexa1}
C_{P2}^{a}=\left(\begin{array}{c}
1\\
1
\end{array}\right)\end{equation}
 However, in the SSG $P\overline{4}2m$ (N. 251), the WP $i$ has also a site-symmetry group isomorphic to point group 2, but WP $i$ is connected to $a$ and $e$ with site-symmetry groups isomorphic to $\overline{4}2m$ and $222$, respectively. Looking at the Table \ref{tab:cex} with entries $G^{q}=6$ and $G^{Q}=48$ on the one hand, and $G^{q}=6$ and $G^{Q}=17$ on the other hand, we see that the two irreps in point group 2 are conjugated by the operations of $\overline{4}2m$ and $222$, so the $C$ matrix of WP $i$ must be extended adding the extra columns due to the exchange term. The two columns added by the connection to $a$ and $e$ are identical (see the last column in Table \ref{tab:cex}), and it is enough to add one of them to the $C$ matrix (because only the linearly independent columns matter the solution of Eq. (\ref{eq:cmat})),
 \begin{equation}
 \label{eqexexa2}
 C_{P\overline{4}2m}^i=\left(\begin{array}{rr}
 1&1\\
 1&-1
 \end{array}\right)\end{equation}
 The indices (\ref{listindices}) given by the $C$ matrices (\ref{eqexexa1}) and (\ref{eqexexa2}) are different.
 
 We have implemented the tools {\color{blue} RSI} (\href{http://www.cryst.ehu.es/cryst/RSI}{www.cryst.ehu.es/cryst/RSI}) and {\color{blue} MagRSI} (\href{http://www.cryst.ehu.es/cryst/MagRSI}{www.cryst.ehu.es/cryst/MagRSI}) in the Bilbao Crystallographic Server for non-magnetic and magnetic groups, respectively. These tools give the $C$ matrix for all the WPs in the chosen SSG together with the RSI indices (\ref{listindices}) derived from the matrix.
 
 \subsection{RSI indices of SSGs in momentum space}
 \label{app:momentum}
 In the previous sections we have developed an algorithm to decide whether a given set of orbitals can be moved from a given specific WP to another WPs connected to the first one. For the present work (see the main text), it allows to elucidate if a material whose band structure corresponds to an insulator without stable topology has a set of orbitals pinned at empty WPs or not, and then determine whether the material is a trivial insulator or an OAI. However, the previous analysis relies in the knowledge of the specific orbitals located at each WP, and this information is not direct because, in general, it involves time-consuming involved calculations based on the whole set of wave-functions in the whole first Brillouin zone, such as sets of Wilson loops. In this section, we use the relationship between the localized electronic orbitals in direct space and the electronic bands (or extended states) in momentum space to simplify the analysis. It must be stressed, however, that the RSI indices calculated in this section are unable to identify the fragile OAIs. Moreover, a result \emph{compatible with} a trivial atomic insulator (not OAI) based on the RSI indices obtained in this section does not mean that the material is a trivial insulator. Other calculations (as sets of Wilson loops) are necessary to discriminate true trivial insulators from materials with no trivial topology not detected by symmetry indicators.
 
 The relation between the localized orbitals around a WP in direct space and the extended electronic states in momentum space using symmetry consideration was firstly introduced by Zak \cite{zak1982band}, and Bacry et al. \cite{Bacry1988} in systems without SOC. The method is based on the site-symmetry approach \cite{Evarestov1997}. The relation is constructed recognizing that an orbital located at a given WP $Q$ transforms under an irrep $\rho_{G^{Q}}$ of the site-symmetry group $G^{Q}$ of that WP. As the site-symmetry group $G^{Q}$ is a finite subgroup of the whole space group $G$ in a non-magnetic system (\ie $G^{Q}\subset G$), the irrep $\rho_{G^{Q}}$ induces a representation in $G$, $\rho_{G^{Q}}\uparrow G$, known as band representation, and it is reducible. Each band representation is a direct sum of the irreps of the space group, and can be denoted as a row vector of integer components that represent the multiplicities of all the irreps assigned to all the $k$-vectors in the first Brillouin zone. In fact, it is enough to consider the multiplicities of the irreps of the $k$-vectors of maximal symmetry (\ie the maximal $k$-vectors) in the space group, because the multiplicities of the irreps at non-maximal $k$-vectors are determined by the former ones through the so-called compatibility relations (see, for example, Ref. \cite{bradlyn_topological_2017} or \cite{elcoro_double_2017}). In the last years, Zak's theory has been extended to systems with SOC and applied to the analysis of the topology of materials. The theoretical background, known as Topological Quantum Chemistry (TQC) was developed in a series of papers \cite{bradlyn_topological_2017,vergniory_graph_2017,elcoro_double_2017,SlagerSymmetry,song_quantitative_2018,cano_building_2018}, where all the details of TQC and examples of application can be found. Afterwards, the concept of band representation was also adapted to magnetic space groups, defining the band co-representations \cite{EvarestovMEBR}. A equivalent theory to TQC in magnetic systems, Magnetic Topological Quantum Chemistry (MTQC) has also been recently developed \cite{MTQC,xu2020high}, together with other alternative methods \cite{watanabe2018structure,peng2021}. 
 
 For our purposes, it is important to realize that every well localized orbital around a given WP $Q$ that transforms under a given irrep $\rho_{G^Q}$ of $G^Q$ induces a band representation $\rho_{G^Q}\uparrow G$ in $G$ described by a set of integers that correspond to the multiplicities of the irreps at high symmetry $k$-vectors (the symmetry data-vector defined in Eq. (\ref{eq:B-vector})). All these multiplicities were tabulated for all the irreps of all WPs in all the 230 space groups both for single-valued (no SOC) and double valued (SOC) irreps and implemented in the tool {\color{blue} BANDREP} (\href{http://www.cryst.ehu.es/cryst/bandrep}{www.cryst.ehu.es/cryst/bandrep}) in the BCS. The band co-representations of the magnetic space groups have been recently implemented in the tool of the BCS {\color{blue} MBANDREP} (\href{http://www.cryst.ehu.es/cryst/mbandrep}{www.cryst.ehu.es/cryst/mbandrep}). The notation and characters of the irreps on magnetic and non-magnetic space groups used in this work have been taken from the mentioned tools.
 
 Using the example of section \ref{RSI:theory}, the list of maximal $k$-vectors of the type-II SSG $P4/m1'$ are: $\Gamma:(0,0,0)$ of littile group $4/m$, $A:(1/2,1/2,1/2)$ of littile group $4/m$, $M:(1/2,1/2,0)$ of littile group $4/m$, $R:(0,1/2,1/2)$ of littile group $2/m$, $X:(0,1/2,0)$ of littile group $2/m$ and $Z:(0,0,1/2)$ of littile group $4/m$. The full set of 20 maximal double valued irreps at these $k$ points are,
 \begin{equation}
 \label{irreps}
 \begin{array}{l}
 \overline{\Gamma}_5\overline{\Gamma}_7,\,\,
 \overline{\Gamma}_6\overline{\Gamma}_8,\,\,
  \overline{\Gamma}_{10}\overline{\Gamma}_{12},\,\,
   \overline{\Gamma}_{11}\overline{\Gamma}_9,\,\,
 \overline{A}_5\overline{A}_7,\,\,
 \overline{A}_6\overline{A}_8,\,\,
 \overline{A}_{10}\overline{A}_{12},\,\,
 \overline{A}_{11}\overline{A}_9,\,\,
 \overline{M}_5\overline{M}_7,\,\,
 \overline{M}_6\overline{M}_8,\,\,
 \overline{M}_{10}\overline{M}_{12},\,\,
 \overline{M}_{11}\overline{M}_9,\\
 \overline{R}_3\overline{R}_4,\,\,
 \overline{R}_5\overline{R}_6,\,\,
 \overline{X}_3\overline{X}_4,\,\,
 \overline{X}_5\overline{X}_6,\,\,
 \overline{Z}_5\overline{Z}_7,\,\,
 \overline{Z}_6\overline{Z}_8,\,\,
 \overline{Z}_{10}\overline{Z}_{12},\,\,
 \overline{Z}_{11}\overline{Z}_9
 \end{array}
   \end{equation}
According to Eq. (\ref{eq:exaind1}), an orbital located at some position in the line $g$ and that transforms under the irrep $\,^2\overline{E}_{1}\,^1\overline{E}_{1}$ of the site-symmetry group $G^g$ induces the direct sum of irreps $\,^2\overline{E}_{1g}\,^1\overline{E}_{1g}\oplus\,^2\overline{E}_{1u}\,^1\overline{E}_{1u}$ of the site-symmetry group $G^a$ of the WP $a$ connected to $g$. In the momentum space, according to {\color{blue} BANDREP} (\href{http://www.cryst.ehu.es/cryst/bandrep}{www.cryst.ehu.es/cryst/bandrep}), the band representation induced by the irrep $\,^2\overline{E}_{1}\,^1\overline{E}_{1}$ at $g$ has the following symmetry-data vector $B$,
\begin{equation}
\label{momentumg}
B(g,\,^2\overline{E}_{1}\,^1\overline{E}_{1})\rightarrow(0, 1, 1, 0, 0, 1, 1, 0, 0, 1, 1, 0, 1, 1, 1, 1, 0, 1, 1, 0)
	\end{equation}
	 Equivalently, BANDREP shows that the multiplicities of the band representations induced by the irreps $\,^2\overline{E}_{1g}\,^1\overline{E}_{1g}$ and $\,^2\overline{E}_{1u}\,^1\overline{E}_{1u}$ of $G^a$ are,
	\begin{eqnarray}
	\label{momentuma1}
	B(a,\,^2\overline{E}_{1g}\,^1\overline{E}_{1g})&\rightarrow&(0, 1, 0, 0, 0, 1, 0, 0, 0, 1, 0, 0, 1, 0, 1, 0, 0, 1, 0, 0)\\
	\label{momentuma2}
	B(a,\,^2\overline{E}_{1u}\,^1\overline{E}_{1u})&\rightarrow&(0, 0, 1, 0, 0, 0, 1, 0, 0, 0, 1, 0, 0, 1, 0, 1, 0, 0, 1, 0)
	\end{eqnarray}
and they fulfill the relation,
\begin{equation}
\label{summomentum}
B(g,\,^2\overline{E}_{1}\,^1\overline{E}_{1})=B(a,\,^2\overline{E}_{1g}\,^1\overline{E}_{1g})+B(a,\,^2\overline{E}_{1u}\,^1\overline{E}_{1u})
\end{equation}
The relation in Eq. (\ref{summomentum}) in momentum space is equivalent to the relation in Eq. (\ref{eq:exaind1}) in direct space, \ie we can split a band induced from orbitals at WP $g$ into bands induced from orbitals at WP $a$.

In the next step we will establish the connection between the local RSI indices at the WPs in the SSG and the indices of the SSG in the momentum space. This allows for an easy identification of an OAI from the band structure. First let us consider a SSG with a single WP as, for instance, the type-II SSG $Pna2_11'$ (N. 33.145) of generators $\{C_{2z}|(0,0,1/2)\}, \{\mathcal{M}_y|(1/2,1/2,0)\}$ and TRS $\mathcal{T}$. The site-symmetry group of the unique WP $a$ is isomorphic to the trivial point group, 1, with only one irrep, $\overline{A}$. Any center of charges at any point in this SSG transforms under this irrep $\overline{A}$ or a direct sum of several copies of $\overline{A}$. In the momentum space, any gapped set of bands has as symmetry data vector $B$ a multiple of the $B$ vector of the band representation induced by $\overline{A}$. This band representation is a \emph{basis} of all the possible gapped sets of bands in SSG $Pna2_11'$. Now we consider another SSG in which, apart from the general position, there are another WPs with site-symmetry groups whose maximal subgroup is the trivial group, as for instance type-I SSG $P2$ (N. 3.1). There are 4 WPs, $a,b,c$ and $d$, with site-symmetry groups isomorphic to point group 2, whose irreps are $\,^2\overline{E}$ and $\,^1\overline{E}$. Looking at Table \ref{tab:ci} we recognize that in these four WPs the $C$ matrix is $C_{G^a}=C_{G^b}=C_{G^c}=C_{G^d}=(1,1)^T$ and there is a $\delta$ index at each WP, $\delta_i=m(\,^1\overline{E})-m(\,^2\overline{E})$. Up to moving as many pairs of $(\,^1\overline{E},\,^2\overline{E})$ orbitals as necessary(possible) from the $a$-$d$ WPs to the general position, any set of orbitals can be expressed as the sum of $|\delta_a|$ orbitals at $a$ ($\,^1\overline{E}$ orbitals  if  $\delta_a>0$ or $\,^2\overline{E}$ orbitals if $\delta_a<0$) plus $|\delta_b|$ orbitals at $b$ ($\,^1\overline{E}$ orbitals  if  $\delta_b>0$ or $\,^2\overline{E}$ orbitals if $\delta_b<0$), etc \ldots and the necessary number of $\overline{A}$ orbitals at the general position.

This argument can  also be extended to SSGs with three types of WPs (general position, planes and lines) or four types  (general position, planes, lines and points). First we move as many orbitals from the WPs $Q_i$ of highest symmetry to the WPs $q_i^{\ell}$ ($\ell$ stands for \emph{line}) whose site-symmetry groups are maximal subgroups of the groups $G^{Q_i}$, next we move as many orbitals as possible to the WPs $q_i^{p}$ ($p$ stands for \emph{plane}) whose site-symmetry groups are maximal subgroups of the groups $G^{q_i^{\ell}}$ and finally to the general position. At the end, in every SSG, any set of orbitals can be expressed as a linear combination of a subset of \emph{basis} orbitals, which consist of one orbital for each local RSI index at every WP and the single orbital that corresponds to the general position. The general algorithm to calculate the RSI indices in the momentum space then consists of these steps:
\begin{itemize}
	\item Identify all the local indices at all WPs following the process described in section (\ref{calcRSI}),
	\begin{equation}
	\label{indices}
	(\delta_1^1,\delta_1^2,\ldots,\delta_2^1,\delta_2^2,\ldots,\delta_j^1,\delta_j^2,\ldots)
	\end{equation}
	The subscript identifies the WP and, for a common subscript, different superscripts identify the different indices at the same WP. This is independent of the band structure of a given material.
	\item For each $\delta_i^j$ index, choose an orbital (or set of orbitals) such that the local RSI indices $\delta_i^j=1$ and $\delta_k^{\ell}=0$ for $i\neq k$ or $j\neq \ell$. As $L_C$ an unimodular matrix, it is always possible to find a column vector $p(i)$ in (\ref{listindices}) such that $L_C^{-1}\cdot p(i)=c(i)$, being $c(i)$ a column vector whose $i^{\textrm{th}}$ component $c(i)_i=1$ and $c(i)_j=0$ for $j\neq i$. $p(i)$ is just the integer column vector $p(i)=L_C\cdot c(i)$. It is not guaranteed that all the components of $p(i)$ are non-negative but, as stressed before, our method does not allow to identify fragile OAIs, so that we can consider negative numbers of orbitals to identify OAIs. When there is a single RSI index at a WP, $\delta_i^j$ with $j=1$, the corresponding orbital that makes $\delta_i^1=1$ can be taken as the representative orbital of the orbitals whose multiplicity is included in the definition of the index $\delta_i^j$, because the condition $\delta_k^j$ for $k\neq i$ is automatically fulfilled. Special care must be taken when there are more than one index in a WP, $\delta_i^j$ with $j=1,2\ldots $ because it must be checked that the chosen orbital (or set of orbitals) gives $\delta_i^j=1$ and $\delta_i^k=0$ for $k\neq j$.
	\item Add to the list of representative orbitals made in the previous step the single orbital that corresponds to the general position. This is necessary to obtain the general solution of the system.
Therefore, we have a list of $n_d$ different basis orbitals; $n_d-1$ is the number of local RSI indices.

\item Using BANDREP for non-magnetic groups or MBANDREP for magnetic groups, take the symmetry data vector $B$ of each orbital chosen in the previous steps, and form the BR matrix,
\begin{equation}
\label{mat:BR}
BR=\left(\begin{array}{cccc}
m_{11}&m_{12}&\ldots&m_{1n_d}\\
m_{21}&m_{22}&\ldots&m_{2n_d}\\
\ldots&\ldots&\ldots&\ldots\\
m_{n_{\rho}1}&m_{n_{\rho}2}&\ldots&m_{n_{\rho}n_d}
\end{array}\right)
\end{equation}
$n_{\rho}$ is the number of irreps at maximal $k$-vectors. Every column in the tatrix (\ref{mat:BR}) is the symmetry data vector of one of the chosen orbitals. 

\item As it has been argued in the paragraph immediately after expression (\ref{summomentum}), any gapped set of orbitals can be expressed as linear combination of the chosen orbitals. If the linear combination does not contain any non-integer coefficient, the set of valence bands below the Fermi level is an insulator without stable topology. If $B$ represents the symmetry data vector of the gapped set of bands and the system does not exhibit stable topology, there must exist an $n_{\rho}$-dimensional integer $p$ vector such that,
\begin{equation}
\label{eq:mom}
BR\cdot p=B
\end{equation}
In general, the set of orbitals chosen to build the $BR$ matrix is an over-complete basis of the set of possible gapped bands without stable topology. In order to analyze the complete sets of solutions of Eq. (\ref{eq:mom}), following the same method used to construct the local indices from Eq. (\ref{eq:cmat}), we perform the Smith decomposition of the integer $n_{\rho}\times n_d$ matrix (\ref{mat:BR}),
\begin{equation}
\label{eq:smith}
BR=L\cdot \Delta\cdot R
\end{equation}
being $L$ and $R$ unimodular matrices of dimensions $n_{\rho}\times n_{\rho}$ and $n_d\times n_d$, respectively, and $\Delta$ being the Smith normal form of dimension $n_{\rho}\times n_d$. This normal form is a diagonal matrix ($\Delta_{ij}=0$ for $i\neq j$) and $L$ and $R$ can be chosen such that $1\le \Delta_{11}\le \Delta_{2}\le\ldots\le \Delta_{rr}$, being $r$ the rank of $\Delta$ and $BR$.
\end{itemize}
Introducing (\ref{eq:smith}) into (\ref{eq:mom}), we get the set of equations,
\begin{equation}
\label{eq:mom2}
\Delta\cdot R\cdot p=L^{-1}\cdot B
\end{equation}
Considering the diagonal form of $\Delta$, we can split the equations (\ref{eq:mom2}) into two subsets,
\begin{equation}
\label{eq:first}
\left(R\cdot p\right)_i=\frac{1}{\Delta_{ii}}\left(L^{-1}\cdot B\right)_i \textrm{ for }i=1,\ldots,r~(\text{r is the rank of the BR})
\end{equation}
and
\begin{equation}
\label{eq:second}
0=\left(L^{-1}\cdot B\right)_i \textrm{ for }i=r+1,\ldots,n_{\rho}
\end{equation}
We are assuming that the symmetry data vector $B$ corresponds to a set of gapped bands without stable topology, so there must exist a solution for the equation system given by (\ref{eq:first}) and (\ref{eq:second}), but the solution is not unique, in general. There is a set of arbitrary $n_{\rho}-r$ integer numbers $k_i$ such that,
\begin{equation}
\label{eq:third}
\left(R\cdot p\right)_i=k_i  \textrm{ for }i=r+1,\ldots,n_{\rho}
\end{equation}
Therefore, the general solution to the equation system (\ref{eq:mom}) is,
\begin{equation}
\label{eq:sol}
p_i=\sum_{j=1}^r\frac{1}{\Delta_{jj}}R_{ij}^{-1}\left(L^{-1}B\right)_j+\sum_{j=r+1}^{n_{\rho}}R_{ij}^{-1}k_{j-r}
\end{equation}
In general, the set of integers $p_i$ in Eq. (\ref{eq:sol}) depend on arbitrary parameters $\{k_i\}$ and the corresponding RSI indices of the SSG are not uniquely defined in the momentum space. However, we can divide the $p_i$ indices into two subsets, depending on the specific form of the $R^{-1}$ matrix. First, let us assume that a given row $i$ of the $R^{-1}$ matrix fulfills the next condition,
\begin{equation}
\label{row:Rmatrix}
R_{ij}^{-1}=0\textrm{ for }j=r+1,\ldots,n_{\rho}
\end{equation}
When this condition is fulfilled, the corresponding integer $p_i$ in (\ref{eq:sol}) does not depend on the $k_i$ and it is a well-defined RSI. The formula of the RSI is
\begin{equation}
\label{eq:solgood}
p_i=\sum_{j=1}^r\frac{1}{\Delta_{jj}}R_{ij}^{-1}\left(L^{-1}B\right)_j
\end{equation}
and its value is obtained from the symmetry data vector $B$. A non-zero value of this index implies a non-zero value of the corresponding local index in the list (\ref{indices}) (the $i^{\textrm{th}}$ index in the list): the symmetry-data vector $B$ of the band structure must contain $p_i$ times the EBR coming from the $i^{\textrm{th}}$ EBR orbital. Therefore, there are at least $p_i$ orbitals that cannot be moved from the corresponding WP to WPs of lower symmetry connected to it. This does not mean that they will be necessarily located at the WP where the $i^{\textrm{th}}$ index is defined, because all the orbitals at this WP can always be moved to another WPs of \emph{higher} symmetry unless the Wyckoff position was already maximal. Therefore, if the WP where the $i^{\textrm{th}}$ index is defined and all the WPs of higher symmetry connected to it are empty, the material is an OAI. There are orbitals necessarily located at empty WPs that cannot be moved away to occupied WPs.

If the row $i$ of the $R^{-1}$ matrix does not fulfill the condition (\ref{row:Rmatrix}), the corresponding index in the list (\ref{indices}) is not well defined (there can be orbitals at the corresponding WP or not, because the arbitrary $k_i$ arbitrary numbers can be adjusted to give $p_i=0$ or $p_i\ne0$). However, combinations of $p_i$ indices, which themselves depend on the arbitrary integer numbers $k_i$, can be independent on the $k_i$. In that case, the combinations of local indices (\ref{indices}) are well defined. If there exists a set of $c_i$ integers such that,
\begin{equation}
\label{eq:cond2}
\sum_{i=1}^{n_d}c_iR_{ij}^{-1}=0\textrm{ for }j=r+1,\ldots,n_{\rho}
\end{equation}
then, the linear combination of integers,
\begin{equation}
\label{eq:solgood2}
P_i=\sum_{i=1}^{n_d}c_ip_i
\end{equation}
does not depend on the arbitrary integers $k_i$ and $P_i$ is a good RSI index which is a linear combination of local RSI indices that defined in Appendix \ref{calcRSI}. In this work, we refer such a $P_i$ index in Eq. (\ref{eq:solgood2}) to as a $Z$-type composite RSI index. In general, a composite RSI index is defined at a set of different Wyckoff positions. In some cases where either Eq. (\ref{row:Rmatrix}) or Eq. (\ref{eq:cond2}) is satisfied, there is another way to construct the composite RSI indices if there exists a linear combination of not well-defined $p_i$ numbers satisfying the following equation,
\begin{equation}
\sum_{i=1}^{n_d}c_iR_{ij}^{-1}\textrm{ mod }N=0\textrm{ for }j=r+1,\ldots,n_{\rho}
\end{equation}
with $N>1$ being an integer number. In those cases, the integer
\begin{equation}
\label{eq:solgood3}
P_i=\sum_{i=1}^{n_d}c_ip_i\textrm{ mod }N
\end{equation}
does not depend on $k_i$ and $P_i$ is a well defined $Z_N$-type composite RSI index in the momentum space. We have checked that, in the 1651 SSGs, $|R_{ij}|=0,1,2,4$ for $j>r$ and any $i$, and therefore there are well-defined indices only with $N=2,4$. It is necessary thus to consider, separately, two sets of equations (\ref{eq:solgood3}), one set for $N=2$ and another one for $N=4$. In the next sections we denote the three types of RSI indices, \ie the $Z$, $Z_2$ and $Z_4$ RSI indices, as $\delta_i(WP)$ ($Z$-type indices), $\eta_i(WP)$ ($Z_2$-type indices) and $\zeta_i(WP)$ ($Z_4$-type indices), where the label $i$ distinguish different indices of the same type in a given Shubnikov group and $WP$ is the list of WPs involved in the definition of the index.

In general, in a SSG there are RSI indices of type (\ref{eq:solgood}), (\ref{eq:solgood2}) and (\ref{eq:solgood3}). In the last two cases, the RSI index defined in momentum space is the combination of several local RSI indices (\ref{indices}) that correspond, in principle, to different WPs. We call these indices \emph{Composite RSI} indices. A non-zero value of a composite index, (\ref{eq:solgood2}) and/or (\ref{eq:solgood3}), means that, at least $P_i$ orbitals must be located at these WPs. If one of the WPs (or one of the WPs of higher symmetry connected to these WPs) is occupied by atoms, the set of bands is compatible with a trivial insulator. If all the involved WPs (and all the WPS of higher symmetry connected to them) are empty, the material is an OAI. Note that the last integer in Eq. (\ref{eq:sol}) for $i=n_{\rho}$ corresponds to the trivial band induced from the general position. An index of type (\ref{eq:solgood2}) or (\ref{eq:solgood3}) with $c_{n_d}\ne0$ must be discarded, because the integer $p_{n_d}$ does not correspond to any local RSI index. A non-zero value of an index that involves the general WP would give the trivial result that some orbitals must be located at the general WP or at WPs connected to it (the complete set of WPs of the Shubnikov group).

Finally, it must also be stressed that not all the indices that can be obtained by relations as (\ref{eq:solgood}), (\ref{eq:solgood2}) and (\ref{eq:solgood3}) are all independent. For instance, it is obvious that, if a set of coefficients $c_i$ in Eq. (\ref{eq:solgood}) gives a well-defined RSI index $P_i$, the set of coefficients $c_i'=2c_i$ gives also a well-defined RSI index $P_i'$, but they are not independent because $P_i'=2P_i$ and it gives no extra information. The final set of RSI indices of a SSG should contain only an independent set of indices.

Together with the local RSI indices at each WP, the new tools {\color{blue} RSI} (\href{http://www.cryst.ehu.es/cryst/RSI}{www.cryst.ehu.es/cryst/RSI})  and {\color{blue} MagRSI}  (\href{http://www.cryst.ehu.es/cryst/MagRSI}{www.cryst.ehu.es/cryst/MagRSI}) in the Bilbao Crystallographic Server give the RSI indices in momentum space in non-magnetic and magnetic space groups, respectively, for double valued irreps (SOC).

In section \ref{example:mom} we give a detailed calculation of the RSI indices in momentum space in SSG \msgsymb{67}{508} (N. 67.508) as example.

\subsection{Example: calculation of the RSI indices in the SSG \msgsymb{67}{508} (N. 67.508)}
\label{example:RSIcalc}
As example of the determination of the RSI indices, in this section we will calculate the RSI indices of the Type IV SSG \msgsymb{67}{508} (N. 67.508) (its generators include the inversion symmetry, $C_{2x}$, $\{C_{2y}|(1/2,0,0)\}$, a fractional translation $\vec\tau=(1/2,1/2,1/2)$ and an anti-unitary translation $\{\mathcal{T}|(0,0,1/2)\}$) both in the direct space and in the momentum space. First, we will calculate the local RSI indices in the direct space at the maximal WPs, then the indices at the non-maximal WPs, and finally we determine the complete set of indices in the momentum space.
\subsubsection{RSI indices at the maximal Wyckoff positions}
The SSG \msgsymb{67}{508} has seven maximal WPs: $8a$:$(1/4,0,0)$, $8b$:$(1/4,0,1/4)$, $8c$:$(0,0,0)$, $8d$:$(0,0,1/4)$, $8e$:$(1/4,1/4,0)$, $8f$:$(1/4,1/4,1/4)$ and $8g$:$(0,1/4,z)$ with site-symmetry groups $G^Q$ isomorphic to the magnetic point groups $222$, $2'2'2$, $2/m$, $2'/m$, $2/m$, $2'/m$ and $mm2$, respectively. Note that the three lattice vectors of a crystal lattice in SSG \msgsymb{67}{508} are $(a,0,0)$, $(0,b,0)$ and $(0,0,c)$ where $a$, $b$ and $c$ are the lattice constants.

{\bf Wyckoff position $8a$}

The site-symmetry group of $8a$:$(1/4,0,0)$ is isomorphic to the magnetic point group $222$ (of generators $C_{2x}$, $C_{2y}$ and $C_{2z}$), whose unique double irrep is $\overline{E}(2)$. The number in parentheses gives the dimension or the irrep. This irrep can be induced from the irreps of the site-symmetry groups of the non-maximal WPs $16h$:$(x,0,0)$, $16l$:$(1/4,0,z)$ and $16m$:$(0,y,z)$ which represent two different lines and a plane, respectively, where the WP $8a$ sits. The site-symmetry groups of $16h$, $16l$ and $16m$ are isomorphic to point groups $2$ ($C_{2x}$), $2$ ($C_{2z}$) and $m$ ($\mathcal{M}_x$), respectively, and the irreps of these magnetic point groups are $\,^1\overline{E}(1)$ and $\,^2\overline{E}(1)$ in the three groups. The subduction relations between the irreps of $8a$ and the irreps of $16h$, $16l$ and $16m$ are exactly the same,
\begin{eqnarray}
\overline{E}\downarrow G^{h,l,m}=\,^1\overline{E}\oplus\,^2\overline{E}
\end{eqnarray}
The induction relations then are,
\begin{eqnarray}
\,^1\overline{E}\uparrow G^{a}=\overline{E}\nonumber\\
\,^2\overline{E}\uparrow G^{a}=\overline{E}
\end{eqnarray}
Each induction relation gives a column in the $C$ matrix with as many rows as irreps in $G^a$ (in our case just one). The $C$ matrix is,
\begin{equation}
\label{eq:cmata}
C_{G^{a}}=\left(\begin{array}{rrrrrr}
1&1&1&1&1&1
\end{array}\right)
\end{equation}
whose Smith decomposition is,
\begin{equation}
C_{G^{a}}=L\cdot\Delta\cdot R=\left(\begin{array}{rr}
1
\end{array}\right)
\cdot
\left(\begin{array}{rrrrrr}
1&0&0&0&0&0
\end{array}\right)
\cdot
\left(
\begin{array}{rrrrrr}
1 & 1 & 1 & 1 & 1 & 1 \\
0 & 1 & 0 & 0 & 0 & 0 \\
0 & 0 & 1 & 0 & 0 & 0 \\
0 & 0 & 0 & 1 & 0 & 0 \\
0 & 0 & 0 & 0 & 1 & 0 \\
0 & 0 & 0 & 0 & 0 & 1 \\
\end{array}
\right)\end{equation}
Following the general procedure explained in section (\ref{app:rsi3D}), as the number of rows is the same as the rank of the $\Delta$ Smith form, there are no indices of the second type in Eq. (\ref{listindices}) at the WP $8a$. Moreover, as there is no element in the diagonal of the $\Delta$ Smith form different from 0 and 1, there are no indices of the first type in Eq. (\ref{listindices}) of Wyckoff position $8a$.

{\bf Wyckoff position $8b$}

The site-symmetry group of $8b$:$(1/4,0,1/4)$ is isomorphic to the magnetic point group $2'2'2$ (of generators $C_{2z}$ and $\mathcal{T}\cdot C_{2x}$) with irreps $\,^1\overline{E}(1)$ (with $C_{2z}$ eigenvalue $i$) and $\,^2\overline{E}(1)$ (with $C_{2z}$ eigenvalue $-i$) that can be induced from the irreps of the site-symmetry groups of the WPs $16i$:$(x,0,1/4)$  with site-symmetry group $2'$ (of generator $\mathcal{T}C_{2x}$) and irrep $\overline{A}(2)$, $16k$:$(1/4,y,1/4)$  with site-symmetry group $2'$ (of generator $\mathcal{T}C_{2y}$) and irrep $\overline{A}(2)$ and $16l$:$(1/4,0,z)$ with site-symmetry group $2$ (of generator $C_{2z}$) and irreps $\,^1\overline{E}(1)$ and $\,^2\overline{E}(1)$.

The subduction relations are:
\begin{eqnarray}
\,^1\overline{E}\downarrow G^{i,k}=\overline{A}\nonumber\\
\,^2\overline{E}\downarrow G^{i,k}=\overline{A}
\end{eqnarray}
and,
\begin{eqnarray}
\,^1\overline{E}\downarrow G^{l}=\,^1\overline{E}\nonumber\\
\,^2\overline{E}\downarrow G^{l}=\,^2\overline{E}
\end{eqnarray}
and the induction relations are,
\begin{equation}
\overline{A}\uparrow G^{b}=\,^1\overline{E}\oplus\,^2\overline{E}
\end{equation}
from irrep $\overline{A}$ in both $16i$ and $16k$ and
\begin{eqnarray}
\,^1\overline{E}\uparrow G^{b}=2\,^1\overline{E}\nonumber\\
\,^2\overline{E}\uparrow G^{b}=2\,^2\overline{E}
\end{eqnarray}
from the irreps at $16l$.

The $C$ matrix is,

\begin{equation}
\label{eq:cmatb}
C_{G^b}=\left(
\begin{array}{cccc}
	1 & 1 & 2 & 0 \\
	1 & 1 & 0 & 2 \\
\end{array}
\right)
\end{equation}
whose Smith decomposition is,

\begin{equation}
C_{G^b}=L\cdot\Delta\cdot R=
\left(
\begin{array}{rr}
1 & 1 \\
1 & 0 \\
\end{array}
\right)\cdot
\left(
\begin{array}{rrrr}
1 & 0 & 0 & 0 \\
0 & 2 & 0 & 0 \\
\end{array}
\right)\cdot
\left(
\begin{array}{rrrr}
1 & 1 & 0 & 2 \\
0 & 0 & 1 & -1 \\
0 & 1 & 0 & 0 \\
0 & 0 & 0 & 1 \\
\end{array}
\right)\end{equation}
The number of rows in $\Delta$ is the same as its rank, so there are no indices of the second type in Eq. (\ref{listindices}) at WP $b$. However, as one of the elements of the diagonal is 2, there is a $Z_2$ index, given by the second row of $L^{-1}$, $(1,-1)$. The index is thus,
\begin{equation}
\label{indexb}
\delta_{b}=m(\,^1\overline{E})-m(\,^2\overline{E})\textrm{ mod 2}
\end{equation}

{\bf Wyckoff position $8c$}

The site-symmetry group of $8c:(0,0,0)$ is isomorphic to the magnetic point group $2/m$ (of generators inversion symmetry and $C_{2x}$) with irreps $^2\overline{E}_g(1)$ (with parity $+1$ and $C_{2x}$ eigenvalue $-i$), $^1\overline{E}_g(1)$ (with parity $+1$ and $C_{2x}$ eigenvalue $+i$), $^2\overline{E}_u(1)$ (with parity $-1$ and $C_{2x}$ eigenvalue $-i$) and  $^1\overline{E}_u(1)$ (with parity $-1$ and $C_{2x}$ eigenvalue $+i$).
These irreps can be induced from the irreps of the site-symmetry groups of the non-maximal WPs $16h$:$(x,0,0)$ and $16m$:$(0,y,z)$ which represent a line and a plane, respectively, where the WP $8c$ sits. $G^{h}$ and $G^{m}$ are isomorphic to $2$ and $m$, respectively, and the irreps of these magnetic point groups are $\,^1\overline{E}(1)$ and $\,^2\overline{E}(1)$ in both cases. The subduction relations between the irreps of $8c$ and the irreps of $16h$ and $16m$ are,

\begin{eqnarray}
\,^1\overline{E}_g\downarrow G^{h}&=&\,^1\overline{E}\nonumber\\
\,^1\overline{E}_u\downarrow G^{h}&=&\,^1\overline{E}\nonumber\\
\,^2\overline{E}_g\downarrow G^{h}&=&\,^2\overline{E}\nonumber\\
\,^2\overline{E}_u\downarrow G^{h}&=&\,^2\overline{E}
\end{eqnarray}
and
\begin{eqnarray}
\,^1\overline{E}_g\downarrow G^{m}&=&\,^1\overline{E}\nonumber\\
\,^1\overline{E}_u\downarrow G^{m}&=&\,^2\overline{E}\nonumber\\
\,^2\overline{E}_g\downarrow G^{m}&=&\,^2\overline{E}\nonumber\\
\,^2\overline{E}_u\downarrow G^{m}&=&\,^1\overline{E}
\end{eqnarray}
The induction relations from the irreps of $G^{h}$ are,
\begin{eqnarray}
\,^2\overline{E}\uparrow G^{c}&=&\,^2\overline{E}_g\oplus\,^2\overline{E}_u\nonumber\\
\,^1\overline{E}\uparrow G^{c}&=&\,^1\overline{E}_g\oplus\,^1\overline{E}_u
\end{eqnarray}
and the induction relations from the irreps of $G^{m}$ are,
\begin{eqnarray}
\,^2\overline{E}\uparrow G^{c}&=&\,^2\overline{E}_g\oplus\,^1\overline{E}_u\nonumber\\
\,^1\overline{E}\uparrow G^{c}&=&\,^1\overline{E}_g\oplus\,^2\overline{E}_u
\end{eqnarray}
The $C$ matrix is,
\begin{equation}
\label{eq:cmatc}
C_{G^c}=\left(
\begin{array}{cccc}
1 & 0 & 1 & 0 \\
0 & 1 & 0 & 1 \\
1 & 0 & 0 & 1 \\
0 & 1 & 1 & 0 \\
\end{array}
\right)
\end{equation}
whose Smith decomposition is,

\begin{equation}
C_{G^c}=L\cdot\Delta\cdot R=
\left(
\begin{array}{rrrr}
1 & 0 & 1 & 0 \\
0 & 1 & 0 & 0 \\
1 & 0 & 0 & 0 \\
0 & 1 & 1 & 1 \\
\end{array}
\right)\cdot
\left(
\begin{array}{rrrr}
1 & 0 & 0 & 0 \\
0 & 1 & 0 & 0 \\
0 & 0 & 1 & 0 \\
0 & 0 & 0 & 0 \\
\end{array}
\right)
\cdot
\left(
\begin{array}{rrrr}
1 & 0 & 0 & 1 \\
0 & 1 & 0 & 1 \\
0 & 0 & 1 & -1 \\
0 & 0 & 0 & 1 \\
\end{array}
\right)
\end{equation}
In this case the number of rows exceeds in 1 the rank of the Smith form, so there is one index of the second type in Eq. (\ref{listindices}) at the WP $8c$, given by the fourth row of the $L^{-1}$ matrix, $(-1,-1,1,1)$,
\begin{equation}
\label{indexc}
\delta_{c}=-m(\,^1\overline{E}_g)-m(\,^1\overline{E}_u)+m(\,^2\overline{E}_g)+m(\,^2\overline{E}_u)
\end{equation}

{\bf Wyckoff position $8d$}

The site-symmetry group of $8d:(0,0,1/4)$ is isomorphic to the magnetic point group $2'/m$ (of generators $\mathcal{T}\cdot C_{2x}$ and $\mathcal{M}_x$) with irrep $\,^2\overline{E}\,^1\overline{E}(2)$.
This irrep can be induced from the irreps of the site-symmetry groups of the non-maximal WPs $16i$:$(x,0,1/4)$ and $16m$:$(0,y,z)$ which represent a line and a plane, respectively, where the WP $8d$ sits. $G^{i}$ and $G^{m}$ are isomorphic to $2'$ and $m$, respectively, with irrep $\overline{A}(1)$ in the first group and  $\,^1\overline{E}(1)$ (with $\mathcal{M}_x$ eigenvalue $+i$) and $\,^2\overline{E}(1)$ (with $\mathcal{M}_x$ eigenvalue $-i$) in the second one. The subduction relations between the irrep of $G^{d}$ and the irreps of $G^{i}$ and $G^{m}$ are,

\begin{eqnarray}
\,^2\overline{E}\,^1\overline{E}\downarrow G^{i}&=&2\,\overline{A}\nonumber\\
\,^2\overline{E}\,^1\overline{E}\downarrow G^{m}&=&\,^1\overline{E}\oplus\,^2\overline{E}
\end{eqnarray}
The induction relation from $16i$ to $8d$ is,
\begin{equation}
\overline{A}\uparrow G^{d}=\,^2\overline{E}\,^1\overline{E}
\end{equation}
and the induction relations from $16m$ to $8d$ are,
\begin{eqnarray}
^1\overline{E}\uparrow G^{d}&=&\,^2\overline{E}\,^1\overline{E}\nonumber\\
^2\overline{E}\uparrow G^{d}&=&\,^2\overline{E}\,^1\overline{E}
\end{eqnarray}
The $C$ matrix is,
\begin{equation}
\label{eq:cmatd}
C_{G^d}=\left(\begin{array}{lll}
1&1&1
\end{array}\right)
\end{equation}
whose Smith decomposition is,
\begin{equation}
C_{G^d}=L\cdot\Delta\cdot R=
\left(
\begin{array}{r}
1 \\
\end{array}
\right)
\cdot
\left(
\begin{array}{rrr}
1 & 0 & 0 \\
\end{array}
\right)
\cdot
\left(
\begin{array}{rrr}
1 & 1 & 1 \\
0 & 1 & 0 \\
0 & 0 & 1 \\
\end{array}
\right)
\end{equation}
The number of rows and the rank of the Smith form are the same and it has no diagonal element different from 0 and 1. There are no RSI indices at the WP $8d$.

{\bf Wyckoff position $8e$}

The site-symmetry group of the WP $8e$ is isomorphic to the site-symmetry group of WP $8c$. The calculation and the result are exactly the same. There is one index of the second type in Eq. (\ref{listindices}) at WP $e$ given by,
\begin{equation}
\label{indexe}
\delta_{e}=-m(\,^1\overline{E}_g)-m(\,^1\overline{E}_u)+m(\,^2\overline{E}_g)+m(\,^2\overline{E}_u)
\end{equation}

{\bf Wyckoff position $8f$}

The site-symmetry group of the WP $8f$ is isomorphic to the site-symmetry group of WP $8d$. The calculation and the result are exactly the same. There are thus no indices at WP $8f$.

\subsubsection{RSI indices at the non-maximal Wyckoff positions}
Apart from the general position, there are 7 non-maximal WPs in this SSG, $16h$:$(x,0,0)$, $16i$:$(x,0,1/4)$, $16j$:$(1/4,y,0)$, $16k$:$(1/4,y,1/4)$, $16l$:$(1/4,0,z)$, $16m$:$(0,y,z)$, and $16n$:$(x,1/4,z)$, with site-symmetry groups isomorphic to the magnetic point groups $2$, $2'$, $2$, $2'$, $2$, $m$ and $m$, respectively. As explained in section \ref{sub:exchange}, the determination of the RSI indices at non-maximal WPs includes the addition, in general, of exchange terms in the $C$ matrix. Therefore, together with the induction-subduction relations between the site-symmetry groups of the WPs and of their subgroups (which are site-symmetry groups of WPs of \emph{lower} symmetry), we have to consider also induction-subduction relations between the site-symmetry groups of the WP and of their supergroups (which are site-symmetry groups of WPs of \emph{higher} symmetry) in order to find the exchange term .

{\bf Wyckoff positions $16h$, $16j$ and $16l$}

These three WPs have site-symmetry groups isomorphic to the point group $2$, and the calculations of the RSIs and the resulting indices at these positions are exactly the same. We choose the WP $16h$ as example. The irreps are $\,^1\overline{E}(1)$ and $\,^2\overline{E}(1)$, and can be induced from the irrep of the site-symmetry group of the general position, with irrep $\overline{A}$. 
The induction relation is,
\begin{equation}
\label{eq:indh}
\overline{A}\uparrow G^{h}=\,^1\overline{E}\oplus\,^2\overline{E}
\end{equation}
This relation gives a column in the $C$ matrix, $(1,1)^T$.
Now we analyze the existence or not of additional columns in the $C$ matrix due to exchange terms. The WP $16h$ is connected to the maximal WP $8a$. Following the notation of the symmetry operations of the double groups introduced in Ref.  \cite{elcoro_double_2017}, the rotational part of the operations of $G^{h}$ are $\{1,\,^d1,2_x,\,^d2_x\}$. If we take the rotational part of a symmetry operation $h$ such that $h\in G^{a}$ but $h\notin G^{h}$ as, for example, the symmetry operation whose rotational part is the 2-fold axis $2_y$, the conjugated elements of $G^{h}$ through $h$ (only the rotational parts are explicitly given),
\begin{eqnarray}
2_y^{-1}\cdot1\cdot2_y&=&1\nonumber\\
2_y^{-1}\cdot\,^d1\cdot2_y&=&\,^d1\nonumber\\
2_y^{-1}\cdot2_x\cdot2_y&=&\,^d2_x\\
2_y^{-1}\cdot\,^d2_x\cdot2_y&=&2_x\nonumber
\end{eqnarray}
The symmetry operations $2_x$ and $^d2_x$ are interchanged through conjugation under $2_y$. If we interchange the traces of the symmetry operations $2_x$ and $^d2_x$ in the point group 2, we get the relations,
\begin{equation}
\label{eq:conj}
\,^1\overline{E}\leftrightarrow\,^2\overline{E}
\end{equation}
Therefore, the two irreps in $G^{h}$ are mutually conjugated under $h=2_y$ (or any other $h\in G^{a}$ but $h\notin G^{h}$). Thus, we have to add a row $(1,-1)^T$ to the $C$ matrix.

We could repeat a similar calculation recognizing that the WP $16h$ is also connected to the maximal WP $8c$ with $G^{c}$ isomorphic to $2/m$. However, the two irreps of  $G^{h}$ are self-conjugated under the symmetry operations $h\in G^{c}$ but $h\notin G^{h}$, so no extra row is added to the $C$ matrix.

The $C$ matrix is thus,
\begin{equation}
\label{eq:cmath}
C_{G^h}=\left(\begin{array}{rr}
1&1\\
1&-1
\end{array}\right)
\end{equation}
whose Smith decomposition is,
\begin{equation}
\label{eq:desch}
C_{G^h}=L\cdot\Delta\cdot R=
\left(
\begin{array}{rr}
1 & 0 \\
1 & -1 \\
\end{array}
\right)
\cdot
\left(
\begin{array}{rr}
1 & 0 \\
0 & 2 \\
\end{array}
\right)
\cdot
\left(
\begin{array}{rr}
1 & 1 \\
0 & 1 \\
\end{array}
\right)
\end{equation}
There are no indices of the second type in Eq. (\ref{listindices}) at this WP, but there is a $Z_2$ index given by the second row of $L^{-1}$, $(1,-1)$,
\begin{equation}
\label{indexh}
\delta_h=m(\,^1\overline{E})-m(\,^2\overline{E})\textrm{ mod 2}
\end{equation}
An equivalent calculation gives $Z_2$ indices at $16j$ and $16l$,
\begin{equation}
\label{indexj}
\delta_j=m(\,^1\overline{E})-m(\,^2\overline{E})\textrm{ mod 2}
\end{equation}
\begin{equation}
\label{indexl}
\delta_l=m(\,^1\overline{E})-m(\,^2\overline{E})\textrm{ mod 2}
\end{equation}

{\bf Wyckoff positions $16i$ and $16k$}

These WPs have site-symmetry groups isomorphic to the point group $2'$, and the calculations of the RSIs and the resulting indices at these positions are exactly the same. We choose the WP $16i$ as example. The unique irrep of $2'$ is $\overline{A}(1)$, which can be induced from the unique irrep $\overline{A}(1)$ of the site-symmetry group of the general position. 
The induction relation is,
\begin{equation}
\label{eq:indi}
\overline{A}\uparrow G^{i}=2\,\overline{A}
\end{equation}
There are no exchange terms in this case because there is a single irrep and, then, it must be self-conjugated under any symmetry operation. The $C$ matrix is thus,
\begin{equation}
\label{eq:cmati}
C_{G^i}=\left(\begin{array}{l}
2
\end{array}\right)
\end{equation}
whose Smith decomposition is trivial,
\begin{equation}
C_{G^h}=L\cdot\Delta\cdot R=
\left(
\begin{array}{c}
1
\end{array}
\right)
\cdot
\left(
\begin{array}{c}
2
\end{array}
\right)
\cdot
\left(
\begin{array}{c}
1
\end{array}
\right)
\end{equation}
As the number of rows and the rank of the Smith form is the same, there are no indices of the second type in Eq. (\ref{listindices}) at these WPs, but there is one $Z_2$ index,
\begin{equation}
\label{indexi}
\delta_i=m(\overline{A})\textrm{ mod 2}
\end{equation}
An equivalent calculation gives another $Z_2$ index at $16k$,
\begin{equation}
\label{indexk}
\delta_k=m(\overline{A})\textrm{ mod 2}
\end{equation}

{\bf Wyckoff positions $16m$ and $16n$}

These WPs have site-symmetry groups isomorphic to the point group $m$, and the calculations of the RSIs and the resulting indices at these positions are exactly the same. We take $16m$ as example to do the calculation. The point group $m$ has the same irreps as the point group $2$ of WP $16h$ with the same labels. Therefore, the determination of the column of the $C$ matrix that correspond to the induction of the irreps from the irrep of the general position is exactly the same as for the WP $16h$ done above. The column in the $C$ matrix is $(1,1)^T$.

In this case, there is also an exchange term (see section \ref{sub:exchange}) noting that the line $8g$ belongs to the plane $16m$. The rotational part of the symmetry elements of the site-symmetry group of WP $16m$ are, $\{1,\,^d1,m_x,\,^dm_x\}$, following the notation in Ref. (\onlinecite{elcoro_double_2017}) for the symmetry operations of double groups, where the action of two point operations $R$ and $^dR$ differ in the spin space (one of them acts as the other one followed by the inversion in the spin space). The conjugation of these elements under $2_z$, the rotational part of a symmetry operation that belongs to $G^g$ but not to $G^m$, gives,
\begin{eqnarray}
2_z^{-1}\cdot1\cdot2_z&=&1\nonumber\\
2_z^{-1}\cdot\,^d1\cdot2_z&=&\,^d1\nonumber\\
2_z^{-1}\cdot m_x\cdot2_z&=&\,^dm_x\\
2_z^{-1}\cdot\,^dm_x\cdot2_z&=&m_x\nonumber
\end{eqnarray}
The symmetry operations $m_x$ and $^dm_x$ are interchanged through conjugation under $2_z$. The interchange of the traces of the symmetry operations $m_x$ and $^dm_x$ in the irreps of the point group $m$ gives the same  conjugation relations between the irreps as in Eq. (\ref{eq:conj}),
\begin{equation}
\,^1\overline{E}\leftrightarrow\,^2\overline{E}
\end{equation}
This relation introduces an extra column in the $C$ matrix, $(1,-1)^T$. Therefore, the whole $C$ matrix is the same as Eq. (\ref{eq:cmath}) with Smith decomposition given by Eq. (\ref{eq:desch}). The resulting $Z_2$ indices are
\begin{equation}
\label{indexm}
\delta_m=m(\,^1\overline{E})-m(\,^2\overline{E})\textrm{ mod 2}
\end{equation}
\begin{equation}
\label{indexn}
\delta_n=m(\,^1\overline{E})-m(\,^2\overline{E})\textrm{ mod 2}
\end{equation}
\subsubsection{RSI indices in momentun space}
\label{example:mom}
Following the general algorithm described in section \ref{app:momentum}, we select a representative band for each RSI index calculated at the WPs and add the band representation induced from the general position $o$. The list of selected bands is,
\begin{equation}
\label{eq:examom}
\begin{array}{llllcccccccc}
&&&&\overline{\Gamma}_5&\overline{\Gamma}_6&\overline{R}_2\overline{R}_2&\overline{S}_2\overline{S}_2&\overline{T}_5\overline{T}_6&\overline{Y}_5&\overline{Y}_6&\overline{Z}_5\overline{Z}_6\\
\delta_b&\rightarrow&\,^1\overline{E}&\rightarrow&1& 1& 1& 1& 1& 1& 1& 1\\
\delta_c&\rightarrow&\,^2\overline{E}_u&\rightarrow&0& 2& 1& 1& 1& 0& 2& 1\\
\delta_e&\rightarrow&\,^2\overline{E}_u&\rightarrow&0& 2& 1& 1& 1& 2& 0& 1\\
\delta_h&\rightarrow&\,^1\overline{E}&\rightarrow&2& 2& 2& 2& 2& 2& 2& 2\\
\delta_i&\rightarrow&\overline{A}&\rightarrow&2& 2& 2& 2& 2& 2& 2& 2\\
\delta_j&\rightarrow&\,^1\overline{E}&\rightarrow&2& 2& 2& 2& 2& 2& 2& 2\\
\delta_k&\rightarrow&\overline{A}&\rightarrow&2& 2& 2& 2& 2& 2& 2& 2\\
\delta_{\ell}&\rightarrow&\,^1\overline{E}&\rightarrow&2& 2& 2& 2& 2& 2& 2& 2\\
\delta_m&\rightarrow&\,^1\overline{E}&\rightarrow&2& 2& 2& 2& 2& 2& 2& 2\\
\delta_n&\rightarrow&\,^1\overline{E}&\rightarrow&2& 2& 2& 2& 2& 2& 2& 2\\
o&\rightarrow&\overline{A}&\rightarrow&4& 4& 4& 4& 4& 4& 4& 4\\
\end{array}
\end{equation}
In the first line of Eq. (\ref{eq:examom}) we have explicitly given the list of irreps at the maximal k-vectors of the SSG \msgsymb{67}{508} (N. 67.508) for reference. The multiplicities of the band representations have been taken from the tool {\color{blue} MBANDREP} (\href{http://www.cryst.ehu.es/cryst/mbandrep}{www.cryst.ehu.es/cryst/mbandrep}) in the BCS.

The $BR$ matrix (\ref{inductionmatrix}), which is the transpose of the matrix outlined by the lists of multiplitities in Eq. (\ref{eq:examom}) and its Smith decomposition (\ref{eq:cdesc}) are,
\begin{scriptsize}
\begin{equation}
BR=
\left(
\begin{array}{ccccccccccc}
1 & 0 & 0 & 2 & 2 & 2 & 2 & 2 & 2 & 2 & 4 \\
1 & 2 & 2 & 2 & 2 & 2 & 2 & 2 & 2 & 2 & 4 \\
1 & 1 & 1 & 2 & 2 & 2 & 2 & 2 & 2 & 2 & 4 \\
1 & 1 & 1 & 2 & 2 & 2 & 2 & 2 & 2 & 2 & 4 \\
1 & 1 & 1 & 2 & 2 & 2 & 2 & 2 & 2 & 2 & 4 \\
1 & 0 & 2 & 2 & 2 & 2 & 2 & 2 & 2 & 2 & 4 \\
1 & 2 & 0 & 2 & 2 & 2 & 2 & 2 & 2 & 2 & 4 \\
1 & 1 & 1 & 2 & 2 & 2 & 2 & 2 & 2 & 2 & 4 \\
\end{array}
\right)
=
\left(
\begin{array}{rrrrrrrr}
1 & 0 & 0 & 0 & 0 & 0 & 0 & 0 \\
1 & 2 & 0 & 0 & 0 & 0 & 1 & 0 \\
1 & 1 & 0 & 0 & 0 & 0 & 0 & 0 \\
1 & 1 & 0 & 1 & 0 & 0 & 0 & 0 \\
1 & 1 & 0 & 0 & 1 & 0 & 0 & 0 \\
1 & 0 & 1 & 0 & 0 & 1 & 0 & 0 \\
1 & 2 & -1 & 0 & 0 & 0 & 0 & 0 \\
1 & 1 & 0 & 0 & 0 & 0 & 0 & 1 \\
\end{array}
\right)
\cdot
\left(
\begin{array}{ccccccccccc}
1 & 0 & 0 & 0 & 0 & 0 & 0 & 0 & 0 & 0 & 0 \\
0 & 1 & 0 & 0 & 0 & 0 & 0 & 0 & 0 & 0 & 0 \\
0 & 0 & 2 & 0 & 0 & 0 & 0 & 0 & 0 & 0 & 0 \\
0 & 0 & 0 & 0 & 0 & 0 & 0 & 0 & 0 & 0 & 0 \\
0 & 0 & 0 & 0 & 0 & 0 & 0 & 0 & 0 & 0 & 0 \\
0 & 0 & 0 & 0 & 0 & 0 & 0 & 0 & 0 & 0 & 0 \\
0 & 0 & 0 & 0 & 0 & 0 & 0 & 0 & 0 & 0 & 0 \\
0 & 0 & 0 & 0 & 0 & 0 & 0 & 0 & 0 & 0 & 0 \\
\end{array}
\right)
\cdot
\left(
\begin{array}{ccccccccccc}
1 & 0 & 0 & 2 & 2 & 2 & 2 & 2 & 2 & 2 & 4 \\
0 & 1 & 1 & 0 & 0 & 0 & 0 & 0 & 0 & 0 & 0 \\
0 & 0 & 1 & 0 & 0 & 0 & 0 & 0 & 0 & 0 & 0 \\
0 & 0 & 0 & 1 & 0 & 0 & 0 & 0 & 0 & 0 & 0 \\
0 & 0 & 0 & 0 & 1 & 0 & 0 & 0 & 0 & 0 & 0 \\
0 & 0 & 0 & 0 & 0 & 1 & 0 & 0 & 0 & 0 & 0 \\
0 & 0 & 0 & 0 & 0 & 0 & 1 & 0 & 0 & 0 & 0 \\
0 & 0 & 0 & 0 & 0 & 0 & 0 & 1 & 0 & 0 & 0 \\
0 & 0 & 0 & 0 & 0 & 0 & 0 & 0 & 1 & 0 & 0 \\
0 & 0 & 0 & 0 & 0 & 0 & 0 & 0 & 0 & 1 & 0 \\
0 & 0 & 0 & 0 & 0 & 0 & 0 & 0 & 0 & 0 & 1 \\
\end{array}
\right)
\end{equation}
\end{scriptsize}
The indices given by equation (\ref{eq:sol}) are, 
\begin{equation}
\label{eq:exammom}
\left(
\begin{array}{c}
\delta_b\\
\delta_c\\
\delta_e\\
\delta_h\\
\delta_i\\
\delta_j\\
\delta_k\\
\delta_{\ell}\\
\delta_m\\
\delta_n\\
-\\
\end{array}\right)
=
\left(
\begin{array}{rrrrrrrr}
1 & 0 & 0 & 0 & 0 & 0 & 0 & 0 \\
-\frac{1}{2} & 0 & 0 & 0 & 0 & 0 & \frac{1}{2} & 0 \\
-\frac{1}{2} & 0 & 1 & 0 & 0 & 0 & -\frac{1}{2} & 0 \\
0 & 0 & 0 & 0 & 0 & 0 & 0 & 0 \\
0 & 0 & 0 & 0 & 0 & 0 & 0 & 0 \\
0 & 0 & 0 & 0 & 0 & 0 & 0 & 0 \\
0 & 0 & 0 & 0 & 0 & 0 & 0 & 0 \\
0 & 0 & 0 & 0 & 0 & 0 & 0 & 0 \\
0 & 0 & 0 & 0 & 0 & 0 & 0 & 0 \\
0 & 0 & 0 & 0 & 0 & 0 & 0 & 0 \\
0 & 0 & 0 & 0 & 0 & 0 & 0 & 0 \\
\end{array}
\right)
\cdot
\left(
\begin{array}{c}
m(\overline{\Gamma}_5)\\
m(\overline{\Gamma}_6)\\
m(\overline{R}_2\overline{R}_2)\\
m(\overline{S}_2\overline{S}_2)\\
m(\overline{T}_5\overline{T}_6)\\
m(\overline{Y}_5)\\
m(\overline{Y}_6)\\
m(\overline{Z}_5\overline{Z}_6)
\end{array}
\right)\hspace{0.5cm}+\hspace{0.5cm}
\left(
\begin{array}{cccccccc}
-2 & -2 & -2 & -2 & -2 & -2 & -2 & -4 \\
0 & 0 & 0 & 0 & 0 & 0 & 0 & 0 \\
0 & 0 & 0 & 0 & 0 & 0 & 0 & 0 \\
1 & 0 & 0 & 0 & 0 & 0 & 0 & 0 \\
0 & 1 & 0 & 0 & 0 & 0 & 0 & 0 \\
0 & 0 & 1 & 0 & 0 & 0 & 0 & 0 \\
0 & 0 & 0 & 1 & 0 & 0 & 0 & 0 \\
0 & 0 & 0 & 0 & 1 & 0 & 0 & 0 \\
0 & 0 & 0 & 0 & 0 & 1 & 0 & 0 \\
0 & 0 & 0 & 0 & 0 & 0 & 1 & 0 \\
0 & 0 & 0 & 0 & 0 & 0 & 0 & 1 \\
\end{array}
\right)
\cdot
\left(
\begin{array}{c}
k_1\\
k_2\\
k_3\\
k_4\\
k_5\\
k_6\\
k_7\\
k_8\\
\end{array}\right)
\end{equation}
where the second term in the right side depends on the $k_i$, $i=1,\ldots,8$ free parameters.
The second and third rows of the matrix of the second term fulfill the condition (\ref{row:Rmatrix}), and $\delta_c$ and $\delta_e$ are good RSI indices,
\begin{eqnarray}
\label{ind1}
	\delta_c&=&-\frac{1}{2}m(\overline{\Gamma}_5)+\frac{1}{2}m(\overline{Y}_6)\\
\label{ind2}
	\delta_e&=&-\frac{1}{2}m(\overline{\Gamma}_5)-\frac{1}{2}m(\overline{Y}_6)
\end{eqnarray}
expressed in momentum space and calculated from the symmetry data vector $B$. Without having linear combinations, no other local index is well defined in this SSG. Now we explore the possibility of having combinations of indices that do not depend on the $k_i$ integers and which do not involve the last row of expression (\ref{eq:exammom}). We find that, once removed the last row of the matrix on the right in Eq. (\ref{eq:exammom}), there are no linear combinations of the type of Eq. (\ref{eq:solgood2}), but there is a trivial linear combination mod 2 which removes the $k_i$-dependence,
\begin{equation}
\label{ind3}
\delta_b'=\delta_b\textrm{ mod 2}=m(\overline{\Gamma}_5)\textrm{ mod 2}
\end{equation}
and another combination mod 4 which also removes the $k_i$-dependence,
\begin{equation}
\label{ind4}
\delta_{b,h,i,j,k,\ell,m,n}=(\delta_b+2\delta_h+2\delta_i+2\delta_j+2\delta_k+2\delta_{\ell}+2\delta_m+2\delta_n)\textrm{ mod 4}=(m(\overline{\Gamma}_5)+4k_8)\textrm{ mod 4}=m(\overline{\Gamma}_5)\textrm{ mod 4}
\end{equation}
This last index is a composite RSI index, as defined in section \ref{app:momentum}. 
Indices (\ref{ind1}), (\ref{ind2}), (\ref{ind3}) and (\ref{ind4}) are the four independent indices of SSG \msgsymb{67}{508} (N. 67.508) in momentum space, denoted as $\delta_1(c)$, $\delta_2(e)$, $\eta_1(b)$ and $\zeta_1(b,h,i,j,k,\ell,m,n)$, respectively, following the convention introduced in section \ref{app:momentum}.

Although very similar, the indices (\ref{ind3}) and (\ref{ind4}) are indicators of different possible types of materials. For instance, let's take a gapped set of bands with odd multiplicity of irrep $\overline{\Gamma}_5$ at the $\Gamma$ point, $m(\overline{\Gamma}_5)=2\mathbb{N}+1$. Both the $Z_2$-type RSI $\eta_1(b)=\delta_b'$ and the $Z_4$-type RSI $\zeta_1(b,h,i,j,k,\ell,m,n)=\delta_{b,h,i,j,k,\ell,m,n}$ are different from 0. If the WP $b$ is occupied, the gapped bands are compatible with a trivial insulator (all the bands that cannot be moved away from WPs $\{b,h,i,j,k,\ell,m,n\}$ can, in principle, be located at $b$). If the WP $b$ is empty, the gapped bands corresponds to an OAI due to the non-zero value of $\eta_1(b)$, independently of the occupancy or not of the WPs $\{h,i,j,k,\ell,m,n\}$.

Now let us assume that the multiplicity of $\overline{\Gamma}_5$ in the set of bands is even, but not multiple of 4, \ie $m(\overline{\Gamma}_5)=4\mathbb{N}+2$. The indices are $\eta_1(b)=0$ and $\zeta_1(b,h,i,j,k,\ell,m,n)=2$. In this case, the condition for a trivial insulator is that, at least, one of the WPs $\{b,h,i,j,k,\ell,m,n\}$ must be occupied. If all of them (and the WPs of higher symmetry connected to them) are empty positions, the material is an OAI.
\LTcapwidth=1.0\textwidth
\renewcommand\arraystretch{1.2}
\begin{longtable}{|l|l|l|l|l|}
\caption[Induction matrices of the 122 site-symmetry groups realized in the 1651 SSGs]{Induction matrices (\ref{inductionmatrix}) of the 122 site-symmetry groups realized in the 1651 SSGs. The first two columns show the number and symbol of the magnetic point group isomorphic to the site-symmetry group $G^{Q}$ of a WP $Q$. The third column gives the list of symbols(numbers) of the magnetic point groups isomorphic to the maximal site-symmetry subgroups $G^{q_i}$ of the WPs $q_i$ connected to $Q$. Subgroups with the same number are non-conjugated subgroups with respect to $G^{Q}$. The fourth column gives the list of irreps $\rho_{G^{Q}}^i$ of $G^{Q}$ and the last column the induction matrix $C_Q$.}
 \label{tab:ci} \\
\hline
\multicolumn{2}{|c|}{$G^{Q}$}&\multicolumn{1}{|c|}{$G^{q_i}$}&\multicolumn{1}{|c|}{$\rho_{G^Q}$}&\multicolumn{1}{|c|}{$C_Q$}\\
\hline
1&$1$&&$\begin{array}{c}
\overline{A}\\
\end{array}$&\\
\hline
2&$11'$&&$\begin{array}{c}
\overline{A}\overline{A}\\
\end{array}$&\\
\hline
3&$\overline{1}$&$1$(1)&$\begin{array}{c}
\overline{A}_{u}\\
\overline{A}_{g}\\
\end{array}$&$\left(
\begin{array}{c}
 1 \\
 1 \\
\end{array}
\right)$\\
\hline
4&$\overline{1}1'$&$11'$(2)&$\begin{array}{c}
\overline{A}_{u}\overline{A}_{u}\\
\overline{A}_{g}\overline{A}_{g}\\
\end{array}$&$\left(
\begin{array}{c}
 1 \\
 1 \\
\end{array}
\right)$\\
\hline
5&$\overline{1}'$&$1$(1)&$\begin{array}{c}
\overline{A}\overline{A}\\
\end{array}$&$\left(
\begin{array}{c}
 2 \\
\end{array}
\right)$\\
\hline
6&$2$&$1$(1)&$\begin{array}{c}
\,^{2}\overline{E}\\
\,^{1}\overline{E}\\
\end{array}$&$\left(
\begin{array}{c}
 1 \\
 1 \\
\end{array}
\right)$\\
\hline
\end{longtable}

\begin{longtable}{|l|l|l|l|l|l|}
\caption[Exchange matrices of the site-symmetry group-subgroup pairs with at least an exchange term]{Exchange matrices (Eq. \ref{exchangematrix}) of the site-symmetry group-subgroup pairs with at least an exchange term. The first two columns show the number and symbol of the magnetic point group isomorphic to the site-symmetry group $G^{q}$ of the WP $q$ of lowest symmetry. The third and forth columns give the number and symbol of the magnetic point group isomorphic to the site-symmetry group $G^{Q}$ of the WP $Q$ of highest symmetry connected to  $q$ such that $G^{q}\subset G^{Q}$. The fourth column gives the list of irreps of $G^{q}$ and the last column the exchange matrix  $C_q^{e}$.}
 \label{tab:cex} \\
\hline
\multicolumn{2}{|c|}{$G^{q}$}&\multicolumn{2}{c|}{$G^{Q}$}&\multicolumn{1}{|c|}{$\rho_{G^q}^i$}&\multicolumn{1}{c|}{$C_q^{e}$}\\
\hline
6&$2$&15&$2/m'$&$\begin{array}{c}
\,^{2}\overline{E}\\
\,^{1}\overline{E}\\
\end{array}$&$\left(
\begin{array}{r}
 1 \\
 -1 \\
\end{array}
\right)$\\
\hline
6&$2$&17&$222$&$\begin{array}{c}
\,^{2}\overline{E}\\
\,^{1}\overline{E}\\
\end{array}$&$\left(
\begin{array}{r}
 1 \\
 -1 \\
\end{array}
\right)$\\
\hline
6&$2$&34&$\overline{4}'$&$\begin{array}{c}
\,^{2}\overline{E}\\
\,^{1}\overline{E}\\
\end{array}$&$\left(
\begin{array}{r}
 1 \\
 -1 \\
\end{array}
\right)$\\
\hline
6&$2$&40&$422$&$\begin{array}{c}
\,^{2}\overline{E}\\
\,^{1}\overline{E}\\
\end{array}$&$\left(
\begin{array}{r}
 1 \\
 -1 \\
\end{array}
\right)$\\
\hline
6&$2$&42&$4'22'$&$\begin{array}{c}
\,^{2}\overline{E}\\
\,^{1}\overline{E}\\
\end{array}$&$\left(
\begin{array}{r}
 1 \\
 -1 \\
\end{array}
\right)$\\
\hline
6&$2$&48&$\overline{4}2m$&$\begin{array}{c}
\,^{2}\overline{E}\\
\,^{1}\overline{E}\\
\end{array}$&$\left(
\begin{array}{r}
 1 \\
 -1 \\
\end{array}
\right)$\\
\hline
\end{longtable}

\section{An overview of the material databases}\label{app:TQCdatabase}

In this Appendix, we introduce the \webTQC~and \webMTQC~that we are relying on for the high-throughput search of OAIs, OOAIs and magnetic OAIs (mOAIs). 

\subsection{Topological quantum chemistry website}\label{app:TQCweb}

The high-throughput searches of 3D paramagnetic OAIs and OOAIs are relying on the database of the \webTQC, which was originally built in Ref. \cite{vergniory_complete_2019} and upgraded in Ref. \cite{Vergniory2021}. 
For each material in this database, the crystal structure was obtained from the \href{http://www2.fiz-karlsruhe.de/icsd_home.html}{Inorganic Crystal Structure Database} (ICSD)~\cite{ICSD}. The {\it ab initio} calculations were done using the Density Functional Theory (DFT)~\cite{Hohenberg-PR64,Kohn-PR65} and performed on the Vienna Ab-initio Simulation Package (VASP)~\cite{vasp1,PhysRevB.48.13115} with the generalized gradient approximation–Perdew, Burke and Ernzerhof (GGA-PBE)-type exchange correlation potential~\cite{PhysRevLett.77.3865} employed.
All the calculations in \webTQC~were done assuming a paramagnetic phase, \ie without the on-site spin polarization. Although the calculations with and without spin-orbit coupling (SOC)~\cite{PhysRevB.62.11556} term were performed for each material, we only consider the results with SOC in this work. Since several ICSDs might correspond to the same compound, an \emph{unique material} is defined as all the ICSD entries with the same space group, chemical formula and topological properties at the Fermi energy. The  \webTQC\ has \TQCDBNbrUniqueMaterials\ unique materials.

The {\it ab initio} calculation with SOC was performed successfully for \TQCDBNbrICSDs~ICSD entries of stoichiometric materials. Symmetry eigenvalues at all the maximal $k$-vectors, namely the symmetry-data-vector, of the occupied Bloch states were calculated using the \href{https://github.com/zjwang11/irvsp/}{VASP2Trace} program \cite{vergniory_complete_2019, gao2020irvsp} and identified using the tool \href{https://www.cryst.ehu.es/cgi-bin/cryst/programs/magnetictopo.pl?tipog=gesp}{Check Topological Mat.} on the \webBCS. Among the \TQCDBNbrICSDs~ICSDs, the symmetry-data-vectors of \TQCDBNbrICSDsTrivial\ ICSDs were finally classified into topologically trivial insulators (\ie LCEBR) at the Fermi level. In this work, we take the \TQCDBNbrICSDsTrivial\ topologically trivial insulators as the initial inputs for the exhaustive catalogues of paramagnetic OAIs and OOAIs. 

\subsection{Topological magnetic materials website}\label{app:MTQCweb}

The primary search of 3D mOAIs relies on the database of the \webMTQC, which was built in Ref. \cite{xu2020high}. The {\it ab initio} calculations in this website were performed on VASP~\cite{vasp1,PhysRevB.48.13115} with the GGA-PBE-type~\cite{PhysRevLett.77.3865} exchange correlation potential employed. Considering the strong correlation of $d$ and $f$ electrons in magnetic compounds, series of Hubbard-$U$ values incorporating SOC~\cite{PhysRevB.62.11556} were added in the calculations. 

For each material, the analysis used as input the experimental crystal and magnetic structures provided in the \webBCSMAG\  database of magnetic structures hosted at the \webBCS \cite{gallego_magndata1,gallego_magndata2}. 
In Ref. \cite{xu2020high}, among \MTQCDBNbrBCSIDs~magnetic material entries, the self consistent calculations were converged for \MTQCDBNbrBCSIDsWithPhaseDiagram ~materials. Using the magnetic version of VASP2trace, namely the \href{https://www.cryst.ehu.es/cgi-bin/cryst/programs/magnetictopo.pl?tipog=gmag}{Mvassp2trace} and the \href{https://www.cryst.ehu.es/cgi-bin/cryst/programs/magnetictopo.pl?tipog=gmag}{Check Topological Magnetic Mat} programs, symmetry-data-vector of the occupied Bloch bands and the topological phase diagram as a function of Hubbard-$U$ were obtained for each magnetic material. 
From Ref.\cite{xu2020high}, \MTQCDBNbrBCSIDsTrivial~magnetic material entries are diagnosed as magnetic topologically trivial insulators (\ie LCEBR) at certain Hubbard-$U$ values in their topological phase diagrams. In this work, we perform a primary search for mOAIs from the \MTQCDBNbrBCSIDsTrivial~magnetic materials which are topologically trivial insulators by implementing some Hubbard-$U$ values.

\section{Implementation of RSIs to the high-throughput search for OAIs and OOAIs}\label{app:highthroughput}

The RSIs,  as defined in Appendix \ref{app:rsi3D} and originally introduced in Ref.~\cite{song2020},  indicate the multiplicities of irreps of a BR pinned at the corresponding Wyckoff positions of a crystal lattice \cite{song2020,song_fragile_2019}. In the following subsections of this Appendix, we detail the methods to classify the OAIs and OOAIs from topologically trivial insulators using the RSI indices.

\subsection{The method of diagnosing OAIs}\label{app:methodOAI}

For a given topologically trivial band structure, its BR can be decomposed into linear combination of EBRs, \ie the BRs induced from the irreps at the maximal Wyckoff positions, with non-negative integer coefficients ~\cite{bradlyn_topological_2017,po2017symmetry,MTQC}. As the BR of a topologically trivial insulator is characterized by the symmetry-data-vector (Eq. \ref{eq:B-vector}), which is provided on the \webTQC~ for \TQCDBNbrICSDsTrivial\ ICSD entries (\TQCDBNbrMaterialsTrivial\ unique materials) being crystalline topologically trivial insulators, we can directly calculate the RSIs of the BR by substituting the symmetry-data-vector into the RSI formula, as defined in Appendix \ref{app:rsi3D} and tabulated on the \webBCS. If all the RSIs of a BR are zero, the BR is compatible with a set of Wannier functions centered at any set of WPs and the compound is clearly an atomic insulator (AI). 
For a BR with non-zero RSIs, we adopt the following strategies to check if the BR is an OAI.

\begin{enumerate}
    \item Starting from the crystal structural file of a topologically trivial insulator used for the {\it ab initio} calculations in the Refs.~\cite{vergniory_complete_2019,Vergniory2021} and the \webTQC, we first identify the Wyckoff positions that are occupied by all the non-equivalent atoms which are not related by any symmetry operation of the corresponding space group.
    \item Then, for each non-zero RSI index, we check if the set of WPs that correspond to this (in general, composite) RSI index are occupied by atoms. If at least one WP of the set is occupied by an atom, the non-zero RSI index is compatible with an atomic insulator. If none of the WPs of the set is occupied, we go to the next step.
    \item For each of the non-zero RSIs defined at empty WPs, as selected in the last step, we check if the corresponding set of WPs are maximal or non-maximal. If all of them are maximal WPs, this RSI indicates an OAI. If the set includes non-maximal WPs, we check whether these non-maximal WPs are connected to the WPs of higher symmetry and occupied by atoms. If all the WPs, which are of higher symmetry and connected to the set of WPs that correspond to the non-zero RSI index, are empty, this RSI indicates an OAI. 
    Otherwise, the non-zero RSI index is compatible with an atomic insulator.
    \item If at least one RSI index indicates an OAI in the previous step, we refer this material to as an OAI.
\end{enumerate}

Using the above method, we have performed the high-throughput search for OAIs from all the topologically trivial insulators provided on the \webTQC, as detailed in the subsection \ref{app:subhighthroughput}.

\subsection{The method of diagnosing orbital-selected OAIs (OOAIs)}\label{app:methodOOAI}

For the non-zero RSIs that do not indicate an OAI phase, as described in the Appendix \ref{app:methodOAI}, we further analyze them in this subsection and check if they indicate an OOAI phase. 

As introduced in Appendix~\ref{app:OAI}, if a BR can be induced from a set of occupied WPs but not from the irreps of the site-symmetry group that correspond to the outer-shell atomic orbitals of the atoms, we refer the BR to as an OOAI. 
In other words, the non-zero-integer RSI at an occupied WP indicates an OOAI if the orbitals involved in the definition of the (in general composite) RSI do not correspond to the outer-shell orbitals of the atoms located at that WP.

In the high-throughput search for OOAIs, for the non-zero (in general, composite) RSI , $\delta$($\{\alpha_i\}$), which is defined at a set of WPs $\{\alpha_i\}$ (where $i=1,2,...,N^{wp}$ and $N^{wp}$ is the number of WPs in the definition of $\delta$), if at least one of the WPs in $\{\alpha_i\}$ is occupied and the RSI value could be contributed by the outer-shell atomic orbitals of the atoms sitting at $\{\alpha_i\}$, this non-zero RSI is trivial and does not indicate an OAI or OOAI.  We adopt the following four steps to enumerate all the trivial RSIs ($\{\Delta(\{\alpha_i\})\}$) which can be contributed by outer-shell electrons of the non-equivalent atoms sitting at the WPs $\{\alpha_i\}$:

\begin{enumerate}
    \item First, we obtain the outer-shell atomic orbitals, which are provided by the pseudo-potential file adopted in the ab-initio calculations, of the non-equivalent atoms sitting at $\{\alpha_i\}$ of the related site symmetry groups $\{G^{\alpha_i}\}$. We assume the set of outer-shell atomic orbitals of the atom sitting at $\alpha_i$ is $L(\alpha_i)=\{l(\alpha_i)\}$, where $l=s,p$ or $d$ represents the $s, p$ or $d$ outer-shell atomic orbital of the non-equivalent atoms sitting at $\alpha_i$. (The outer-shell atomic orbitals of all the related atoms are tabulated in Table \ref{tab:eleconfig}.)
    \item Then, we calculate the set of irreps, $\rho(\alpha_i)=\{\rho^i_k\}$($k=1,2,...,N^{i,irrep}$) of the site-symmetry group $G^{\alpha_i}$ given by the orbitals in $L(\alpha_i)$, where $N^{i,irrep}$ is the number of such irreps. (Note that some irreps in $\rho(\alpha_i)$ could be the same. For example, the closed-shell $p$ orbitals sitting at the site of point group $\bar 1$ split into three $\bar A_u \bar A_u$ irreps, of which each $\bar A_u \bar A_u$ irrep has a unique $k$ index.
    The set of irreps given by the $s,p$ or $d$ atomic orbitals under different site symmetry groups are tabulated in Table \ref{tab:3dpgirrep}.)
    \item We exhaust all the possible combinations of irreps in $\{\rho^i_k\}$($i=1,2,...,N^{wp},k=1,2,...,N^{i,irrep}$) and calculate the corresponding induced BRs of each combination, $\{B_m\}$, where $m=1,2,...,N^B$ and $N^B$ is the number of possible combinations. Note that each combination corresponds to a possible valence state of the atoms in $\{\alpha_i\}$.
    \item For each BR in $\{B_m\}$, we calculate its related RSI $\Delta_m(\{\alpha_i\})$ at the WPs $\{\alpha_i\}$ and refer the RSI to as a trivial RSI.
\end{enumerate}
If the RSI is not trivial, \ie $\delta(\{\alpha_i\})\notin \{\Delta_m(\{\alpha_i\})\}$, we refer the topologically trivial insulator to as an OOAI.

\begin{table}
\centering
\caption[Outer-shell electronic configuration of each type of atom]{Outer-shell electronic configuration, provided in the pseudo potential file (\ie \emph{POTCAR}) and adopted in the {\it ab-initio} calculations in Refs. \cite{vergniory_complete_2019, Vergniory2021} and on the \webTQC, of each type of atom.}
\label{tab:eleconfig}
\begin{tabular}{cc|cc|cc|cc|cc|cc|cc|cc}
\hline
Atom & $L$ &  Atom & $L$ &  Atom & $L$ &  Atom & $L$ & Atom & $L$ &  Atom & $L$ & Atom & $L$ & Atom & $L$ \\
\hline
Ag &s,d &Al &s,p &Ar &s,p &As &s,p &Au &s,d &B &s,p &Ba &s,p,s &Be &s \\ 
Bi &s,p &Br &s,p &C &s,p &Ca &s,p,s &Cd &s,d &Cl &s,p &Co &d,s & Cr &d,s \\ 
Cs &s,p,s &Cu &d,p &F &s,p &Fe &d,s &Ga &s,p &Ge &s,p &H &s & He &s \\ 
Hf &s,d &Hg &s,d &I &s,p &In &s,p &Ir &s,d &K &p,s &La &s,d & Li &s \\ 
Mg &p,s &Mn &d,s &Mo &s,d &N &s,p &Na &s &Nb &s,d &Ne &s,p &Ni &d,s \\ 
O &s,p &Os &s,d &P &s,p &Pb &s,p &Pd &s,d &Pt &s,d &Rb &p,s &Re &s,d \\ 
Rh &s,d &Ru &s,d &S &s,p &Sb &s,p &Sc &s,d &Se &s,p &Si &s,p &Sn &s,p \\ 
Sr &s,p,s &Ta &s,d &Tc &s,d &Te &s,p &Ti &d,s &V &p,d,s &W &s,d &Xe &s,p \\ 
Y &s,d &Zn &d,p &Zr &s,d & & & & & & & & & \\
\hline
\end{tabular}
\end{table}

\LTcapwidth=1.0\textwidth
\renewcommand\arraystretch{1.1}
\begin{table}
\centering
\caption[Induced irreps from atomic orbitals under the 32 double point groups]{Induced irreps from $s$, $p$ or $d$ atomic orbitals under the 32 double point groups.}
\label{tab:3dpgirrep}
\begin{tabular}{cc|c|c|c}
\hline
PG & (Symbol) & $s$ & $p$($p_x,p_y,p_z$) & $d$($d_{xy},d_{yz},d_{zx},d_{x^2-y^2},d_{z^2}$) \\
\hline
1& $1$ & $\bar {A}\bar {A}$  &$3\bar {A}\bar {A}$  &$5\bar {A}\bar {A}$   \\
2& $\bar 1$ & $\bar {A}_{g}\bar {A}_{g}$  &$3\bar {A}_{u}\bar {A}_{u}$  &$5\bar {A}_{g}\bar {A}_{g}$   \\
3& $2$ & $^1\bar {E}^2\bar {E}$  &$3^1\bar {E}^2\bar {E}$  &$5^1\bar {E}^2\bar {E}$   \\
4& $m$ & $^1\bar {E}^2\bar {E}$  &$3^1\bar {E}^2\bar {E}$  &$5^1\bar {E}^2\bar {E}$   \\
5& $2/m$ & $^1\bar {E}_{g}^2\bar {E}_{g}$  &$3^1\bar {E}_{u}^2\bar {E}_{u}$  &$5^1\bar {E}_{g}^2\bar {E}_{g}$   \\
6& $222$ & $\bar {E}$  &$3\bar {E}$  &$5\bar {E}$   \\
7& $mm2$ & $\bar {E}$  &$3\bar {E}$  &$5\bar {E}$   \\
8& $mmm$ & $\bar {E}_{g}$  &$3\bar {E}_{u}$  &$5\bar {E}_{g}$   \\
9& $4$ & $^1\bar {E}_{1}^2\bar {E}_{1}$  &$2^1\bar {E}_{1}^2\bar {E}_{1}$ , $^1\bar {E}_{2}^2\bar {E}_{2}$  &$2^1\bar {E}_{1}^2\bar {E}_{1}$ , $3^1\bar {E}_{2}^2\bar {E}_{2}$   \\
10& $\bar 4$ & $^1\bar {E}_{1}^2\bar {E}_{1}$  &$^1\bar {E}_{1}^2\bar {E}_{1}$ , $2^1\bar {E}_{2}^2\bar {E}_{2}$  &$2^1\bar {E}_{1}^2\bar {E}_{1}$ , $3^1\bar {E}_{2}^2\bar {E}_{2}$   \\
11& $4/m$ & $^1\bar {E}_{1g}^2\bar {E}_{1g}$  &$^1\bar {E}_{2u}^2\bar {E}_{2u}$ , $2^1\bar {E}_{1u}^2\bar {E}_{1u}$  &$2^1\bar {E}_{1g}^2\bar {E}_{1g}$ , $3^1\bar {E}_{2g}^2\bar {E}_{2g}$   \\
12& $422$ & $\bar {E}_{1}$  &$\bar {E}_{2}$ , $2\bar {E}_{1}$  &$3\bar {E}_{2}$ , $2\bar {E}_{1}$   \\
13& $4mm$ & $\bar {E}_{1}$  &$\bar {E}_{2}$ , $2\bar {E}_{1}$  &$3\bar {E}_{2}$ , $2\bar {E}_{1}$   \\
14& $\bar 42m$ & $\bar {E}_{1}$  &$2\bar {E}_{2}$ , $\bar {E}_{1}$  &$3\bar {E}_{2}$ , $2\bar {E}_{1}$   \\
15& $4/mmm$ & $\bar {E}_{1g}$  &$2\bar {E}_{1u}$ , $\bar {E}_{2u}$  &$3\bar {E}_{2g}$ , $2\bar {E}_{1g}$   \\
16& $3$ & $^1\bar {E}^2\bar {E}$  &$2^1\bar {E}^2\bar {E}$ , $\bar {E}\bar {E}$  &$3^1\bar {E}^2\bar {E}$ , $2\bar {E}\bar {E}$   \\
17& $\bar 3$ & $^1\bar {E}_{g}^2\bar {E}_{g}$  &$2^1\bar {E}_{u}^2\bar {E}_{u}$ , $\bar {E}_{u}\bar {E}_{u}$  &$2\bar {E}_{g}\bar {E}_{g}$ , $3^1\bar {E}_{g}^2\bar {E}_{g}$   \\
18& $32$ & $\bar {E}_{1}$  &$^1\bar {E}^2\bar {E}$ , $2\bar {E}_{1}$  &$2^1\bar {E}^2\bar {E}$ , $3\bar {E}_{1}$   \\
19& $3m$ & $\bar {E}_{1}$  &$^1\bar {E}^2\bar {E}$ , $2\bar {E}_{1}$  &$2^1\bar {E}^2\bar {E}$ , $3\bar {E}_{1}$   \\
20& $\bar 3m$ & $\bar {E}_{1g}$  &$^1\bar {E}_{u}^2\bar {E}_{u}$ , $2\bar {E}_{1u}$  &$3\bar {E}_{1g}$ , $2^1\bar {E}_{g}^2\bar {E}_{g}$   \\
21& $6$ & $^1\bar {E}_{3}^2\bar {E}_{3}$  &$^1\bar {E}_{1}^2\bar {E}_{1}$ , $2^1\bar {E}_{3}^2\bar {E}_{3}$  &$2^1\bar {E}_{1}^2\bar {E}_{1}$ , $^1\bar {E}_{2}^2\bar {E}_{2}$ , $2^1\bar {E}_{3}^2\bar {E}_{3}$   \\
22& $\bar 6$ & $^1\bar {E}_{3}^2\bar {E}_{3}$  &$^1\bar {E}_{1}^2\bar {E}_{1}$ , $2^1\bar {E}_{2}^2\bar {E}_{2}$  &$2^1\bar {E}_{1}^2\bar {E}_{1}$ , $^1\bar {E}_{2}^2\bar {E}_{2}$ , $2^1\bar {E}_{3}^2\bar {E}_{3}$   \\
23& $6/m$ & $^1\bar {E}_{3g}^2\bar {E}_{3g}$  &$2^1\bar {E}_{3u}^2\bar {E}_{3u}$ , $^1\bar {E}_{1u}^2\bar {E}_{1u}$  &$2^1\bar {E}_{1g}^2\bar {E}_{1g}$ , $2^1\bar {E}_{3g}^2\bar {E}_{3g}$ , $^1\bar {E}_{2g}^2\bar {E}_{2g}$   \\
24& $622$ & $\bar {E}_{1}$  &$\bar {E}_{3}$ , $2\bar {E}_{1}$  &$\bar {E}_{2}$ , $2\bar {E}_{3}$ , $2\bar {E}_{1}$   \\
25& $6mm$ & $\bar {E}_{1}$  &$\bar {E}_{3}$ , $2\bar {E}_{1}$  &$\bar {E}_{2}$ , $2\bar {E}_{3}$ , $2\bar {E}_{1}$   \\
26& $\bar 6m2$ & $\bar {E}_{1}$  &$2\bar {E}_{2}$ , $\bar {E}_{3}$  &$\bar {E}_{2}$ , $2\bar {E}_{3}$ , $2\bar {E}_{1}$   \\
27& $6/mmm$ & $\bar {E}_{1g}$  &$\bar {E}_{3u}$ , $2\bar {E}_{1u}$  &$2\bar {E}_{3g}$ , $\bar {E}_{2g}$ , $2\bar {E}_{1g}$   \\
28& $23$ & $\bar {E}$  &$^1\bar {F}^2\bar {F}$ , $\bar {E}$  &$2^1\bar {F}^2\bar {F}$ , $\bar {E}$   \\
29& $m\bar 3$ & $\bar {E}_{g}$  &$^1\bar {F}_{u}^2\bar {F}_{u}$ , $\bar {E}_{u}$  &$2^1\bar {F}_{g}^2\bar {F}_{g}$ , $\bar {E}_{g}$   \\
30& $432$ & $\bar {E}_{1}$  &$\bar {F}$ , $\bar {E}_{1}$  &$\bar {E}_{2}$ , $2\bar {F}$   \\
31& $\bar 43m$ & $\bar {E}_{1}$  &$\bar {E}_{2}$ , $\bar {F}$  &$\bar {E}_{2}$ , $2\bar {F}$   \\
32& $m\bar 3m$ & $\bar {E}_{1g}$  &$\bar {E}_{1u}$ , $\bar {F}_{u}$  &$2\bar {F}_{g}$ , $\bar {E}_{2g}$ \\
\hline
\end{tabular}
\end{table}

Using the above method, we have performed the high-throughput search for OOAIs from all the paramagnetic topologically trivial insulators provided on the \webTQC, as detailed in the subsection \ref{app:subhighthroughput}. Note that a topologically trivial insulator can be diagnosed both as an OAI and as an OOAI.

\subsection{High-throughput searches for OAIs, mOAIs and OOAIs}\label{app:subhighthroughput}

Using the methods in Appendices \ref{app:methodOAI} and \ref{app:methodOOAI}, we have performed the high-throughput searches for the OAIs, OOAIs and mOAIs from \TQCDBNbrICSDsTrivial\ ICSD entries (\TQCDBNbrMaterialsTrivial\ unique materials) for the paramagnetic topologically trivial insulators available on the \webTQC~and \MTQCDBNbrBCSIDsTrivial~magnetic topologically trivial insulators on the \webMTQC, respectively. We found \TQCDBNbrICSDsOAI\ ICSDs (\TQCDBNbrICSDsOAIPercent\  of the topologically trivial ICSDs) are OAIs, including \TQCDBNbrICSDsOAIIndirectGap~OAIs with finite indirect gap and \TQCDBNbrICSDsOAIOnlyDirectGap~OAIs without an indirect gap but a direct gap, \TQCDBNbrICSDsOOAI~ICSDs (\TQCDBNbrICSDsOOAIPercent\  of the topologically trivial ICSDs) tagged as OOAIs and \MTQCDBNbrBCSIDsMOAI~mOAIs.  A complete summary of these figures is tabulated for all the three types of materials in Appendix~\ref{app:3DOAI}. In Table~\ref{tab:statisticspersg}, we provide the statistics of OAIs and OOAIs per SG  as found in the high-throughput search. For sake of completeness, we point out that we provided the statistics for the filling-enforced OAIs (feOAIs), a special types of OAIs, in Ref. \cite{xu2021filling}.

\begin{longtable}{|c|c|c|c|c|c|c|c|c|}
\caption[Statistics of OAIs per SG]{Statistics of OAIs, OOAIs and fe-OAIs per space group (SG). For each SG, we provide the total number of unique materials in this SG (second column) and the number of trivial insulator (third column). The fourth column gives the number of OAIs and the fifth column, the number of OAIs with a non-zero indirect gap (IGap). Similar figures are provided in columns 6-7 and 8-9 for OOAIs and fe-OAIs, respectively. All percentages are defined with respect to the number of unique materials in the SG.}\label{tab:statisticspersg}\\
\hline
SG & Nbr materials & Nbr trivial & Nbr OAIs & \begin{tabular}{c}Nbr OAIs\\ with IGap\end{tabular} & Nbr OOAIs & \begin{tabular}{c}Nbr OOAIs\\ with IGap \end{tabular} & Nbr fe-OAIs& \begin{tabular}{c}Nbr fe-OAIs\\ with IGap\end{tabular} \\
\hline
1 & 151 & 142 {\tiny(94.0\%)} & ---  & ---  & ---  & ---  & ---  & ---  \\ 
2 & 1574 & 1276 {\tiny(81.1\%)} & 149 {\tiny(9.5\%)} & 126 {\tiny(8.0\%)} & ---  & ---  & 92 {\tiny(5.8\%)} & 77 {\tiny(4.9\%)} \\ 
3 & 13 & 13 {\tiny(100\%)} & ---  & ---  & ---  & ---  & ---  & ---  \\ 
4 & 228 & 220 {\tiny(96.5\%)} & ---  & ---  & ---  & ---  & ---  & ---  \\ 
5 & 156 & 141 {\tiny(90.4\%)} & 6 {\tiny(3.9\%)} & 3 {\tiny(1.9\%)} & ---  & ---  & 6 {\tiny(3.9\%)} & 3 {\tiny(1.9\%)} \\ 
6 & 28 & 27 {\tiny(96.4\%)} & ---  & ---  & ---  & ---  & ---  & ---  \\
\hline
\end{longtable}

\begin{figure}
\centering\includegraphics[width=1.0\textwidth]{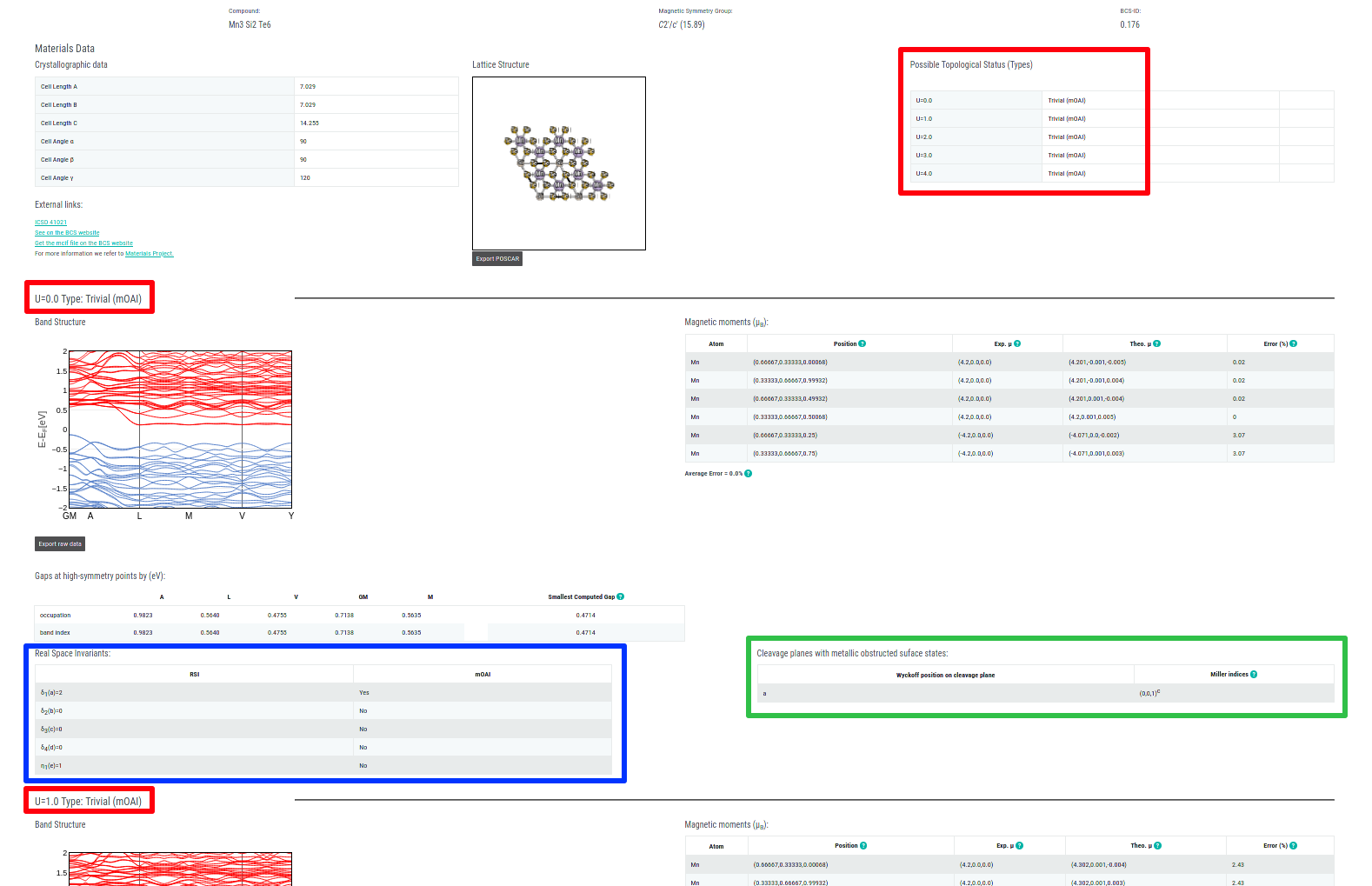}
\caption{An example, based on ${\rm Mn}_3 {\rm Si}_2 {\rm Te}_6$ [\bcsidweblong{0.176}, \msgsymbnum{15}{89}], showing how the mOAI information is displayed on the \webMTQC. When appropriate, a tag ``mOAI" is added to the topological property summary (red box) and the title of each Hubbard-$U$ value. The RSIs are provided in the table ``Real Space Invariants" (blue box), with a clear indication pointing out those leading to an mOAI. If available, the Miller indices of the cleavage planes are given in the table ``Cleavage planes with metallic obstructed surface states" (green box).}\label{fig:mtqcweb}
\end{figure}

\begin{figure}
\centering\includegraphics[width=1.0\textwidth]{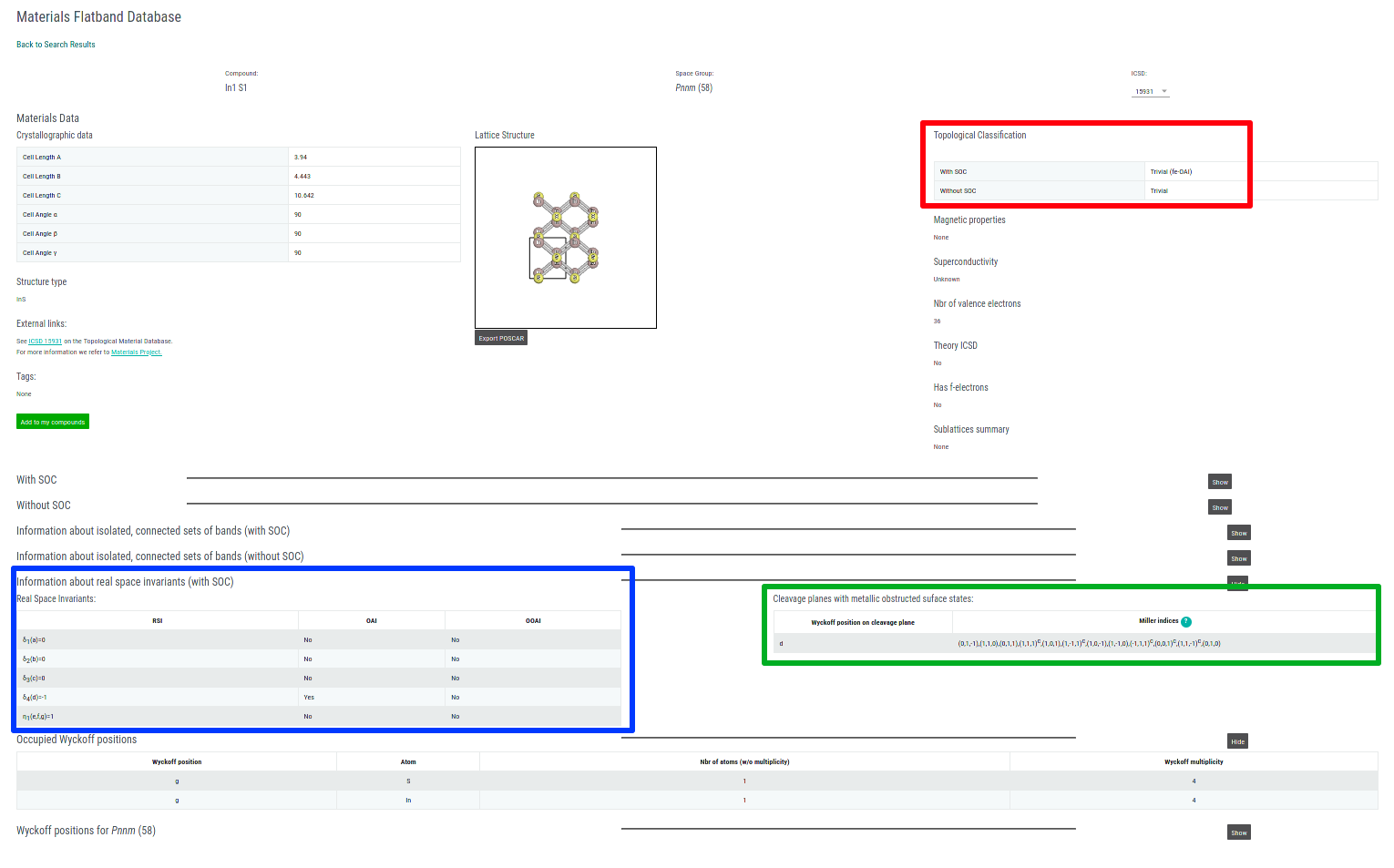}
\caption{An example, based on ${\rm In} {\rm S}$ [\icsdweb{15931}, SG 58 (\sgsymb{58})], showing how the OAI (or OOAI) information is displayed on the \webflatband. When appropriate, a tag ``OAI", ``OOAI" or ``fe-OAI" is added to the topological classification summary (red box). The RSIs are provided in the table ``Real Space Invariants" (blue box), with a clear indication pointing out those leading to an OAI or OOAI. If available, the Miller indices of the cleavage planes are given in the table ``Cleavage planes with metallic obstructed surface states" (green box).}\label{fig:flatbandweb}
\end{figure}

For each compound we have analyzed, we provide online information on our different websites. The mOAI details are available on the \webMTQC. An example of how the data are displayed is shown in Fig.~\ref{fig:mtqcweb}. In particular for every Hubbard-$U$ value that was considered, we give the RSI values, pointing out those leading to an mOAI, and the Miller indices of the cleavage planes exhibiting metallic obstructed surface states (if they are known). About the OAI, fe-OAI and OOAI, the \webTQC\ only indicates for each ICSD if it hosts such a type of insulator and we provide an external link to the full data. These latest are located on the \webflatband, as exemplified in Fig.~\ref{fig:flatbandweb}. Like the mOAI, we give the list of RSI values and when relevant, the Miller indices related to the metallic obstructed surface state.

\section{The RSI calculations of InS, Mn$_3$Si$_2$Te$_6$, La$_2$Ti$_2$O$_7$ and 2H-MoS$_2$,  }\label{app:rsical}

In this section, using the symmetry-data-vectors as tabulated in the \webTQC, we calculate the nontrivial RSIs indicating an OAI or OOAI phase of the materials \ce{InS}, \ce{Mn3Si2Te6}, \ce{la2Ti2O7} and 2H-\ce{MoS2}.

\subsection{RSI indices of \ce{InS}}
The symmetry-data-vector of \ce{InS} is,
\begin{equation}
\begin{aligned}
B=&(10\bar \Gamma_5, 8\bar \Gamma_6, 9\bar R_3\bar R_4, 4\bar S_5\bar S_5, 5\bar S_6\bar S_6, 5\bar T_2\bar T_5, 5\bar T_3\bar T_4, 4\bar T_6\bar T_9, 4\bar T_7\bar T_8, 4\bar U_2\bar U_3, 4\bar U_4\bar U_5, 5\bar U_6\bar U_7, 5\bar U_8\bar U_9, 9\bar X_3\bar X_4, 9\bar Z_3\bar Z_4, 9\bar Z_3\bar Z_4 )
\end{aligned}
\label{eq:B15931}
\end{equation}
By substituting the $B$ vector in Eq.~(\ref{eq:B15931}) into all the RSI indices of SG 58 (See the BCS tools \href{http://www.cryst.ehu.es/cryst/RSI}{RSIsg}), it shows that \ce{InS} is an OAI indicated by a non-zero-integer $Z$-type RSI which is defined at the empty sites of Wyckoff position $2d$, namely the $\delta_4(d)$ index, 
\begin{equation}
\begin{aligned}
\delta_4(d)=&-\frac{1}{4}[m(\bar \Gamma_5)+2m(\bar S_6\bar S_6) -2m(\bar T_6\bar T_9)+2m(\bar U_6\bar U_7) -2m(\bar Z_3\bar Z_4)] = -1
\end{aligned}
\label{eq:rsi15931}
\end{equation}
where $m(\bar X_i)$ is the multiplicity of the irrep $\bar X_i$ at the $X$ momenta. Hence, the Wyckoff position $2d$ is the OWCC of this OAI.

\subsection{RSI indices of \ce{Mn3Si2Te6}}
The symmetry-data-vector of \ce{Mn3Si2Te6} is,
\begin{equation}
\begin{aligned}
B=&(81\bar \Gamma_2, 85\bar \Gamma_3, 83\bar Z_2 \bar Z_3, 83\bar U_2,83\bar U_3, 83\bar T_2, 83\bar T_3, 83\bar R_2 \bar R_3, 81\bar X_2, 85\bar X_3, 81\bar Y_2, 85\bar Y_3, 81\bar V_2, 85\bar V_3)
\end{aligned}
\label{eq:B0.176}
\end{equation}
By substituting the $B$ vector in Eq.~(\ref{eq:B0.176}) into the formula of RSIs of SSG 15.89, we found that \ce{Mn3Si2Te6} is a mOAI indicated by a $Z$-type magnetic RSI at the empty Wyckoff position $4a$, \ie
\begin{equation}
\begin{aligned}
\delta_1(a)=&\frac{1}{4}[m(\bar \Gamma_3)-2 m(\bar Z_2 \bar Z_3)-m(\bar U_2) - m(\bar U_3) +2 m(\bar X_3)+m(\bar V_3)]=2
\end{aligned}
\label{eq:rsi0.176}
\end{equation}
where $m(\bar X_i)$ is the multiplicity of irrep $\bar X_i$ at the momenta $X$.

\subsection{RSI indices of \ce{La2Ti2O7}}
From the TQCDB, the BR of \ce{La2Ti2O7} is characterized by the symmetry-data-vector,
\begin{equation}
    \begin{aligned}
    B=&(6\bar \Gamma_6,6\bar \Gamma_7,7\bar \Gamma_8,7\bar \Gamma_9,10\bar \Gamma_{10},13\bar \Gamma_{11},12\bar L_4\bar L_5,11\bar L_6\bar L_7, 26\bar L_8,23\bar L_9,18\bar W_3\bar W_4,18\bar W_5\bar W_6,36\bar W_7,36\bar X_5)
    \end{aligned}
    \label{eq:B164027}
\end{equation}
Using the $B$ vector in Eq.~(\ref{eq:B164027}), we calculate all the RSIs defined in SG 227 and found that three $Z$-type RSIs have non-zero-integer values,
\begin{equation}
    \begin{aligned}
    \delta_3(c)=&-\frac{1}{2}[m(\bar \Gamma_6)-m(\bar \Gamma_9)-m(\bar L_4 \bar L_5) +m(\bar L_6 \bar L_7)]=1 
    \end{aligned}
    \label{eq:3rsi164027}
\end{equation}
\begin{equation}
    \begin{aligned}
    \delta_4(c)=&-\frac{1}{4}[m(\bar \Gamma_6)+2m(\bar \Gamma_8)+m(\bar \Gamma_9)-2m(\bar \Gamma_{11}) +3m(\bar L_4 \bar L_5)+m(\bar L_6 \bar L_7)-2m(\bar L_8)]=1
    \end{aligned}
    \label{eq:4rsi164027}
\end{equation}
\begin{equation}
    \begin{aligned}
    \delta_7(e)=&2m(\bar \Gamma_6)+2m(\bar \Gamma_8)-2m(\bar \Gamma_9)+m(\bar \Gamma_{11}) -m(\bar L_4 \bar L_5)-m(\bar L_6 \bar L_7)=2
    \end{aligned}
    \label{eq:7rsi164027}
\end{equation}
where $m(\bar X_i)$ is the multiplicity of irrep $\bar X_i$ at the momenta $X$.

\subsection{RSI indices of \ce{MoS2}}
2H-\ce{MoS2} [\icsdweb{105091}] has the symmetries of the SG 194 (\sgsymb{194}) with the Mo and S atoms occupying the respective $2c$ and $4f$ Wyckoff positions. From the \webTQC, \ce{MoS2} is a topologically trivial insulator with indirect band gap of 0.7 eV. 
The symmetry-data-vector of \ce{MoS2} is,
\begin{equation}
    \begin{aligned}
    B=&(2\bar \Gamma_7,3\bar \Gamma_8,4\bar \Gamma_9,2\bar \Gamma_{10},4\bar \Gamma_{11},3\bar \Gamma_{12},2\bar A_4\bar A_5,7\bar A_6, 3\bar H_4 \bar H_5, 3\bar H_6 \bar H_7, 6\bar H_8, 6\bar H_9, 6\bar K_7, 5\bar K_8, 7\bar K_9, 9\bar L_3\bar L_4, 9\bar M_5, 9\bar M_6)
    \end{aligned}
    \label{eq:MoS2105091}
\end{equation}
By substituting the $B$ vector in Eq.~(\ref{eq:MoS2105091}) into the formula of RSIs of SG 194, we found that \ce{MoS2} is an OAI indicated by a $Z$-type RSI at the empty Wyckoff position $2b$, \ie
\begin{equation}
\begin{aligned}
\delta_3(b)=&\frac{1}{3}[m(\bar \Gamma_7)+m(\bar \Gamma_{10})-m(\bar \Gamma_{12})+2 m(\bar A_6)-m(\bar H_4 \bar H_5)-m(\bar H_6 \bar H_7)-2m(\bar K_8)]=1
\end{aligned}
\label{eq:rsi105091}
\end{equation}
where $m(\bar X_i)$ is the multiplicity of irrep $\bar X_i$ at the momenta $X$. Using the method in Appendix~\ref{app:filling_anomaly}, we find the cleavage plane of Miller index $(100)$ exhibits the metallic OSSs.

\section{Filling anomaly and obstructed surface states of OAIs}\label{app:filling_anomaly}

For an OAI material with non-zero RSIs defined at the empty sites, there must be Wannier functions centered at the empty positions, which are referred to as the obstructed Wannier charge centers(OWCCs). 
So it is possible to have a cleavage Miller plane that cuts through \emph{at least} one of the OWCCs but away from all the atoms, which results in the obstructed surface states (OSSs) in the bulk band gap. If the cleaved terminations of the finite-size crystal of an OAI are related by the elements of the site-symmetry group at the OWCC, the OSSs are \emph{filling anomaly}~\cite{song_d-2-dimensional_2017,PhysRevB.95.035421,benalcazar2019quantization,schindler2019fractional}, \ie by counting the number of valence electrons the OSSs have to be partially filled. 

To be specific, for a given OAI material with occupied WPs $\{{\alpha_i}^{occ}\}$ and a non-zero composite RSI defined at a set of Wyckoff positions $\{{\alpha_j}^{owcc}\}$, which are also referred to as the OWCCs, the cleavage plane with normal vector of Miller index $(h,k,l)$, which is of the conventional unit cell, has OSSs, if it satisfies the two following criteria: 
\begin{enumerate}
    \item[I] The cleavage plane has to be away from all the atoms of coordinates $\{(X(a_i),Y(a_i),Z(a_i))\}$ , where $i=1,2,...,N^{atom}$ and $N^{atom}$ is the number of atoms in a supercell of a large enough size, in this work, $5\times5\times5$. 
    \item[II] The cleavage plane cuts through \emph{all} the Wyckoff positions in $\{{\alpha_j}^{owcc}\}$ , where $j=1,2,...,N^{wp}$ and $N^{wp}$ is the number of non-equivalent WPs in the definition of the composite RSI.
\end{enumerate}

For a 2D cleavage plane of coordinates $(x,y,z)$ and Miller index ($h,k,l$), the criteria-I is characterized by the following inequality,

\begin{equation}
\begin{aligned}
&(h\vec{b}_1+k\vec{b}_2+l\vec{b}_3)\cdot[x-{X(a_i)},y-{Y(a_i)},z-{Z(a_i)}]\ne 0, & i=1,...,N^{atom},  &  \forall x, y, z 
\end{aligned}
\label{eq:millerindex}
\end{equation}
where $\vec{b}_1$, $\vec{b}_2$ and $\vec{b}_3$ are the three reciprocal lattice vectors of the material

While in the criteria-II, to identify if the cleavage plane of Miller index $(h,k,l)$ is cutting through a WP $\alpha_j^{owcc}$ in the OWCCs set $\{\alpha_j^{owcc}\}$, the plane should satisfy one of the following criteria depends on the types of the WP $\alpha_j^{owcc}$,

\begin{enumerate}

\item If the WP $\alpha_j^{owcc}$ is a center of coordinates ($X(\alpha_j^{owcc}),Y(\alpha_j^{owcc}),Z(\alpha_j^{owcc})$), the definition of the cleavage plane $z(x,y)$ should satisfy the following equation,
\begin{equation}
(h\vec{b}_1+k\vec{b}_2+l\vec{b}_3)\cdot[x-{X(\alpha_j^{owcc})},y-{Y(\alpha_j^{owcc})},z-{Z(\alpha_j^{owcc})}]=0 \\
\label{eq:millerpoint}
\end{equation}

\item If the WP $\alpha_j^{owcc}$ is defined on a line with one degree of freedom along the lattice direction $[n_1,n_2,n_3]$, the normal vector of the cleavage plane (\ie $(h\vec{b}_1+k\vec{b}_2+l\vec{b}_3)$) should be perpendicular to the Wyckoff line defined by the WP $\alpha_j^{owcc}$ and cuts through the line, \ie it satisfies the following equations,
\begin{equation}
\begin{aligned}
&(h\vec{b}_1+k\vec{b}_2+l\vec{b}_3)\cdot(n_1\vec{c}_1+n_2\vec{c}_2+n_3\vec{c}_3)=0, \\
&(h\vec{b}_1+k\vec{b}_2+l\vec{b}_3)\cdot[x-{X(\alpha_j^{owcc})},y-{Y(\alpha_j^{owcc})},z-{Z(\alpha_j^{owcc})}]=0 \\
\end{aligned}
\label{eq:millerline}
\end{equation}
where ($X(\alpha_j^{owcc}),Y(\alpha_j^{owcc}),Z(\alpha_j^{owcc})$) is any point on the Wyckoff line, $(\vec{c}_1,\vec{c}_2,\vec{c}_3)$ are the three lattice vectors of the unit cell.

\item  If the WP $\alpha_j^{owcc}$ is defined on a plane of norm vector along the lattice direction $[p_1,p_2,p_3]$, the cleavage plane should be parallel to the Wyckoff plane defined by WP $\alpha_j^{owcc}$ and cuts through the Wyckoff plane, \ie it satisfies the following equations,
\begin{equation}
\begin{aligned}
&(h\vec{b}_1+k\vec{b}_2+l\vec{b}_3)\times(p_1\vec{c}_1+p_2\vec{c}_2+p_3\vec{c}_3)=0, \\
&(h\vec{b}_1+k\vec{b}_2+l\vec{b}_3)\cdot(x-{X(\alpha_j^{owcc})},y-{Y(\alpha_j^{owcc})},z-{Z(\alpha_j^{owcc})})=0 \\
\end{aligned}
\label{eq:millerplane}
\end{equation}
where ($X(\alpha_j^{owcc}),Y(\alpha_j^{owcc}),Z(\alpha_j^{owcc})$) is any point on the Wyckoff plane.
\end{enumerate}
If each of the WPs in $\{\alpha_j^{owcc}\}$ satisfies one of the above three criteria, \ie Eqs. (\ref{eq:millerpoint}), (\ref{eq:millerline}) and (\ref{eq:millerplane}), the cleavage plane of Miller index $(h,k,l)$ satisfies the criteria-II.

For each of the \TQCDBNbrICSDsOAI\ paramagnetic OAIs and \MTQCDBNbrBCSIDsMOAI\ mOAIs obtained in the high-throughput search, we have identified its possible cleavage planes, which satisfy both criteria-I and II, that host metallic OSSs. The results are tabulated in the Tables \ref{tab:oaiindirect}, \ref{tab:oaidirect} and \ref{tab:mOAIs} in Appendix \ref{app:OAIlist}.

In the following subsections, we select seven paramagnetic OAIs and one mOAI to analyze their RSIs and showcase the OSSs. 
The eight materials are the paramagnetic OAIs \ce{NbBr2O} (\icsdweb{416669}), \ce{CaIn2P2} (\icsdweb{260562}), \ce{InSe} (\icsdweb{185172}), \ce{Hg2IO} (\icsdweb{33275}), \ce{PtSbSi} (\icsdweb{413194}), \ce{Nb3Br8} (\icsdweb{25766}), \ce{B12} (\icsdweb{431636}) and the anti-ferromagnetic mOAI \ce{CsFe2Se3} (\bcsidweblong{1.26}).

\subsection{NbBr$_2$O}\label{app:NbBr2O}

\begin{figure}[htbp]
\centering\includegraphics[width=6.5in]{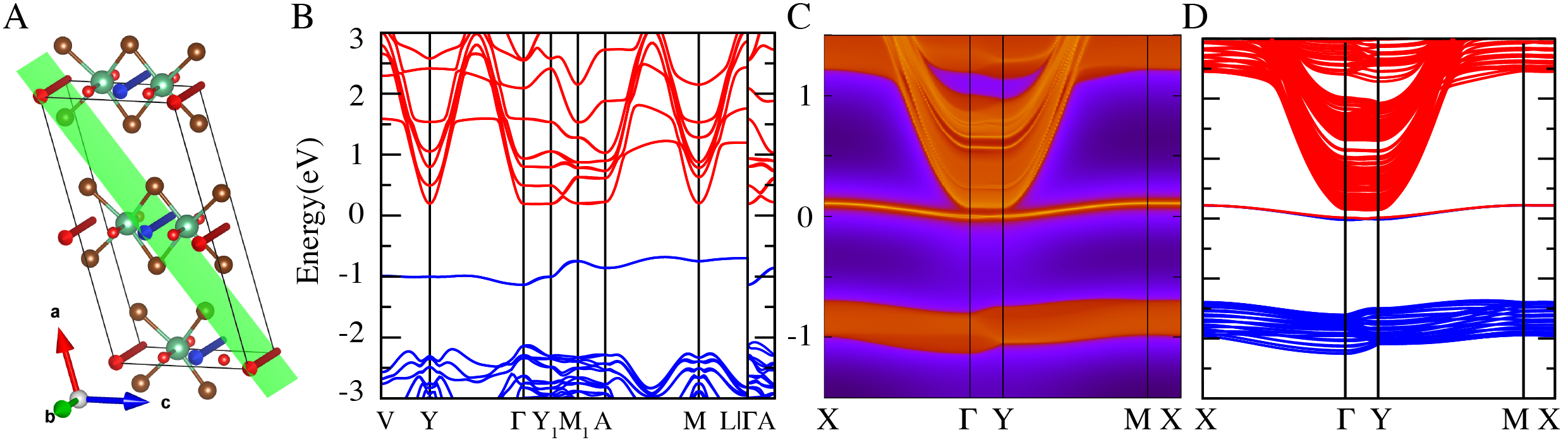}
\caption{(A) Crystal structure of \ce{NbBr2O}, where the red and blue lines indicate the positions of Wyckoff letter $a$ and $b$, which are identified as the OWCCs. The green plane with Miller index $(101)$ cuts through the OWCCs and exhibits the OSSs.
(B) Bulk band structure along the high-symmetry paths of \ce{NbBr2O}, where the blue lines represent the valence bands (VB) and the red lines represent the conduction bands (CB). (C) Band structure of a semi-infinite slab structure with the $(010)$ cleavage plane as defined in (A). (D) Same as (C) but of a finite-size slab structure. In (D), the valence and conduction bands are represented by the blue and red lines, respectively. }
\label{fig:NbBr2O}
\end{figure}

As shown in Fig. \ref{fig:NbBr2O}(A), \ce{NbBr2O} of \icsdweb{416669} adopts a monoclinic lattice with SG 5 (\sgsymb{5}). All atoms occupy the positions of Wyckoff letter $2c$. The band structure, as calculated in the \webTQC\ and shown in Fig. \ref{fig:NbBr2O}(B), indicates that \ce{NbBr2O} is a topologically trivial insulator in terms of irreps (\ie LCEBR) with large indirect band gap of $0.86 eV$. The BR of \ce{NbBr2O} at the maximal $k$-vectors is,
\begin{equation}
B=33(\bar \Gamma_3\bar \Gamma_4,\bar A_3\bar A_4,\bar M_3\bar M_4,\bar Y_3\bar Y_4,\bar L_2\bar L_2,\bar V_2\bar V_2) \label{eq:B416669}
\end{equation}
which can be decomposed into the linear combination of EBRs of SG 5 (\sgsymb{5}), \ie $B=^1\bar{E}^2\bar{E}@a \oplus ^1\bar{E}^2\bar{E}@b$. 
As the EBRs $^1\bar E ^2\bar E@a$ and $^1\bar E ^2\bar E@b$ have exactly the same symmetry eigenvalues at the maximal $k$-vectors, the decomposition of the BR is not unique. Therefore, different decompositions give different values of the local RSIs at $a$ and $b$ and hence the RSI defined in the momentum space is not well defined at either $a$ or $b$ individually. As developed in Appendix \ref{app:rsi3D} and provided on the \webBCS, for SG 5, there is one composite $Z_2$-type RSI, namely the $\eta$ index, defined at the set of Wyckoff positions $(a,b)$,

\begin{equation}
\eta_1(a,b)=m(\bar \Gamma_3\bar \Gamma_4)~mod~2
\label{eq:rsi416669}
\end{equation}
where $m(\bar \Gamma_3\bar \Gamma_4)$ is the multiplicity of the irreducible representation $\bar \Gamma_3\bar \Gamma_4$ at the $\Gamma$ point.
When $\eta_1(a,b)=1$, either the $a$ or $b$ position must have an odd number of irreps $^1\bar{E} {}^2\bar{E}$.

By substituting the symmetry-data-vector $B$ into Eq. \ref{eq:rsi416669}, we obtain the RSI index $\eta_1(a,b)=1$.
As both of the Wyckoff positions $a$ and $b$ are empty, \ce{NbBr2O} is an OAI indicated by the non-zero $Z_2$-type RSI $\eta_1(a,b)$. 
By using the criteria in Eqs. \ref{eq:millerindex}-\ref{eq:millerplane}, we find only the cleavage planes of Miller indices $(101)$ and $10\bar1$ could cut through both of the lines defined by the Wyckoff positions $a$ and $b$, both of which are defined on lines along the $[010]$ lattice direction, but away from all the occupied sites, hence the $(101)$ and $(10\bar1)$ cleavage planes could exhibit the OSSs when it cuts through the Wyckoff positions both $a$ and $b$.
As schematically shown in Fig. \ref{fig:NbBr2O}(A), the positions defined by $a$ and $b$ are represented by the red and blue arrow lines, respectively, and the cleavage plane with OSSs is represented by the green plane. 

In Fig. \ref{fig:NbBr2O}(C) and (D), we have calculated the surface states of \ce{NbBr2O} on the $(101)$ surface. 
Fig.~\ref{fig:NbBr2O}(C) shows the band structure of a semi-infinite slab structure calculated with the Green's function method~\cite{sancho1985highly,WU2017}. The surface states are lying in between the conduction and the valence bands of the bulk states. To clarify the charge neutrality point of the band structure, we have also performed the surface states calculation of a finite-size slab structure. As shown in Fig.~\ref{fig:NbBr2O}(D), by counting the number of valence bands, it shows that the two connected bands of surface states near the Fermi level are half filled, giving a metallic configuration with filling anomaly on the surface.

\subsection{CaIn$_2$P$_2$}\label{app:CaIn2P2}

\begin{figure}[htbp]
\centering\includegraphics[width=6.5in]{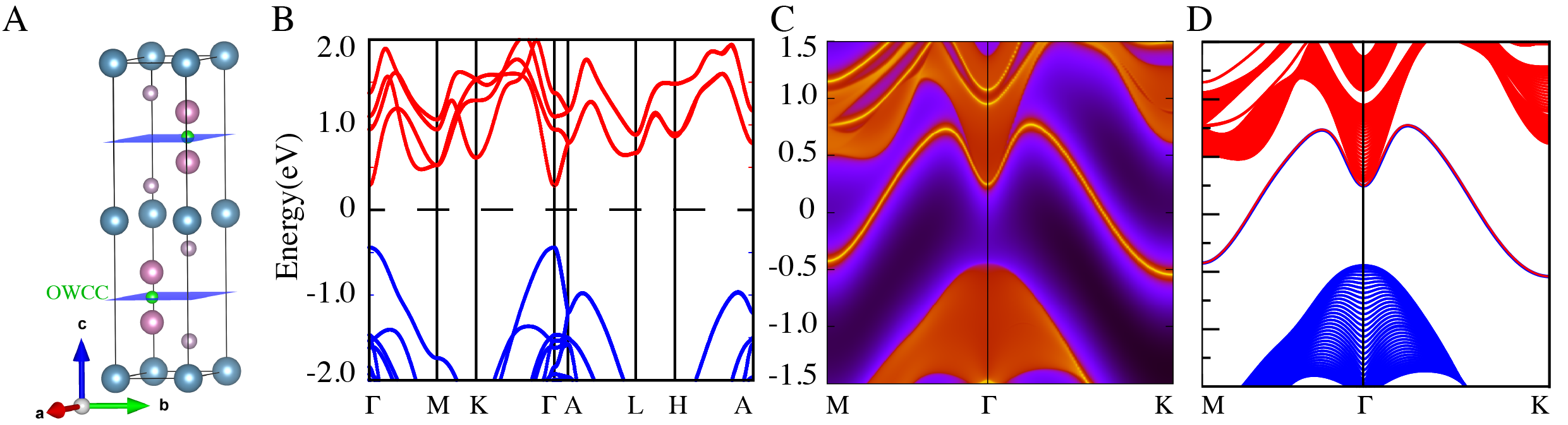}
\caption{(A) Crystal structure of \ce{CaIn2P2}, where the green spheres represent the positions of the OWCCs. The blue planes with Miller index $(001)$ cuts through the OWCCs and exhibits the OSSs.
(B)Bulk band structure along the high-symmetry paths of \ce{CaIn2P2}, where the blue lines represent the VBs and the red lines represent the CBs. (C) Band structure of a semi-infinite slab structure with the $(001)$ cleavage plane as defined in (A). (D) Same as (C) but of a finite-size slab structure. In (D), the valence and conduction bands are represented by the blue and red lines, respectively.
}\label{fig:CaIn2P2}
\end{figure}

\ce{CaIn2P2} with~\icsdweb{260562} adopts a hexagonal lattice with SG 194 ($P6_3/mmc$). As shown in Fig.~\ref{fig:CaIn2P2}(A), Ca atoms occupy the Wyckoff position $a$. Both In and P atoms occupy the Wyckoff position $f$. The band structure, as calculated on the \webTQC\ and shown in Fig.~\ref{fig:CaIn2P2}(B), indicates that \ce{CaIn2P2} is a topologically trivial insulator (\ie LCEBR) with a direct band gap of $0.708 eV$. 
The BR of \ce{CaIn2P2} at the maximal $k$-vectors is:

\begin{equation}
B=(2\bar\Gamma_7,3\bar\Gamma_8,4\bar\Gamma_9,2\bar\Gamma_{10},4\bar\Gamma_{11},3\bar\Gamma_{12},2\bar A_4\bar A_5,7\bar A_6,3\bar H_4\bar H_5,4\bar H_6\bar H_7,6\bar H_8,5\bar H_9,7\bar K_7,6\bar K_8,5\bar K_9,9\bar L_3\bar L_4,9\bar M_5,9\bar M_6) \label{eq:B260562}
\end{equation}

By substituting the $B$ vector given by Eq. \ref{eq:B260562} in the expressions of the RSI indices of SG 194, as defined in Appendix~\ref{app:rsi3D}, we find that \ce{CaIn2P2} is an OAI indicated by a $Z$-type RSI, namely a $\delta$ index, at the empty WP $d$,
\begin{equation}
\delta_5(d)=\frac{1}{3}[-2m(\bar\Gamma_7)-2m(\bar\Gamma_{10})+m(\bar\Gamma_{11})-m(\bar\Gamma_{12})+3m(\bar A_4\bar A_5)-m(\bar A_6) -m(\bar H_4\bar H_5)+2m(\bar H_6\bar H_7)+m(\bar K_8)]=1 \label{eq:rsi260562}
\end{equation}
where $m(\bar X_i)$ is the multiplicity of $\bar X_i$ at $X$ point.

By using the criteria in Eqs. \ref{eq:millerindex} and \ref{eq:millerpoint}, we find the cleavage plane with Miller index $(001)$ has OSSs when it cuts through the Wyckoff position $d$, which is a center. As there always exist other atoms on the top (or bottom) of the OWCC at $d$ along the $[001]$ direction, all the cleavage planes parallel to $[001]$ direction are impossible to avoid cutting through the atoms. In this case, there must have atoms co-planar with the OWCCs on the top \emph{or} bottom surface (not both of them) and there has no filling anomaly. 

As schematically shown in Fig.~\ref{fig:CaIn2P2}(A), the atoms at WP $d$ are represented by the green spheres and the $(001)$ cleavage planes with OSSs are indicated by the blue planes. 

Similarly, in Fig.~\ref{fig:CaIn2P2}(C) and (D), we have calculated the surface states of \ce{CaIn2P2} on the $(001)$ surface. Fig.~\ref{fig:CaIn2P2}(C) shows the band structure of a semi-infinite slab structure calculated with the Green's function method. The surface states locate between the gap of the conduction bands and the valence bands of bulk states. To clarify the charge neutrality point of the band structure, we have also performed the surface states calculation of a finite-size slab. As shown in Fig.~\ref{fig:CaIn2P2}(D), by counting the number of valence bands,  we find the two connected bands of surface states near the Fermi level are half filled and hence it is a metallic configuration with filling anomaly on the surface.

\subsection{InSe}\label{app:InSe}

\begin{figure}[htbp]
\centering\includegraphics[width=6.5in]{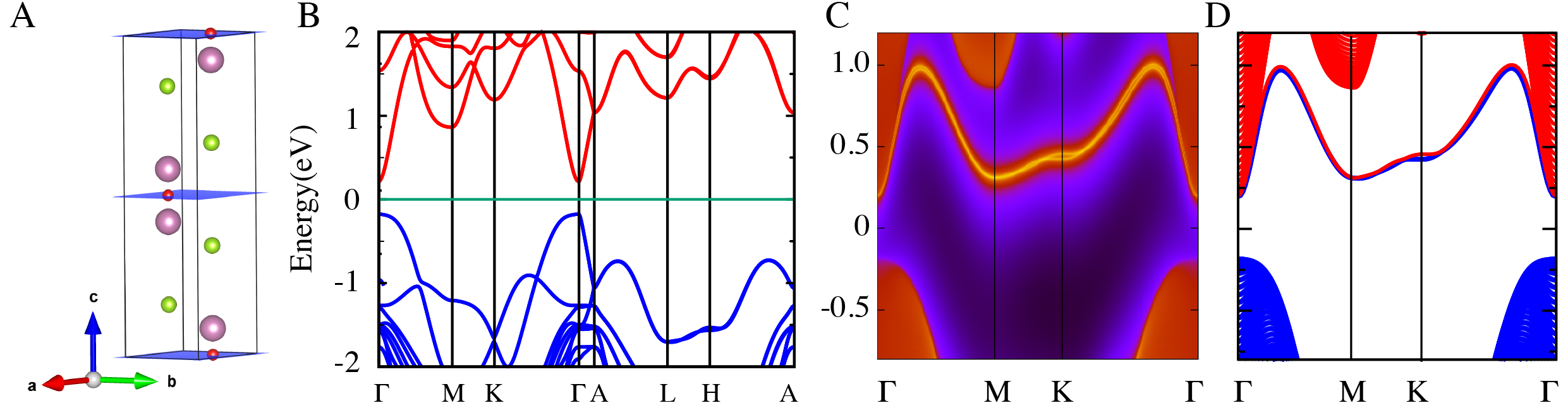}
\caption{(A) Crystal structure of \ce{InSe}, where the red spheres represent the positions of the OWCCs. The blue planes with Miller index $(001)$ cuts through the OWCCs and exhibits the OSSs.
(B)Bulk band structure along the high-symmetry paths of \ce{InSe}, where the blue lines represent the VBs and the red lines represent CBs. (C) Band structure of a semi-infinite slab structure with the $(001)$ cleavage plane as defined in (A). (D) Same as (C) but of a finite-size slab structure. In (D), the valence and conduction bands are represented by the blue and red lines, respectively.}\label{fig:InSe}
\end{figure}

Similar to the structure of the previous example, \ce{CaIn2P2}, the compound \ce{InSe} with~\icsdweb{185172} adopts a hexagonal lattice of SG 194 (\sgsymb{194}). As shown in Fig.~\ref{fig:InSe}(A), both In and Se atoms occupy the Wyckoff position $f$. The band structure, as calculated in the \webTQC~and in Fig.~\ref{fig:InSe}(B), indicates that \ce{InSe} is a topologically trivial insulator with a direct band gap of $0.479 eV$. 
The BR of \ce{InSe} is:
\begin{equation}
    B=(2\bar\Gamma_7,3\bar\Gamma_8,4\bar\Gamma_9,2\bar\Gamma_{10},4\bar\Gamma_{11},3\bar\Gamma_{12},2\bar A_4\bar A_5,7\bar A_6,4\bar H_4\bar H_5,3\bar H_6\bar H_7,5\bar H_8,6\bar H_9,7\bar K_7,6\bar K_8,5\bar K_9,9\bar L_3\bar L_4,9\bar M_5,9\bar M_6) \label{eq:B185172}
\end{equation}

By substituting the $B$ vector in Eq. \ref{eq:B185172} to the RSI indices of SG 194, as defined in Appendix~\ref{app:rsi3D}, we find \ce{InSe} is an OAI indicated by a $Z$-type RSI, namely a $\delta$ index, at the empty Wyckoff position $c$,

\begin{equation}
\delta_4(c)=\frac{1}{3}[-2m(\bar\Gamma_7)-2m(\bar\Gamma_{10})+m(\bar\Gamma_{11})-m(\bar\Gamma_{12})+3m(\bar A_4\bar A_5)-m(\bar A_6) +2m(\bar H_4\bar H_5)-m(\bar H_6\bar H_7)+m(\bar K_8)]=1 \label{eq:rsi185172}
\end{equation}
where $m(\bar X_i)$ is the multiplicity of $\bar X_i$ at $X$ point.

Similar to the case in \ce{CaIn2P2}, using the criteria in Eqs. \ref{eq:millerindex} and \ref{eq:millerpoint}, we find the cleavage plane with Miller index $(001)$ has OSSs when it cuts through the Wyckoff position $c$. As schematically shown in Fig.~\ref{fig:InSe}(A), the positions defined by $c$ are represented by the red spheres and the cleavage plane with OSSs is indicated by the blue planes. Similarly, the OSSs of \ce{InSe} on the $(001)$ surface are shown in  the surface states calculations in Fig.~\ref{fig:InSe}.

\subsection{Hg$_2$IO}\label{app:Hg2IO}

\begin{figure}[htbp]
\centering\includegraphics[width=6.5in]{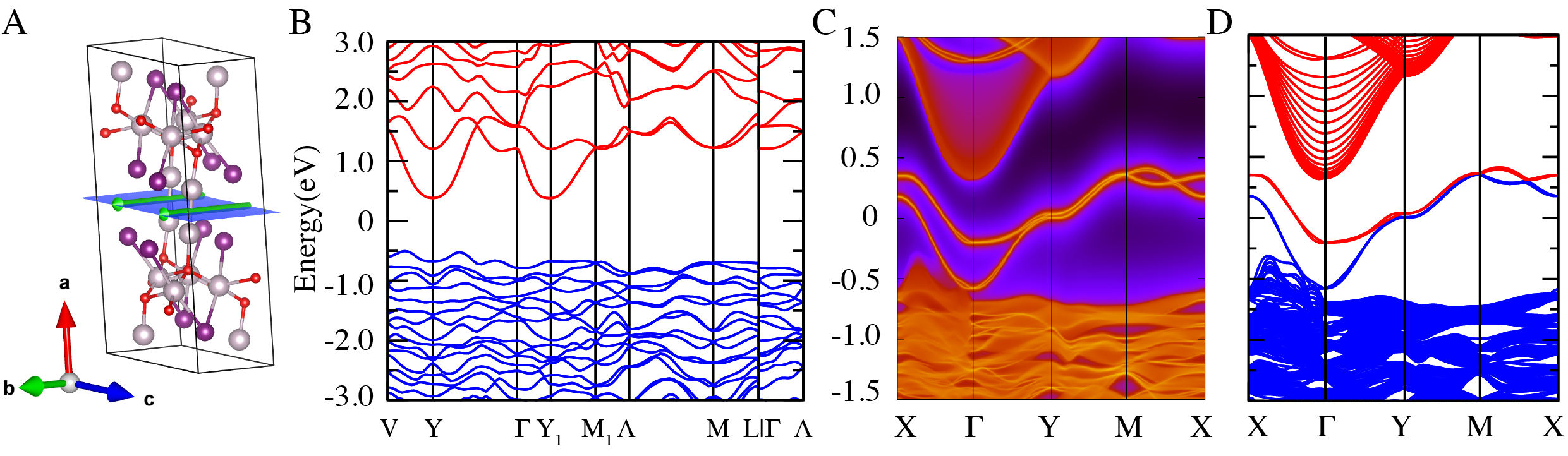}
\caption{(A) Crystal structure of \ce{Hg2IO}, where the green lines represent the positions of OWCCs. The blue planes with Miller index $(001)$ cuts through the OWCCs and exhibits the OSSs.
(B)Bulk band structure along the high-symmetry paths of \ce{Hg2IO}, where the blue lines represent the VBs and the red lines represent CBs. (C) Band structure of a semi-infinite slab structure with the $(001)$ cleavage plane as defined in (A). (D) Same as (C) but of a finite-size slab structure. In (D), the valence and conduction bands are represented by the blue and red lines, respectively.}\label{fig:Hg2IO}
\end{figure}

As shown in Fig.~\ref{fig:Hg2IO}(A) and (B), \ce{Hg2IO} (\icsdweb{33275}) with SG 15 (\sgsymb{15}) is a topologically trivial band insulator with an indirect band gap 0.983eV. All atoms occupy the Wyckoff position $f$. 
The BR of \ce{Hg2IO} at the maximal high-symmetry $\bf k$ points is:
\begin{equation}
B=37(\bar \Gamma_3\bar \Gamma_4, \bar \Gamma_5\bar \Gamma_6, \bar A_2\bar A_2, \bar L_2\bar L_2, \bar L_3\bar L_3, \bar M_2\bar M_2, \bar V_2\bar V_2, \bar V_3\bar V_3, \bar Y_3\bar Y_4, \bar Y_5\bar Y_6)\label{eq:B33275}
\end{equation}

By substituting the $B$ vector into the formula of RSI indices in SG 15, as defined in Appendix~\ref{app:rsi3D}, it shows that \ce{Hg2IO} is an OAI indicated by a $Z_2$-type RSI, namely an $\eta$ index, at the empty Wyckoff position $e$,

\begin{equation}
\eta_1(e)=\frac{1}{2}[m(\bar \Gamma_3\bar \Gamma_4)+m(\bar L_2\bar L_2)+3m(\bar L_3\bar L_3)+m(\bar Y_3\bar Y_4)]~mod~2=1 \label{eq:rsi33275}\end{equation}
where $m(\bar X_i)$ is the multiplicity of the irrep $\bar X_i$ at the $X$ point.

As the OWCC $e$ is defined on lines in the $[010]$ direction, using the criteria in Eqs. \ref{eq:millerindex} and \ref{eq:millerline}, we find the cleavage plane with Miller index $(001)$, as indicated by the blue plane in Fig.~\ref{fig:Hg2IO}(A), has OSSs. 

Similarly, in Fig.~\ref{fig:Hg2IO}(C) and (D), we have calculated the surface states of \ce{Hg2IO} on the $(001)$ surface. Fig.~\ref{fig:Hg2IO}(C) shows the band structure of a semi-infinite slab structure calculated with the Green's function method. The surface states locate between the   conduction bands and the valence bands of bulk states. A finite-size slab calculation in Fig.~\ref{fig:Hg2IO}(D) shows that the two connected surface bands near the Fermi level are half filled and hence it is a metallic configuration with filling anomaly on the surface.

\subsection{PtSbSi}\label{app:PtSbSi}

\begin{figure}[htbp]
\centering\includegraphics[width=6.5in]{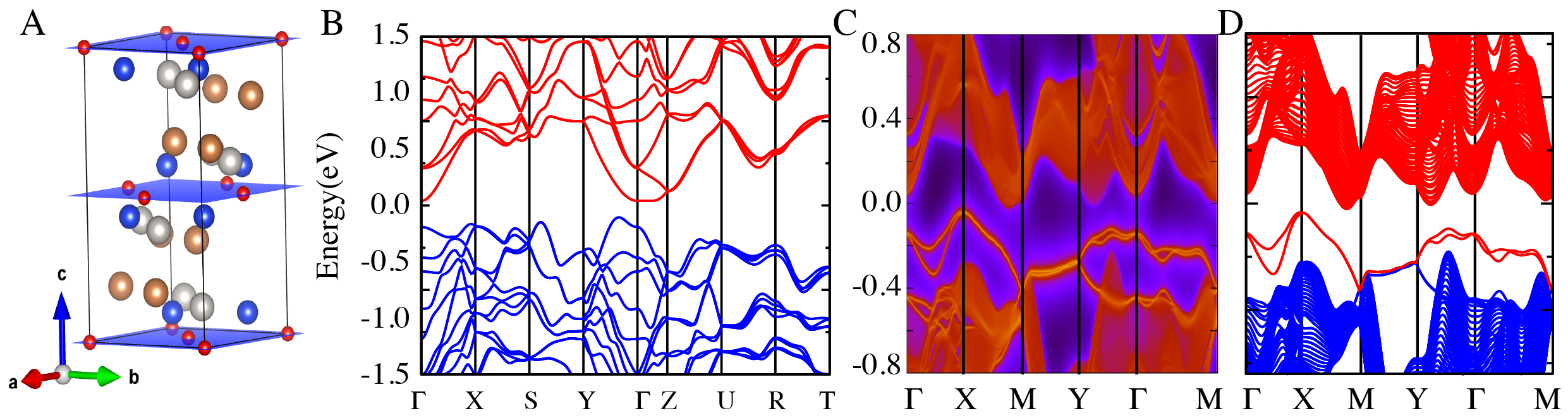}
\caption{(A) Crystal structure of \ce{PtSbSi}, where the red spheres represent the positions of the OWCCs. The blue planes with Miller index $(001)$ cuts through the OWCCs and exhibits the OSSs.
(B)Bulk band structure along the high-symmetry paths of \ce{PtSbSi}, where the blue lines represent the VBs and the red lines represent CBs. (C) Band structure of a semi-infinite slab structure with the $(001)$ cleavage plane as defined in (A). (D) Same as (C) but of a finite-size slab structure. In (D), the valence and conduction bands are represented by the blue and red lines, respectively.}\label{fig:PtSbSi}
\end{figure}

As shown in Fig.~\ref{fig:PtSbSi}(A) and (B), \ce{PtSbSi} (\icsdweb{413194}) with SG 61 ($Pbca$) is a topologically trivial insulator with an indirect band gap $0.199 eV$. All atoms occupy the Wyckoff position $c$. 
The BR of \ce{PtSbSi} is,
\begin{equation}
\begin{aligned}
&B=(40\bar \Gamma_5,36\bar \Gamma_6,10\bar R_2\bar R_2,10\bar R_3\bar R_3,10\bar R_4\bar R_4,10\bar R_5\bar R_5,9\bar R_6\bar R_6,9\bar R_7\bar R_7,9\bar R_8\bar R_8,9\bar R_9\bar R_9, \\
&19\bar S_3\bar S_3,19\bar S_4\bar S_4,19\bar T_3\bar T_3,19\bar T_4\bar T_4,19\bar U_3\bar U_3,19\bar U_4\bar U_4,38\bar X_3\bar X_4,38\bar Y_3\bar Y_4,38\bar Z_3\bar Z_4)
\end{aligned}
\label{eq:B413194}
\end{equation}
By substituting the $B$ vector into the formula of RSI indices in SG 61, as defined in Appendix~\ref{app:rsi3D}, it shows that \ce{PtSbSi} is an OAI indicated by a $Z$-type RSI, namely a $\delta$ index, at the empty Wyckoff position $a$,

\begin{equation}
\delta_1(a)=-\frac{1}{4}m(\bar \Gamma_6)+m(\bar R_2\bar R_2)=1
\label{eq:rsi413194}
\end{equation}
where $m(\bar X_i)$ is the multiplicity of $\bar X_i$ at $X$ point.

Similar to the case in \ce{CaIn2P2}, using the criteria in Eqs. \ref{eq:millerindex} and \ref{eq:millerpoint}, we find the cleavage planes with Miller indices $(010)$ and $(001)$, as indicated by the blue plane in Fig.~\ref{fig:PtSbSi}(A), has OSSs. 

In Fig.~\ref{fig:PtSbSi}(C) and (D), we calculate the surface states of \ce{PtSbSi} on the $(001)$ surface. Fig.~\ref{fig:PtSbSi}(C) shows the band structure of a semi-infinite slab structure calculated with the Green's function method. The surface states locate between the conduction bands and the valence bands of bulk states. A finite-size slab calculation in Fig.~\ref{fig:PtSbSi}(D) shows that the two connected surface bands near the Fermi level are half filled and hence are metallic.

\subsection{Nb$_3$Br$_8$}\label{app:Nb3Br8}

\begin{figure}[htbp]
\centering\includegraphics[width=6.5in]{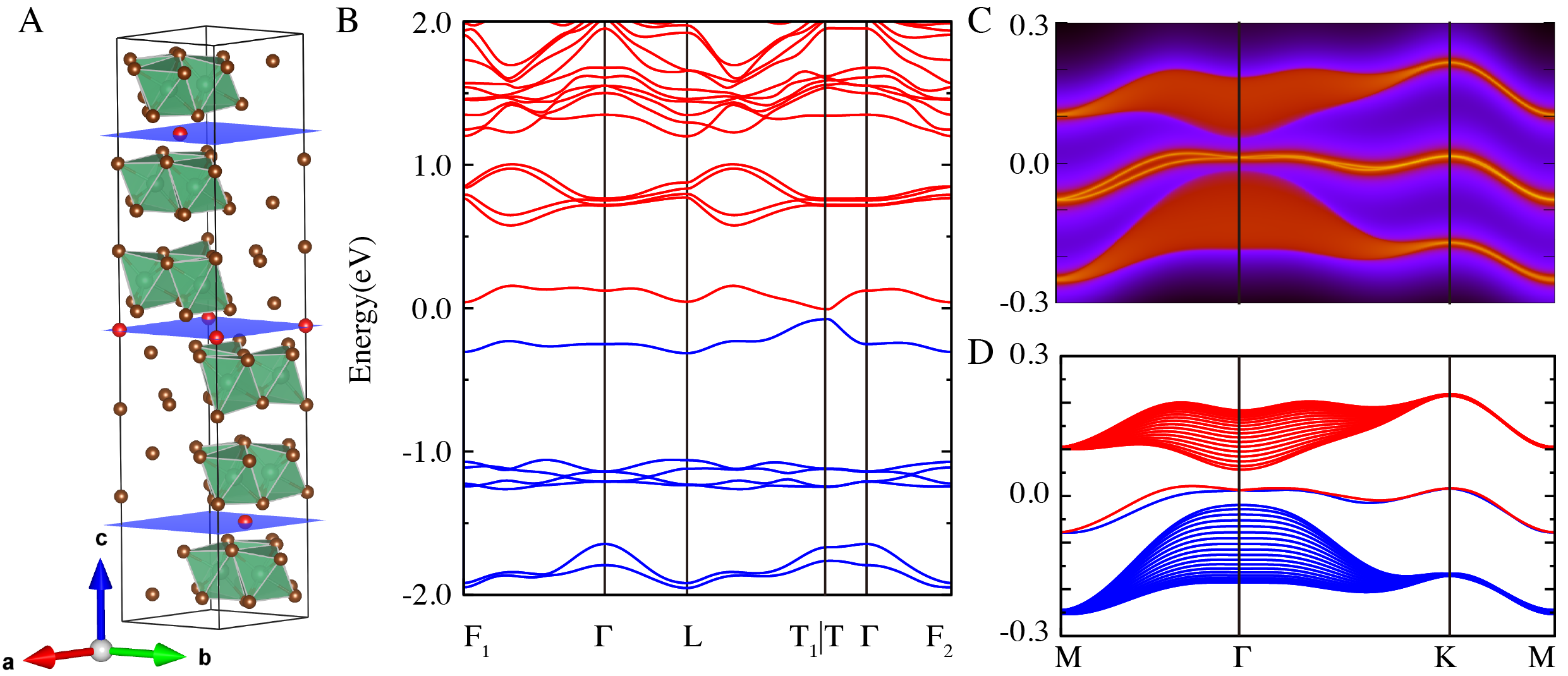}
\caption{(A) Crystal structure of \ce{Nb3Br8}, where the red spheres represent the positions of the OWCCs. The blue planes with Miller indices $(001)$ are the cleavage surfaces cutting through the OWCCs and exhibit OSSs. (B) Bulk band structure along the high-symmetry paths of \ce{Nb3Br8}, where the blue and red lines represent VBs and CBs, respectively. (C) Band structure of a semi-infinite slab structure with the $(001)$ cleavage plane as defined in (A). (D) Same as (C) but of a finite-size slab structure. In (D), the valence and conduction bands are represented by the blue and red lines, respectively.}
\label{fig:Nb3Br8}
\end{figure}

As shown in Fig. \ref{fig:Nb3Br8}(A), \ce{Nb3Br8} (\icsdweb{25766}) has a layered structure with SG 166 (\sgsymb{166}). The Wyckoff positions $6c$ and $18h$ are occupied by the respective Nb and Br atoms. The band structure, as calculated in the \webTQC~and shown in Fig. \ref{fig:Nb3Br8}(B), indicates \ce{Nb3Br8} is a topologically trivial insulator with a small indirect band gap of $0.069 eV$. The BR of \ce{Nb3Br8} at the maximal high-symmetry $\bf k$ points is,
\begin{equation}
B=(15\bar \Gamma_4\bar \Gamma_5,15\bar \Gamma_6\bar \Gamma_7,33\bar \Gamma_8,32\bar \Gamma_9,48\bar F_3\bar F_4,47\bar F_5\bar F_6, 47\bar L_3\bar L_4,48\bar L_5\bar L_6, 15\bar T_4\bar T_5,15\bar T_6\bar T_7,32\bar T_8,33\bar T_9)
\label{eq:B25766}
\end{equation}
where $m(\bar X_i)$ is the multiplicity of $\bar X_i$ at $X$ point.
By substituting the $B$ vector into the formula of RSI indices in SG 166, as defined in Appendix~\ref{app:rsi3D}, it shows that \ce{Nb3Br8} is an OAI indicated by a $Z$-type RSI, namely a $\delta$ index, at the empty Wyckoff position $b$,

\begin{equation}
\delta_4(b)=\frac{1}{2}[-m(\bar \Gamma_6\bar \Gamma_7)+m(\bar \Gamma_9)+m(\bar F_5\bar F_6)-m(\bar L_5\bar L_6)+m(\bar T_6\bar T_7)-m(\bar T_9)]=-1 \label{eq:rsi25766}
\end{equation}
where $m(\bar X_i)$ is the multiplicity of $\bar X_i$ at $X$ point.

Similar to the case in \ce{CaIn2P2}, using the criteria defined in Eqs. \ref{eq:millerindex} and \ref{eq:millerpoint}, we find the cleavage plane with Miller index $(001)$, as indicated by the blue planes in Fig.~\ref{fig:Nb3Br8}(A), has OSSs. 

In Fig.~\ref{fig:Nb3Br8}(C) and (D), we calculate the surface states of \ce{Nb3Br8} on the $(001)$ surface. Fig.~\ref{fig:Nb3Br8}(C) shows the band structure of a semi-infinite slab structure calculated with the Green's function method. The surface states locate between the conduction bands and the valence bands of bulk states. A finite-size slab calculation in Fig.~\ref{fig:Nb3Br8}(D) shows that the two connected surface bands near the Fermi level are half filled and hence it is a metallic configuration with filling anomaly on the surface.

\subsection{B$_{12}$}\label{app:B12}

\begin{figure}[htbp]
\centering\includegraphics[width=6.5in]{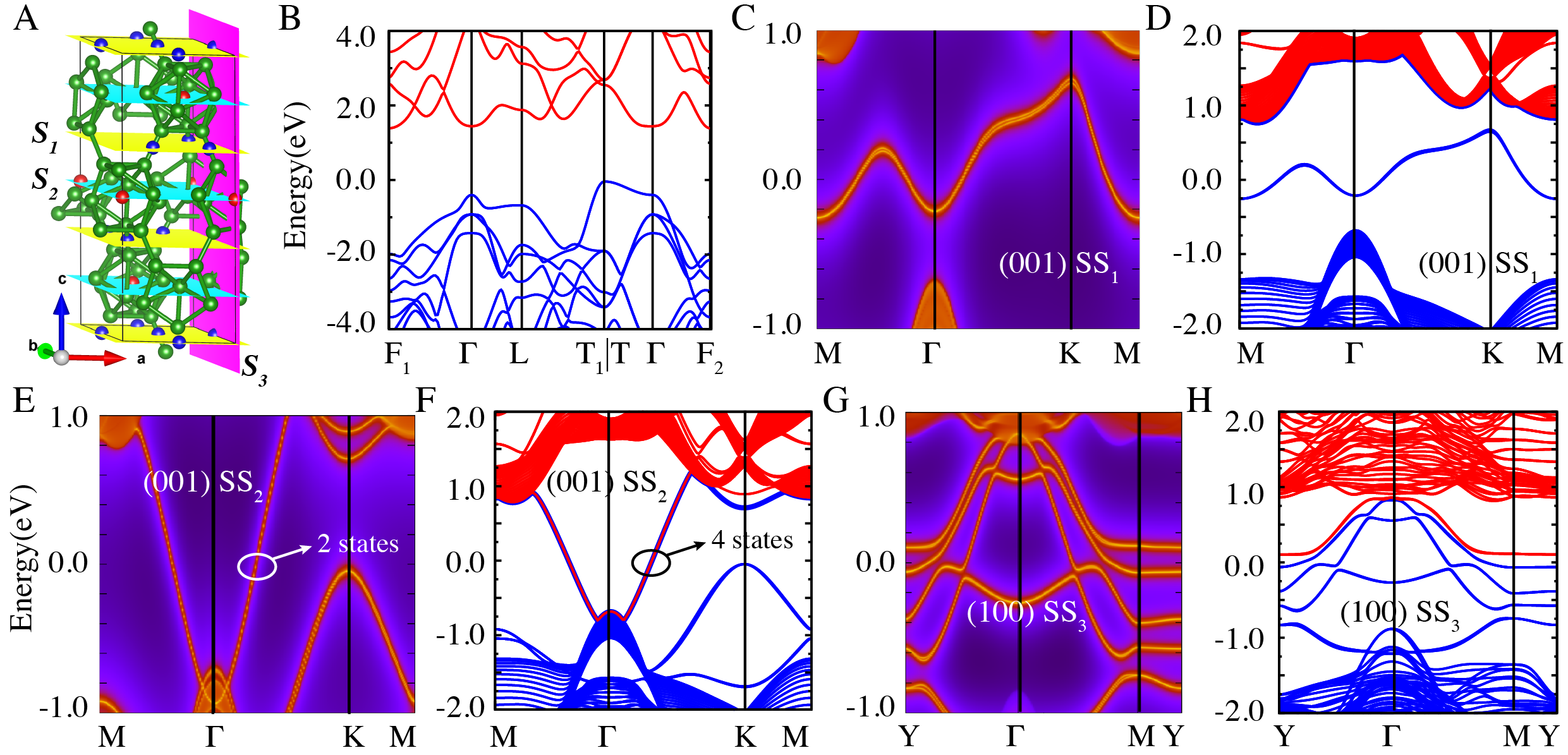}
\caption{(A) Crystal structure of \ce{B12}, where the red and blue spheres represent the positions of the OWCCs at $b$ and $e$, respectively. The Wyckoff position $c$ is represented as line segments on the $c$-axis. The yellow, cyan and pink planes cutting through them are the cleavage planes $S_1$, $S_2$ and $S_3$, which exhibit the OSSs. (B) Bulk band structure along the high-symmetry paths of \ce{B12}, where the blue and red lines represent VBs and CBs, respectively; (C) and (D) show the surface states of a semi-infinite and a finite-size slab structure with cleavage plane $S_1$. Similarly, (E) and (F) are the surface states on the cleavage plane $S_2$, (G) and (H) are the surface states on the cleavage plane $S_3$.}
\label{fig:B12}
\end{figure}

As shown in Fig. \ref{fig:B12}(A), \ce{B12} with~\icsdweb{431636} and SG 166 ($R\bar3m$) is an elementary compound with occupied Wyckoff position $18h$ occupied. The band structure shown in Fig. \ref{fig:B12}(B) indicates that \ce{B12} is a topologically trivial insulator with an indirect band gap of $1.149 eV$. The BR of \ce{B12} at the maximal high-symmetry $\bf k$ points is,

\begin{equation}
B=(3\bar \Gamma_4\bar \Gamma_5,2\bar \Gamma_6\bar \Gamma_7,7\bar \Gamma_8,6\bar \Gamma_9,8\bar F_3\bar F_4,10\bar F_5\bar F_6, 9\bar L_3\bar L_4,9\bar L_5\bar L_6, 3\bar T_4\bar T_5,2\bar T_6\bar T_7,8\bar T_8,5\bar T_9) \label{eq:B431636}
\end{equation}
where $m(\bar X_i)$ is the multiplicity of $\bar X_i$ at $X$ point.
By substituting the $B$ vector into the formula of RSI indices in SG 166, as defined in Appendix~\ref{app:rsi3D}, it shows that \ce{B12} is an OAI indicated by three different $Z$-type RSIs, namely $\delta$ indices, at the empty Wyckoff positions $b$, $c$
 and $e$,

\begin{equation}
\begin{aligned}
&\delta_4(b)=\frac{1}{2}[-m(\bar \Gamma_6\bar \Gamma_7)+m(\bar \Gamma_9)+m(\bar F_5\bar F_6)-m(\bar L_5\bar L_6)+m(\bar T_6\bar T_7)-m(\bar T_9)]=1 \\
&\delta_5(c)=2m(\bar \Gamma_6\bar \Gamma_7)+3m(\bar \Gamma_8)+2m(\bar \Gamma_9)-2m(\bar L_3\bar L_4)-2m(\bar L_5\bar L_6)=1 \\
&\delta_7(e)=\frac{1}{4}[m(\bar \Gamma_6\bar \Gamma_7)+m(\bar \Gamma_9)-m(\bar F_5\bar F_6)-m(\bar L_5\bar L_6)+m(\bar T_6\bar T_7)+m(\bar T_9)]=-1
\end{aligned}
\label{eq:rsi431636}
\end{equation}
where $m(\bar X_i)$ is the multiplicity of $\bar X_i$ at $X$ point. The maximal Wyckoff positions $b$ and $e$ are isolated points, as indicated by the respective red and blue spheres in Fig.~\ref{fig:B12}(A).
The non-maximal position $c$ represents a line parallel to the $[001]$ direction.

Using the criteria defined in Eqs. \ref{eq:millerindex}, \ref{eq:millerpoint} and \ref{eq:millerline},  we find three non-equivalent cleavage planes of lower Miller indices that exhibit the OSSs. They are the $(001)$ plane (\ie the $S_1$ plane in Fig. \ref{fig:B12}) cutting through the Wyckoff position $e$, the $(001)$ plane (\ie the $S_2$ plane in Fig. \ref{fig:B12}) cutting through the Wyckoff position $b$, and the $(100)$ plane (\ie the $S_3$ plane in Fig. \ref{fig:B12}) cutting through the Wyckoff position $b, e$ and $c$. In Fig.~\ref{fig:B12}(C)-(H), we show the result of the calculation of the surface states of the above three cleavage planes using both the semi-finite and finite-size slab structures. 
On the $S_1$ plane, irreps at the three OWCCs on the surface contribute three branches of disconnected surface bands, where each branch is two-fold degenerate at the time-reversal invariant $k$ points. In Fig. \ref{fig:B12}(C) and (D), only one branch of the surface bands is in the bulk band gap and another two branches are merged with the bulk conduction bands.
On the $S_2$ plane, there is only one OWCC which contributes with one branch of surface states in the bulk band gap and it's half-filled. While on the $S_3$ plane, it cuts through four OWCCs, including one OWCC at $b$, one OWCC at $e$ and two OWCCs at $c$, which contribute with four branches of surface bands. 

\subsection{CsFe$_2$Se$_3$}\label{app:CsFe2Se3}

\begin{figure}[htbp]
\centering\includegraphics[width=6.0in]{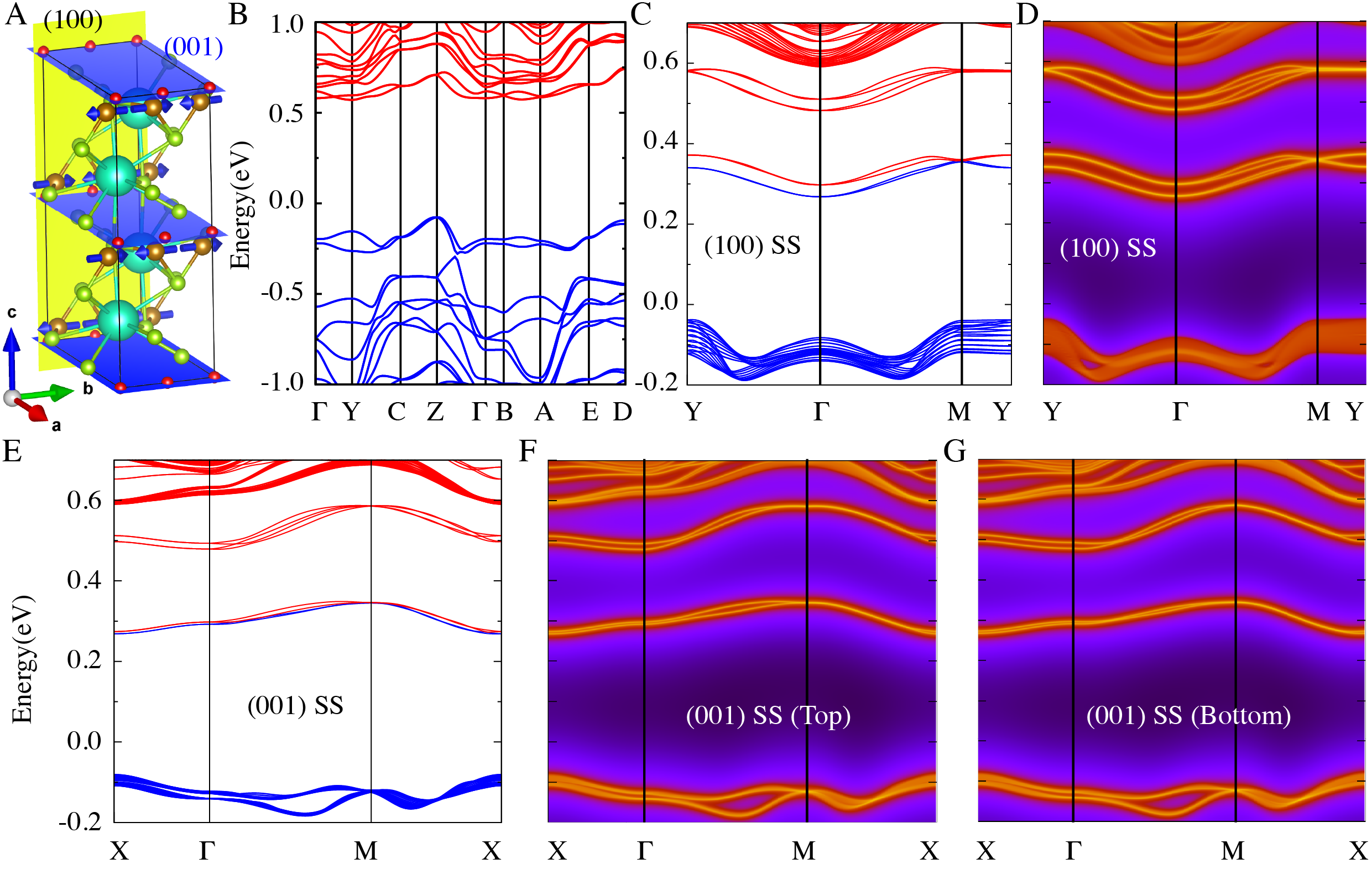}
\caption{(A) Crystal structure of \ce{CsFe2Se3}, where the red spheres indicate the positions of the OWCCs. The blue and yellow planes represent the respective $(001)$ and $(100)$ cleavage planes that  cut through the OWCCs and exhibit the OSSs. (B) Bulk band structure along the high-symmetry paths of \ce{CeFe2Se2}, where the blue and red lines represent the VBs and CBs, respectively. 
(C) and (D) show the surface states of a semi-infinite and a finite-size slab structures of  $(100)$ cleavage plane.
(E) is the surface states calculation of the finite-size slab structure of $(001)$ cleavage plane. (F) and (G) are the surface states calculations of the semi-infinite slab of $(001)$ cleavage plane on the respective top and bottom terminations.}
\label{fig:CsFe2Se3}
\end{figure}

As shown in Fig. \ref{fig:CsFe2Se3}(A), \ce{CsFe2Se3} with~\bcsidweblong{1.26} and MSG 14.82 (\msgsymb{14}{82}) is an anti-ferromagnetic material. Cs and Fe atoms occupy the magnetic Wyckoff positions $4e$ and $8f$, respectively. Three non-equivalent Se atoms occupy three different $4e$ positions. The band structure, as calculated in the \webMTQC ~and shown in Fig. \ref{fig:Nb3Br8}(B), indicates that \ce{CsFe2Se3} is a magnetic topologically trivial insulator. The BR of \ce{CsFe2Se3} at the maximal high-symmetry $\bf k$ points is,
\begin{equation}
B=(56\bar \Gamma_3\bar \Gamma_4,54\bar \Gamma_5\bar \Gamma_6,110\bar A_2,110\bar B_2,55\bar C_2\bar C_2,55\bar D_3\bar D_6, 55\bar D_4\bar D_5,55\bar E_3\bar E_6, 55\bar E_4\bar E_5,56\bar Y_3\bar Y_4,54\bar Y_5\bar Y_6, 55\bar Z_2\bar Z_2)
\label{eq:CsFe2Se3}
\end{equation}
where $m(\bar X_i)$ is the multiplicity of $\bar X_i$ at $X$ point.
By substituting the $B$ vector into the formula of RSI indices of MSG 14.82, as defined in Appendix~\ref{app:rsi3D}, it shows that \ce{CsFe2Se3} is a mOAI indicated by a $Z$-type RSIs, namely a $\delta$ index, at the empty Wyckoff position $4c$,

\begin{equation}
\delta_1(c)=\frac{1}{2}[-m(\bar \Gamma_5\bar \Gamma_6)-m(\bar A_2)+m(\bar Y_3\bar Y_4)+2m(\bar Z_2\bar Z_2)]=1
\label{eq:rsi126}
\end{equation}
where $m(\bar X_i)$ is the multiplicity of $\bar X_i$ at $X$ point.

The atoms at WP $4c$ are indicated by red spheres in Fig.\ref{fig:CsFe2Se3}(A). 
Using the criteria defined in Eqs. \ref{eq:millerindex} and \ref{eq:millerpoint}, we find that the cleavage planes with Miller indices $(001)$ and $(100)$ are identified to exhibit the OSSs. Note that the magnetic space group on the $(100)$ plane is MSG 7.28 (\msgsymb{7}{28}) and each band has degeneracy two at the $Y$ and $M$ points on the $(100)$ plane. In Fig.~\ref{fig:CsFe2Se3}(C)-(D), we have calculated and plotted the surface states of $(100)$ plane, where the four OWCCs contribute with eight surface bands on the top or bottom surface.
In Fig.~\ref{fig:CsFe2Se3}(E)-(G), we calculate and plot the surface states on the $(001)$ plane, which cuts through two OWCCs as shown in Fig.~\ref{fig:CsFe2Se3}(A). The two OWCCs contribute with four surface bands on both the top and bottom $(001)$ surfaces. As the corresponding MSG on the $(001)$ cleavage plane is MSG 4.9 (\msgsymb{4}{9}), the surface bands at the $M$ point have two-fold degeneracy.

\section{Methods and extended data of the electrochemical catalytic measurements on 2H-MoS$_2$}\label{app:experiments}

\subsection{Experimental Methods}

2H-\ce{MoS2} single crystals were purchased from HQ Graphene. All the electrochemical HER catalytic measurements were performed on an Autolab PGSTAT302N electrochemistry workstation an Ar saturated 0.5 M \ce{H2SO4} solution. An Ag/AgCl (3 M KCl) electrode and graphite rod as the reference electrode and counter electrode, respectively. The bulk single crystal of \ce{MoS2} is attached to a Ti wire with silver paint and served as the working electrode. Linear sweep voltammograms were recorded with a scan rate of 1 $mV/S$. The electrochemical impedance spectroscopy (EIS) measurements were conducted from 100 kHz to 0.1 Hz. The amplitude of the sinusoidal potential signal was 10 mV. All potentials were referenced to a reversible hydrogen electrode according to $E(vs~RHE)=E(vs~Ag/AgCl)+(0.210+0.059\times pH)V$.

\subsection{Extended Data}
\begin{figure}
\centering\includegraphics[width=3.4in]{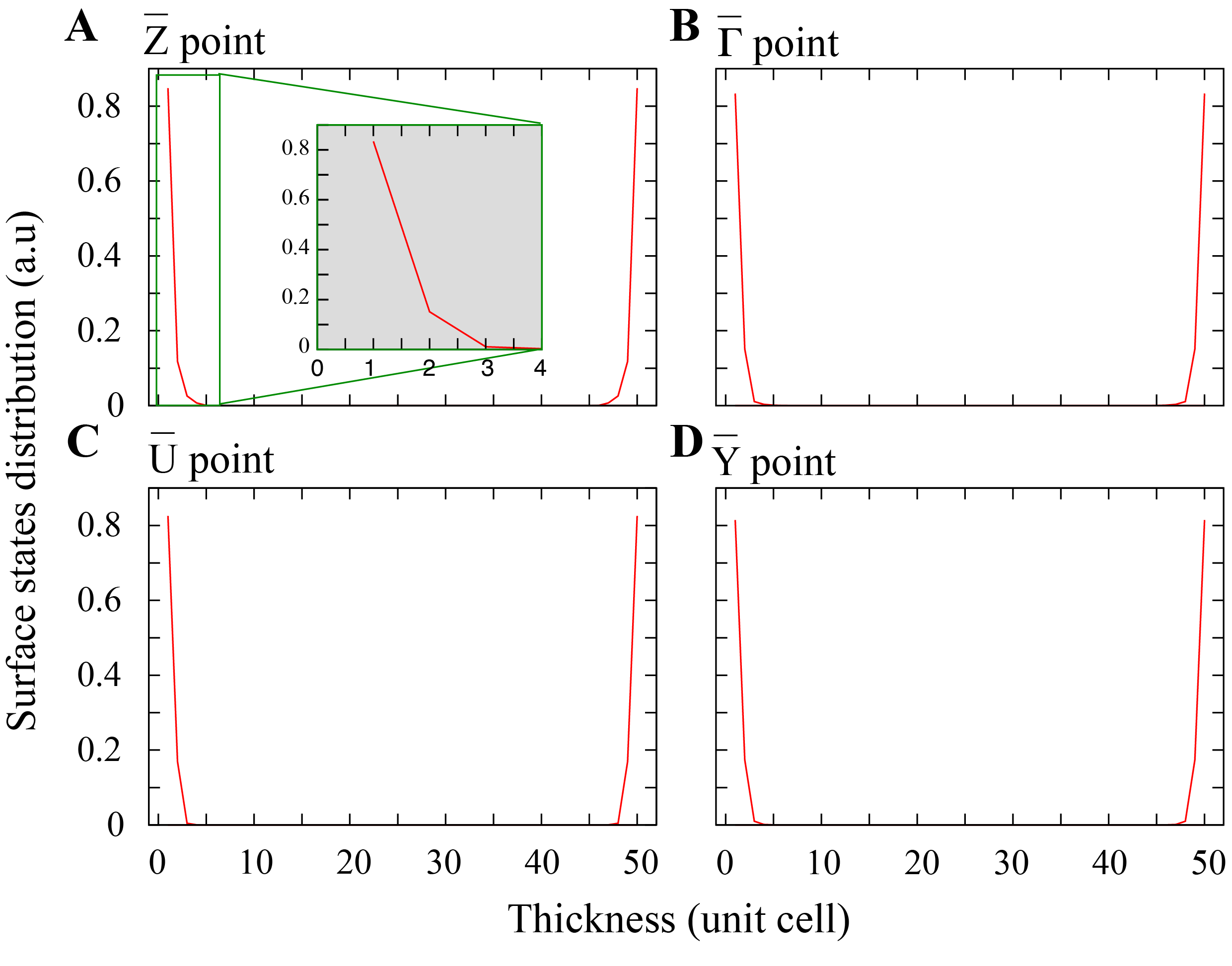}
\caption{The decay of OSSs at (A) $\bar{Z}$ (B) $\bar{\Gamma}$ (C) $\bar{U}$ and (D) $\bar{Y}$ points, respectively, for the 2H-\ce{MoS2}. The slab is with a thickness of 50 unit cells. The first unit cell and the 50th unit cells are the left and right terminations, respectively. The obstructed surface states are mainly localized at the first three unit cells close to the surface, as presented in the inset of (A).}\label{fig:figs1}
\end{figure}

\begin{figure}
\centering\includegraphics[width=2.0in]{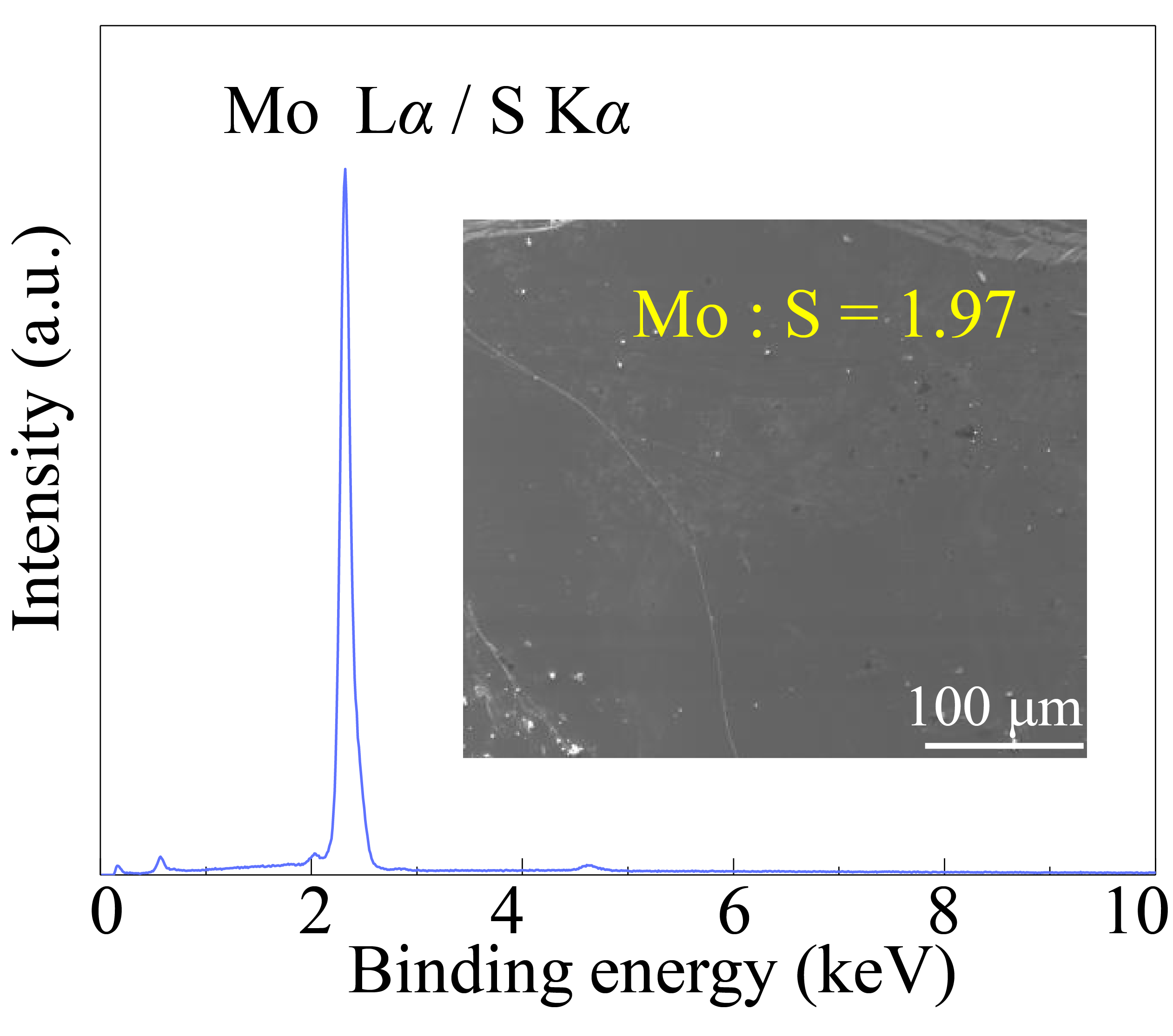}
\caption{Energy-dispersive X-ray spectroscopy and SEM image of the 2H-\ce{MoS2} bulk single crystal. The atomic concentration of Mo and S is very close to the stoichiometric concentration of \ce{MoS2}. }\label{fig:figs2}
\end{figure}

\begin{figure}
\centering\includegraphics[width=3.4in]{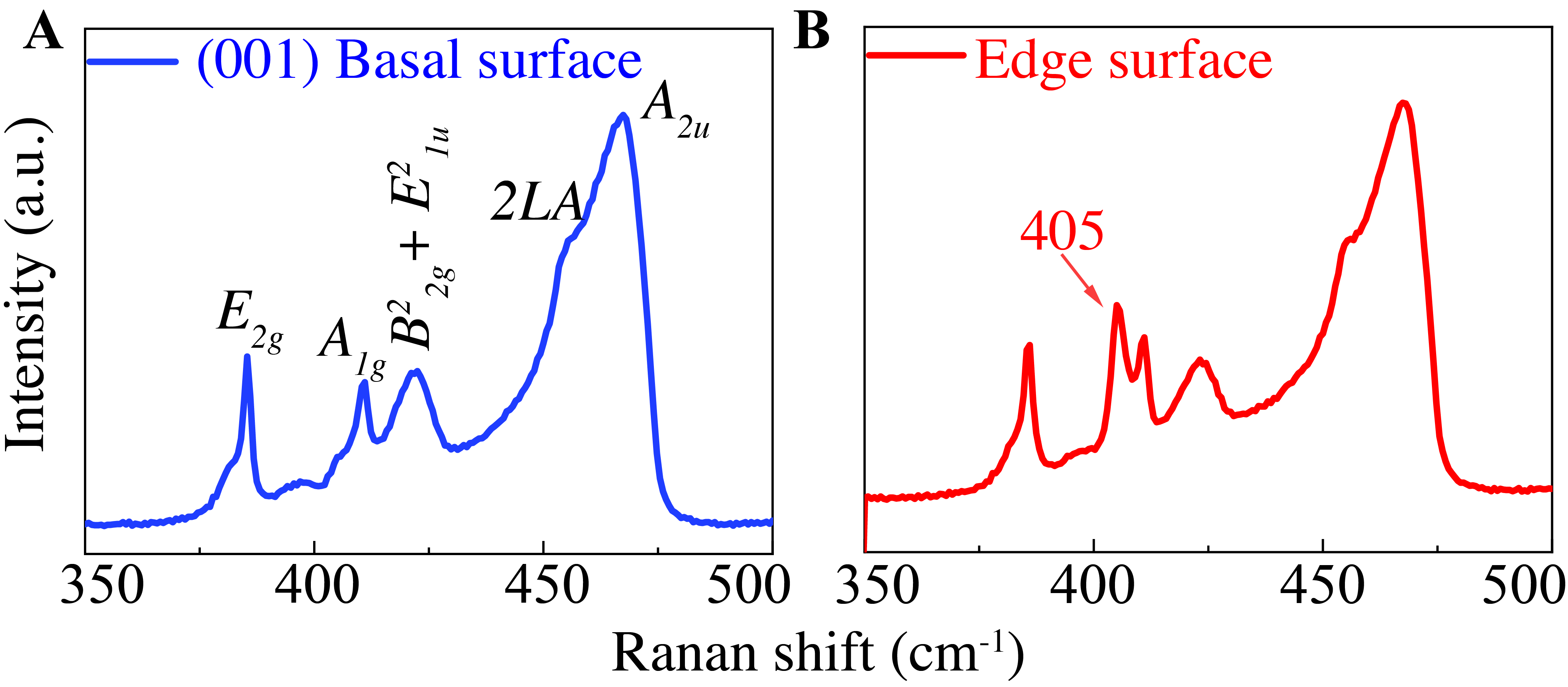}
\caption{Raman spectra of the 2H-\ce{MoS2} bulk crystal recorded on the (A) $(001)$ basal plane and (B) side surfaces. The appearance of a peak below the $A_{1g}$ peak (about 405 cm$^{-1}$) may be related to side surfaces with lower asymmetry. All the other peaks can be indexed to the 2H-\ce{MoS2}.}\label{fig:figs3}
\end{figure}

\begin{figure}
\centering\includegraphics[width=2.0in]{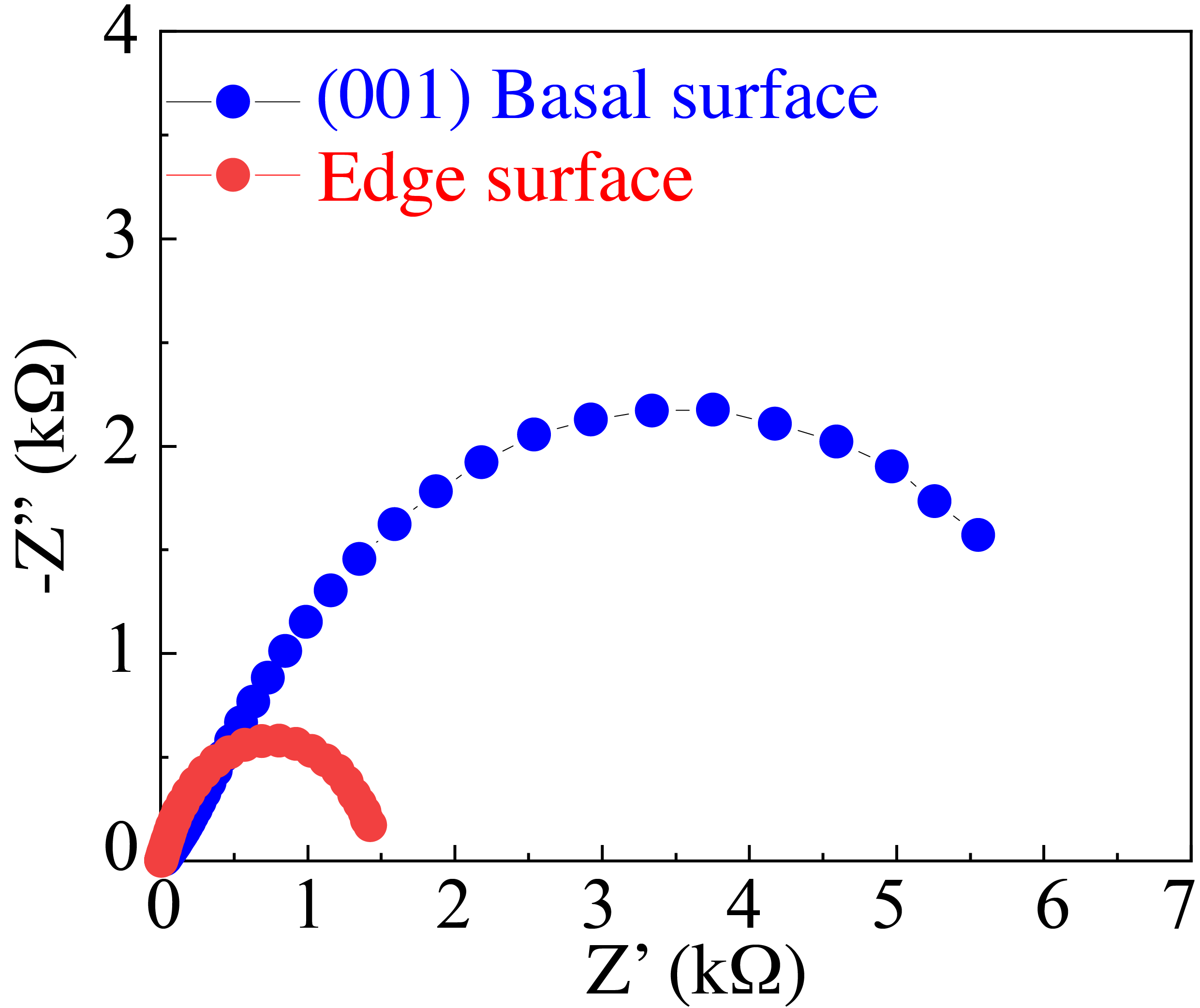}
\caption{Electrochemical impedance spectra of the 2H-\ce{MoS2} single crystal when exposing the $(001)$ basal plane and edge side surfaces. The semi-circle corresponds to the charge transfer kinetics in the hydrogen reduction process. The small radius suggests a low interfacial charge transfer resistance due to the increased conductivity at the 2H-\ce{MoS2} side surfaces.}\label{fig:figs4}
\end{figure}

There are four extended figures for the ``proof of principle'' catalytic experiments on 2H-\ce{MoS2}.
In Fig.~\ref{fig:figs1}, by calculating a slab structure with a thickness of 50 unit cell, we analyze the decay of the OSSs on the (100) surface.
In In Figs.~\ref{fig:figs2} and \ref{fig:figs3}, the crystals of \ce{MoS2} are confirmed to be the stoichiometric 2H phase by energy-dispersive spectroscopy and Raman spectroscopy, respectively.
In Fig.~\ref{fig:figs4}, the electrochemical impedance spectroscopy demonstrated a much lower charge transfer resistance for the edge sites than for the basal plane, suggesting enhanced electron transfer kinetics on the edge surface due to the highly conducting metallic surface states \cite{li2015charge}.

\section{List of three dimensional obstructed atomic insulators and orbital selected obstructed atomic insulators}\label{app:3DOAI}

In this Appendix, we detail the material lists of 3D OAIs, OOAIs and mOAIs, as obtained from the high-throughput searches described in Appendix \ref{app:highthroughput}. For each material, we provide its basic band and electronic properties and the RSIs that indicate the obstructed phases. 

\subsection{List of 3D paramagnetic OAIs}\label{app:OAIlist}

Among the \TQCDBNbrICSDs\ stoichiometric materials with distinct ICSD entries that successfully computed on the \webTQC, there are \TQCDBNbrICSDsTrivial\ ICSDs (\TQCDBNbrMaterialsTrivial\ unique materials) classified into symmetry-indicated topologically trivial insulators (\ie LCEBR). By applying the RSIs to the BR of each LCEBR material, we have finally diagnosed \TQCDBNbrICSDsOAI\ ICSDs (\TQCDBNbrMaterialsOAI\ unique materials) as OAIs, whose BRs \emph{can not} be induced from the irreps sitting at the occupied Wyckoff positions.

Among the OAI materials, \TQCDBNbrICSDsOAIIndirectGap\ ICSDs (\TQCDBNbrMaterialsOAIIndirectGap\ unique materials) are fully gapped along all the high-symmetry lines in the BZ with finite indirect band gap and the other \TQCDBNbrICSDsOAIOnlyDirectGap\ ICSD entries (\TQCDBNbrMaterialsOAIOnlyDirectGap\ unique materials) only have a finite direct band gap but no indirect band gap.
The OAIs with and without indirect band gap are tabulated in the respective Tables~\ref{tab:oaiindirect} and~\ref{tab:oaidirect}. In the tables, we group together all the ICSD entries of the  unique material (as defined in Refs.\cite{vergniory_complete_2019,Vergniory2021} and in Appendix~\ref{app:TQCdatabase}). For each OAI of a given unique material, we provide the chemical formula with a direct link to the \webTQC\ entry, the space group, both the indirect and direct band gaps, whether the material is identified as a feOAI in the Ref~\cite{xu2021filling}, the chemical properties, all the related ICSD entries, all the RSIs that indicate the OAI phase, the non-equivalent occupied Wyckoff positions and the corresponding atoms.
For most materials, we also provide the Miller indices of the cleavage planes that have metallic OSSs obtained using the method in Appendix~\ref{app:filling_anomaly}. To facilitate the potential application of OAIs to asymmetric electrocatalysis \cite{noyori2002asymmetric,list2000proline,ahrendt2000new}, we have also indicated the OAIs in chiral space groups and the cleavage planes with 2D chiral plane groups.

\LTcapwidth=1.0\textwidth
\renewcommand\arraystretch{1.0}
\begin{longtable*}{|@{\hskip0.1pt}c@{\hskip0.1pt}|@{\hskip0.1pt}c@{\hskip0.1pt}|@{\hskip0.1pt}c@{\hskip0.1pt}|@{\hskip0.1pt}c@{\hskip0.1pt}|@{\hskip0.1pt}c@{\hskip0.1pt}|@{\hskip0.1pt}c@{\hskip0.1pt}|@{\hskip0.1pt}c@{\hskip0.1pt}|@{\hskip0.1pt}c@{\hskip0.1pt}|@{\hskip0.1pt}c@{\hskip0.1pt}|@{\hskip0.1pt}c@{\hskip0.1pt}|}
\caption[List of OAIs with finite indirect band gap]{List of OAIs with finite indirect band gap. For each material, in the first row, the table provides the chemical formula (Formula) with a direct link to the \webTQC\ entry, the space group symbol and number (SG), indirect band gap (IGap), direct band gap (DGap), whether the material is identified as a feOAI in the Ref~\cite{xu2021filling}, the flags indicating if the material is ``high-quality'' (HQ) or theory (th) compound and if the material compound contains elements with $f$ electron (f), all the related ICSD entries (each with a direct to the full RSI information), all the RSIs that indicate the OAI phase, the non-equivalent occupied Wyckoff positions (Occ. WP) and the corresponding atoms. For each material, since the indirect band gap might slightly varies from one ICSD entry to another, IGap is the one of the first ICSD entry and is the minimal indirect band gap for all entries. In addition, we provide in parentheses the indirect band gap for each ICSD. For most materials, we also provide the Miller indices of the cleavage planes that have metallic OSSs below the first row, which are obtained using the method in Appendix \ref{app:filling_anomaly}. Each cleavage plane is defined by the Wyckoff letter of the OWCC, which is on the plane, and the Miller index of the plane. Moreover, in the ``ICSDs'' column, the flag `$*$' (`$\dagger$') indicates that the RSIs (cleavage planes) for an ICSD entry differ from the RSIs (cleavage planes) of the first ICSD entry listed for the unique material. These potential differences are usually due to slightly different crystal structures of these ICSD entries. The superscript `$c$' on each Miller index indicates that the 2D MSG of the related surface is a chiral group.}\label{tab:oaiindirect} \\
\hline
\end{longtable*}

\LTcapwidth=1.0\textwidth
\renewcommand\arraystretch{1.0}
\begin{longtable*}{|@{\hskip0.1pt}c@{\hskip0.1pt}|@{\hskip0.1pt}c@{\hskip0.1pt}|@{\hskip0.1pt}c@{\hskip0.1pt}|@{\hskip0.1pt}c@{\hskip0.1pt}|@{\hskip0.1pt}c@{\hskip0.1pt}|@{\hskip0.1pt}c@{\hskip0.1pt}|@{\hskip0.1pt}c@{\hskip0.1pt}|@{\hskip0.1pt}c@{\hskip0.1pt}|@{\hskip0.1pt}c@{\hskip0.1pt}|}
\caption[List of OAIs with zero minimal indirect band gap]{List of OAIs with zero minimal indirect band gap. A material is considered as having a zero minimal indirect band gap as soon as one of its ICDS entries has a zero indirect gap. For each material, in the first row, the table provides the chemical formula (Formula) with a direct link to the \webTQC\ entry, the space group symbol and number (SG), direct band gap (DGap), whether the material is identified as a feOAI in the Ref~\cite{xu2021filling}, the flags indicating if the material is ``high-quality'' (HQ) or theory (th) compound and if the material compound contains elements with $f$ electron (f), all the related ICSD entries (each with a direct to the full RSI information), all the RSIs that indicate the OAI phase, the non-equivalent occupied Wyckoff positions (Occ. WP) and the corresponding atoms. Note that for some materials, \emph{a fraction of their ICSD entries might still have an indirect band gap}, since the indirect band gap might slightly varies from one ICSD entry to another. For that reason, we provide in parentheses the indirect band gap for each ICSD. For most materials, we also provide the Miller indices of the cleavage planes that have metallic OSSs below the first row, which are obtained using the method in Appendix \ref{app:filling_anomaly}. Each cleavage plane is defined by the Wyckoff letter of the OWCC, which is on the plane, and the Miller index of the plane. Moreover, in the ``ICSDs'' column, the flag `$*$' (`$\dagger$') indicates that the RSIs (cleavage planes) for an ICSD entry differ from the RSIs (cleavage planes) of the first ICSD entry listed for the unique material. These potential differences are usually due to slightly different crystal structures of these ICSD entries. The superscript `$c$' on each Miller index indicates that the 2D MSG of the related surface is a chiral group.}\label{tab:oaidirect} \\
\hline
\end{longtable*}

\subsection{List of 3D OOAIs}\label{app:OOAIlist}

Using the method of diagnosing OOAIs, as detailed in the Appendix~\ref{app:methodOOAI}, we have identified \TQCDBNbrICSDsOOAI\ ICSD entries (\TQCDBNbrMaterialsOOAI\ unique materials) from the \webTQC\ to be OOAIs, of which each is indicated by \emph{at least} one non-zero RSI defined at the occupied sites. Among them, \TQCDBNbrICSDsOOAIIndirectGap\ ICSD entries (\TQCDBNbrMaterialsOOAIIndirectGap\ unique materials) have a finite indirect band gap and \TQCDBNbrICSDsOOAIandOAI\ ICSD entries (\TQCDBNbrMaterialsOOAIandOAI\ unique materials) are also diagnosed as OAIs in the Tables~\ref{tab:oaiindirect} and~\ref{tab:oaidirect} of Appendix~\ref{app:OAIlist}.

In Table~\ref{tab:ooailist} dedicated to OOAI, for each ICSD entry we tabulate its space group (SG), Chemical formula, indirect band gap (IGap), direct band gap (Gap), the occupied Wyckoff positions (WP) where the RSIs indicating an OOAI phase are defined, the atoms that occupy these WPs, and the RSIs indicating the OOAI phase.

\LTcapwidth=1.0\textwidth
\renewcommand\arraystretch{1.0}
\begin{longtable*}{|c|c|c|c|c|c|c|c|c|c|}
\caption[List of unique materials that are OOAIs]{List of unique materials from the \webTQC\ that are OOAIs. In this table, we provide the chemical formula (second column) with a direct link to the \webTQC\ entry, space group (third column), indirect band gap (IGap, fourth column), direct band gap (DGap, fifth column). The sixth column (OAI) indicates if the compound is also an OAI (as listed in Appendix \ref{app:highthroughput}) or a feOAI. All the ICSD numbers related to the unique materials are given in the seventh column, each with a direct to the full RSI information. Since the IGap might differ from one ICSD entry to another, we provide in parentheses the IGap for each ICSD. Moreover, a star indicates that the RSI for an ICSD entry might different from (but is equivalent to) the RSI of the first ICSD entry listed for the unique material. Finally in the last three columns, we provide the site of orbital-selected obstructed center (WP) and its associated site symmetry (site sym.) in parentheses, the atom sitting at the WP (Atom) and its outer-shell electronic configuration (orb.) in parentheses, and the RSI of the WP which indicates an OOAI.\label{tab:ooailist}}\\
\hline
No. & Formula & SG & \begin{tabular}{c}IGap\\ (eV)\end{tabular} & \begin{tabular}{c}DGap\\ (eV)\end{tabular} & OAI & ICSDs & \begin{tabular}{c}Wyckoff\\ (site sym.)\end{tabular} & \begin{tabular}{c}Atom\\ (orb.)\end{tabular}& RSI \\
\hline
{\scriptsize{1}} & {\scriptsize{ \icsdwebdirectlink{155716}{$\rm{K} \rm{Mn} (\rm{H} \rm{P}_{3} \rm{O}_{10})$}}} & {\scriptsize{\sgsymb{15} (15)}} & {\scriptsize{0.000}} & {\scriptsize{0.009}} & {\scriptsize{---}}  & {\tiny{ \flatwebdirectlink{155716}{155716}(0.000) \flatwebdirectlink{161904}{161904}(0.000)}} &  {\scriptsize{ $c (\bar{1})$}}  & {\scriptsize{$ \rm{H} (\rm{s})$}} & {\scriptsize{$\delta_{3}(c)=2$ }} \\
\hline
\end{longtable*}

\subsection{List of 3D mOAIs}\label{app:mOAIlist}

Among the \MTQCDBNbrBCSIDsWithPhaseDiagram\ stoichiometric materials with distinct BCSIDs on the \webMTQC, there are \MTQCDBNbrBCSIDsTrivial\ BCSIDs classified as symmetry-indicated topologically trivial insulators (\ie LCEBR). By applying the RSI indices of MSGs to the BRs of each LCEBR-type magnetic material of different Hubbard-$U$ values, we diagnosed \MTQCDBNbrBCSIDs\ BCSIDs as  mOAIs, whose BRs cannot be induced from the irreps sitting at the occupied Wyckoff positions in the magnetic crystal strucutures.

In Table~\ref{tab:mOAIs}, for each mOAI we tabulate its  chemical formula, MSG, BCSID, number of valence electrons in one magnetic unit cell as adopted in the ab-initio calculations, the occupied WPs, all the  Hubbard-$U$ values leading to an LCEBR topology in the topological phase diagram of this material \cite{xu2020high}. For each Hubbard-$U$ values of the LCEBR topology, we tabulate the indirect band gap and the RSIs that indicate an mOAI phase.
Similar to the OAIs in Tables~\ref{tab:oaiindirect} and~\ref{tab:oaidirect}, we also provide the Miller indices of the cleavage planes which exhibit OSSs.

\LTcapwidth=1.0\textwidth
\renewcommand\arraystretch{1.0}
\begin{longtable*}{|c|c|c|c|c|c|c|c|c|}
\caption[List of unique materials that are mOAIs]{The list of magnetic OAIs on the \webMTQC. In the table, we provide in columns 2 to 5,the chemical formula (Formula), identification number (BCSID) of the material on the MAGNDATA database of \webBCS\ (including a direct link to the \webMTQC),  magnetic space group (MSG) ,the number of valence electrons (Nele) in one primitive unit cell of the magnetic material. The sixth, seventh and eighth columns provide the Hubbard-$U$ value as adopted in the first-principle calculations, the indirect gap(IGap), and the real space invariants (RSIs) indicating the OAI phase (Obstructed RSIs), respectively. The last column gives the list of occupied Wyckoff positions (WP$_{occ}$). For most materials, we also provide the Miller indices of the cleavage planes that have metallic OSSs below the first row, which are obtained using the method in Appendix \ref{app:filling_anomaly}. Each cleavage plane is defined by the Wyckoff letter of the OWCC, which is on the plane, and the Miller index of the plane. The superscript `$c$' on each Miller index indicates that the 2D MSG of the related surface is a chiral group.}\label{tab:mOAIs} \\
\hline
 & Formula & BCSID & MSG & Nele  & U (eV) & IGap (eV) & Obstructed RSIs & WP$_{occ}$\\
\hline
 \multirow{2}{*}{{\tiny{1}}} & \multirow{2}{*}{{\tiny{ $\rm{Lu} \rm{Fe}_{2} \rm{O}_{4}$}}} & \multirow{2}{*}{{\tiny{ \bcsidwebshort{1.0.7}}}} & \multirow{2}{*}{{\tiny{\msgsymb{12}{62} (12.62)}}} & \multirow{2}{*}{{\tiny{183}}} & {\tiny{0.0}}  & {\tiny{0.000}} & {\tiny{ $\delta_{4}(d)=1$}}  & {\tiny{$a(\rm{Lu}),i(\rm{O}),i(\rm{Fe})$}} \\
\cline{6-8}
  &  &  &  &  & {\tiny{2.0}}  & {\tiny{0.009}} & {\tiny{ $\delta_{4}(d)=1$}}  & {\tiny{$j(\rm{O}),j(\rm{Fe}),j(\rm{Lu})$}} \\\cline{2-9}
 & \multicolumn{8}{c|}{\tiny{$d: \; (1,-2,0)^c,(0,0,1)^c$}} \\ \hline 
 \multirow{5}{*}{{\tiny{2}}} & \multirow{5}{*}{{\tiny{ $\rm{Cs} \rm{Fe}_{2} \rm{Se}_{3}$}}} & \multirow{5}{*}{{\tiny{ \bcsidwebshort{1.26}}}} & \multirow{5}{*}{{\tiny{\msgsymb{14}{82} (14.82)}}} & \multirow{5}{*}{{\tiny{220}}} & {\tiny{0.0}}  & {\tiny{0.647}} & {\tiny{ $\delta_{1}(c)=1$}}  & \multirow{5}{*}{{\tiny{$e(\rm{Se}),e(\rm{Cs}),f(\rm{Fe})$}}} \\
\cline{6-8}
  &  &  &  &  & {\tiny{1.0}}  & {\tiny{0.933}} & {\tiny{ $\delta_{1}(c)=1$}}  &  \\
\cline{6-8}
  &  &  &  &  & {\tiny{2.0}}  & {\tiny{1.162}} & {\tiny{ $\delta_{1}(c)=1$}}  &  \\
\cline{6-8}
  &  &  &  &  & {\tiny{3.0}}  & {\tiny{1.212}} & {\tiny{ $\delta_{1}(c)=1$}}  &  \\
\cline{6-8}
  &  &  &  &  & {\tiny{4.0}}  & {\tiny{1.140}} & {\tiny{ $\delta_{1}(c)=1$}}  &  \\\cline{2-9}
 & \multicolumn{8}{c|}{\tiny{$c: \; (1,-1,0)^c,(1,1,0)^c$}} \\ \hline 
 \multirow{5}{*}{{\tiny{3}}} & \multirow{5}{*}{{\tiny{ $\rm{Mn}_{3} \rm{Si}_{2} \rm{Te}_{6}$}}} & \multirow{5}{*}{{\tiny{ \bcsidwebshort{0.176}}}} & \multirow{5}{*}{{\tiny{\msgsymb{15}{89} (15.89)}}} & \multirow{5}{*}{{\tiny{}}} & {\tiny{0.0}}  & {\tiny{0.254}} & {\tiny{ $\delta_{1}(a)=2$}}  & {\tiny{$e(\rm{Mn}),f(\rm{Si}),f(\rm{Mn})$}} \\
\cline{6-8}
  &  &  &  &  & {\tiny{1.0}}  & {\tiny{0.493}} & {\tiny{ $\delta_{1}(a)=2$}}  & {\tiny{$f(\rm{Te})$}} \\
\cline{6-8}
  &  &  &  &  & {\tiny{2.0}}  & {\tiny{0.686}} & {\tiny{ $\delta_{1}(a)=2$}}  &  \\
\cline{6-8}
  &  &  &  &  & {\tiny{3.0}}  & {\tiny{0.842}} & {\tiny{ $\delta_{1}(a)=2$}}  &  \\
\cline{6-8}
  &  &  &  &  & {\tiny{4.0}}  & {\tiny{0.863}} & {\tiny{ $\delta_{1}(a)=2$}}  &  \\\cline{2-9}
 & \multicolumn{8}{c|}{\tiny{$a: \; (0,0,1)^c$}} \\ \hline 
 {\tiny{4}} & {\tiny{ $\rm{Dy}_{2} \rm{Se} \rm{O}_{2}$}} & {\tiny{ \bcsidwebshort{1.212}}} & {\tiny{\msgsymb{15}{90} (15.90)}} & {\tiny{116}} & {\tiny{0.0}}  & {\tiny{0.000}} & {\tiny{ $\delta_{3}(f)=1$}}  & {\tiny{$b(\rm{Se}),i(\rm{O}),i(\rm{Dy})$}} \\\cline{2-9}
 & \multicolumn{8}{c|}{\tiny{$f: \; (1,0,0)^c,(0,1,0)^c$}} \\ \hline 
 \multirow{2}{*}{{\tiny{5}}} & \multirow{2}{*}{{\tiny{ $\rm{Ag}_{2} \rm{Ni} \rm{O}_{2}$}}} & \multirow{2}{*}{{\tiny{ \bcsidwebshort{1.49}}}} & \multirow{2}{*}{{\tiny{\msgsymb{15}{90} (15.90)}}} & \multirow{2}{*}{{\tiny{300}}} & {\tiny{3.0}}  & {\tiny{0.000}} & {\tiny{ $\delta_{2}(d)=-1$}}  & {\tiny{$c(\rm{Ni}),h(\rm{Ni}),i(\rm{O})$}} \\
\cline{6-8}
  &  &  &  &  & {\tiny{4.0}}  & {\tiny{0.000}} & {\tiny{ $\delta_{2}(d)=-1$}}  & {\tiny{$i(\rm{Ag}),j(\rm{O}),j(\rm{Ag})$}} \\\cline{2-9}
 & \multicolumn{8}{c|}{\tiny{$d: \; (1,1,0)^c$}} \\ \hline 
 \multirow{2}{*}{{\tiny{6}}} & \multirow{2}{*}{{\tiny{ $\rm{Pr} \rm{Ag}$}}} & \multirow{2}{*}{{\tiny{ \bcsidwebshort{1.150}}}} & \multirow{2}{*}{{\tiny{\msgsymb{53}{334} (53.334)}}} & \multirow{2}{*}{{\tiny{48}}} & \multirow{2}{*}{{\tiny{0.0}}}  & \multirow{2}{*}{{\tiny{0.000}}} & {\tiny{ $\delta_{2}(b)=1$}}  & \multirow{2}{*}{{\tiny{$a(\rm{Ag}),c(\rm{Pr})$}}} \\
  &  &  &  &  & &  & {\tiny{ $\delta_{4}(d)=-1$}}  &  \\ \hline 
 {\tiny{7}} & {\tiny{ $\rm{Ca} \rm{Co}_{2} \rm{P}_{2}$}} & {\tiny{ \bcsidwebshort{1.252}}} & {\tiny{\msgsymb{59}{416} (59.416)}} & {\tiny{76}} & {\tiny{0.0}}  & {\tiny{0.000}} & {\tiny{ $\delta_{1}(k)=1$}}  & {\tiny{$a(\rm{Ca}),g(\rm{P}),h(\rm{Co})$}} \\\cline{2-9}
 & \multicolumn{8}{c|}{\tiny{$k: \; (1,0,0)^c,(0,1,0)^c$}} \\ \hline 
 \multirow{2}{*}{{\tiny{8}}} & \multirow{2}{*}{{\tiny{ $\rm{Mn}_{5} \rm{Si}_{3}$}}} & \multirow{2}{*}{{\tiny{ \bcsidwebshort{1.88}}}} & \multirow{2}{*}{{\tiny{\msgsymb{60}{431} (60.431)}}} & \multirow{2}{*}{{\tiny{308}}} & {\tiny{2.0}}  & {\tiny{0.000}} & {\tiny{ $\delta_{1}(a)=1$}}  & {\tiny{$c(\rm{Si}),c(\rm{Mn}),e(\rm{Mn})$}} \\
\cline{6-8}
  &  &  &  &  & {\tiny{3.0}}  & {\tiny{0.000}} & {\tiny{ $\delta_{1}(a)=1$}}  & {\tiny{$g(\rm{Si}),g(\rm{Mn})$}} \\ \hline 
 {\tiny{9}} & {\tiny{ $\rm{Eu} \rm{Fe}_{2} \rm{As}_{2}$}} & {\tiny{ \bcsidwebshort{2.1}}} & {\tiny{\msgsymb{61}{439} (61.439)}} & {\tiny{220}} & {\tiny{0.0}}  & {\tiny{0.000}} & {\tiny{ $\delta_{2}(b)=1$}}  & {\tiny{$a(\rm{Eu}),e(\rm{Fe}),f(\rm{As})$}} \\ \hline 
 \multirow{2}{*}{{\tiny{10}}} & \multirow{2}{*}{{\tiny{ $\rm{Cr}_{2} \rm{As}$}}} & \multirow{2}{*}{{\tiny{ \bcsidwebshort{1.130}}}} & \multirow{2}{*}{{\tiny{\msgsymb{62}{450} (62.450)}}} & \multirow{2}{*}{{\tiny{116}}} & {\tiny{0.0}}  & {\tiny{0.000}} & {\tiny{ $\delta_{1}(d)=1$}}  & \multirow{2}{*}{{\tiny{$a(\rm{Cr}),b(\rm{Cr}),b(\rm{As})$}}} \\
\cline{6-8}
  &  &  &  &  & {\tiny{3.0}}  & {\tiny{0.000}} & {\tiny{ $\delta_{1}(d)=1$}}  &  \\\cline{2-9}
 & \multicolumn{8}{c|}{\tiny{$d: \; (1,0,0)^c,(0,1,0)^c,(0,0,1)^c$}} \\ \hline 
 \multirow{2}{*}{{\tiny{11}}} & \multirow{2}{*}{{\tiny{ $\rm{U}_{2} \rm{Ni}_{2} \rm{Sn}$}}} & \multirow{2}{*}{{\tiny{ \bcsidwebshort{1.200}}}} & \multirow{2}{*}{{\tiny{\msgsymb{63}{466} (63.466)}}} & \multirow{2}{*}{{\tiny{296}}} & {\tiny{0.0}}  & {\tiny{0.000}} & {\tiny{ $\delta_{3}(e)=1$}}  & {\tiny{$f(\rm{Sn}),g(\rm{U}),h(\rm{Ni})$}} \\
  &  &  &  &  & &  &  & {\tiny{$i(\rm{U}),j(\rm{Ni})$}} \\\cline{2-9}
 & \multicolumn{8}{c|}{\tiny{$e: \; (0,1,1)^c,(1,0,1)^c$}} \\ \hline 
 \multirow{2}{*}{{\tiny{12}}} & \multirow{2}{*}{{\tiny{ $\rm{Np} \rm{Ga}_{5} \rm{Rh}$}}} & \multirow{2}{*}{{\tiny{ \bcsidwebshort{1.262}}}} & \multirow{2}{*}{{\tiny{\msgsymb{63}{466} (63.466)}}} & \multirow{2}{*}{{\tiny{190}}} & {\tiny{0.0}}  & {\tiny{0.000}} & {\tiny{ $\delta_{3}(e)=-1$}}  & {\tiny{$a(\rm{Np}),b(\rm{Ga}),d(\rm{Rh})$}} \\
  &  &  &  &  & &  &  & {\tiny{$m(\rm{Ga})$}} \\ \hline 
 {\tiny{13}} & {\tiny{ $\rm{Er}_{2} \rm{Ni}_{2} \rm{In}$}} & {\tiny{ \bcsidwebshort{1.195}}} & {\tiny{\msgsymb{63}{467} (63.467)}} & {\tiny{356}} & {\tiny{0.0}}  & {\tiny{0.000}} & {\tiny{ $\eta_{1}(f)=1$}}  & {\tiny{$e(\rm{In}),i(\rm{Ni}),j(\rm{Er})$}} \\\cline{2-9}
 & \multicolumn{8}{c|}{\tiny{$f: \; (1,1,0)^c$}} \\ \hline 
 \multirow{2}{*}{{\tiny{14}}} & \multirow{2}{*}{{\tiny{ $\rm{Ce} \rm{B}_{6}$}}} & \multirow{2}{*}{{\tiny{ \bcsidwebshort{3.13}}}} & \multirow{2}{*}{{\tiny{\msgsymb{64}{479} (64.479)}}} & \multirow{2}{*}{{\tiny{480}}} & \multirow{2}{*}{{\tiny{0.0}}}  & \multirow{2}{*}{{\tiny{0.000}}} & {\tiny{ $\delta_{1}(b)=1$}}  & {\tiny{$i(\rm{B}),j(\rm{B}),k(\rm{Ce})$}} \\
  &  &  &  &  & &  & {\tiny{ $\delta_{2}(c)=-1$}}  & {\tiny{$l(\rm{B})$}} \\\cline{2-9}
 & \multicolumn{8}{c|}{\tiny{$b: \; (1,1,0)^c,(-1,1,0)^c$}} \\\cline{2-9}
 & \multicolumn{8}{c|}{\tiny{$c: \; (1,1,0)^c,(-1,1,0)^c$}} \\ \hline 
 \multirow{3}{*}{{\tiny{15}}} & \multirow{3}{*}{{\tiny{ $\rm{Mn}_{3} \rm{Ni}_{20} \rm{P}_{6}$}}} & \multirow{3}{*}{{\tiny{ \bcsidwebshort{1.145}}}} & \multirow{3}{*}{{\tiny{\msgsymb{64}{480} (64.480)}}} & \multirow{3}{*}{{\tiny{778}}} & \multirow{2}{*}{{\tiny{4.0}}}  & \multirow{2}{*}{{\tiny{0.000}}} & {\tiny{ $\delta_{2}(b)=-1$}}  & {\tiny{$a(\rm{Mn}),f(\rm{Mn}),g(\rm{P})$}} \\
  &  &  &  &  & &  & {\tiny{ $\delta_{3}(e)=-1$}}  & {\tiny{$h(\rm{P}),i(\rm{P}),m(\rm{Ni})$}} \\
  &  &  &  &  & &  &  & {\tiny{$n(\rm{Ni}),o(\rm{Ni}),p(\rm{Ni})$}} \\\cline{2-9}
 & \multicolumn{8}{c|}{\tiny{$b: \; (1,1,1)^c$}} \\\cline{2-9}
 & \multicolumn{8}{c|}{\tiny{$e: \; (1,1,1)^c$}} \\ \hline 
 {\tiny{16}} & {\tiny{ $\rm{Ba} \rm{Fe}_{2} \rm{As}_{2}$}} & {\tiny{ \bcsidwebshort{1.16}}} & {\tiny{\msgsymb{64}{480} (64.480)}} & {\tiny{96}} & {\tiny{4.0}}  & {\tiny{0.000}} & {\tiny{ $\delta_{3}(e)=1$}}  & {\tiny{$a(\rm{Ba}),f(\rm{Fe}),g(\rm{As})$}} \\\cline{2-9}
 & \multicolumn{8}{c|}{\tiny{$e: \; (1,0,1)^c$}} \\ \hline 
 \multirow{2}{*}{{\tiny{17}}} & \multirow{2}{*}{{\tiny{ $\rm{Ce} \rm{Rh}_{2} \rm{Si}_{2}$}}} & \multirow{2}{*}{{\tiny{ \bcsidwebshort{1.188}}}} & \multirow{2}{*}{{\tiny{\msgsymb{64}{480} (64.480)}}} & \multirow{2}{*}{{\tiny{100}}} & \multirow{2}{*}{{\tiny{2.0}}}  & \multirow{2}{*}{{\tiny{0.000}}} & {\tiny{ $\delta_{2}(b)=-1$}}  & \multirow{2}{*}{{\tiny{$a(\rm{Ce}),f(\rm{Rh}),g(\rm{Si})$}}} \\
  &  &  &  &  & &  & {\tiny{ $\delta_{3}(e)=2$}}  &  \\\cline{2-9}
 & \multicolumn{8}{c|}{\tiny{$e: \; (0,1,0)^c,(1,0,0)^c$}} \\ \hline 
 \multirow{2}{*}{{\tiny{18}}} & \multirow{2}{*}{{\tiny{ $\rm{Ca} \rm{Fe}_{2} \rm{As}_{2}$}}} & \multirow{2}{*}{{\tiny{ \bcsidwebshort{1.52}}}} & \multirow{2}{*}{{\tiny{\msgsymb{64}{480} (64.480)}}} & \multirow{2}{*}{{\tiny{96}}} & {\tiny{3.0}}  & {\tiny{0.000}} & {\tiny{ $\delta_{3}(e)=1$}}  & \multirow{2}{*}{{\tiny{$a(\rm{Ca}),f(\rm{Fe}),g(\rm{As})$}}} \\
\cline{6-8}
  &  &  &  &  & {\tiny{4.0}}  & {\tiny{0.000}} & {\tiny{ $\delta_{3}(e)=1$}}  &  \\\cline{2-9}
 & \multicolumn{8}{c|}{\tiny{$e: \; (0,1,1)^c$}} \\ \hline 
 \multirow{3}{*}{{\tiny{19}}} & \multirow{3}{*}{{\tiny{ $\rm{Gd}_{2} \rm{Cu} \rm{O}_{4}$}}} & \multirow{3}{*}{{\tiny{ \bcsidwebshort{1.104}}}} & \multirow{3}{*}{{\tiny{\msgsymb{66}{500} (66.500)}}} & \multirow{3}{*}{{\tiny{118}}} & {\tiny{0.0}}  & {\tiny{0.000}} & {\tiny{ $\delta_{2}(b)=-3$}}  & {\tiny{$a(\rm{Cu}),c(\rm{O}),f(\rm{O})$}} \\
\cline{6-8}
  &  &  &  &  & {\tiny{1.0}}  & {\tiny{0.000}} & {\tiny{ $\delta_{2}(b)=-3$}}  & {\tiny{$g(\rm{Gd})$}} \\
\cline{6-8}
  &  &  &  &  & {\tiny{2.0}}  & {\tiny{0.051}} & {\tiny{ $\delta_{2}(b)=-3$}}  &  \\ \hline 
 \multirow{3}{*}{{\tiny{20}}} & \multirow{3}{*}{{\tiny{ $\rm{Pr}_{2} \rm{Cu} \rm{O}_{4}$}}} & \multirow{3}{*}{{\tiny{ \bcsidwebshort{1.106}}}} & \multirow{3}{*}{{\tiny{\msgsymb{66}{500} (66.500)}}} & \multirow{3}{*}{{\tiny{134}}} & \multirow{2}{*}{{\tiny{0.0}}}  & \multirow{2}{*}{{\tiny{0.000}}} & {\tiny{ $\delta_{2}(b)=-3$}}  & {\tiny{$a(\rm{Cu}),c(\rm{O}),f(\rm{O})$}} \\
  &  &  &  &  & &  & {\tiny{ $\delta_{3}(e)=-1$}}  & {\tiny{$g(\rm{Pr})$}} \\
\cline{6-8}
  &  &  &  &  & {\tiny{2.0}}  & {\tiny{0.000}} & {\tiny{ $\delta_{2}(b)=-3$}}  &  \\ \hline 
 {\tiny{21}} & {\tiny{ $\rm{Ce} \rm{Mg} \rm{Pb}$}} & {\tiny{ \bcsidwebshort{1.142}}} & {\tiny{\msgsymb{67}{510} (67.510)}} & {\tiny{136}} & {\tiny{0.0}}  & {\tiny{0.000}} & {\tiny{ $\delta_{1}(c)=6$}}  & {\tiny{$d(\rm{Mg}),h(\rm{Ce}),h(\rm{Pb})$}} \\\cline{2-9}
 & \multicolumn{8}{c|}{\tiny{$c: \; (0,1,0)^c$}} \\ \hline 
 \multirow{3}{*}{{\tiny{22}}} & \multirow{3}{*}{{\tiny{ $\rm{Ni} \rm{Cr}_{2} \rm{O}_{4}$}}} & \multirow{3}{*}{{\tiny{ \bcsidwebshort{0.4}}}} & \multirow{3}{*}{{\tiny{\msgsymb{70}{530} (70.530)}}} & \multirow{3}{*}{{\tiny{128}}} & {\tiny{2.0}}  & {\tiny{0.571}} & {\tiny{ $\eta_{2}(e,f)=1$}}  & \multirow{3}{*}{{\tiny{$b(\rm{Ni}),c(\rm{Cr}),h(\rm{O})$}}} \\
\cline{6-8}
  &  &  &  &  & {\tiny{3.0}}  & {\tiny{1.126}} & {\tiny{ $\eta_{2}(e,f)=1$}}  &  \\
\cline{6-8}
  &  &  &  &  & {\tiny{4.0}}  & {\tiny{1.675}} & {\tiny{ $\eta_{2}(e,f)=1$}}  &  \\ \hline 
 \multirow{2}{*}{{\tiny{23}}} & \multirow{2}{*}{{\tiny{ $\rm{La} \rm{Fe} \rm{As} \rm{O}$}}} & \multirow{2}{*}{{\tiny{ \bcsidwebshort{1.125}}}} & \multirow{2}{*}{{\tiny{\msgsymb{73}{553} (73.553)}}} & \multirow{2}{*}{{\tiny{144}}} & {\tiny{1.0}}  & {\tiny{0.000}} & {\tiny{ $\delta_{1}(c)=1$}}  & {\tiny{$a(\rm{O}),b(\rm{Fe}),g(\rm{As})$}} \\
\cline{6-8}
  &  &  &  &  & {\tiny{3.0}}  & {\tiny{0.515}} & {\tiny{ $\delta_{1}(c)=1$}}  & {\tiny{$g(\rm{La})$}} \\\cline{2-9}
 & \multicolumn{8}{c|}{\tiny{$c: \; (1,1,0)^c$}} \\ \hline 
 \multirow{2}{*}{{\tiny{24}}} & \multirow{2}{*}{{\tiny{ $\rm{Yb} \rm{Co}_{2} \rm{Si}_{2}$}}} & \multirow{2}{*}{{\tiny{ \bcsidwebshort{1.176}}}} & \multirow{2}{*}{{\tiny{\msgsymb{73}{553} (73.553)}}} & \multirow{2}{*}{{\tiny{200}}} & {\tiny{0.0}}  & {\tiny{0.000}} & {\tiny{ $\delta_{1}(c)=1$}}  & {\tiny{$a(\rm{Co}),b(\rm{Co}),g(\rm{Yb})$}} \\
  &  &  &  &  & &  &  & {\tiny{$n(\rm{Si})$}} \\ \hline 
 \multirow{4}{*}{{\tiny{25}}} & \multirow{4}{*}{{\tiny{ $\rm{Ce} \rm{Sb} \rm{Te}$}}} & \multirow{4}{*}{{\tiny{ \bcsidwebshort{1.271}}}} & \multirow{4}{*}{{\tiny{\msgsymb{130}{432} (130.432)}}} & \multirow{4}{*}{{\tiny{92}}} & {\tiny{0.0}}  & {\tiny{0.000}} & {\tiny{ $\delta_{3}(d)=1$}}  & \multirow{4}{*}{{\tiny{$a(\rm{Sb}),c(\rm{Te}),c(\rm{Ce})$}}} \\
\cline{6-8}
  &  &  &  &  & {\tiny{2.0}}  & {\tiny{0.000}} & {\tiny{ $\delta_{3}(d)=1$}}  &  \\
\cline{6-8}
  &  &  &  &  & {\tiny{4.0}}  & {\tiny{0.000}} & {\tiny{ $\delta_{3}(d)=1$}}  &  \\
\cline{6-8}
  &  &  &  &  & {\tiny{6.0}}  & {\tiny{0.000}} & {\tiny{ $\delta_{3}(d)=1$}}  &  \\\cline{2-9}
 & \multicolumn{8}{c|}{\tiny{$d: \; (1,0,0)^c$}} \\ \hline 
 \multirow{2}{*}{{\tiny{26}}} & \multirow{2}{*}{{\tiny{ $\rm{U} \rm{Ga}_{5} \rm{Ni}$}}} & \multirow{2}{*}{{\tiny{ \bcsidwebshort{1.254}}}} & \multirow{2}{*}{{\tiny{\msgsymb{140}{550} (140.550)}}} & \multirow{2}{*}{{\tiny{190}}} & {\tiny{6.0}}  & {\tiny{0.000}} & {\tiny{ $\delta_{5}(d)=1$}}  & {\tiny{$a(\rm{U}),b(\rm{Ni}),c(\rm{Ga})$}} \\
  &  &  &  &  & &  &  & {\tiny{$i(\rm{Ga})$}} \\ \hline 
 \multirow{5}{*}{{\tiny{27}}} & \multirow{5}{*}{{\tiny{ $\rm{Co} \rm{Al}_{2} \rm{O}_{4}$}}} & \multirow{5}{*}{{\tiny{ \bcsidwebshort{0.58}}}} & \multirow{5}{*}{{\tiny{\msgsymb{141}{556} (141.556)}}} & \multirow{5}{*}{{\tiny{78}}} & {\tiny{0.0}}  & {\tiny{0.846}} & {\tiny{ $\delta_{3}(b)=-1$}}  & \multirow{5}{*}{{\tiny{$a(\rm{Co}),d(\rm{Al}),h(\rm{O})$}}} \\
\cline{6-8}
  &  &  &  &  & {\tiny{1.0}}  & {\tiny{1.784}} & {\tiny{ $\delta_{3}(b)=-1$}}  &  \\
\cline{6-8}
  &  &  &  &  & {\tiny{2.0}}  & {\tiny{2.479}} & {\tiny{ $\delta_{3}(b)=-1$}}  &  \\
\cline{6-8}
  &  &  &  &  & {\tiny{3.0}}  & {\tiny{3.165}} & {\tiny{ $\delta_{3}(b)=-1$}}  &  \\
\cline{6-8}
  &  &  &  &  & {\tiny{4.0}}  & {\tiny{3.842}} & {\tiny{ $\delta_{3}(b)=-1$}}  &  \\\cline{2-9}
 & \multicolumn{8}{c|}{\tiny{$b: \; (1,1,1)^c$}} \\ \hline 
 {\tiny{28}} & {\tiny{ $\rm{Tb} \rm{Mg}_{3}$}} & {\tiny{ \bcsidwebshort{1.189}}} & {\tiny{\msgsymb{167}{108} (167.108)}} & {\tiny{86}} & {\tiny{0.0}}  & {\tiny{0.002}} & {\tiny{ $\delta_{5}(e)=1$}}  & {\tiny{$a(\rm{Tb}),b(\rm{Mg}),c(\rm{Mg})$}} \\\cline{2-9}
 & \multicolumn{8}{c|}{\tiny{$e: \; (1,1,0)^c$}} \\ \hline 
 \multirow{6}{*}{{\tiny{29}}} & \multirow{6}{*}{{\tiny{ $\rm{Ni} \rm{S}_{2}$}}} & \multirow{6}{*}{{\tiny{ \bcsidwebshort{0.150}}}} & \multirow{6}{*}{{\tiny{\msgsymb{205}{33} (205.33)}}} & \multirow{6}{*}{{\tiny{112}}} & \multirow{2}{*}{{\tiny{1.0}}}  & \multirow{2}{*}{{\tiny{0.000}}} & {\tiny{ $\delta_{5}(b)=-1$}}  & \multirow{6}{*}{{\tiny{$a(\rm{Ni}),c(\rm{S})$}}} \\
  &  &  &  &  & &  & {\tiny{ $\delta_{6}(b)=-1$}}  &  \\
\cline{6-8}
  &  &  &  &  & \multirow{2}{*}{{\tiny{2.0}}}  & \multirow{2}{*}{{\tiny{0.000}}} & {\tiny{ $\delta_{5}(b)=-1$}}  &  \\
  &  &  &  &  & &  & {\tiny{ $\delta_{6}(b)=-1$}}  &  \\
\cline{6-8}
  &  &  &  &  & \multirow{2}{*}{{\tiny{3.0}}}  & \multirow{2}{*}{{\tiny{0.040}}} & {\tiny{ $\delta_{5}(b)=-1$}}  &  \\
  &  &  &  &  & &  & {\tiny{ $\delta_{6}(b)=-1$}}  &  \\\cline{2-9}
 & \multicolumn{8}{c|}{\tiny{$b: \; (1,1,1)^c$}} \\ \hline 
 \multirow{10}{*}{{\tiny{30}}} & \multirow{10}{*}{{\tiny{ $\rm{Mn} \rm{Te}_{2}$}}} & \multirow{10}{*}{{\tiny{ \bcsidwebshort{0.20}}}} & \multirow{10}{*}{{\tiny{\msgsymb{205}{33} (205.33)}}} & \multirow{10}{*}{{\tiny{100}}} & \multirow{2}{*}{{\tiny{0.0}}}  & \multirow{2}{*}{{\tiny{0.000}}} & {\tiny{ $\delta_{5}(b)=-1$}}  & \multirow{10}{*}{{\tiny{$a(\rm{Mn}),c(\rm{Te})$}}} \\
  &  &  &  &  & &  & {\tiny{ $\delta_{6}(b)=-1$}}  &  \\
\cline{6-8}
  &  &  &  &  & \multirow{2}{*}{{\tiny{1.0}}}  & \multirow{2}{*}{{\tiny{0.107}}} & {\tiny{ $\delta_{5}(b)=-1$}}  &  \\
  &  &  &  &  & &  & {\tiny{ $\delta_{6}(b)=-1$}}  &  \\
\cline{6-8}
  &  &  &  &  & \multirow{2}{*}{{\tiny{2.0}}}  & \multirow{2}{*}{{\tiny{0.216}}} & {\tiny{ $\delta_{5}(b)=-1$}}  &  \\
  &  &  &  &  & &  & {\tiny{ $\delta_{6}(b)=-1$}}  &  \\
\cline{6-8}
  &  &  &  &  & \multirow{2}{*}{{\tiny{3.0}}}  & \multirow{2}{*}{{\tiny{0.311}}} & {\tiny{ $\delta_{5}(b)=-1$}}  &  \\
  &  &  &  &  & &  & {\tiny{ $\delta_{6}(b)=-1$}}  &  \\
\cline{6-8}
  &  &  &  &  & \multirow{2}{*}{{\tiny{4.0}}}  & \multirow{2}{*}{{\tiny{0.393}}} & {\tiny{ $\delta_{5}(b)=-1$}}  &  \\
  &  &  &  &  & &  & {\tiny{ $\delta_{6}(b)=-1$}}  &  \\\cline{2-9}
 & \multicolumn{8}{c|}{\tiny{$b: \; (1,1,1)^c$}}  \\ \hline
\end{longtable*}

\end{document}